\documentclass[aps,pra,twocolumn,preprintnumbers,amsmath,amssymb,longbibliography]{revtex4-1}

\usepackage{graphics}
\usepackage{amsmath}
\usepackage{graphicx}
\usepackage{amsfonts}
\usepackage{amssymb}
\usepackage{float}
\usepackage{longtable}
\usepackage{epsfig}
\usepackage{latexsym}
\usepackage{theorem}
\usepackage{bbm}
\usepackage{bm}
\usepackage{verbatim}
\usepackage{psfrag}
\usepackage{multirow}
\usepackage{pifont}
\usepackage{ifthen}
\usepackage{color}
\usepackage{subfigure}
\usepackage{color}

\def\EQ#1{\begin{eqnarray}#1\end{eqnarray}}

\newcommand{\beq}{\begin{equation}}
\newcommand{\eeq}{\end{equation}}
\newcommand{\bea}{\begin{eqnarray}}
\newcommand{\eea}{\end{eqnarray}}

\newcommand{\ros}{\color{black}}

\def\({\left(}
\def\){\right)}
\def\[{\left[}
\def\]{\right]}
\def \X {\mathbb{X}}
\def \Y {\mathbb{Y}}
\def \Z {\mathbb{Z}}
\def \I {\mathbb{I}}

\newcommand{\inprod}[2]{\langle #1 | #2 \rangle}

\newcommand{\ket}[1]{|#1 \rangle}

\newcommand{\proj}[1]{|#1\rangle\!\langle#1|}
\newcommand{\id}{\openone}
\newcommand{\tr}{\text{Tr}}

\newcommand{\ot}{\otimes}

\newcommand{\eps}{\varepsilon}
\newcommand{\corr}{\textrm{corr}}
\newcommand{\secr}{\textrm{secr}}

\begin{document}

\title{Advances in Quantum Cryptography}
\author{S.
Pirandola$^{1,2}$, U. L. Andersen$^{3}$, L. Banchi$^{4}$, M.
Berta$^{5}$, D. Bunandar$^{2}$, R. Colbeck$^{6}$, D.
Englund$^{2}$, T. Gehring$^{3}$, C. Lupo$^{7}$, C.
Ottaviani$^{1}$, J. Pereira$^{1}$, M. Razavi$^{8}$, J. S.
Shaari$^{9,10}$, M. Tomamichel$^{11}$, V. C. Usenko$^{12}$, G.
Vallone$^{13}$, P. Villoresi$^{13}$, P. Wallden$^{14}$}

\affiliation{}

\affiliation{$^{1}\mbox{Computer Science and York Centre for
Quantum Technologies, University of York, York YO10 5GH, UK}$}
\affiliation{$^{2}\mbox{Research Laboratory of Electronics,
Massachusetts Institute of Technology (MIT), Cambridge,
Massachusetts 02139, USA}$} \affiliation{$^{3}\mbox{Center for
Macroscopic Quantum States (bigQ), Department of Physics, }$}
\affiliation{$\mbox{Technical University of Denmark, Fysikvej,
2800 Kgs. Lyngby, Denmark}$} \affiliation{$^{4}\mbox{Department of
Physics and Astronomy, University of Florence, via G. Sansone 1,
I-50019 Sesto Fiorentino (FI), Italy}$}
\affiliation{$^{5}\mbox{Department of Computing, Imperial College,
Kensington, London SW7 2AZ, UK}$}
\affiliation{$^{6}\mbox{Department of Mathematics, University of
York, York YO10 5DD, UK}$} \affiliation{$^{7}\mbox{Department of
Physics and Astronomy, University of Sheffield, Sheffield S3 7RH,
UK}$} \affiliation{$^{8}\mbox{School of Electronic and Electrical
Engineering, University of Leeds, Leeds, LS2 9JT, UK}$}
\affiliation{$^{9}\mbox{Faculty of Science, International Islamic
University Malaysia (IIUM),}$} \affiliation{$\mbox{Jalan Sultan
Ahmad Shah, 25200 Kuantan, Pahang, Malaysia}$}
\affiliation{$^{10}\mbox{Institute of Mathematical Research
(INSPEM), University Putra Malaysia, 43400 UPM Serdang, Selangor,
Malaysia}$} \affiliation{$^{11}\mbox{Centre for Quantum Software
and Information, School of Software,}$} \affiliation{$\mbox{
University of Technology Sydney, Sydney NSW 2007, Australia}$}
\affiliation{$^{12}\mbox{Department of Optics, Palacky University,
17. listopadu 50, 772 07 Olomouc, Czech Republic}$}
\affiliation{$^{13}\mbox{Dipartimento di Ingegneria
dell'Informazione, Universit\'{a} degli Studi di Padova,}$}
\affiliation{$\mbox{via Gradenigo 6B, 35131 Padova, Italy}$}
\affiliation{$^{14}\mbox{School of Informatics, University of
Edinburgh, 10 Crichton Street, Edinburgh EH8 9AB, UK}$}

\begin{abstract}
Quantum cryptography is arguably the fastest growing area in
quantum information science. Novel theoretical protocols are
designed on a regular basis, security proofs are constantly
improving, and experiments are gradually moving from
proof-of-principle lab demonstrations to in-field implementations
and technological prototypes. In this review, we provide both a
general introduction and a state of the art description of the
recent advances in the field, both theoretically and
experimentally. We start by reviewing protocols of quantum key
distribution based on discrete variable systems. Next we consider
aspects of device independence, satellite challenges, and high
rate protocols based on continuous variable systems. We will then
discuss the ultimate limits of point-to-point private
communications and how quantum repeaters and networks may overcome
these restrictions. Finally, we will discuss some aspects of
quantum cryptography beyond standard quantum key distribution,
including quantum data locking and quantum digital signatures.
\end{abstract}

\maketitle

\tableofcontents



\section{Introduction}

Quantum information~\cite{NielsenChuang,Bengtsson,Hayashi,
Watrous,SamRMPm,RMPwee,HolevoBOOK,sera,AdessoR,hybrid,hybrid2} is
the core science behind the so-called second quantum
revolution~\cite{qrev,qrev2}, or quantum 2.0. This is the rapid
development of new disrupting technologies that are based on the
most powerful features and resources of quantum mechanics, such as
quantum entanglement~\cite{entREVIEW},
teleportation~\cite{teleBENNETT,teleCV,Samtele2,telereview}, and
the no-cloning theorem~\cite{noclone1,noclone2}. In this context,
quantum computing~\cite{NielsenChuang} has recently gained a lot
of momentum, also thanks to the involvement of multinational
corporations in competition to develop the first large quantum
computer. In particular, superconducting chips based on Josephson
junctions~\cite{superQC} are rapidly scaling up their number of
qubits and soon may start to factorize non-trivial integers by
using Shor's algorithm~\cite{ShorAlgo1,ShorAlgo2}. The threat for
the Rivest-Shamir-Adleman (RSA) protocol~\cite{RSA} and the other
public key cryptosystems not only comes from quantum computing but
also from potential advances in number theory, where an efficient
factorization algorithm might be found for classical Turing
machines (e.g., already in 2004 the test of primality has become
polynomial, thanks to the Agrawal-Kayal-Saxena
algorithm~\cite{AKS}).

An important point to understand is that the fragility of current
classical cryptosystems not only is a potential threat for the
present, but a more serious and realistic threat for the future.
Today, eavesdroppers may intercept cryptograms that they are not
able to decrypt. However, they may store these encrypted
communications and wait for their decryption once a sufficiently
large quantum computer is technologically available (or a new
classical algorithm is discovered). This means that the
confidentiality of messages may have a very limited lifespan.
Following Michele Mosca~\cite{Mosca}, we may write a simple
inequality. Let us call $x$ the \textit{security shelf-life} which
is the length of time (in years) we need the classical
cryptographic keys to be secure. Then, let us call $y$ the
\textit{migration time} which is the time (in years) needed to
adapt the current classical infrastructure with quantum-secure
encryption. Finally, let us call $z$ the \textit{collapse time}
which is the time (in years) for a large quantum computer to be
built. If $x+y>z$ then \textquotedblleft
worry\textquotedblright~\cite{Mosca}.

It is therefore clear that suitably countermeasures are necessary.
One approach is known as post-quantum cryptography. This is the
development of novel classical cryptosystems which are robust to
factorization and other quantum algorithms. This is certainly one
option but it does not completely solve the problem. The point is
that there may be undiscovered quantum algorithms (or undiscovered
classical ones) that might easily break the security of the new
cryptosystems. In other words, post-quantum cryptography is likely
to offer only a partial and temporary solution to the problem. By
contrast, quantum key distribution (QKD) offers the ultimate
solution: restoring security and confidentiality by resorting to
unbreakable principles of nature, such as the uncertainty
principle or the monogamy of
entanglement~\cite{rev1,rev2,rev3,rev4}.

Even though QKD offers the ultimate solution to the security
problem, its ideal implementation is hard to implement in practice
and there are a number of open problems to be addressed. One the
one side, fully-device independent QKD
protocols~\cite{MayersYao,BHK} (discussed in
Section~\ref{sec:DIQKD}) provide the highest level of quantum
security but they are quite demanding to realize and are
characterized by extremely low secret key rates. On the other
hand, more practical QKD protocols assume some level of trust in
their devices, an assumption that allows them to achieve
reasonable rates, but this also opens the possibility of dangerous
side-channel attacks.

Besides a trade-off between security and rate, there is also
another important one which is between rate and distance. Today,
we know that there is a fundamental limit which restricts any
point to point implementation of QKD. Given a lossy link with
transmissivity $\eta$, two parties cannot distribute
more than the secret key capacity of the channel, which is $-\mathrm{log}%
_{2}(1-\eta)$~\cite{PLOB}, i.e., a scaling of $1.44\eta$ secret
bits per channel use at long distance. Ideal implementations of
QKD protocols based on continuous-variable
systems~\cite{SamRMPm,sera} and Gaussian states~\cite{RMPwee} may
approach this capacity~\cite{ReverseCAP}, while those based on
discrete variables falls below by additional factors. In order to
overcome this limit and enable long-distance high-rate
implementations of QKD, we need the develop quantum
repeaters~\cite{Briegel,Dur,DLCZ_01} and quantum
networks~\cite{qNetworkBook}. In this way, we may achieve better
long-distance scalings and further boost the rates by resorting to
more complex routing strategies. The study of quantum repeaters
and secure QKD networks is one of the hottest topics
today~\cite{net2006,net2006p,netO1,netO2,netO3,netO4,netO5,netO6,DurPirker2018,DurPirker2019}.

The present review aims at providing an overview of the most
important and most recent advances in the field of quantum
cryptography, both theoretically and experimentally. After a brief
introduction of the general notions, we will review the main QKD
protocols based on discrete- and continuous-variable systems. We
will consider standard QKD, device-independent and
measurement-device independent QKD. We will discuss the various
levels of security for the main communication channel, from
asymptotic security proofs to analyzes accounting for finite-size
effects and composability aspects. We will also briefly review
quantum hacking and side-channel attacks. Then, we will present
the most recent progress in the exploration of the ultimate limits
of QKD. In particular, we will discuss the secret key capacities
associated with the most important models of quantum channels over
which we may implement point-to-point QKD protocols, and their
extension to quantum repeaters and networks. Practical aspects of
quantum repeaters will then be thoroughly discussed. Finally, we
will treat topics beyond QKD, including quantum data locking,
quantum random number generators, and quantum digital signatures.


\section{Basic notions in quantum key distribution}\label{BasicQKDSection}

\subsection{Generic aspects of a QKD protocol}

In our review we consider both discrete-variable systems, such as
qubits or other quantum systems with finite-dimensional Hilbert
space, and continuous-variable systems, such as bosonic modes of
the electromagnetic field which are described by an
infinite-dimensional Hilbert space. There a number of reviews and
books on these two general areas (e.g., see Refs.
\cite{NielsenChuang,RMPwee}). Some of the concepts are repeated in
this review but we generally assume basic knowledge of these
systems. Here we mention some general aspects that apply to both
types of systems.

A generic ``prepare and measure'' QKD protocol can be divided in
two main steps:\ quantum communication followed by classical
postprocessing. During quantum communication the sender (Alice)
encodes instances of a random classical variable $\alpha$ into
non-orthogonal quantum states. These states are sent over a
quantum channel (optical fiber, free-space link) controlled by the
eavesdropper (Eve), who tries to steal the encoded information.
The linearity of quantum mechanics forbids to perform perfect
cloning~\cite{noclone1,noclone2}, so that Eve can only get partial
information while disturbing the quantum signals. At the output of
the communication channel, the receiver (Bob) measures the
incoming signals and obtains a random classical variable $\beta$.
After a number of uses of the channel, Alice and Bob share raw
data described by two correlated variables $\alpha$ and $\beta$.

The remote parties use part of the raw data to estimate the
parameters of the channel, such as its transmissivity and noise.
This stage of parameter estimation is important in order to
evaluate the amount of post-processing to extract a private shared
key from the remaining data. Depending on this information, they
in fact perform a stage of error correction, which allows them to
detect and eliminate errors, followed by a stage of privacy
amplification that allows them to reduce Eve's stolen information
to a negligible amount. The final result is the secret key.

Depending on which variable is guessed, we have direct or reverse
reconciliation. In direct reconciliation, it is Bob that
post-process its outcomes in order to infer Alice's encodings.
This procedure is usually assisted by means of forward CC from
Alice to Bob. By contrast, in reverse reconciliation, it is Alice
who post-process her encoding variable in order to infer Bob's
outcomes. This procedure is usually assisted by a final round of
backward CC from Bob to Alice. Of course, one may more generally
consider two-way procedures where the extraction of the key is
helped by forward and feedback CCs, which may be even interleaved
with the various communication rounds of the protocol.

Let us remark that there may also be an additional post-processing
routine, called sifting, where the remote parties communicate in
order to agree instances while discard others, depending on the
measurement bases they have independently chosen. For instance
this happens in typical DV protocols, where the $Z$-basis is
randomly switched with the $X$-basis, or in CV protocols where the
homodyne detection is switched between the $q$ and the $p$
quadrature.

Sometimes QKD protocols are formulated in entanglement-based
representation. This means that Alice' preparation of the input
ensemble of states is replaced by an entangled state $\Psi_{AB}$
part of which is measured by Alice. The measurement on part $A$
has the effect to conditionally prepare a state on part $B$. The
outcome of the measurement is one-to-one with the classical
variable encoded in the prepared states. This representation is
particularly useful for the study of QKD protocols, so that their
prepare and measure formulation is replaced by an
entanglement-based formulation for assessing the security and
deriving the secret key rate.

\subsection{Asymptotic security and eavesdropping strategies}

The asymptotic security analysis is based on the assumption that
the parties exchange a number $n$ $\gg1$ (ideally infinite) of
signals. The attacks can then be divided in three classes of
increasing power: Individual, collective, and general-coherent. If
the attack is individual, Eve uses a fresh ancilla to interact
with each transmitted signal and she performs individual
measurements on each output ancillary systems. The individual
measurements can be done run-by-run or delayed at the end of the
protocol, so that Eve may optimize over Alice and Bob's CC (also
known as delayed-choice strategy). In the presence of an
individual attacks, we have three classical variables for Alice,
Bob and Eve, say $\alpha$, $\beta$ and $\gamma$. The asymptotic
key rate is then given by the difference of the mutual
information~\cite{cover} $I$ among the various parties according
to Csiszar and Korner's classical theorem~\cite{Czisar}. In direct
reconciliation (DR), we have the key rate
\begin{equation}
R_{\mathrm{DR}}:=I(\alpha : \beta)-I(\alpha:\gamma),
\label{KidealDR}
\end{equation}
where $I(\alpha:\beta):= H(\alpha)-H(\alpha | \beta)$ with $H$
being the Shannon entropy and $H(|)$ its conditional version. In
reverse reconciliation (RR), we have instead
\begin{equation}
R_{\mathrm{RR}}:=I(\alpha : \beta)-I(\beta:\gamma),
\label{KidealRR}
\end{equation}

If the attack is collective then Eve still uses a fresh ancilla
for each signal sent but now her output ancillary systems are all
stored in a quantum memory which is collectively measured at the
end of the protocol after Alice and Bob's CCs. In this case, we
may compute a lower bound to the key rate by replacing Eve's
mutual information with Eve's Holevo information on the relevant
variable. In direct reconciliation, one considers Eve's ensemble
of output states conditioned to Alice's variable $\alpha$, i.e.,
$\{\rho_{\mathrm{E}|\alpha}, P(\alpha) \}$ where $P(\alpha)$ is
the probability of the encoding $\alpha$. Consider then Eve's
average state $\rho_{\mathrm{E}}:= \int d\alpha P(\alpha)
\rho_{\mathrm{E}|\alpha}$. Eve's Holevo information on $\alpha$ is
equal to
\begin{equation}
I(\alpha : E):= S(\rho_{\mathrm{E}}) - \int d\alpha P(\alpha)
S(\rho_{\mathrm{E}|\alpha}),
\end{equation}
where $S(\rho):=-\mathrm{Tr}(\rho \mathrm{log}_{2}\rho)$ is the
von Neumann entropy. In reverse reconciliation, Eve's Holevo
information on $\beta$ is given by
\begin{equation}
I(\beta : E):= S(\rho_{\mathrm{E}}) - \int d\beta P(\beta)
S(\rho_{\mathrm{E}|\beta}),
\end{equation}
where $\rho_{\mathrm{E}|\beta}$ is Eve's output state conditioned
to the outcome $\beta$ with probability $P(\beta)$. Thus, we may
write the two key rates~\cite{winter}
\begin{align}
& R_{\mathrm{DR}}:=I(\alpha : \beta)-I(\alpha: E), \\
&  R_{\mathrm{RR}}:=I(\alpha : \beta)-I(\beta: E).
\end{align}

In a general-coherent attack, Eve's ancillae and the signal
systems are collectively subject to a joint unitary interaction.
The ancillary output is then stored in Eve's quantum memory for
later detection after the parties' CCs. In the asymptotic
scenario, it has been proved~\cite{renner2} that this attack can
be reduced to a collective one by running a random symmetrization
routine which exploits the quantum de Finetti
theorem~\cite{renner1,renner2,renner-cirac}. By means of random
permutations, one can in fact transform a general quantum state of
$n$ systems into a tensor product $\rho^{\otimes n}$, which is the
structure coming from the identical and independent interactions
of a collective attack.

\subsection{Finite-size effects}

Finite-size effects come into place when the number of signal
exchanged $n$ is not so large to be considered to be infinite (see
\ref{sectionMB_MT} for more details). If we assume that the
parties can only exchange a finite number of signals, them the key
rate must be suitably modified and takes the form
\begin{equation}
K_{c}:=\xi I(\alpha : \beta)- I_{\mathrm{E}}-\Delta(n,\epsilon).
\end{equation}
Here $\xi$ accounts for non-ideal reconciliation efficiency of
classical protocols of error correction and privacy amplification,
while $\Delta (n,\epsilon)$ represents the penalty to pay for
using the Holevo quantity $I_{\mathrm{E}}=I(\alpha:\mathrm{E})$ or
$I(\beta:\mathrm{E})$ in the non-asymptotic context. An important
point is the computation of $\Delta(n,\epsilon)$ which is function
of the number of signals exchanged $n$, and of composite
$\epsilon$-parameter that contains contributions from the
probability that the protocol aborts, the probability of success
of the error correction, parameter estimation etc. This is related
to the concept of composability that we briefly explain in the
next section. Composable security proofs are today known for both
discrete- and continuous-variable QKD
protocols~\cite{renner-scarani,Sheridan,renner-comp,furrer-comp,furrer-comp2,leverrier-comp,lupo-PE,lupo-comp}.

\subsection{Composable security of QKD}\label{sec:compos} Cryptographic tasks often
form parts of larger protocols. Indeed the main reason for our
interest in QKD is that secure communication can be built by
combining key distribution with the one-time pad protocol. If two
protocols are proven secure according to a composable security
definition, then the security of their combination can be argued
based on their individual functionalities and \emph{without} the
need to give a separate security proof for the combined protocol.
Since individual cryptographic tasks are often used in a variety
of applications, composability is highly desirable.  Furthermore,
the early security proofs for QKD~\cite{Mayers2,ShorPreskill} did
not use a composable definition and were consequently shown to be
inadequate (even when combined with the one-time pad)~\cite{KRBM}.

The concept of composability was first introduced in classical
cryptography~\cite{Canetti,Can01,PW00,PW} before being generalized
to the quantum setting~\cite{Ben-OrMayers,BHLMO,Unruh}.  A new
security definition was developed~\cite{RennerKoenig,RennerPhD}
that is composable in the required sense and is the basis of the
accepted definition, which we discuss here.  The main idea behind
a composable security definition is to define an ideal protocol,
which is secure by construction, and then show that the real
implementation is virtually indistinguishable from the ideal in
\emph{any} situation.  Therefore, in effect it takes into account
the worst possible combined protocol for the task in question.  To
think about this concretely, it is often phrased in terms of a
game played by a distinguisher whose task it is to guess whether
Alice and Bob are implementing the real protocol or the ideal. The
distinguisher is permitted to do anything that an eavesdropper
could in a real implementation of the protocol.  They are also
given access to the outputs of the protocol, but not to any data
private to Alice and Bob during the protocol (e.g., parts of any
raw strings that are not publicly announced).

Coming up with a reasonable ideal for a general cryptographic task
is not usually straightforward because the ideal and real
protocols have to be virtually indistinguishable even after
accounting for all possible attacks of an adversary.  However, in
the case of key distribution it is relatively straightforward. The
ideal can be phrased in terms of a hypothetical device that
outputs string $S_A$ to Alice and $S_B$ to Bob (each having $n$
possible values) such that
\begin{equation}\label{eq:rhoI}
\rho^I_{S_AS_BE}=\frac{1}{n}\sum_{x=0}^{n-1}\proj{x}\ot\proj{x}\ot\rho_E\,.
\end{equation}
This captures that Alice's and Bob's strings are identical and
uncorrelated with $E$ (which represents all of the systems held by
Eve).  These conditions are often spelled out separately:
\begin{enumerate}
\item $P(S_A\neq S_B)_{\rho^I}=0$ (correctness, i.e., Alice and Bob have identical outputs).
\item $\rho^I_{S_AE}=n^{-1}\id_n\ot\rho_E$ (the output string is secret).
\end{enumerate}

The ideal protocol then says perform the real protocol and if it
does not abort, replace the output with one from this hypothetical
device with the same length. It may seem strange that the ideal
involves running the real. However, if the ideal protocol just
said use the hypothetical device, a distinguisher could readily
distinguish it from the real protocol by blocking the quantum
channel between Alice and Bob.  This would force the real protocol
to abort, while the ideal would not.  By defining the ideal using
the real protocol, both protocols abort with the same probability
for any action of the distinguisher.

From the point of view of the distinguisher, the aim is to
distinguish two quantum states: those that the protocol outputs in
the real and ideal case.  The complete output of the real protocol
(taking into account the possibility of abort) can be written
$$\sigma^R_{S_AS_BE}=p(\perp)\proj{{\perp}}\ot\proj{{\perp}}\ot\rho_E^{{\perp}}+p(\bar{\perp})\rho^R_{S_AS_BE}\,,$$
where
$$\rho^R_{S_AS_BE}=\sum_{xy}P_{XY}(x,y)\proj{x}\ot\proj{y}\ot\rho_E^{x,y}$$
is the state conditioned on the real protocol not aborting,
$\ket{{\perp}}$ as a special symbol representing abort (this is
orthogonal to all the $\ket{x}$ or $\ket{y}$ terms in the sum),
$p({\perp})$ and $p(\bar{\perp})=1-p({\perp})$ are the
probabilities of abort and not abort respectively. (Note that any
information sent over the authenticated public channel that Eve
could listen in on during the implementation is included in $E$.)
The output of the ideal is instead
$$\sigma^I_{S_AS_BE}=p(\perp)\proj{{\perp}}\ot\proj{{\perp}}\ot\rho_E^{{\perp}}+p(\bar{\perp})\rho^I_{S_AS_BE}\,,$$
with $\rho^I_{S_AS_BE}$ defined in Eq.~\eqref{eq:rhoI}.

The measure of distinguishability for these is the trace distance
$D$~\cite{NielsenChuang}. This has the operational meaning that,
given either $\sigma^R_{S_AS_BE}$ or $\sigma^I_{S_AS_BE}$ with
50\% chance of each, the optimal probability of guessing which is
\beq
p_{\mathrm{guess}}=\frac{1}{2}[1+D(\sigma^R_{S_AS_BE},\sigma^I_{S_AS_BE})],
\eeq which accounts for any possible quantum strategy to
distinguish them. If the distance is close to zero, then the real
protocol is virtually indistinguishable from the real.
Quantitatively, if
$D(\sigma^R_{S_AS_BE},\sigma^I_{S_AS_BE})\leq\eps$ for all
possible strategies an eavesdropper could use, then the protocol
is said to be $\eps$-secure.  The analogue of this definition for
probability distributions was used in~\cite{bcktwo} to prove
security of a QKD protocol against an adversary limited only by
the no-signalling principle.  However, it is more common to
express security in another way as described below.

By using properties of the trace distance it can be shown that the
probability of successfully distinguishing can be bounded by the
sum of contributions from the two conditions stated
previously~\cite{Portmann14}.  These are usually called the
\emph{correctness error}
$$\eps_{\corr}=p(\bar{\perp})P(S_A\neq S_B)_{\rho^R}\,,$$
and the \emph{secrecy error},
$$\eps_{\secr}=p(\bar{\perp})D(\rho^R_{S_AE},n^{-1}\id_n\ot\rho_E)\,.$$
The correctness error is the probability that the protocol outputs
different keys to Alice and Bob.  The secrecy error is the
probability that the key output to Alice can be distinguished from
uniform given the system $E$.  In security proofs it is often
$\eps_{\corr}$ and $\eps_{\secr}$ that are computed.


\section{Overview of DV-QKD}\label{mainDVsection}

DV protocols can be seen as the earliest (and possibly the
simplest) form of QKD. Despite the development of the famous BB84
protocol with its name accorded based on a 1984 paper~\cite{BB84},
the first ideas for the use of quantum physics in the service of
security can be traced as far back as the early 70s (A detailed
history on the beginnings of quantum cryptography can be found in
Ref.~\cite{history}). Wiesner was then toying with the idea of
making bank notes that would resist counterfeit~\cite{history}.
The first paper published on quantum cryptography, on the other
hand was in 1982 \cite{1stBB84}. In this section we give a brief
description of DV protocols for QKD. It is instructive to
introduce some preliminary notation which will be useful in the
subsequent sections. The reader expert in quantum information may
skip most of the following notions.

\subsection{Preliminary notions}

Recall that a qubit is represented as a vector in a bidimensional
Hilbert space, which is drawn by the following basis vectors:
\begin{align}
& \ket{0} \equiv\binom{1}{0}, & \ket{1} \equiv\binom{0}{1}.
\label{eq:basis_states}
\end{align}
Any pure qubit state can thus be expressed as a linear
superposition of these basis states,
\begin{equation}
\ket{\psi} = \alpha\ket{0}+\beta\ket{1} = \cos(\theta/2)
\ket{0}+e^{i\phi} \sin(\theta/2) \ket{1},\label{eq:qubit}
\end{equation}
with $\theta\in(0,\pi)$, $\phi\in(0,2\pi)$ and $i$ the imaginary
unit. This state can be pictorially represented as a vector in the
so-called ``Bloch sphere''. When $\theta=0$ or $\theta=\pi$, we
recover the basis states $\ket{0}$ and $\ket{1}$, respectively,
which are placed at the poles of the sphere. When $\theta=\pi/2$,
the qubit pure state is a vector lying on the equator of the
sphere. Here we can identify the four vectors aligned along the
$\hat{x}$ and $\hat{y}$ axes, which are obtained in correspondence
of four specific values of $\phi$, i.e., we have:
\begin{align}
\phi & =0 : ~~~~~\ket{+} = \frac{1}{\sqrt{2}}\binom{1}{1},\label{eq:x0}\\
\phi & =\pi: ~~~~~\ket{-} = \frac{1}{\sqrt{2}}\binom{1}{-1},\label{eq:x1}\\
\phi & =\pi/2 : ~~\ket{+i} = \frac{1}{\sqrt{2}}\binom{1}{i},\label{eq:y0}\\
\phi & =3\pi/2: \ket{-i} = \frac{1}{\sqrt{2}}\binom{1}{-i}.\label{eq:y1}%
\end{align}
These four states are particularly important in QKD as they are
associated with the popular BB84 protocol~\cite{BB84}.

The basis vectors in Eq.~\eqref{eq:basis_states} are eigenstates
of the Pauli matrix \beq \sigma_{z}=\( \begin{matrix}
1 & 0\\
0 & -1
\end{matrix}\), \eeq which we shall simply refer to as the ``$\Z$ basis'', as it is
customary in QKD. Similarly, the states in Eqs.~\eqref{eq:x0} and
\eqref{eq:x1} are eigenstates of the Pauli matrix \beq
\sigma_{x}=\(
\begin{matrix}
0 & 1\\
1 & 0
\end{matrix} \), \eeq known as the $\X$ basis, and the states in Eqs.~\eqref{eq:y0} and
\eqref{eq:y1} are eigenstates of \beq \sigma_{y}=\( \begin{matrix}
0 & -i\\
i & 0
\end{matrix} \), \eeq known as the $\Y$ basis. {It is worth noting that each of these
pairs of eigenstates forms a basis which are mutually unbiased to
one another, referred to as mutually unbiased bases (MUB).
Formally, two orthogonal basis of a $d$-dimensional Hilbert space,
say $\{|\psi_{1},...,\psi_{d}\}$ and
$\{|\psi_{1},...,\psi_{d}\}$, are mutually unbiased if $|\langle\psi_{i}%
|\phi_{j}\rangle|^{2}=1/d$ for any $i$ and $j$. Measuring a state
from one MUB in another would thus produce either one of the
eigenstates with equal probability. 

Using the three Pauli matrices and the bidimensional identity
matrix \beq \I= \(
\begin{matrix}
1 & 0\\
0 & 1
\end{matrix} \) ,\eeq it is possible to write the most generic state of a qubit in
the form of a density operator, \beq \rho= \frac{1}{2} I+
\underline{n} \cdot \underline{\sigma}, \eeq with $\underline{n}$
the Bloch vector and $\underline{\sigma}=\{\sigma_{x}, \sigma_{y},
\sigma_{z}\}$. This notation comes handy when the qubit states are
mixed, which can be described with a vector $\underline{n}$ whose
modulo is less than 1, as opposed to pure states, for which
$|\underline{n}|=1$.

To give a physical meaning to the representation of a qubit, we
can interpret the qubit state in Eq.~\eqref{eq:qubit} as the
polarization state of a photon. In this case, the Bloch sphere is
conventionally called the Poincar\'{e} sphere, but its meaning is
unchanged. The basis vectors on the poles of the Poincar\'{e}
sphere are usually associated with the linear polarization states
$\ket{H}=\ket{0}$ and $\ket{V}=\ket{1}$, where $H$ and $V$ refer
to the horizontal or vertical direction of oscillation of the
electromagnetic field, respectively, with respect to a given
reference system. The $\X$ basis states are also associated with
linear polarization but along diagonal ($\ket{D}=\ket{+}$) and
anti-diagonal ($\ket{A}=\ket{-}$) directions. Finally, the $\Y$
basis states are associated with right-circular
($\ket{R}=\ket{+i}$) and left-circular ($\ket{L}=\ket{-i}$)
polarization states. Any other state is an elliptical polarization
state and can be represented by suitably choosing the parameters
$\theta$ and $\phi$.

It is worth noting that polarization can be cast in one-to-one
correspondence with another degree of freedom of the photon which
is particularly relevant from an experimental point of view. This
is illustrated in Fig.~\ref{fig:Bennett_interf}. The light source
emits a photon that is split into two arms by the first
beam-splitter (BS). The transmission of this BS represents the
angle $\theta$ of the Bloch sphere. More precisely, if $r$ and $t$
are the reflection and transmission coefficients of the BS,
respectively, such that $|r|^{2}+|t|^{2}=1$, we can write
$r=\cos(\theta/2)$ and $t=e^{i\phi} \sin(\theta/2)$ so to recover
Eq.~\eqref{eq:qubit}. If the BS is 50:50, then $\theta=\pi/2$ and
the state after the BS becomes \beq \ket{\psi} = \frac
{1}{\sqrt{2}} \( \ket{0}+e^{i\phi}\ket{1}\). \label{eq:qubit2}
\eeq The phase $\phi$ now has a clear physical meaning, i.e., it
represents the relative electromagnetic phase between the upper
and lower arms of the interferometer in
Fig.~\ref{fig:Bennett_interf}. This phase can be modified by
acting on the phase shifters in Fig.~\ref{fig:Bennett_interf} and
this is one of the most prominent methods to encode and decode
information in QKD. In fact, it is fair to say that the vast
majority of QKD experiments were performed using either the
polarization or the relative phase to encode information.

\begin{figure}[h]
\centering \vspace{-1.8cm}
\includegraphics[width=\columnwidth]{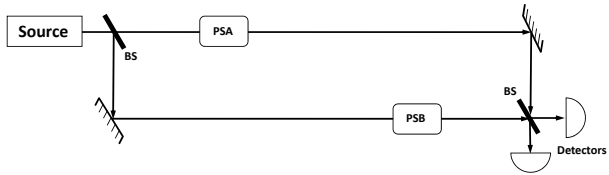}
\vspace{-2.4cm} \caption{Fundamental phase-based interferometer.
BS: beam-splitter; PSA: phase shift Alice; PSB: phase shift Bob.}
\label{fig:Bennett_interf}%
\end{figure}

As we well know, from a historical perspective, the first QKD
protocols were introduced using DVs, especially polarization. This
remains even today the simplest way to describe an otherwise
complex subject. The seminal BB84 protocol~\cite{BB84} was
described using polarization. In 1991 Ekert
suggested a scheme, the ``E91''~\cite{Eke91}, that for the first
time exploits entanglement for cryptographic purposes. The
conceptual equivalence of this scheme with the BB84 protocol was
demonstrated in 1992 by Bennett, Brassard and Mermin~\cite{BBM92},
who also proposed a simplified version of the E91 later called
``BBM92'' or more simply ``EPR scheme''. However, this supposed
equivalence cannot be taken strictly as it can be shown that the
entangled based protocol of E91 can provide device independent
security, which is impossible for the BB84 using separable states
even in a noise free scenario \cite{acin}.} A few years later, Lo
and Chau first~\cite{LC99} and Shor and Preskill
later~\cite{SP00}, will exploit this equivalence between the
prepare-and-measure BB84 and the entanglement-based BBM92 to
demonstrate the unconditional security of the BB84 protocol.
Another important protocol, the ``B92''~\cite{B92}, was proposed
in 1992 by Bennett, showing that QKD can be performed with even
only two non-orthogonal states. In the next sections, we will
describe these protocols and the advances over them in more
detail.

\subsection{Prepare and measure protocols}

In this section, we outline the most intuitive DV-QKD protocols,
generally denoted ``prepare-and-measure''. Here, the transmitting
user, Alice, prepares the optical signals by encoding on them a
discrete random variable, e.g., a bit. The optical signals are
then sent to the receiving user, Bob, who measures them in order
to retrieve the information sent by Alice. In describing the
protocols in this category, we will often use a single-photon
description highlighting the protocol's `in principle' workings,
even if in practice true single-photon sources are not yet widely
available.\newline

\subsubsection{BB84 protocol}

In the BB84 protocol, Alice (the transmitter) prepares a random
sequence of four states in two complementary bases. These are
usually chosen as $\ket{0}$, $\ket{1}$ ($\Z$ basis), $\ket{+}$,
$\ket{-}$ ($\X$ basis). However, other choices are possible,
including the four states in Eqs.~\eqref{eq:x0}-\eqref{eq:y1}. The
users associate a binary 0 (a binary 1) with the non-orthogonal
states $\ket{0}$ and $\ket{+}$ ($\ket{1}$ and $\ket{-}$). The
non-orthogonality condition guarantees that an eavesdropper cannot
clone or measure the prepared states with perfect fidelity. This
is true because the no cloning theorem assures that she cannot
replicate a particle of unknown state~\cite{noclone1,noclone2}.
This implies that she cannot perfectly retrieve the information
encoded by Alice and that her action causes a disturbance on the
quantum states that can be detected by the legitimate users. The
states prepared by Alice are sent to Bob (the receiver), who
measures them in one of the two bases $\Z$ or $\X$, selected at
random. If, for a particular photon, Bob chooses the same basis as
Alice, then in principle, Bob should measure the same bit value as
Alice and thus he can correctly infer the bit that Alice intended
to send. If he chose the wrong basis, his result, and thus the bit
he reads, will be random.

When the quantum communication is over, Bob notifies Alice over a
public channel what basis he used to measure each photon, for each
of the photons he detected. Alice reports back her bases and they
discard all the events corresponding to different bases used.
Provided no errors occurred or no one manipulated the photons, the
users should now both have an identical string of bits which is
called ``sifted key''.

At this point, Alice and Bob test their key by agreeing upon a
random subset of the bits to compare their results. If the bits
agree, they are discarded and the remaining bits form the shared
secret key. In the absence of noise or any other measurement
error, a disagreement in any of the bits compared would indicate
the presence of an eavesdropper on the quantum channel.

For the sake of clarity, we shall describe how an eavesdropper can
gain any information while inducing noise. This is really simple
quantum mechanics. Let us assume, for simplicity, that Eve makes a
measurement to project the state of the photon onto one given by
$|\theta\rangle=\cos(\theta/2) \ket{0}+e^{i\phi} \sin(\theta/2)
\ket{1}$ and a state orthogonal to it, $|\theta^{\bot}\rangle$.
She can infer Alice's state $|a\rangle$, after Alice's disclosure
on the public channel of the basis used by Bayes theorem,
\begin{align}
\label{bayes}\text{Pr}(|a\rangle||\theta\rangle)=\dfrac{\text{Pr}%
(|\theta\rangle||a\rangle)\text{Pr}(|a\rangle)}{\text{Pr}(|\theta
\rangle||a\rangle)\text{Pr}(|a\rangle)+\text{Pr}(|\theta\rangle||a^{\bot
}\rangle)\text{Pr}(|a^{\bot}\rangle)}%
\end{align}
As $\text{Pr}(|a\rangle)=1/2$ and Alice's disclosure limits the
possible states to only $|a\rangle$ and $|a^{\bot}\rangle$ for a
run, the above simplifies to
$\text{Pr}(|a\rangle||\theta\rangle)=\text{Pr}(|\theta
\rangle||a\rangle)$. In order to have an idea of Eve's information
gain, let us consider a specific example~\cite{rev1}; assuming
$|a\rangle=|0\rangle$. It can be easily shown that
$\text{Pr}(|0\rangle||\theta\rangle)=|\langle0|\theta
\rangle|^{2}=\cos^{2}{(\theta/2)}$. Her uncertainty,
$H_{E}^{\mathbb{Z}}$, on Alice's encoding is given by Shannon's
binary entropic function as
\begin{align}
H_{E}^{\mathbb{Z}}=-\cos^{2}{(\theta/2)}\log_{2}\[{\cos^{2}{(\theta/2)}}\] \nonumber \\
-\sin^{2}{(\theta/2)}\log_{2}\[{\sin^{2}{(\theta/2)}}\].
\end{align}

If Alice had used the $\mathbb{X}$ basis, a similar calculation
would have given
$\text{Pr}(|+\rangle||\theta\rangle)=|\langle+|\theta\rangle
|^{2}=(1+\sin{\theta}\cos{\phi})/2$ and Eve's corresponding
uncertainty, $H_{E}^{\mathbb{X}}$ is
\begin{align}
H_{E}^{\mathbb{X}}=-\dfrac{1+\sin{\theta}\cos{\phi}}{2}\log_{2} \( {\dfrac{1+\sin{\theta}\cos{\phi}}{2}} \) \nonumber \\
-\dfrac{1-\sin{\theta}\cos{\phi}}{2}\log_{2} \(
{\dfrac{1-\sin{\theta} \cos{\phi}}{2}} \).
\end{align}
The first thing to note is that, to have zero uncertainty for
$H_{E}^{\mathbb{Z}}$ is to set $\theta=0$ corresponding to a
measurement in the $\mathbb{Z}$ basis (mathematically, it is
certainly possible to set $\theta=\pi$, however this simply means
that $|\theta\rangle\equiv|1\rangle$ and the measurement basis is
still $\mathbb{Z}$). However, this forces a maximal uncertainty
for $H_{E}^{\mathbb{X}}$ i.e. when Alice uses the $\mathbb{X}$
basis. On the other hand, minimizing the uncertainty of
$H_{E}^{\mathbb{X}}$ (e.g. by setting $\theta=\pi/2$ and $\phi=0$)
would maximize $H_{E}^{\mathbb{Z}}$. This is certainly in line
with the use of MUBs where maximizing the information gain when
measuring in one basis maximizes the uncertainty for the
complimentary basis. The only way out, i.e. to minimize both
uncertainty is to use two different measurement bases
corresponding to Alice's choices of bases; this can be chosen
randomly and the events where the choices do not match would be
discarded. This is precisely Bob's situation! Hence we can see how
Alice and Bob can actually share maximal information in principle
as they discard the runs where their bases do not match. Eve on
the other hand does not have that luxury, as she would only have
her bases match Alice's half the time and her information gain is
$0.5$. This is the most basic \textit{intercept-resend} attack
strategy.

Now, let us consider what happens after Eve makes a measurement.
Quantum mechanics tells us that her measurement would project
Alice's state into an eigenstate of her measurement basis, and she
would thus forward to Bob the state $|\theta\rangle$. Bob on the
other hand, when measuring in the same basis as Alice (when she
sends $|a\rangle$) would thus register an error
($|a^{\bot}\rangle$) with probability $|\langle
a^{\bot}|\theta\rangle |^{2}=\sin^{2}{(\theta/2)}$ in those
instances. Hence, if Eve uses the $\mathbb{Z}$ basis for
measurement while Alice and Bob's are $\mathbb{X}$ the error rate
in these instances becomes $1/2$. However, as she would be able to
guess correctly half the time, the error rate is halved and on
average, the users will detect an error with a probability of
25\%.

Obviously this choice of $\theta$ need not be limited to $0$ or
$\phi/2$. A well known example is when $\theta=\pi/4$, a
measurement in the so called Breidbart basis. This would make
$H_{E}^{\mathbb{Z}}=H_{E}^{\mathbb{X}}$. The calculation for the
error that Bob would note is straightforward. Let us take the case
when Alice and Bob uses the $\mathbb{Z}$ basis. When Eve projects
her the qubit into the state $|\theta\rangle$, this happens with
probability $\cos^{2}{(\pi/8)}$. Bob gets an erroneous result with
probability $\sin^{2}{(\pi/8)}$. In the instance Eve projects onto
$|\theta^{\bot}\rangle$ which happens with probability
$\sin^{2}{(\pi/8)}$, Bob registers an error with probability
$\cos^{2}{(\pi/8)}$. The error rate thus becomes
$2\cos^{2}{(\pi/8)}\sin^{2}{(\pi/8)}=0.25$. A similar calculation
can be done for the case when Alice uses the $\mathbb{X}$ basis to
also yield an error rate of 25\%.

In a noiseless scenario, the presence of an error would reveal
with certainty the presence of an eavesdropper. In this case the
users can abort the whole communication, discard their key and
start a new communication. However, in realistic situation, noise
is always present given imperfection of physical implementations.
It is tempting to imagine that one can characterize the errors on
the physical channel and then any `extra' error can be assumed due
to Eve. However, assuming Eve can actually substitute the channel
with a perfect noiseless one, Alice and Bob would not be able to
distinguish between errors that are genuine (i.e. not due to Eve)
or errors due to her meddling. A pessimistic stand is to assume
\textit{all} errors due to Eve. Aborting the protocol every time
an error is detected would translate into Alice and Bob never able
to establish a secure key. Thus the trick is not so much in
detecting an eavesdropper, rather, given the presence of an
eavesdropper, how can one still distill a secret key.

When noise is present, the users can detect an error even if Eve
is not on the line. In this case they run an error correction
algorithm followed by a compression algorithm called privacy
amplification (PA). The amount of PA necessary is estimated by the
users starting from the percentage of errors measured in their
experiment, the so-called ``quantum bit error rate'' (QBER). Hence
the search for an ultimate security proof is simply the search for
the best strategy Eve can employ to achieve the highest
information gain given the amount of QBER detected.

A general attack strategy an eavesdropper can consider is to
attach an ancilla, $|\mathcal{E}\rangle$, (a quantum system
possibly higher dimension than a qubit) to Alice's qubit and let
them interact in the hope of gleaning some information. This
interaction (with Alice's state in the computational basis) can be
written as
\begin{align}
U|0\rangle|\mathcal{E}\rangle=\sqrt{F_{0}}|0\rangle|\mathcal{E}_{00}%
\rangle+\sqrt{D_{0}}|1\rangle|\mathcal{E}_{01}\rangle,\\
U|1\rangle|\mathcal{E}\rangle=\sqrt{F_{1}}|1\rangle|\mathcal{E}_{11}%
\rangle+\sqrt{D_{1}}|0\rangle|\mathcal{E}_{10}\rangle,
\end{align}
with $|\mathcal{E}_{ij}\rangle$ being Eve's possible ancillary
states after the interaction. These equations literally mean that
when Alice sends a $|0\rangle$ ($|1\rangle$) state, Bob has a
probability $F_{0}$ ($F_{1}$) of getting the right result when
measuring in the $\mathbb{Z}$ basis and $D_{0}$ ($D_{1}$)
otherwise.

There are two points worth noting here; firstly, the Stinespring
dilation theorem allows us to limit our consideration of Eve's
ancillae to a four dimensional quantum system or two qubits.
Secondly, given linearity, the interaction with Eve's ancillae can
also be written directly for Alice's $\mathbb{X}$ basis, thus
defining the QBER in that basis. In order to ensure that the QBER
in both bases $\mathbb{Z}$ and $\mathbb{X}$ are equal, the overlap
between Eve's ancillary states must be defined accordingly. We
begin by rewriting the above equations more concisely as
\begin{align}
\label{nonortho}U|a\rangle|\mathcal{E}\rangle=\sqrt{F_{a}}|a\rangle
|\mathcal{E}_{aa}\rangle+\sqrt{D_{a}}|a^{\bot}\rangle|\mathcal{E}_{aa^{\bot}%
}\rangle,
\end{align}
where $|a\rangle\in\{|0\rangle,|1\rangle,|+\rangle,|-\rangle\}$
and $\langle a|a^{\bot}\rangle=0$. Unitarity of $U$ ensures
\begin{align}
\langle\mathcal{E}_{aa}|\mathcal{E}_{aa}\rangle=F_{a},\\~\langle\mathcal{E}%
_{aa^{\bot}}|\mathcal{E}_{aa^{\bot}}\rangle=D_{a},\\~\langle\mathcal{E}%
_{aa}|\mathcal{E}_{aa^{\bot}}\rangle=0,
\end{align}
and $F_{a}+D_{a}=1$. Imposing the symmetry of errors in both bases
leads to
\begin{align}
\langle\mathcal{E}_{aa}|\mathcal{E}_{a^{\bot}a^{\bot}}\rangle=F_{a}\cos
{x},\\ ~\langle\mathcal{E}_{aa}|\mathcal{E}_{a^{\bot}a}\rangle=0,\\
\langle\mathcal{E}_{aa^{\bot}}|\mathcal{E}_{a^{\bot}a}\rangle=D_{a}
\cos {(y)},
\end{align}
implying the QBER
\begin{align}
\label{D}D_{a}=\dfrac{1-\cos x}{2-\cos x+\cos y}.
\end{align}
This is the essence of a \textit{symmetric} attack \cite{SP} which
can be seen as a contraction of the Bloch sphere by $F_{a}-D_{a}$.

Assume that Eve keeps her ancillary system in a quantum memory and
waits for Alice and Bob to end all the classical communication
related with the reconciliation of the bases (sifting). In this
way she can distinguish between her ancillary
states given by $|\mathcal{E}_{a a}\rangle$ and $|\mathcal{E}_{a^{\bot}%
a^{\bot}}\rangle$. Then assume that she can also perform a joint
measurement on her entire quantum memory, a scenario known as
`collective attack'. In such a case, Eve's amount of information
is upper bounded by the Holevo information
\begin{align}
\chi=S(\rho_{E})-\dfrac{S[\rho_{E}(a)]+S[\rho_{E}(a^{\bot})]}{2},
\end{align}
where $S(\cdot)$ is the von Neumann entropy, and $\rho_{E}(a)$
($\rho _{E}(a^{\bot})$) is Eve's state for Alice's $|a\rangle$
($|a^{\bot}\rangle$). In the presence of this symmetric collective
attack, it can be shown that the secret key rate is then given
by~\cite{SP}
\begin{align}
\label{piran}R_{\mathrm{BB84}}=1-S(\rho_{E})=1-2H_2{(D_{a})},
\end{align}
where the binary Shannon entropy $H_2$ is computed over the QBER
$D_{a}$. As a result, a key can be extracted for a QBER with a
value no greater than approximately $11\%$.

This security threshold value of $11\%$ is exactly the same as the
one that is found by assuming the most general `coherent attack'
against the protocol, where all the signal states undergo a joint
unitary interaction together with Eve's ancillae, and the latter
are jointly measured at the end of protocol. In this general case
the security proof was provided by Shor and Preskill~\cite{SP00}.

The main idea to show the unconditional security of the BB84
protocol is based on the reduction of a QKD protocol into an
entanglement distillation protocol (EDP). Given a set of
non-maximally entangled pairs, the EDP is a procedure to
\textit{distill} a smaller number of entangled pairs with a higher
degree of entanglement using only local operations and classical
communication (LOCC). In some ways, employing this for a security
proof for QKD actually makes perfect sense as it involves the two
parties ending with a number of maximally entangled pairs. Given
the monogamous nature of entanglement, no third party can be privy
to any results of subsequent measurements the two make.

In particular, Shor and Preskill~\cite{SP00} showed that EDP can
be done using quantum error correction codes, namely the
Calderbank-Shor-Steane (CSS) code~\cite{NielsenChuang} which has
the interesting property which decouples phase errors from bit
errors. This allows for corrections to be made independently. In
this way, one can show that the key generation rate becomes \beq
R_{\mathrm{BB84}}=1-H_2(e_{b})-H_2(e_{p}) \eeq where $e_{b}$ and
$e_{p}$ are bit and phase error rates with $e_{b}=e_{p}$. This
results in the same formula of Eq. (\ref{piran}). It is simple to
see that $R=0$ for $e_{b}\approx11\%$.

\subsubsection{Six-state protocol}

The BB84 protocol has also been extended to use six states in
three bases to enhance the key generation rate and the tolerance
to noise \cite{bruss}. 6-state BB84 is identical to BB84 except,
as its name implies, rather than using two or four states, it uses
six states on three bases $\X$, $\Y$ and $\Z$. This creates an
obstacle to the eavesdropper who has to guess the right basis from
among three possibilities rather than just two of the BB84. This
extra choice causes the eavesdropper to produce a higher rate of
error, for example, $1/3$ when attacking all qubits with a simple
IR strategy; thus becoming easier to detect.

One can extend the analysis of Eve's symmetric collective attack
to the 6-state BB84 by considering a third basis for Eq. (\ref{D})
which immediately sets a further constraint on Eve's ancillary
state; i.e. $\cos{y}=0$ (Eve's states $|
\mathcal{E}_{aa^{\bot}}\rangle$ and
$|\mathcal{E}_{a^{\bot}a}\rangle$ are orthogonal). The new QBER
$D^{\prime}_{a}$ is then given by
\begin{align}
\label{D6}D_{a}^{\prime}=\dfrac{1-\cos x}{2-\cos x},
\end{align}
as also noted in~\cite{hbg} (but reported in terms of the fidelity
rather than the QBER). Assuming a symmetric collective
attack~\cite{SP}, a similar calculation to the one for BB84 gives
the following secret key rate for the 6-state protocol as
\begin{align}
R_{\mathrm{6-state}}=1 & +\dfrac{3D_{a}^{\prime}}{2}\log_{2}\dfrac{D_{a}^{\prime}}{2}\\
& +\left(  1-\dfrac{3D_{a}^{\prime}}{2}\right)  \log_{2}\left(
1-\dfrac {3D_{a}^{\prime}}{2}\right).
\end{align}
This rate exactly coincides with the unconditional key rate,
proven against coherent attacks, and gives a security threshold
value of about $12.6\%$ slightly improving that of the BB84
protocol.

Before moving on, it is worth noting that the symmetric attacks
described in both the BB84 protocol as well as the 6-state
protocol are equivalent to the action of quantum cloning machines
(QCMs)~\cite{clone}. Notwithstanding the no-cloning theorem, QCMs
imperfectly clone a quantum state, producing a number of copies,
not necessarily of equal fidelity. QCMs which result in copies
that have the same fidelity are referred to as symmetric. In the
case of the BB84, the states of interest come from only 2 MUBs,
hence the relevant QCM would be the \textit{phase covariant} QCM
which clones all the states of the equator defined by two MUBs
(the term `phase covariant' comes from the original formulation of
the QCM cloning states of the form
$(|0\rangle+e^{i\phi}|1\rangle)/\sqrt{2}$ independently of
$\phi$~\cite{brussclone}; this QCM thus copies equally well the
states from the $\mathbb{X}$ and $\mathbb{Y}$ bases). As for the
6-state protocol, the relevant QCM is universal, meaning that it
imperfectly clones all states from 3 MUBs with the same fidelity.

\subsubsection{B92 protocol}

In 1992, Charles Bennett proposed what is arguably the simplest
protocol of QKD, the ``B92''~\cite{B92}. It uses only two states
to distribute a secret key between the remote parties. This is the
bare minimum required to transmit one bit of a cryptographic key.
More precisely, in the B92 protocol, Alice prepares a qubit in one
of two quantum states, $\ket{\psi_0}$ and $\ket{\psi_1}$, to which
she associates the bit values 0 and 1, respectively. The state is
sent to Bob, who measures it in a suitable basis, to retrieve
Alice's bit. If the states $\ket{\psi_0}$, $\ket{\psi_1}$ were
orthogonal, it is always possible for Bob to deterministically
recover the bit. For instance, if $\ket{\psi_0}=\ket{0}$ and
$\ket{\psi_1}=\ket{1}$, Bob can measure the incoming states in the
$\Z$ basis and recover the information with 100\% probability.

However, Bob's ability to retrieve the information without any
ambiguity also implies that Eve can do it too. She will measure
the states midway between Alice and Bob, deterministically
retrieve the information, prepare new states identical to the
measured ones, and forward them to Bob, who will never notice any
difference from the states sent by Alice. Orthogonal states are
much alike classical ones, that can be deterministically measured,
copied and cloned. Technically, the orthogonal states are
eigenstates of some common observable, thus measurements made
using that observable would not be subjected to any uncertainty.
The no-cloning theorem~\cite{noclone1,noclone2} does not apply to
this case.

By contrast, measurements will be bounded by inherent
uncertainties if Alice encodes the information in two
non-orthogonal states, for example the following ones:
\begin{align}
& \ket{\psi_0} = \ket{0}, & \ket{\psi_1} =
\ket{+}~,~\inprod{\psi_0}{\psi_1}=s\neq0.\label{eq:B92_states}%
\end{align}
As Bennett showed in his seminal paper~\cite{B92}, any two
non-orthogonal states, even mixed, spanning disjoint subspaces of
the Hilbert space can be used. In the actual case, the scalar
product $s$ is optimized to give the best performance of the
protocol. For the states in Eq.~\eqref{eq:B92_states}, this
parameter is fixed and amounts to $1/\sqrt{2}$; i.e. the states
are derived from bases which are mutually unbiased one to the
other. Given the complementary nature of the observables involved
in distinguishing between these states, neither Bob nor Eve can
measure or copy the states sent by Alice with a 100\% success
probability. However, while Alice and Bob can easily overcome this
problem (as described in the following) and distil a common bit
from the data, Eve is left with an unsurmountable obstacle, upon
which the whole security of the B92 protocol is based.

In B92, Bob's decoding is peculiar and worth describing. It is a
simple example of ``unambiguous state discrimination''
(USD)~\cite{Chefles,QHT2}. To explain it, it is useful to remember
that the state $\ket{0}$ ($\ket{+}$) is an eigenstate of $\Z$
($\X$) and that $\ket{\pm} = (\ket{0}\pm\ket{1})/\sqrt{2}$, as it
is easy to verify from Eqs.~\eqref{eq:basis_states}, \eqref{eq:x0}
and \eqref{eq:x1}. Suppose first that Alice prepares the state
$\ket{\psi_0}$. When Bob measures it with $\Z$, he will obtain
$\ket{0}$ with probability $100\%$ whereas when he measures it
with $\X$, he will obtain either $\ket{+}$ or $\ket{-}$ with
probability 50\%. In particular, there is one state that Bob will
never obtain, which is $\ket{1}$. Now suppose that Alice prepares
the other state of B92, $\ket{\psi_1}$. Bob will still measure in
the same bases as before but in this case, if we repeat the
previous argument, we conclude that Bob can never obtain the state
$\ket{-}$ as a result. See the table below for a schematic
representation of Bob's outcomes and their probabilities
($\mathrm{Pr}$) depending on Alice's encoding state and Bob's
chosen basis for measurement.

\beq
\begin{tabular}
[c]{c|c|c|c}%
bit & Alice & Bob ($\mathbb{Z}$) & Bob
($\mathbb{X}$)\\\hline\hline
0 & $\left\vert 0\right\rangle $ & \multicolumn{1}{|l|}{$%
\begin{array}
[c]{c}%
\left\vert 0\right\rangle ,~\mathrm{Pr}=1\\
\left\vert 1\right\rangle ,~\mathrm{Pr}=0
\end{array}
$} & \multicolumn{1}{|l}{$%
\begin{array}
[c]{c}%
\left\vert +\right\rangle ,~\mathrm{Pr}=1/2\\
\left\vert -\right\rangle ,~\mathrm{Pr}=1/2
\end{array}
$}\\\hline
1 & $\left\vert +\right\rangle $ & \multicolumn{1}{|l|}{$%
\begin{array}
[c]{c}%
\left\vert 0\right\rangle ,~\mathrm{Pr}=1/2\\
\left\vert 1\right\rangle ,~\mathrm{Pr}=1/2
\end{array}
$} & \multicolumn{1}{|l}{$%
\begin{array}
[c]{c}%
\left\vert +\right\rangle ,~\mathrm{Pr}=1\\
\left\vert -\right\rangle ,~\mathrm{Pr}=0
\end{array}
$}
\end{tabular}
\eeq

\noindent From the table it is clear that, for the conditional
probability $p(A|B)$ of guessing Alice's encoding $A$ given Bob's
outcome $B$, we may write
\begin{align}
\text{Pr}(|+\rangle||1\rangle)=\text{Pr}(|0\rangle||-\rangle)=1.
\end{align}
In other words, Bob can logically infer that when he detects
$\ket{1}$, Alice must have prepared the state $\ket{+}$, so he
decodes the bit as `1', whereas when he detects $\ket{-}$, Alice
must have prepared the state $\ket{0}$ so he decodes the bit as
`0'. Whenever he detects any other state, Bob is unsure of Alice's
preparation and the users decide to simply discard these
``inconclusive'' events from their records.

This way, using this sort of ``reversed decoding'', which is
typical of USD, and his collaboration with Alice, Bob manages to
decode the information encoded by Alice. {\ros Despite the fact
that USD can also be used by Eve, the unconditional security of
the B92 protocol was rigorously proven in~\cite{TKI03} for a
lossless scenario and then extended to a lossy, more realistic,
case in~\cite{TL04}, under the assumption of single photons
prepared by Alice. This assumption is not necessary in the B92
version with a strong reference pulse, which has been proven
secure in~\cite{Koa04}. Remarkably, this particular scheme has
been shown to scale linearly with the channel transmission at long
distance, a desirable feature in QKD. Two interesting variants of
this scheme appeared in~\cite{Tam08} and~\cite{TLK+09}, which
allow for a much simpler implementation. 

Generally speaking, the performance of the B92 protocol is not as
good as that of BB84. The presence of only two linearly
independent states makes it possible for the eavesdropper to
execute a powerful USD measurement on the quantum states prepared
by Alice. This makes the B92 very loss dependent and reduces its
tolerance to noise from a depolarizing channel to about
3.34\%~\cite{TKI03}. This value is much smaller than the one
pertaining to the BB84 protocol, which is 16.5\%~\cite{SP00} (it
should be stressed here that these values refer to the
depolarizing parameter $p$ for a depolarizing channel acting on a
state $\rho$ as $(1-p)\rho+p/3\sum_{i}\sigma_{i}\rho \sigma_{i}$
with $\sigma_{i}$ as the Pauli matrices).

However, it was recently shown that the B92 can be made
loss-tolerant if Alice prepares a pair of uninformative states in
addition to the usual B92 states, while leaving Bob's setup
unchanged~\cite{LDGT09}. This is due to the fact that the two
extra states make the B92 states linearly dependent, thus
preventing the possibility of a USD measurement by Eve. The
existence of the uninformative states paved the way to a
device-independent entanglement-based description of the B92
protocol~\cite{LVG+12}, which was not previously available. In
this description, Eve herself can prepare a non-maximally
entangled state and distribute it to Alice and Bob. By measuring
in suitable bases, Alice and Bob can test the violation of the
Clauser-Horne inequality~\cite{CH74}, a special form of Bell
inequality, thus guaranteeing the security of the protocol from
any attack allowed by quantum mechanics, irrespective of the
detailed description of the hardware. Despite the radically
different security proof used~\cite{MPA11}, the tolerance to the
noise from a depolarizing channel was found to be 3.36\%,
remarkably close to the value of the standard prepare-and-measure
B92 protocol.

Before concluding, it is worth mentioning that both the
prepare-and-measure B92~\cite{B92} and the entanglement-based
B92~\cite{LVG+12} have a clear advantage in the implementation, as
experimentally shown in~\cite{LKDG+10}. The asymmetry of the B92
states allows for an automatic feedback that can keep distant
systems aligned without employing ad-hoc resources at no extra
cost for the key length.

\subsection{Practical imperfections and countermeasures}

\subsubsection{PNS attacks}

DV-QKD protocols are ideally defined on qubits (or qudits) for
which security analysis drawn are based on. Moving from a
theoretical protocol where single qubits are used to carry one bit
of classical information to a practical implementation should in
principle require the most faithful adaptation possible. However
in practice, perfect single-photon sources are generally not
available and there is some probability for a source to emit
multiple photons with identical encodings in a given run of the
QKD protocol. This can be a security vulnerability to an
eavesdropper who employs the photon number splitting (PNS)
attack~\cite{hut,lut}. The essential idea behind the attack is
that Eve can perform a quantum non-demolition measurement to
determine the number of photons in a run and when it is greater
than $1$, she could steal one of the excess photons while
forwarding the others to Bob. In this way Bob would not be able to
detect her presence while she lies in wait for Alice's basis
revelation to make sharp measurements of the stolen photons and
obtain perfect information of the multi-photon runs. The case for
single photons can be attacked using the ancillary assisted attack
strategy described earlier.

A weak coherent laser source is commonly used to implement DV-QKD
protocols. Such a source generates a pulse having a finite
probability of multiple photons with the number of photons $n$
described by the Poisson distribution. Thus, the probability for a
pulse sent to contain a number of photons $n$ is given by
\begin{align}
\label{poisson}\text{Pr}(n)=\dfrac{\mu}{n!}\exp{(-\mu)}%
\end{align}
where $\mu$ being the average photon number per pulse. It is not
difficult to imagine that an eavesdropper's information gain thus
increases with $\mu$. Intuitively, one can imagine that as secret
bits can only be derived from the single photon pulses, the number
of bits from the multi-photon pulses needs to be subtracted from
the total signal gain for the raw key. Writing $p$ as the fraction
of signals detected by Bob, we write the minimum fraction for
single photon pulses as $\mathcal{P}$, where
$\mathcal{P}=[p-\text{Pr}(n>1)]/p$.

Considering the case of Eve committing to an individual attack
strategy, the QBER estimated as $e$, need to be rescaled to
$e^{\prime}=e/\mathcal{P}$ for PA purposes with the assumption
that all errors stem from Eve's attack of single photon pulses.
Let $Q$ be the ratio of the total bits for a raw key over the
total signals detected. The effective key rate is then given by
\begin{align}
\label{lutk}R_{m}=[\mathcal{P}(1-r_{\mathrm{PA}} )-H_2(e)]Q
\end{align}
where $r_{\mathrm{PA}}$ is the rate for PA and $H_2(e)$ is the
rate due to error correction procedure (note that the QBER need
not to be rescaled for error correction). For an individual attack
strategy, the PA rate is given by
$r_{PA}=\log_{2}(1+4e^{\prime}-4e^{\prime^{2}})$, for
$e^{\prime}\le0.5$ and $1$ otherwise.

In a realistic setup, considerations for dark counts (Bob's
detector clicking for vacuum pulses) must also be taken into
account. It should also be noted that for a BB84 setup, Eq.
(\ref{lutk}) would be multiplied further by a factor of half to
reflect the instances where Alice's and Bob's choice of
measurement bases coincide. A full fledge treatment for the above
can be referred to in~\cite{lut}. While the PNS attack decreases
the secure key generation rate drastically, the decoy state method
and the SARG04 protocol are possible approaches to solve these
issues.

\subsubsection{Decoy States}

As we have seen in the previous section, practical implementations
which include multi photon pulses is detrimental to the key rate
when the legitimate parties cannot distinguish between photons
detected from single or multi photon pulses. At best, they can try
to determine the fraction of single photon pulses received and
have privacy amplification done only on that. The case of the
individual attack strategy by an eavesdropper was studied
in~\cite{lut} with the key rate reflected by equation
(\ref{lutk}). In Ref.~\cite{GLLP}, commonly referred to as `GLLP',
a more general scenario was considered. In some sense, this can be
understood as a generalization to the EDP based security proof of
BB84 to include non-single photon sources.

In a nutshell, it demonstrates that it is sufficient to consider
the PA rate for single-photon pulses when considering a string
derived from both single and multi-photon pulses. Thus, the key
rate is essentially given by equation (\ref{lutk}) except for
$\mathcal{P}$ and $r_{\mathrm{PA}}$ terms substituted with $Q_{1}$
and $H_{2}(e_{1})$, where $Q_{1}$ and $e_{1}$ are the gain and the
QBER corresponding to single-photon pulses. Hence, if one can
determine accurately the gain (which could be greater than
$\mathcal{P}$) as well as the amount of error relevant to the
single photon pulses (which may be less than $e^{\prime}$), then
the PA rate may be reduced and the key rate would definitely
receive a boost. This is where the decoy state method comes in.
First introduced in Ref.~\cite{decoy1}, it was shown to be
practically useful in Ref.~\cite{decoyWang}, where the method was
studied assuming three different intensities under finite-size
effects (see also Ref.~\cite{Wang4decoy}). It was further
developed and worked on in Refs.~\cite{decoy,decoyfin}. The decoy
states technique has enabled QKD to be executed over distances
beyond a hundred kilometers despite the imperfections in
implementation.

In a practical setup, Bob's gain is a weighted average of all
detected photons (including the empty pulses) and can be written
as
\begin{align}
\label{Qm}Q_{m}=\sum_{i=0}^{\infty}Y_{i}\exp{(-\mu)}\dfrac{\mu^{i}}{i!}%
\end{align}
where $Y_{i}$ is the probability that Bob detects conclusively an
$i$-photon pulse sent by Alice. The QBER also has contribution
from multi-photon pulses, and can be written as
\begin{align}
\label{Em}E_{m}=\dfrac{1}{Q_{m}}\sum_{i=0}^{\infty}Y_{i}e_{i}\exp{(-\mu
)}\dfrac{\mu^{i}}{i!}%
\end{align}
Let us note that Alice and Bob can only determine the values of
$Q_{m}$ and $E_{m}$ in an actual implementation and do not have
any information of the values of $Y_{i}$. However, if one
considers using differing values of the light intensity, $\mu$,
then one can have a system of linear equations (based on Eq.
(\ref{Qm}) for varying $\mu$) with the solution set $\{Y_{0},
Y_{1},...\}$. Another system of linear equations based on Eq.
(\ref{Em}) provides the solution set $\{e_{0}, e_{1},...\}$.

It is now a rather straightforward matter to determine $Q_{1}$:
Alice could send to Bob photon pulses with varying intensities of
light. Only one intensity is preferred for key bits while the
other intensities would be used as decoy, i.e. \textit{decoy
states}. As these would be done randomly, Eve would not know which
photons would be used for key purposes and which were the decoys.
Linear algebra then should provide a standard method to derive the
solution sets and Alice and Bob would know exactly the values for
$Y_{1}$ and $e_{1}$. These values are the most pertinent.

All seems well except for the fact that the value of $i$ in
equations (\ref{Qm}) and (\ref{Em}) runs from $0$ to infinite!
This literally means that to have a precise value for $Y_{1}$ and
$e_{1}$, Alice should in principle, use and infinite number of
decoy states. This was in fact the consideration done in
Ref.~\cite{decoy}. Using parameters of an experimental setup, they
demonstrated how the decoy state technique could benefit the BB84
protocol by extending the distance. Nevertheless, given the fact
that the event of producing a pulse with higher number of photons
is less likely compared to one with lesser in number, a finite
number of decoy states can be quite sufficient.
Ref.~\cite{decoyfin} showed that an implementation using only two
decoy states (with the sum of the intensities lower than that of
the signal state) is sufficient and in the limit of vanishing
intensities of the decoys, the values estimated for $Y_{1}$ and
$e_{1}$ asymptotically approaches the ideal infinite decoy case.
See also Sec.~4.3 of Ref.~\cite{decoyREV} for a brief review on
decoy states.

\subsubsection{SARG04 protocol}

While the decoy state technique mitigates the problem of PNS
attacks by introducing new elements to the BB84 protocol, a more
subtle approach was introduced in 2004 by Scarani, Acin, Ribordy,
and Gisin, ``SARG04''~\cite{sarg}, a variant of the BB84 at the
classical communication stage. The PNS attack thrives on the
information revealed regarding the basis used in prepare and
measure protocols like BB84. Thus, a natural way against such an
attack would be to discount such an element from the protocol. The
SARG04 protocol shares the first step of photon transmission with
BB84: Alice sends one of four states selected randomly from 2
MUBs, $\Z$ or $\X$, and Bob performs a measurement with the two
bases. In the second step however, when Alice and Bob determine
for which bits their bases matched, Alice does not directly
announce her bases but a pair of non-orthogonal states, one of
which being used to encode her bit.

The decoding is similar to that of the B92 protocol; it is a
procedure of USD between states in the announced pair. For
example, assume Alice transmits $\ket{0}$ and Bob measures it with
the basis $\X$. Alice would announce the set
$\{\ket{0},\ket{+}\}$. If Bob's measurement results in $\ket{+}$,
then Bob cannot infer Alice's state conclusively as the output
$\ket{+}$ could have resulted from either $\ket{0}$ or $\ket{+}$
as input. In such a case, the particular run would be discarded.
If the result was $\ket{-}$ instead, then it is stored for post
processing because it could have only resulted from the
measurement of the $\ket{0}$ state. Since the two states in a set
are non-orthogonal, the PNS attack cannot provide Eve with perfect
information on the encoded bit.

The SARG04 protocol has been shown to be secure up to QBER values
of $9.68\%$ and $2.71\%$ for single photon and double photon
pulses respectively~\cite{sargsecure} using the EDP type proof. It
is worth noting that similar modification to the classical phase
of the six state protocol can be done to give a `six-state SARG04'
where key bits can be derived from even $4$ photon pulse. This is
secure for QBER values of $11.2\%,5.60\%,2.37\%$ and $0.788\%$ for
$1,2,3$ and $4$ photon pulses respectively. See also the recent
analysis in Ref.~\cite{sargZBC}. \newline

\subsection{Entanglement-based QKD}

\subsubsection{E91 protocol}\label{EkertSECTION}

In 1991, Artur Ekert developed a new approach to QKD by
introducing the E91 protocol~\cite{Eke91}. The security of the
protocol is guaranteed by a Bell-like test to rule out Eve. The
E91 considers a scenario where there is a single source that emits
pairs of entangled particles, each described by a Bell state, in
particular the singlet state
$|\Psi\rangle=(|01\rangle-|10\rangle)/\sqrt{2}$. The twin
particles could be polarized photons, which are then separated and
sent to Alice and Bob, each getting one half of each pair. The
received particles are measured by Alice and Bob by choosing a
random basis, out of three possible bases. These bases are chosen
in accordance to a Clauser, Horne, Shimony and Holt (CHSH)
test~\cite{CHSH}. Explicitly, the angles chosen by Alice are \beq
a_{1}=0,~~a_{2}=\pi/4,~~a_{3}=\pi/2, \eeq corresponding to the
bases $\mathbb{Z}$, $(\mathbb{X}+\mathbb{Z})/\sqrt{2}$ and
$\mathbb{X}$, respectively. Bob's on the other hand chooses \beq
b_{1}=\pi/4,~~b_{2}=\pi/2,~~b_{3}=3\pi/4, \eeq corresponding to
$(\mathbb{X}+\mathbb{Z})/\sqrt{2}$, $\mathbb{X}$ and
$(\mathbb{X}-\mathbb{Z})/\sqrt{2}$.

As in BB84, they would discuss in the clear which bases they used
for their measurements. Alice and Bob use the instances where they
chose different basis to check the presence of Eve. By disclosing
the data related to these instances they check the violation of
the CHSH quantity
\begin{align}
E= \langle a_{1}b_{1} \rangle - \langle a_{1}b_{3} \rangle +
\langle a_{3}b_{1} \rangle + \langle a_{3}b_{3} \rangle
\end{align}
where $\langle a_{i}b_{j} \rangle$ represents the expectation
value when Alice measures using $a_{i}$ and Bob, $b_{j}$. If the
inequality $-2\leq E\leq2$ holds, it would indicate either that
the received photons are not truly entangled (which could be due
to an attempt to eavesdrop) or that there is some problem with the
measurement device. By contrast, if everything works perfectly and
there is no eavesdropper, Alice and Bob expected value of $E$ is
the maximal violation $-2\sqrt{2}$. One way of looking at it is by
writing the state of entangled photons subjected to a depolarizing
channel, resulting into the isotropic mixed state
\begin{align}
\rho_{\Psi}=p|\Psi\rangle\langle\Psi|+(1-p)\mathbb{I}_{4}/4,
\end{align}
with probability $p$. It can be shown that the CHSH test has
maximal violation $-2\sqrt{2}$ provided that $p=1$, i.e., for an
unperturbed `Eve-less' channel.

In the case of maximal violation of the CHSH test, Alice and Bob
are sure that their data is totally decoupled from any potential
eavesdropper. From the instances where they chose the same bases,
they therefore process their perfectly anti-correlated results
into a shared private key. While QKD generally capitalizes on the
no-cloning theorem and the inability of perfectly distinguishing
between two non-orthogonal states, the essential feature of the
E91 protocol is its use of the nonlocal feature of entangled
states in quantum physics. Eve's intervention can be seen as
inducing elements of physical reality which affects the
non-locality of quantum mechanics.

\subsubsection{BBM92 protocol}

The BBM92 protocol~\cite{BBM92} was, in some sense, aimed as a
critic to E91's reliance on entanglement for security. Building
upon E91 with a source providing each legitimate party with halves
of entangled pairs, BBM92 works more efficiently by having both
the legitimate parties each measure in only two differing MUBs
instead of the three bases of E91. The two MUBs can be chosen to
be the same as that of BB84. By publicly declaring the bases,
Alice and Bob select the instances where they chose the same basis
to obtain correlated measurement results, from which a secret key
can be distilled. A sample is then disclosed publicly to check for
errors and evaluate the amount of eavesdropped information.

The idea is that Eve cannot become entangled to Alice's and Bob's
qubits while not causing any error in their measurements. This
points out to the claim that there is no need for the legitimate
parties to commit to a Bell test. The similarity between BBM92 and
BB84 is obvious. If Alice possesses the source, her measurement
(in a random basis) would prepare the state to be sent to Bob in
one of the 4 possible of the BB84 states. Hence, without a Bell
test, we are essentially left with BB84. There is no way of
telling whether Alice started off by measuring part of a Bell
state or by preparing a qubit state using a random number
generator. This observation is at the basis of the
entanglement-based representation of prepare and measure
protocols, which is a powerful theoretical tool in order to prove
the security of QKD protocols.

Using or not entangled pairs in a QKD protocol is
non-consequential in the context of standard eavesdropping on the
main communication channel. However, it is also important to note
that a protocol with a Bell test provides a higher level of
security because it allows to relax the assumption that the
legitimate parties have control over the other degrees of freedom
of the quantum signals. This makes way for the most pessimistic
security definition, i.e., device-independent security, a topic to
be delved into later. The security analysis of entanglement-based
QKD protocols is still the subject of very active research, with
recent investigations and simplified proofs based on entanglement
distillation protocols~\cite{Pirker2017,Pirker2019}.

\subsection{Two-way quantum communication}

Quantum cryptographic protocols making a bidirectional use of
quantum channels started with the introduction of deterministic
protocols for the purpose of secure direct
communication~\cite{felbinger,Qin,FG} and later evolved into more
mature schemes of two-way QKD~\cite{FGD,LM05}. A defining feature
of these protocols is that encodings are not based on preparing a
quantum state but rather applying a unitary transformation, by one
party (often Alice) on the traveling qubit sent by another party
(Bob) in a bidirectional communication channel. The initial idea
of direct communication aimed at allowing two parties to
communicate a message secretly, without the need of first
establishing a secret key. However the reality of noisy channels
would render any such direct communication between parties invalid
or very limited. For this reason, two-way protocols for direct
communication were soon replaced by QKD versions, with appropriate
security proofs~\cite{hua}.

\subsubsection{Ping pong protocol}

The ping pong direct communication protocol~\cite{felbinger}
derives its name from the to and from nature of the traveling
qubits between the communicating parties in the protocol. The
\textit{ping} comes from Bob
submitting to Alice half of a Bell pair he had prepared, $|\Psi_{+}%
\rangle=(|01\rangle+|10\rangle)/\sqrt{2}$, and the \textit{pong}
is Alice's submitting of the qubit back to Bob. With probability
$c$, Alice would measure the received qubit in the $\mathbb{Z}$
basis; otherwise, she would operate on it with either the identity
$\mathbb{I}$ with probability $p_{0}$ or the $\sigma_{z}$ Pauli
operator with probability $1-p_{0}$, re-sending the qubit back to
Bob. The former is the case where she could check for disturbance
in the channel and is referred to as the \textit{control mode}
(CM), while the latter is the essential encoding feature of the
protocol and referred to as the \textit{encoding mode} (EM).

The operations in EM flip between two orthogonal Bell states as
$\mathbb{I}$ retains $|\Psi_{+}\rangle$, while $\sigma_{z}$
provides
\begin{align}
\mathbb{I}\otimes\sigma_{z} |\Psi_{+}\rangle =|\Phi_{-}\rangle :=
(|00\rangle -|11\rangle)/\sqrt{2}.
\end{align}
This allows Bob to distinguish between them and infer Alice's
encoding perfectly. The details of the CM is as follows: Alice
measures the received qubit in the $\mathbb{Z}$ basis and
announces her result over a public channel. Bob then measures his
half of the (now disentangled) Bell pair and can determine if Eve
had interacted with the traveling qubit. It should be noted that,
in this protocol, Alice is not expected to resend anything to Bob
in CM. See Fig.~\ref{fig:pp} for a schematic representation.

\begin{figure}[h] \centering
\vspace{-1.3cm}
\includegraphics[width=0.9\columnwidth]{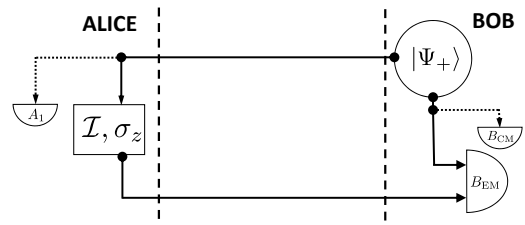}
\vspace{-1.3cm} \caption{A schematic of the ping pong protocol.
Part of a Bell pair $\Psi_{+}$ is sent by Bob to Alice, while the
other part is kept. If Alice chooses the EM (solid lines), she
performs either $\mathcal{I}$ or $\sigma_{z}$ on the received
qubit, which is then sent back to Bob. Finally, Bob performs a
Bell detection on the received and kept qubits. If Alice chooses
the CM (dotted lines), she measures the incoming qubit in the
$\mathbb{Z}$ basis ($A_{1}$), and informs Bob who also
measures its kept qubit in the same basis ($B_{cm}$).}
\label{fig:pp}
\end{figure}

By using the instances in CM, the parties may check the presence
or not of Eve. In particular, Eve's action goes undetected only
with an exponentially decreasing probability in the number of bits
gained. Therefore for long enough communication, its presence is
almost certainly discovered and the protocol aborted. If not
present, then Alice's message is privately delivered to Bob via
the EM instances with a sufficient degree of privacy. The message
that Alice transmits to Bob is not subject to any form of further
processing.

Note that direct private communication is very fragile and easily
fails in realistic conditions where noise on the line is
inevitable and, therefore, the presence of Eve must always be
assumed as worst-case scenario. Note that a similar severe
limitation also affects schemes of quantum direct communication in
continuous-variable systems~\cite{CVdirectCOMM1,CVdirectCOMM2}. In
particular, the ping pong protocol is also subject to a powerful
denial-of-service attack~\cite{QinC} which can be partially
mitigated if Alice returns the qubit to Bob in CM. Finally, note
that the protocol can be easily extended~\cite{fpp} to include all
the Pauli operators plus the identity, therefore doubling of the
communication capacity, resembling the superdense coding scenario.

\subsubsection{Two-way QKD protocols}

Two-way protocols for QKD do not need to use entanglement as in
the ping
pong protocol. According to Refs.~\cite{FGD, LM05}, Bob prepares a state $%
|a\rangle $ randomly selected from the two MUBs $\mathbb{X}$ and
$\mathbb{Z}$ to be sent to Alice. In EM, Alice encodes a bit using
either the identity (corresponding to bit value `0') and $i\sigma
_{y}$ (corresponding to bit value `1'), i.e., \beq
\mathbb{I}|a\rangle =|a\rangle ~,~i\sigma _{y}|a\rangle =|a^{\bot
}\rangle \eeq where $|a^{\bot }\rangle $ is the state orthogonal
to $|a\rangle $. The qubit is then sent back to Bob who measures
it in the same preparation basis. With some probability, Alice
chooses the CM where the incoming qubit is instead measured, and
another qubit is prepared and sent back to Bob for
his measurement. This `double check' was specifically introduced in Ref.~%
\cite{LM05} known as the LM05 protocol. This clearly increases the
detection performance of the protocol. For instance, given an
attack scenario where Eve measures the traveling qubits in either
of the two MUBs $\mathbb{Z}$ and $\mathbb{X}$, the probability of
detecting her is $37.5\%$.

Security proofs are based on the fact that Eve is forced to attack
both the forward and backward paths~\cite{qipj}. In general, from
the CM, Alice and Bob derive the amount of noise in the channels,
which determines how much PA has to be performed in the
post-processing. By disclosing part of the data in EM, they can
also estimate the amount of error correction to be performed.
Practical implementations of the protocol were already carried out
as early as 2006 in Ref.~\cite{Cere} as well as Refs.~\cite{Khir1,
Khir2,Rupesh}. We now discuss basic eavesdropping strategies.

\subsubsection{Intercept-resend strategy }

The simplest attack scheme is IR where Eve measures the traveling
qubit in both channels with a basis of her choice (randomly
selected between the same bases used by Bob). As she would
effectively prepare the traveling qubit into her basis of choice
by virtue of a projective measurement, she plays the role of Bob
and would be able to ascertain Alice's encoding perfectly. In
LM05, she would introduce errors $1/4$ of the time in each path.
This strategy leads to a security threshold of $11.9\%$, in terms
of maximal error (detected in CM) before no key is distillable.

It is worth noting that this attack results in an asymmetry
between Alice-Eve's and Bob-Eve's mutual information. While Eve
attempts to estimate Alice's encoding by inferring the evolution
of the state of the traveling qubit, her estimation of the result
of Bob's final measurement is another matter entirely. This leads
to the idea that Alice and Bob could actually consider doing a
reverse reconciliation (RR) procedure for distilling a key, where
Alice would correct her bits to guess Bob's string. In RR, the
security threshold is increased to $25\%$.

\subsubsection{Non-orthogonal attack strategies}

Here Eve would attach an ancilla to the traveling qubit in the
forward path and another in the backward path with the most
optimal possible interaction between them to glean the maximal
amount of information while minimizing the disturbance on the
channel. In this way, the security threshold for LM05 is about
$10\%$ in DR, while remaining $25\%$ in RR. A specific sub-optimal
version of this attack is the DCNOT attack strategy, where Eve's
ancilla is a qubit, used in the forward as well as the backward
path. The unitary transformation used by Eve in both paths would
be the same CNOT gate (hence the name \textit{double CNOT attack}
or DCNOT).

Let us write Alice's encoding as $U$ which acts on a qubit in the
computational basis as $U|i\rangle \rightarrow |i\oplus j\rangle $ where $%
\oplus $ is the addition modulo $2$ operation and $i,j=0,1$. The
action of the CNOT gates together with Alice's encoding $U_{A}$
can be written as follows: \beq \text{CNOT}(U_{A}\otimes
I)\text{CNOT}|i\rangle |0\rangle _{E} = |j\rangle |j\oplus
i\rangle _{E} \eeq where qubits with subscript $E$ refers to Eve's
ancillae. We see that Eve's qubit would record the evolution of
Bob's qubit. This is not at all
surprising as the CNOT gate allows for the perfect copying of states of the $%
\mathbb{Z}$ basis.

The case where Bob uses the $\mathbb{X}$ basis is no hindrance
either to Eve. Despite the fact that a CNOT between a qubit in the
$\mathbb{X}$ basis (as control qubit) and one in $\mathbb{Z}$ (for
target) would entangle the qubits, a subsequent CNOT would serve
to disentangle them.
\begin{align}
& \text{CNOT}(U_{A}\otimes I)\text{CNOT}~\dfrac{|0\rangle \pm |1\rangle }{%
\sqrt{2}}|0\rangle _{E}  \nonumber \\
& = U_{A}\left( \dfrac{|0\rangle \pm |1\rangle }{\sqrt{2}}\right)
\otimes |j\rangle _{E}  \label{dcnot2}
\end{align}%
The attack would leave no trace of an eavesdropper in EM while she
gains all the information. The attack is however very noisy and
easily detectable in CM with an error rate of $25\%$. If Eve
attacks a fraction $f$ of the runs, then her information gain is
$f$ with an error rate of $f/4$.

\subsubsection{Further considerations}

A general security proof for two-way DV-QKD was reported in Ref.
\cite{hua}
but methods employed led to an over-pessimistic estimation of the key rate ($%
1.7\%$ for LM05). On the other hand, the approach of
Ref.~\cite{norm} based on entropic bounds does not directly apply
to two-way QKD\ protocols based on unitary encodings. A tight
security proof is therefore still very much an open problem. A
number of eavesdropping strategies and technical issues have been
also described in Refs.~\cite{ipe,woj}, and the performance
against lossy channels have been thoroughly studied in
Refs.~\cite{OSID,jsb,dcnot}, where the key rate of the LM05 has
been compared with that of the BB84 at the same distances.

Two-way QKD protocols were also extended to considering
non-orthogonal
unitaries~\cite{dar2,jss,non,bisio}. For instance, the encoding unitaries $%
\mathbb{I}$ and the $(\mathbb{I}-i\sigma _{y})/\sqrt{2}$ were
considered
by Ref.~\cite{EPL}, while Ref.~\cite{jss} exploited the notion of \textit{%
mutually unbiased unitary-operator bases} (MUUB)~\cite{jrm}.
Another development has been the extension of the LM05 from two to
three MUBs (similar to the extension of BB84 to the six-state
protocol). The improvement in security provided by the protocol
known as 6DP~\cite{6DP} by making use of three MUBs instead of
only two is expected. However the extension to include the third
MUB is non-trivial given the no-go theorem which forbids the
flipping of an arbitrary state selected from 3 MUBs (see also
Ref.~\cite{blind}). This can be seen as follows: if we assume the
existence of a unitary transformation $U_{f}$ that flips between
the
orthogonal state of the $\mathbb{Z}$ basis, which can be written as $%
U_{f}|0\rangle = -|1\rangle $ and $U_{f}|1\rangle = |0\rangle $.
The negative phase factor in the first equation is necessary to
ensure $U_{f}$ also flips between the states in the $\mathbb{X}$
basis. However, $U_{f}$ would not flip between the states in the
$\mathbb{Y}$ basis,
\begin{align}
U_{f}(|0\rangle +i|1\rangle )/\sqrt{2}& = (-|1\rangle +i|0\rangle
)/\sqrt{2}  \nonumber \\
& :=(|0\rangle +i|1\rangle )/\sqrt{2}.
\end{align}


\section{Device-independent QKD}\label{sec:DIQKD}

\subsection{Introduction}\label{sec:di_intro} A security proof
for a QKD protocol is a mathematical theorem based on particular
assumptions.  These assumptions might encode that the devices work
in a particular way, e.g., that Alice generates a $\ket{0}$ state
and sends it to Bob, who measures in the $\{\ket{0},\ket{1}\}$
basis.  Although we have rigorous security proofs for QKD
protocols, finding devices satisfying the assumptions of these
proofs is difficult.  Any features of the real devices not modeled
in the security proof could compromise security, and there are
cases where this has happened in actual implementations
(e.g.~\cite{GLLSKM,XQL,LWWESM,WKRFNW}).  Attacks that exploit
features not modeled in the security proof are known as
\emph{side-channel attacks}.

Identified side-channel attacks can be patched sending the hacker
back to the drawing board. This leads to a technological arms race
between the hackers and protocol designers and a sequence of
(hopefully) increasingly secure protocols.  Device-independent
protocols provide a way to break out of this hack-and-patch cycle
with respect to side-channel attacks on the devices.  They are
able to do so because they make no assumptions about how the
devices used in the protocols operate in their security
proofs---instead, security follows from the classical input-output
behavior, which is tested in the protocol.  In this way, a device
independent protocol checks that the devices are functioning
sufficiently well \emph{during the protocol}.  This has a second
advantage: in standard QKD protocols with trusted devices, in
principle a user should check the functionality of their devices
regularly to ensure their behavior is still in line with the
assumptions of the security proof. This is a technically
challenging task and not one that can be expected of an average
user.  By contrast, in a device-independent protocol, no
sophisticated testing is needed to detect devices that are not
functioning sufficiently well (although, technical know-how is
needed to fix them).

At first it may seem intuitive that this is an impossible task:
how can we put any constraints on the workings of a device without
probing its internal behavior?  In particular, is it possible to
test the input-output behavior and ensure that the outputs of a
device could not have been pre-determined by its manufacturer?  In
fact, the intuition that this is impossible is correct if there is
only one device.  However, with two or more devices, this can be
done, thanks to Bell's theorem.  The basic idea is that if two
devices are unable to communicate, are given random inputs and
their input-output behavior gives rise to a distribution that
violates a Bell inequality, then their outputs could not have been
pre-determined and hence are a suitable starting point to generate
a key.  Because this idea is central to device-independence we
will elaborate on it first before discussing DI-QKD protocols.

\subsection{The link between Bell violation and
unpredictability}\label{sec:Bell_unpred} Consider two parties,
Alice and Bob, each of whom have a device. Alice and Bob are each
able to make one of two inputs to their device and obtain one of
two outputs.  Quantum mechanically, these devices may be set up to
measure halves of a pair of entangled qubits, with the inputs
corresponding to the choice of basis. Crucially, although this may
be what honest parties should do to set up their devices, for the
security argument, no details of the setup are required.  In order
to describe the behavior of such devices we will use the following
notation.  Alice's input is modeled by a binary random variable
$A$ and Bob's by $B$ and their respective outputs are binary
random variables $X$ and $Y$. It is convenient to use the
following tables to represent the conditional distribution
$P_{XY|AB}$ as a $4\times 4$ matrix:
\bigskip

\begin{tabular}{ccc|cc|cc|}
$P_{XY|AB}$&&$B$&\multicolumn{2}{|c|}{0}&\multicolumn{2}{|c|}{1}\\
&&$Y$&0&1&0&1\\
$A$&$X$&&&&&\\
\hline
\multirow{2}{*}{0}&0&&$P_{00|00}$&$P_{01|00}$&$P_{00|01}$&$P_{01|01}$\\
&1&&$P_{10|00}$&$P_{11|00}$&$P_{10|01}$&$P_{10|01}$\\
\hline
\multirow{2}{*}{1}&0&&$P_{00|10}$&$P_{01|10}$&$P_{00|11}$&$P_{01|11}$\\
&1&&$P_{10|10}$&$P_{11|10}$&$P_{10|11}$&$P_{11|11}$\\
\hline
\end{tabular}
\bigskip

Suppose now that Alice and Bob's devices behave according to a
particular distribution $P_{XY|AB}$ and imagine an eavesdropper
holding some additional information about the devices and for ease
of this exposition, let us assume that this information is
classical and use the random variable $Z$ to describe it. This
classical information tells Eve additional information about what
is happening.  One can think of this in the following way: Eve
supplies devices that behave according to $P_{XY|AB}^z$, but picks
$z$ with probability $p_z$ such that from Alice and Bob's point of
view the device behavior is the same, i.e.,
\begin{equation}\label{eq:decomp}
P_{XY|AB}=\sum_zp_zP_{XY|AB}^z\,.
\end{equation}

If the devices are used in such a way that each device cannot
access the input of the other then they must act in a local manner
($P_{X|AB}^z=P^z_{X|A}$ and $P_{Y|AB}^z=P^z_{Y|B}$).  The question
of interest is then whether Eve could have supplied deterministic
devices giving rise to the observed distribution.  This can be
stated mathematically as the question whether $P_{XY|AB}$ can be
written in the form~\eqref{eq:decomp} with
$P_{XY|AB}^z=P^z_{X|A}P^z_{Y|B}$ and
$P^z_{X|A=a}(x),P^z_{Y|B=b}(y)\in\{0,1\}$ for all
$x,y,a,b\in\{0,1\}$. In other words, is $P_{XY|AB}$ a convex
combination of the 16 local deterministic distributions
$$\left(\begin{array}{cc|cc}1&0&1&0\\0&0&0&0\\\hline
          1&0&1&0\\0&0&0&0\end{array}\right),\ \left(\begin{array}{cc|cc}1&0&0&1\\0&0&0&0\\\hline
          1&0&0&1\\0&0&0&0\end{array}\right),\ldots,\left(\begin{array}{cc|cc}0&0&0&0\\0&1&0&1\\\hline
          0&0&0&0\\0&1&0&1\end{array}\right)\,?$$
If not, then at least some of the time Eve must be sending a
distribution $P^z_{XY|AB}$ to which she doesn't know either
Alice's or Bob's outcome after later learning their inputs.

A Bell inequality is a relation satisfied by all local
correlations (i.e., all $P_{XY|AB}$ that can be written as a
convex combination of local deterministic distributions).  The
CHSH inequality can be expressed in this notation as $\langle
C,P\rangle\leq 2$, where $P=P^z_{XY|AB}$,
$$C=\left(\begin{array}{cc|cc}1&-1&1&-1\\-1&1&-1&1\\\hline
            1&-1&-1&1\\-1&1&1&-1\end{array}\right)$$
and $\langle C,P\rangle=\tr(C^TP)$ is the Hilbert-Schmidt inner
product. Bell's theorem states that there are quantum correlations
that violate this inequality. To describe these we introduce a
class of distributions parameterized in terms of $\eps\in[0,1/2]$
as follows
\begin{equation}\label{eq:Peps}
P_\eps:=\left(\begin{array}{cc|cc}\frac{1}{2}-\eps&\eps&\frac{1}{2}-\eps&\eps\\
            \eps&\frac{1}{2}-\eps&\eps&\frac{1}{2}-\eps\\
            \hline
            \frac{1}{2}-\eps&\eps&\eps&\frac{1}{2}-\eps\\
            \eps&\frac{1}{2}-\eps&\frac{1}{2}-\eps&\eps\end{array}\right).
\end{equation}

Define the state
$\ket{\psi_\theta}:=\cos\frac{\theta}{2}\ket{0}+\sin\frac{\theta}{2}\ket{1}$.
Then assume that Alice and Bob measure the two halves of the
maximally-entangled state $\frac{1}{\sqrt{2}}(\ket{00}+\ket{11})$
in the following bases:
\begin{eqnarray}
\{\ket{\psi_0},\ket{\psi_\pi}\}&~~\mathrm{for}\ A=0,\nonumber\\
\{\ket{\psi_{\pi/2}},\ket{\psi_{3\pi/2}}\}&~~\mathrm{for}\ A=1,\nonumber\\
\{\ket{\psi_{\pi/4}},\ket{\psi_{5\pi/4}}&~~\mathrm{for}\ B=0,\nonumber\\
\{\ket{\psi_{3\pi/4}},\ket{\psi_{7\pi/4}}\}&~~\mathrm{for}\
B=1.\label{eq:mmts}
\end{eqnarray}
This gives rise to a distribution of the form $P_\eps$ as in
Eq.~\eqref{eq:Peps} where \beq
\eps=\frac{1}{2}\sin^2\frac{\pi}{8}=\frac{1}{8}(2-\sqrt{2})=:\eps_{\text{QM}},
\eeq which leads to $\displaystyle\langle
C,P_{\eps_{\text{QM}}}\rangle=2\sqrt{2}$, i.e., the maximal
violation of the CHSH inequality. Recall that the Tsirelson's
bound~\cite{Cirelson} states that if $P$ is quantum-correlated
then $\langle C,P\rangle\leq2\sqrt{2}$.

One way to think about how random the outcomes are is to try to
decompose this distribution in such a way as to maximize the local
part. For $0\leq\eps\leq 1/8$, this is achieved using the
following decomposition whose optimality can be verified using a
linear program
\begin{align}
  P_\eps:=&\eps\left[
      \left(\begin{array}{cc|cc}1&0&1&0\\0&0&0&0\\\hline1&0&1&0\\0&0&0&0\end{array}\right)
     +\left(\begin{array}{cc|cc}1&0&1&0\\0&0&0&0\\\hline0&0&0&0\\1&0&1&0\end{array}\right)
     +\left(\begin{array}{cc|cc}0&1&1&0\\0&0&0&0\\\hline0&0&0&0\\0&1&1&0\end{array}\right)\right. \nonumber \\&\left.
     +\left(\begin{array}{cc|cc}1&0&0&1\\0&0&0&0\\\hline1&0&0&1\\0&0&0&0\end{array}\right)
     +\left(\begin{array}{cc|cc}0&0&0&0\\0&1&1&0\\\hline0&0&0&0\\0&1&1&0\end{array}\right)
     +\left(\begin{array}{cc|cc}0&0&0&0\\1&0&0&1\\\hline1&0&0&1\\0&0&0&0\end{array}\right)\right. \nonumber
  \\&\left.
      +\left(\begin{array}{cc|cc}0&0&0&0\\0&1&0&1\\\hline0&1&0&1\\0&0&0&0\end{array}\right)
      +\left(\begin{array}{cc|cc}0&0&0&0\\0&1&0&1\\\hline0&0&0&0\\0&1&0&1\end{array}\right)\right] \nonumber \\
  &+(1-8\eps)\left(\begin{array}{cc|cc}\frac{1}{2}&0&\frac{1}{2}&0\\0&\frac{1}{2}&0&\frac{1}{2} \\\hline
          \frac{1}{2}&0&0&\frac{1}{2}\\0&\frac{1}{2}&\frac{1}{2}&0\end{array}\right).
\end{align}
If Eve used this decomposition she would be able to guess Alice's
outcome with probability
$8\eps+\frac{1}{2}(1-8\eps)=\frac{1}{2}+4\eps$. Thus, Alice's
outcome would have some randomness with respect to Eve.


We note however that while the first eight terms in this
decomposition are local, the last is a maximally non-local
distribution~\cite{KT,Cirelson93}, often called a
PR-box~\cite{PR}. This is well-known not to be realizable in
quantum theory.  The stated strategy is hence not available to an
eavesdropper limited by quantum mechanics.  To analyze the case of
a quantum-limited eavesdropper, we also have to ensure that
$P^z_{XY|AB}$ is quantum-realizable for all $z$. It is not easy to
do this in general, but in the case where $A$, $B$, $X$ and $Y$
are binary it can be shown that it is sufficient to consider
qubits~\cite{Cirelson93}. For other cases, there is a series of
increasingly tight outer approximations to the quantum set that
can be tested for using semidefinite programs~\cite{NPA}.

Considering a quantum-limited eavesdropper reduces Eve's power and
hence leads to more randomness in the outcomes.  For a
distribution of the form $P_\eps$ for
$\eps_{\text{QM}}\leq\eps\leq1/8$, for instance, Eve can do a
quantum decomposition as follows:
\begin{align}
  P_\eps:=&\frac{\eps-\eps_{\text{QM}}}{1-8\eps_{\text{QM}}}\left[
      \left(\begin{array}{cc|cc}1&0&1&0\\0&0&0&0\\\hline1&0&1&0\\0&0&0&0\end{array}\right)
     +\left(\begin{array}{cc|cc}1&0&1&0\\0&0&0&0\\\hline0&0&0&0\\1&0&1&0\end{array}\right)\right.\nonumber \\&\left.
     +\left(\begin{array}{cc|cc}0&1&1&0\\0&0&0&0\\\hline0&0&0&0\\0&1&1&0\end{array}\right)
     +\left(\begin{array}{cc|cc}1&0&0&1\\0&0&0&0\\\hline1&0&0&1\\0&0&0&0\end{array}\right)
     +\left(\begin{array}{cc|cc}0&0&0&0\\0&1&1&0\\\hline0&0&0&0\\0&1&1&0\end{array}\right)\right. \nonumber \\&\left.
     +\left(\begin{array}{cc|cc}0&0&0&0\\1&0&0&1\\\hline1&0&0&1\\0&0&0&0\end{array}\right)
      +\left(\begin{array}{cc|cc}0&0&0&0\\0&1&0&1\\\hline0&1&0&1\\0&0&0&0\end{array}\right)
      +\left(\begin{array}{cc|cc}0&0&0&0\\0&1&0&1\\\hline0&0&0&0\\0&1&0&1\end{array}\right)\right] \nonumber \\&+\frac{1-8\eps}{1-8\eps_{\text{QM}}}P_{\eps_{\text{QM}}}\,,
\end{align}
allowing her to predict the outcome correctly with probability 
$8\frac{\eps-\eps_{\text{QM}}}{1-8\eps_{\text{QM}}}+\frac{1-8\eps}{2(1-8\eps_{\text{QM}})}$.

The argument just given is intended to give an intuition to the
idea of why violating a Bell inequality means that there is some
randomness in the outcomes. However, knowing that there is some
randomness is not enough; we also need to know how much key can be
extracted from the raw data.

\subsection{Quantitative bounds}\label{sec:quant} Given a pair of
uncharacterized devices we would like to know how much secure key
we can extract from their outputs. Because the devices are
uncharacterized, we need to test their behavior. Such a test
involves repeatedly making random inputs to the devices and
checking some function of the chosen inputs and the device
outputs.  For convenience, in this section we will mostly consider
the average CHSH value.  Conditioned on this test passing, the
protocol will go on to extract key.

We would like a statement that says that for any strategy of Eve
the probability that both the average CHSH value is high and the
key extraction fails is very small.  For this to be the case we
need to connect the CHSH value with the amount of extractable key.
Since key is shared randomness, before considering sharing we can
ask how much randomness can Alice extract from her outcomes for a
given CHSH value. For a cq-state (i.e., a state of the form
$\rho_{AE}=\sum_{\bf x}P_{\bf X}({\bf x})\proj{{\bf
x}}\ot\rho_E^{{\bf
    x}}$,
where ${\bf X}$ denotes a string of many values), this can be
quantified by the (smooth) min-entropy~\cite{RennerPhD}
$S_{\mathrm{min}}(\textbf{X}|E)$ of Alice's string ${\bf X}$
conditioned on $E$.

This is a difficult quantity to evaluate, in part because of the
lack of structure. In fact, Eve's behavior need not be identical
on every round and she need not make measurements round by round,
but can keep her information quantum. However, a simpler
round-by-round analysis in which the conditional von Neumann
entropy is evaluated can be elevated to give bounds against the
most general adversaries via the entropy accumulation theorem
(EAT)~\cite{DFR,ARV}. The basic idea is that, provided the
protocol proceeds in a sequential way, then the total min-entropy
of the complete output of $n$ rounds conditioned on $E$ is (up to
correction factors of order $\sqrt{n}$) at least $n$ times the
conditional von Neumann entropy of one round evaluated over the
average CHSH value.

The evaluation of the conditional von Neumann entropy as a
function of the CHSH value was done in~\cite{ABGMPS}. There it was
shown that for any density operator $\rho_{ABE}$, if the observed
distribution $P_{XY|AB}$ has CHSH value $\langle
C,P\rangle=\beta\in[2,2\sqrt{2}]$, then the conditional von
Neumann entropy satisfies the bound
\begin{equation}\label{eq:chsh_ent}
S(X|E)\geq
1-H_2\left[\frac{1}{2}\left(1+\sqrt{(\beta/2)^2-1}\right)\right]\,,
\end{equation}
where $H_2(...)$ is the binary Shannon entropy. Combining this
with the EAT, we obtain a quantitative bound on the amount of
uniform randomness that can be extracted from Alice's outcomes of
roughly $n$ times this.

The bound~\eqref{eq:chsh_ent} is obtained by using various
technical tricks specific to the CHSH scenario.  For general
non-local games/device measurements we do not know of good ways to
obtain tight bounds on the conditional von Neumann entropy.
Instead, a typical way to obtain a bound is to note that
$S(X|E)\geq S_{\min}(X|E)$, and that $S_{\min}(X|E)$ can be
bounded via a hierarchy of semi-definite programs~\cite{NPA,NPA2},
as discussed in~\cite{HR,BRC}.  However, the bounds obtained in
this way are fairly loose and it is an open problem to find good
ways to improve them.

\subsection{Protocols for DI-QKD}
\subsubsection{The setup for DI-QKD}\label{sec:DIQKDsetup}
As mentioned in Section~\ref{sec:di_intro}, use of
device-independence eliminate security flaws due to inadequate
modeling of devices. There are  nevertheless, a number of other
assumptions we make in this scenario (note that these assumptions
are also made in the trusted-devices case):
\begin{enumerate}
\item\label{ass:1} Alice and Bob have secure laboratories and control over all
  channels connecting their laboratory with the outside
  world. (Without this assumption, the untrusted devices could
    simply broadcast their outputs to the adversary outside the
    laboratory, or Eve could send a probe into the laboratory to
    inspect any secret data.) For any devices in their labs, Alice and
  Bob can prevent unwanted information flow between it and any other
  devices.
\item Each party has a reliable way to perform classical information
  processing.
\item Alice and Bob can generate perfectly random (and private) bits
  within their own laboratories.
\item Alice and Bob are connected by an authenticated classical
  channel on which an adversary could listen without detection.
\item Alice and Bob are also connected by an insecure quantum
  channel on which an adversary can intercept and modify signals in
  any way allowed by quantum mechanics.
\end{enumerate}

Security is proven in a composable way (cf.\ Section
~\ref{sec:compos}) allowing a key output by the protocol to be
used in an arbitrary application.  Note that because the protocol
is device-independent, the prolonged security of any output relies
on the devices not being reused~\cite{bckone} in subsequent
protocols (note that the same devices can be used many times
within a run of the protocol), although modified protocols to
mitigate this problem have been proposed~\cite{bckone}.

\subsubsection{The spot-checking CHSH QKD
protocol}\label{sec:spotCHSHQKD} A protocol acts as a filter.  It
is a procedure that can be fed by a set of devices such that bad
devices lead to an ``abort'' with high probability, and good
devices lead to success with high probability.  There are many
possible types of protocol; we will describe a specific protocol
here, based on the CHSH game with spot-checking.

The protocol has parameters $\alpha\in(0,1)$, $n\in\mathbb{N}$,
$\beta\in(2,2\sqrt{2}]$, $\delta\in(0,2(\sqrt{2}-1))$, which are
to be chosen by the users before it commences.
\begin{enumerate}
\item \label{st:1} Alice uses a preparation device to generate an
  entangled pair.  She keeps one half and sends the other to
  Bob. This step and the subsequent one refer to
    the generation, sending and storage of an entangled state, but for
    security Alice and Bob do not rely on this taking place correctly
    (if the state created is not of high enough quality the protocol
    should abort).
\item Bob stores it and reports its receipt to Alice.
\item Alice picks a random bit $T_i$, where $T_i=0$ with probability
  $1-\alpha$ and $T_i=1$ with probability $\alpha$. She sends $T_i$ to
  Bob over the authenticated classical channel.
\item \label{st:4} If $T_i=0$ (corresponding to no test) then Alice and Bob each
  make some fixed inputs (choices of bases) into their devices, $A_i=0$ and $B_i=2$ and
  record the outcomes, $X_i$ and $Y_i$.\\
  If $T_i=1$ (corresponding to a test) then Alice and Bob each
  independently pick uniformly random inputs $A_i\in\{0,1\}$ and
  $B_i\in\{0,1\}$ to their devices and record the outcomes, $X_i$ and
  $Y_i$.
\item Steps \ref{st:1}--\ref{st:4} are repeated $n$ times, increasing
  $i$ each time.
\item For all the rounds with $T_i=1$, Bob sends his inputs and
  outputs to Alice who computes the average CHSH value (assigning $+1$
  or $-1$ in accordance with the entries of matrix $C$). If this
  value is below $\beta-\delta$, Alice announces that the protocol aborts.
\item If the protocol does not abort, Alice and Bob use the rounds
  with $T_i=0$ to generate a key using error correction and privacy
  amplification over the authenticated classical channel. The EAT tells them how much key can be extracted, subject
  to adjustments for the communicated error correction information.
\end{enumerate}

To explain the structure of the protocol it is helpful to think
about an ideal implementation.  In this, the preparation device
generates a maximally entangled state
$\frac{1}{\sqrt{2}}(\ket{00}+\ket{11})$ and for $A,B\in\{0,1\}$
the measurements are as described in~\eqref{eq:mmts}. Furthermore,
for $B=2$, the measurement is in the
$\{\ket{\psi_0},\ket{\psi_\pi}\}$ basis, i.e., the same basis as
for $A=0$.  If $\alpha$ is chosen to be small, on most of the
rounds both parties measure in the $\{\ket{0},\ket{1}\}$ basis
which should give perfectly correlated outcomes, suitable for key.
However, on some of the rounds (those with $T_i=1$), a CHSH test
is performed, in order to keep the devices honest.  These are the
spot-checks that give the protocol its name.  The parameter
$\beta$ is the expected CHSH value of the setup ($\beta=2\sqrt{2}$
in the ideal implementation) and $\delta$ is some tolerance to
statistical fluctuations.

The probability that an ideal implementation with no eavesdropping
leads to an abort is called the \emph{completeness error}.  Using
the implementation given above, this occurs when statistical
fluctuations cause devices with an expected CHSH value of $\beta$
to produce a value below $\beta-\delta$.  An ideal implementation
behaves in an i.i.d.\ way and hence standard statistical bounds
imply that the completeness error is exponentially small in the
number of rounds.

It is worth making some remarks about the protocol.
\begin{enumerate}
\item It is important that the preparation device is unable to access
  information from Alice's measurement device, even though these may
  be in the same lab (if access were granted, the preparation device
  could send previous measurement results to Eve via the quantum
  channel).
\item The choice $T_i$ needs to be communicated after the state is
  shared (otherwise Eve can choose whether to intercept and modify the
  quantum state depending on whether or not a test will be
  performed). This requires Alice and Bob to have a (short-lived)
  quantum memory; without such a memory, Alice and Bob could instead
  use some pre-shared randomness to make these choices and then
  consider the modified protocol to be one for \emph{key
    expansion}. For reasonable parameter ranges, this would still lead
  to expansion, because $\alpha$ can be low and so a small amount of
  pre-shared key is needed to jointly choose the values of $\{T_i\}$.
\item Bob's device can tell when it is being used to generate key
  ($B_i=2$).  Crucially though, Alice's device cannot (Alice's device
  learns only $A_i$ and not the value of $T_i$), and it is this that
  forces her device to behave honestly; not doing so will lead to her
  getting caught out if the round is a test.  If Bob's device does not
  behave close enough to the way it should in the case $B_i=2$, then
  the protocol will abort during error correction step.
\end{enumerate}

There are many other possible protocols, but they follow the same
basic idea of generating shared randomness while occasionally
doing tests based on some non-local game, estimate the amount of
min-entropy that any devices that pass the tests with high
probability must give and then using classical protocols to
eliminate errors and remove any information Eve may have through
privacy amplification.

\subsection{Historical remarks} Using violation of a Bell
inequality as part of a key distribution protocol goes back to the
Ekert protocol~\cite{Eke91}, and many device-independent protocols
can be seen as a development of this. However, Ekert's work didn't
envisage foregoing trust on the devices, and the idea behind this
came many years later under the name of
self-checking~\cite{MayersYao}.  The first protocol with a full
security proof was that of Barrett, Hardy and Kent~\cite{BHK}, and
their protocol is even secure against eavesdroppers not limited by
quantum theory, but by some hypothetical post-quantum theory,
provided it is no-signalling. However, it has the drawback of a
negligible key rate and the impracticality of needing as many
devices as candidate entangled pairs to ensure all of the required
no-signalling conditions are met.  Following this were several
works that developed protocols with reasonable key rates, proving
security against restricted
attacks~\cite{AGM,ABGMPS,PABGMS,McKague} with as many devices as
candidate entangled states~\cite{HR,HRW,MPA11,MRC}.  Later proofs
avoided such restrictions~\cite{bcktwo,RUV,VV2,MS1}, but still
were not able to tolerate reasonable levels of either noise or had
poor rates (or both). Using the EAT~\cite{DFR} leads to a
reasonable rate and noise tolerance~\cite{ARV}, and better rates
still can be derived from recent strengthened versions of the
EAT~\cite{dupuis18}.

\smallskip

\subsection{Putting DI-QKD protocols into
practice} Although device-independence in principle allows for
stronger security, adopting it in practice is more challenging
than ordinary QKD.  This is because it is difficult to generate
correlations that violate a Bell inequality at large separations.
Using photons is a natural way to quickly distribute entanglement.
However, detecting single photons is difficult.  In a
device-dependent QKD protocol such as BB84, failed detection
events slow down the generation of key, but it is possible to
post-select on detection; in a device-independent protocol, below
a certain detection threshold, no key can be securely generated.
This is because post-selecting on detection events leads to the
possibility that the post-selected events appear to be non-local
when they are in fact not.  To treat this problem, suppose that
each detector detects a photon with probability $\eta\in[0,1]$.  A
distribution of the form $P_\eps$ from Eq.~\eqref{eq:Peps} will
become
\begin{widetext}
\begin{equation}\label{eq:Peps_eta}
P_{\eps,\eta}:=\left(\begin{array}{ccc|ccc}\eta^2(\frac{1}{2}-\eps)&\eta^2\eps&\frac{\eta(1-\eta)}{2}&\eta^2(\frac{1}{2}-\eps)&\eta^2\eps&\frac{\eta(1-\eta)}{2}\\
            \eta^2\eps&\eta^2(\frac{1}{2}-\eps)&\frac{\eta(1-\eta)}{2}&\eta^2\eps&\eta^2(\frac{1}{2}-\eps)&\frac{\eta(1-\eta)}{2}\\\frac{\eta(1-\eta)}{2}&\frac{\eta(1-\eta)}{2}&(1-\eta)^2&\frac{\eta(1-\eta)}{2}&\frac{\eta(1-\eta)}{2}&(1-\eta)^2\\
            \hline
            \eta^2(\frac{1}{2}-\eps)&\eta^2\eps&\frac{\eta(1-\eta)}{2}&\eta^2\eps&\eta^2(\frac{1}{2}-\eps)&\frac{\eta(1-\eta)}{2}\\
            \eta^2\eps&\eta^2(\frac{1}{2}-\eps)&\frac{\eta(1-\eta)}{2}&\eta^2(\frac{1}{2}-\eps)&\eta^2\eps&\frac{\eta(1-\eta)}{2}\\\frac{\eta(1-\eta)}{2}&\frac{\eta(1-\eta)}{2}&(1-\eta)^2&\frac{\eta(1-\eta)}{2}&\frac{\eta(1-\eta)}{2}&(1-\eta)^2\end{array}\right)\,,
\end{equation}
\end{widetext}
where the third outcome corresponds to a no-detection event.
Post-selecting on both detectors clicking recovers the
distribution $P_\eps$, but it can be the case that $P_\eps$ is not
a convex combination of local deterministic distributions, but
that $P_{\eps,\eta}$ is.  To avoid this, the experimental
conditions need to be such that the distribution \emph{including
no-click events} has no deterministic decomposition.  In the
terminology of Bell experiments, this is referred to as closing
the \emph{detection
  loophole}.  
For the distribution $P_{\eps,\eta}$ given above, this loophole is
closed provided $\eta>2/(3-8\eps)$.  Note that for $\eta\leq2/3$
this cannot be satisfied for any $\eps$.  Hence, for protocols
based on CHSH, $2/3$ is a lower bound on the detection efficiency
required. This is known as Eberhard's bound~\cite{Eberhard}.

Another loophole that is of interest for Bell experiments is the
\emph{locality loophole}, which is closed by doing measurements at
space-like separation. The desire to close this loophole comes
from a concern that the devices are able to talk to each other
during the measurements, and, in particular, that one device is
able to learn the measurement choice of the other, which makes it
trivial to violate a Bell inequality in a classical deterministic
way.  It was a longstanding technical problem to simultaneously
close the locality and detection loopholes, a feat that was only
recently achieved~\cite{Hensen,Giustina,Shalm}.  In the context of
DI-QKD, however, it is not necessary to close the locality
loophole (although it does not hurt).  The reason is that for QKD
it is necessary that Alice's and Bob's lab are secure
(Assumption~\ref{ass:1} above).  If their devices could
communicate with each other during the measurements then this
assumption is broken, and it makes little sense to allow
communication between devices without allowing it from the devices
to Eve.


\subsection{Measurement device independence (MDI)}\label{MDISection}

In DI-QKD one avoids the formulation of a mathematical model
describing the devices involved in the experiment and aims at
proving the security of the communication protocol only from the
collected data. This is possible because only a purely quantum
experiment can provide data that violate Bell inequalities. This
approach is conceptually powerful but limited in terms of
attainable key rates. Here we review the main ideas of measurement
device independent (MDI) QKD~\cite{BP,Lo}. This is a framework in
which no assumptions is made on the detectors involved in the QKD
protocols, which can be operated by a malicious eavesdropper.
In a typical MDI-QKD protocol, both trusted users Alice and Bob
send quantum signals to a central receiver (also called relay).
The assumptions are that Alice and Bob have perfect control on the
quantum state they prepare and send through the quantum channels.
On the other hand, no assumption is made on the central relay,
which can be under control of Eve. In this way one does not need
to bother about the trustfulness of any detector or in general of
any measurement device. Although at first sight it may seem
impossible to extract any secrecy at all from such a scheme, it is
indeed possible to exploit this MDI scheme to generate secret key
at a nonzero rate.

In a simple (idealized) scheme of MDI-QKD, Alice and Bob locally
prepare single-photon states with either rectilinear $\{
|H\rangle, |V\rangle\}$ or diagonal $\{ |D\rangle, |A\rangle\}$
polarization. These states are sent to a central relay that is
assumed under control of Eve. Notice that initially the states
sent to Eve are statistically independent. Any possible physical
transformation may affect the signals traveling through the
quantum channels that connect Alice and Bob to the central relay.
Also, Eve can apply any measurement on the received signals, or
she can store them in a long term quantum memory. However, to
explain the working principle of MDI-QKD let us assume for a
moment that the channels from the trusted users to Eve are
noiseless, and that Eve performs a Bell detection on the incoming
signals. These assumptions will be relaxed later. Moreover, we
require that Eve publicly announces the outcome $\alpha= 0,1,2,3$
of the Bell detection.

The ideal Bell detection is a measurement with four POVM elements,
$\Lambda _{\alpha}:= \sigma_{\alpha}|\beta\rangle\langle\beta|
\sigma_{\alpha}$, where $|\beta\rangle= 2^{-1/2}\left( |HH\rangle+
|VV\rangle\right)  $ is a maximally entangled state, and
$\sigma_{\alpha}$ are the Pauli operators (including the
identity), $\sigma_{0} = |H\rangle\langle H| + |V\rangle\langle
V|$, $\sigma_{1} = |H\rangle\langle V| + |V\rangle\langle H|$,
$\sigma_{2} = i |H\rangle\langle V| - i |V\rangle\langle H|$,
$\sigma_{3} = |H\rangle\langle H| - |V\rangle\langle V|$. Note
that, if both Alice and Bob encode information in the rectilinear
basis, then they know that their encoded bit values are the same
if the outcome is $\alpha=0$ or $\alpha= 3$, otherwise they know
that they are opposite if $\alpha=1$ or $\alpha= 2$. Therefore,
Bob can obtain Alice's bit by flipping (or not flipping) his local
bit according to the value of $\alpha$. Similar is the situation
if Alice and Bob use the diagonal basis, as depicted in Table
\ref{TableMDI}. If the parties choose different bases, they simply
discard their data.

The above example shows that the Bell detection performed by the
relay can induce (or post-select) strong correlations between the
bits locally prepared by the trusted users, after they sift their
data according to the choice of local polarization basis. In other
words, ideal Bell detection simulates a virtual noiseless
communication channel connecting the two honest users. Notice that
the output of the Bell detection contains information about the
identity (or non-identity) of the pair of bit values encoded by
Alice and Bob (after sifting) but does not contain any information
about the actual bit values.

\begin{center}
\begin{table}[ptb]%
\begin{tabular}
[c]{|c||c|c|} & $\{ |H\rangle,|V\rangle\}$ & $\{
|D\rangle,|A\rangle\}$\\\hline\hline $\alpha=0$ & $-$ &
$-$\\\hline $\alpha=1$ & bit flip & $-$\\\hline $\alpha=2$ & bit
flip & bit flip\\\hline $\alpha=3$ & $-$ & bit flip
\end{tabular}
\caption{The table shows the rules for bit-flipping according to
the result
$\alpha=0,1,2,3$ of Bell detection and the sifted basis choice.}%
\label{TableMDI}%
\end{table}
\end{center}

While a noiseless communication channel and an ideal Bell
detection do not introduce any error, noise in the communication
channels and non-ideal Bell measurement introduce an error rate in
this virtual communication channel connecting Alice and Bob.
Standard parameter estimation procedures can then be applied to
estimate the QBER and then provide a lower bound on the secret key
rate. As a matter of fact Eve can apply any quantum operation at
the relay, however she is expected to declare one value of
$\alpha=0,1,2,3$ for any pair of signals received, otherwise the
trusted users will abort the protocol.

To move from this abstract mathematical model towards experimental
implementations, one shall replace single-photon states with
phase-randomized attenuated coherent states. Moreover, with a
linear optics implementation one can only realize two of the four
POVM elements of Bell detection, this does not affect the working
principle of MDI-QKD and has only the effect of introducing a
non-deterministic element that reduces the secret key rate. In
this context, a practical design for DV MDI-QKD has been proposed
in Ref.~\cite{Lo}, where it has been shown that it can be
implemented using decoy states along the same methodology of the
BB84 protocol. We remind that decoy states are crucial for DV QKD
to overcome photon number splitting attacks that the eavesdropper
can implement in realistic cases, where the signals emitter does
not generate truly single photon states. In DV MDI-QKD, both Alice
and Bob emits pulses randomly changing the intensity and revealing
it, publicly, only after the quantum communication has been
concluded. This avoid that Eve may adapt her attacks.

In the protocol described in Ref.~\cite{Lo}, the parties generate
weak coherent pulses passing through two distinct polarization
modulators, which operate randomly and independently. After this
step, the signals are sent through two intensity modulators, which
generate the decoy states. The protocol proceeds with the Bell
measurement realized by the relay. The signals are mixed in a
$50:50$ beam splitter, and the outputs processed by two polarizing
beam splitters (PBS), filtering the input photons into states
$|H\rangle$ or $|V\rangle$, and finally detected by two pairs of
single-photon detectors. The Bell measurement is successful when
two of the four detectors click.

Assuming that the rectilinear basis is used to generate the key
the asymptotic
key rate is given by the following expression~\cite{Lo}%
\begin{align}
&R_{\mathrm{decoy-MDI}}=P_{\mathrm{rect}}^{11}Y_{\mathrm{rect}}^{11}-P_{\mathrm{rect}}
^{11}Y_{\mathrm{rect}}^{11}H_{2}(e_{\mathrm{diag}}^{11}) \nonumber \\
&-G_{\mathrm{rect}}~\delta(Q_{\mathrm{rect}}),\label{RATE-DV-MDI}%
\end{align}
where
$P_{\mathrm{rect}}^{11}=\mu_{A}\mu_{B}\exp[-(\mu_{A}+\mu_{B})]$ is
the joint probability that both emitters generate single-photon
pulses, with $\mu_{A}$ and $\mu_{B}$ describing the intensities
(or the mean photon number) of the photon sources of Alice and
Bob, respectively. The quantity $Y_{\mathrm{rect}}^{11}$ gives the
gain, while $e_{\mathrm{diag}}^{11}$ is the QBER when Alice and
Bob correctly send single-photon pulses. The function $H_{2}(x)$
is the binary Shannon entropy. The gain $G_{\mathrm{rect}}$ and
the QBER $Q_{\mathrm{rect}}$ account for the cases where the
parties sent more than one photon. In particular,
$\delta(x)=f(x)~H_{2}(x)$ gives the leak of information from
imperfect error correction, with $f\geq1$ being the efficiency of
classical error correction codes.

In ideal conditions of perfect transmission and perfect single
photon sources, the key rate of Eq.~(\ref{RATE-DV-MDI}) would be
just the gain $Y_{\mathrm{rect}}^{11}$. By contrast, assuming more
realistic conditions the key rate is re-scaled by the probability
$P_{\mathrm{rect}}^{11}$ and reduced subtracting a term
proportional to information lost to perform privacy amplification
[second term on the right-side of Eq.~(\ref{RATE-DV-MDI})], and
error correction (third term). The security of decoy-state DV
MDI-QKD, including finite-size effects, has been assessed in
Ref.~\cite{FS-DV-MDI}. See also Refs.~\cite{MDIW1,MDIW2} for
practical decoy-state analyzes of the MDI-QKD protocol.

The above example is a special case of a general approach that
protect QKD from side-channel attacks on the measurement devices.
In the more general framework introduced by Ref.~\cite{BP}, each
honest user prepares a bipartite quantum state and sends one
subsystem to the relay. The state received by the relay has thus
the form $\rho_{AA^{\prime}}\otimes\rho_{B^{\prime}B}$, where the
system $A$, $B$ are those retained by the Alice and Bob,
respectively. A generic operation applied by the relay is
described by a quantum instrument
\cite{Ziman} characterized by a set of operators $\Lambda_{A^{\prime}%
B^{\prime}\rightarrow E}^{z}$. This includes a measurement with
outcome $z$ and storage of information in a quantum memory $E$. If
Eve applies the measurement and then announces the outcome $z$,
for any given value of $z$ the correlations between Alice, Bob,
and Eve, are described by the tripartite state \beq
\rho_{AEB}^{z}=\frac{1}{p(z)}\left(  I_{A}\otimes I_{B}\otimes
\Lambda_{A^{\prime}B^{\prime}\rightarrow E}^{z}\right)  (\rho_{AA^{\prime}%
}\otimes\rho_{B^{\prime}B}), \eeq where
$p(z)=\mathrm{Tr}(\Lambda_{A^{\prime }B^{\prime}\rightarrow
E}^{z}\rho_{A^{\prime}}\rho_{B^{\prime}})$.

The conditional state $\rho_{AB}^{z}$ is no more factorized and
exhibits correlations between Alice and Bob. To extract secret
bits from such a state Alice and Bob must apply local measurements
to obtain a tri-partite classical-quantum state $\rho_{XEY}^{z}$
for any given value of $z$. The asymptotic secret key rate is
obtained from the expression of the mutual information between
Alice and Bob averaged over $z$,
$I_{AB}=\sum_{z}p(z)I(X;Y)_{\rho^{z}}$ minus the average Holevo
information between Alice and Eve,
$\chi_{AE}=\sum_{z}p(z)I(X;E)_{\rho^{z}}$ (in the case of direct
reconciliation). The general approach of Ref.~\cite{BP} not only
provides a security proof for DV MDI-QKD schemes but also sets the
basis for an extension to CV systems, later realized in
Ref.~\cite{RELAY}.

\subsection{Twin-field QKD}\label{TFQKDSection}

In the MDI-QKD protocol, the idea is to use a middle relay that
may be untrusted, i.e., run by Eve. This is a very first step
towards the end-to-end principle of networks which assumes a
scenario with unreliable middle nodes. On the other hand, despite
MDI-QKD employs a relay, it is not able to beat the PLOB bound for
point-to-point QKD~\cite{PLOB}. This limitation has been recently
lifted by the introduction of a more efficient protocol called
``twin-field'' (TF) QKD~\cite{TFQKD}. The TF-QKD protocol has led
to further theoretical investigations~\cite{TFQKD2} and a number
of TF-inspired variants, including the phase-matching (PM)
protocol~\cite{PMQKD} (see also Ref.~\cite{PMQKD2}), the ``sending
or not sending'' (SNS) version of
TF-QKD~\cite{SNSwang,SNSpractical,SNSpractical2}, recently
improved into the active odd-parity pair (AOPP)
protocol~\cite{SNSAOPP}, and the no-phase-postselected TF (NPPTF)
protocol~\cite{Cuo3,Cuo1,Cuo2} (see also
Refs.~\cite{NPPTF2,NPPTF3}).

In the TF-QKD protocol, Alice and Bob send two phase-randomized
optical fields (dim pulses) to the middle relay (Charlie/Eve) to
produce a single-photon interference to be detected by a
single-photon detector, whose outcomes are publicly declared. The
term \textit{twin} derives from the fact that the electromagnetic
phases of the optical fields should be sufficiently close in order
to interfere. More precisely, Alice and Bob send to the relay
pulses whose intensity $\mu_i$ (for $i=A$ or $B$) is randomly
selected between three possible values. Then, they respectively
choose phases $\psi_A$ and $\psi_B$ as
$\psi_i=(\alpha_i+\beta_i+\delta_{i})\oplus 2\pi$, where
$\alpha_i\in\{0,\pi\}$ encodes a bit, $\beta_i\in\{0,\pi/2\}$
determines the basis, and the final term $\delta_i$ is randomly
selected from $M$ slices of the interval $[0,2\pi)$, so that it
takes one of the values $2\pi k/M$ for $k=\{0,1...,M-1\}$.

To ensure only phases close enough are selected, after disclosing
on an authenticated channel, Alice and Bob only accept the same
choices of the slice, i.e., the instances for $\delta_A =
\delta_B$. These pulses interfere at the relay interfere
constructively (or destructively). Then, Alice announces the basis
she used $\beta_A$ and the intensity $\mu_A$ for each instance.
The key is extracted from the basis $\beta_A=\beta_B=0$ and for
one of the intensities. In fact, a bit $\alpha_A$ can be shared
between Alice and Bob by considering the absolute difference
between $\alpha_A$ and $\alpha_B$ to be equal to $0$ or $\pi$
(depending on the relay's announcement). The rest of the results
can be used for other purposes, including estimating error rates
as well as decoy-state parameters.

\begin{figure}
\vspace{+.2cm}
\includegraphics[width=0.45\textwidth]{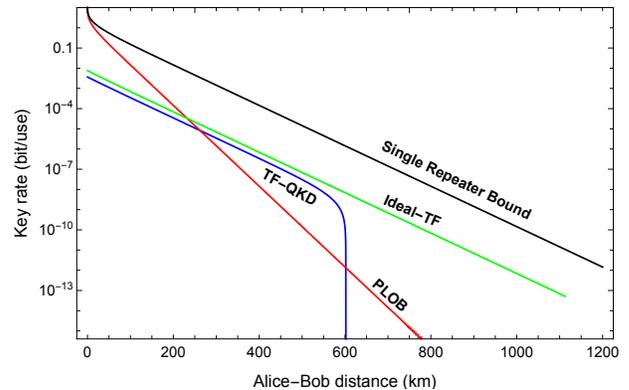}
\caption{Key rate of the TF-QKD protocol~\cite{TFQKD} versus
Alice-Bob total distance in standard optical fiber (0.2 dB/km),
assuming ideal (green line) and realistic (blue line) conditions.
For the realistic key rate we assume $10^{-8}$ dark count
probability per detector, $75\%$ loss at the relay, $50\%$
detector efficiency, and error correction efficiency of $1.1$. For
the ideal key rate we consider no dark counts and perfect detector
efficiency. For more details and other assumptions on these key
rates see Ref.~\cite{TFQKD}. We also plot the point-to-point
repeaterless PLOB bound~\cite{PLOB} and the single-repeater
bound~\cite{net2006,net2006p}. We can see that the PLOB is
violated, showing that the TF-QKD protocol is equivalent to an
active repeater. At the same time, it cannot beat the secret-key
capacity of an ideal repeater.} \label{TFQKDpic}
\end{figure}

Note that the twin pulses are in principle set by requiring
$\delta_A$ to be as close as possible to $\delta_B$, and the
nonzero difference between them introduces an intrinsic QBER. The
two become \textit{identical} provided that $M$ is infinitely
large. Realistically, a finite but large value of $M$ can be used
though this decreases the probability of matching two phase
slices. An estimation made in Ref.~\cite{TFQKD} gives the optimal
value $M=16$ with a QBER of $\approx 1.28\%$.

In Ref.~\cite{TFQKD}, the authors considered a restricted scenario
where the `global phase' does not leak any useful information to
Eve, giving a key rate
\begin{eqnarray}
R_{\mathrm{TF}}(\mu,L)=\dfrac{d}{M}R(\mu,L/2)
\end{eqnarray}
where $R(\cdot)$ is the secret key rate of an efficient BB84
protocol~\cite{effBB84} with tagging argument~\cite{tag}, and
$\mu,L$ are the intensity and the distance respectively. Later,
Ref.~\cite{Heid} considered a collective attack where Eve makes
use of identical beam splitters set along each path connecting
Alice and Bob to the relay. While this attack considerably
increases Eve's gain, the key rate scaling $O(\sqrt{\eta})$
remains unchanged. As a matter of fact, using the TF-QKD protocol
(and the PM-QKD protocol) over a communication line with total
Alice-Bob's transmissivity $\eta$, not only is the PLOB bound
beaten, but the rate performance is also not so far from the
single-repeater bound of
$-\mathrm{log}_2(1-\sqrt{\eta})$~\cite{net2006,net2006p}. See
Fig.~\ref{TFQKDpic}.

Yet other variants of the TF-QKD protocol have been
proposed~\cite{TFQKDvar1,TFQKDvar2,TFQKDvar3} and experimental
implementations have been carried
out~\cite{TREL19,Cuo4,ZhongTF19,ZHC+19}. In particular, the
proof-of-concept experiment in Ref.~\cite{TREL19} has recently
overcome the PLOB bound for the first time, a result previously
thought to be out of the reach of present technology.

\section{Experimental DV-QKD protocols}
The original BB84 protocol requires perfect single photon sources
which emit only one photon at a time. Since these sources are
notoriously hard to build they have been replaced by coherent
state sources which are heavily attenuated to a fraction of a
photon per pulse. However, these sources lead to security concerns
due to the probability to have more than a photon per pulse and a
photon splitting attack has been proposed and demonstrated to
exploit the wrong assumption in the security proofs. As described
before a rigorous security~\cite{GLLP,Inamori2007} analysis has
been proposed with the idea of estimating the ratio of secure
signals from which the secure bits are distilled by
post-processing. For practical sources the bounds found in the
security analysis are not tight leading to a degradation of system
performance. To circumvent this problem several novel protocols
with different encoding schemes have been proposed and in the
following sections we explain the development of their
implementations in detail. Despite the different encoding schemes
all DV QKD system have single photon detectors in common to detect
the arriving states. To achieve high key rates high count rates
and, thus, low dead times are necessary. Extremely long distances
require however low dark count rates.

\subsection{Detector technology}
At the receiver side the arriving photon pulses are processed by
e.g.\ beam splitters, interferometers or a like to decode the
information encoded in various degrees of freedoms. After optical
processing the photons are detected by single photon detectors
which set limits on the achievable performance.

Indium Gallium Arsenide (InGaAs) avalanche photodiodes detect
single photons by generating a strong electron avalanche at the
absorption of a photon when operated with a reverse voltage above
the breakdown voltage. However, the strong avalanche current can
lead to trapped electron charges in defects. Spontaneously
released they trigger a second avalanche pulse, a so-called
afterpulse. A common approach to suppress the afterpulse is
gating. To further suppress this afterpulse and to allow for
gating frequencies beyond 1 GHz, a self-differentiating technique
was introduced to detect much weaker avalanches~\cite{Yuan2007}.
Operating at $-30^\circ$\,C the APD was gated at 1.25 GHz,
obtaining a count rate of 100 MHz with an detection efficiency of
$10.8 \%$, an afterpulse probability of about $6 \%$ and a dark
count rate of about 3 kHz.

To achieve higher quantum efficiencies and in particular lower
dark count rates, superconducting nanowire single photon detectors
(SNPDs) have been developed. They consist of a nanometer thick and
hundreds of nanometer wide nanowire with a length of hundreds of
micrometers. Compactly patterned in a meander structure they fill
a square or circular area on the chip. The nanowire is cooled
below its superconducting critical temperature and a bias current
just below the superconducting critical current is applied. An
incident photon breaks up Cooper pairs in the nanowire which
lowers the superconducting critical current below the bias current
which produces a measurable voltage pulse. A recent
development~\cite{Caloz2018} shows dark count rates of 0.1 Hz, low
jitter of 26 ps and a quantum efficiency of $80\,\%$ at a
temperature of 0.8 K. SNSPDs have been integrated into photonic
circuits~\cite{You2011, Rath2015}.

\subsection{Decoy state BB84}

As described before decoy state QKD severely increases security
and distance for attenuated coherent laser pulse sources and is
much more practical in comparison to single photon sources. The
first implementation was performed in 2006 with one decoy state by
modifying a commercial two-way idQuantique system~\cite{Zhao2006}.
In the two-way protocol with phase encoding Bob sent bright laser
pulses to Alice who after attenuating them to the single photon
level and applying a phase shift sent them back to Bob for
measurement. The intensity of the pulses was randomly modulated by
an acousto-optical modulator inserted into Alice's station to
either signal state or decoy state level before sending the pulses
back to Bob. Shortly later the same group implemented a two decoy
state protocol with an additional vacuum state to detect the
background and dark count detection probability~\cite{Zhao2006a}.

The demonstration of two-decoy states BB84 in a one-way QKD system
was reported by three groups at the same time in 2007. In
Ref.~\cite{Rosenberg2007} phase encoding was employed and secure
key generation was shown over a distance of 107\,km using optical
fiber on a spool in the lab. Including finite statistics in the
parameter estimation, a secret key rate of 12\,bit/s was achieved.
To generate the decoy states pulses from a DFB laser diode at a
repetition rate of 2.5\,MHz were amplitude modulated with an
amplitude modulator. For detection single-photon sensitive
superconducting transition-edge detectors were employed.

The second group demonstrated two-decoy state QKD over a 144\,km
free-space link with 35\,dB attenuation between the canary islands
La Palma and Tenerife~\cite{Schmitt-Manderbach2007}. Here, the
BB84 states were polarization encoded. Four 850\,nm laser diodes
oriented at $45^\circ$ relative to the neighbouring one were used
in the transmitter. At a clock rate of 10\,MHz one of them emitted
a 2\,ns pulse. The decoy states of high intensity were generated
at random times by two laser diodes emitting a pulse at the same
time, while for the vacuum state no pulse was emitted. The
receiver performed polarization analysis using polarizing beam
splitters and four avalanche photo detectors. A secure key rate of
12.8 bit/s was achieved.

The third group used polarization encoding and demonstrated secret
key generation over 102\,km of fiber~\cite{Peng2007}. The
transmitter consisted of 10 laser diodes each of which produced
1\,ns pulses at the central wavelength of 1550\,nm with a
repetition rate of 2.5\,MHz. Four laser diodes were used for
signal and high intensity decoy state generation, respectively,
using a polarization controller to transform the output
polarization of a laser diode to the respective polarization of
one of the four BB84 states. Two additional laser diodes were used
for calibrating the two sets of polarization basis which was
performed in a time multiplexed fashion. The outputs of the 10
laser diodes were routed to a single optical fiber using a network
of multiple beam splitters and polarization beam splitters. An
additional dense wavelength division multiplexing filter ensured
that the wavelengths of the emitted photons was equal. The
receiver consisted of two single photon detectors and a switch to
randomly choose one polarization basis.

Using advances in InGaAs avalanche photon detection (APD)
operating in self-differencing mode~\cite{Yuan2007} GHz clocked
decoy state QKD was demonstrated in 2008~\cite{Dixon2008}. A
self-differencing circuit can sense smaller avalanche charges
thereby reducing after pulse probability and thus dead time. The
demonstrated QKD system clocked at 1.036\,GHz was based on a phase
encoded GHz system implementing the BB84 protocol~\cite{Yuan2008}
and used two decoy states generated by an intensity modulator.
Dispersion shifted single mode fiber was employed since for
channel lengths over 65\,km fiber chromatic dispersion must be
compensated for in standard SMF28 single mode fiber.

In the standard BB84 protocol Bob measures in the wrong basis
$50\,\%$ of the time. Moreover, in decoy state BB84, it is
advantageous to send the states with higher intensity more often
than the others. To increase the usable signal generation rate an
efficient version with asymmetric bases choice and highly
unbalanced intensities was introduced, with an implementation
reported in~\cite{Lucamarini2013a}. They prove the protocol's
composable security for collective attacks and improved parameter
estimation with a numerical optimization technique. Based on phase
encoding the GHz system achieved a secure key rate of 1.09 MBit/s
in contrast to 0.63 MBit/s for the standard protocol over 50\,km
of fiber. Its experimental implementation is depicted in
Fig.~\ref{fig:dvqkdexp}a.

Composable security against coherent attacks was only achieved
recently. Ref.~\cite{Xu2015} describes an experiment demonstrating
it with a modified two-way commercial plug-and-play QKD system
where the authors also included imperfect state generation.
Security against coherent attacks was furthermore demonstrated in
\cite{Frohlich2017} with a one-way phase-encoding system. With the
latter system the authors achieved a distance in ultra-low loss
fiber (0.18\,dB/km) of 240\,km. Using APDs with a detection
efficiency of 10\,\% a dark count rate of 10 counts/s was achieved
at $-60^\circ$C reached with a thermal-electrical cooler.

The current distance record of 421 km ultra low loss optical fiber
(0.17 dB/km) was achieved simplified BB84 scheme with a one-decoy
state~\cite{Boaron2018}. The distance record was achieved by
optimizing the individual components and simplifying the protocol.
The system was clocked at 2.5 GHz and used efficient
superconducting detectors (about 50\,\%) with a dark count rate
below 0.3 Hz. The protocol was based on a scheme with three states
using time bin encoding. Two states were generated in the Z basis,
a weak coherent pulse in the first or the second time bin,
respectively. The third state, a state in the X basis, was a
superposition of two pulses in both time bins. While the Z basis
states were used to estimate the leaked information to the
eavesdropper, the X basis state was used to generate the raw key.
The experimental setup is shown in Fig.~\ref{fig:dvqkdexp}b.

\subsection{Differential phase shift QKD}
Differential phase shift QKD encodes information into the
differential phase shift of two sequential pulses. The first QKD
system employing this encoding technique was reported in 2004 over
20\,km fiber~\cite{Honjo2004}. A continuous-wave laser diode from
an external-cavity laser was intensity modulated at 1\,GHz to
carve 125\,ps long pulses. Afterwards a phase modulator was used
to modulate the phase of each pulse randomly by 0 or $\pi$. An
attenuator attenuated the beam to 0.1 photon per pulse. At the
receiver side the differential phase between two sequential pulses
was measured with an unbalanced Mach-Zehnder interferometer. The
incoming pulses were split 50:50 and before recombination at
another 50:50 splitter, one arm was delayed by the interval of
time between two pulses. The two outputs of the unbalanced
Mach-Zehnder interferometer were detected by gated avalanche
single photon detectors. The Mach-Zehnder interferometer was as
waveguides and the arm length difference could be controlled
thermally.

Using superconducting single photon detectors and a 10\,GHz clock
frequency keys were distributed over 200\,km dispersion shifted
fiber~\cite{Takesue2007}. In a different experiment, a secure bit
rate in the MBit/s range was achieved over 10\,km by using a
2\,GHz pulse train with 70\,ps long pulses~\cite{Zhang2009}. At
the receiver after the unbalanced Mach-Zehnder interferometer the
photons were upconverted in a nonlinear process and detected by a
Silicon avalanche photo diode which enabled count rates of 10\,MHz
with a low timing jitter.

High-rates of 24\,kbit/s over 100\,km were achieved using 2\,GHz
sinusoidally gated avalanche photo diodes and the important
influence of laser phase noise has been studied~\cite{Honjo2011}.
Using a Michelson interferometer with unequal arm length based on
a beam splitter and two Faraday mirrors and superconducting
detectors at the receiver the maximum transmission distance has
been boosted to 260\,km in standard telecom fiber~\cite{Wang2012}.
Its experimental implementation is depicted in
Fig.~\ref{fig:dvqkdexp}c.

The DPS-QKD protocol has been tested in the Tokyo QKD
network~\cite{Sasaki2011,Shimizu2014}.

\subsection{Coherent one-way}
The first proof-of-principle implementation of the COW protocol
has been reported in 2005~\cite{Stucki2005}. A 1550\,nm
continuous-wave laser beam was intensity modulated to generate the
quantum or decoy states and a variable attenuator attenuates the
beam to the single photon level. Bits were encoded into arrival
time by two consecutive pulses: A vacuum state followed by a
coherent state represented bit 0, a coherent state followed by a
vacuum state represented bit 1. The decoy state was represented by
two coherent states. On the receiver side the beam was split by a
tap coupler (tapping e.g.\, 10\,\%). While the highly transmittive
output was detected by a single photon detector, the tap was
injected into an interferometer with asymmetric arms which
interfered the two pulses. One output of the interferometer was
measured by a single photon detector and the measurement outcomes
were used to calculate the visibility to check channel
disturbances. The unbalanced interferometer was implemented as
Michelson interferometer by using a 3\,dB coupler and two Faraday
mirrors.

Running at a high clock speed of 625\,MHz a fully automated system
was built and demonstrated over 150\,km in deployed telecom
fiber~\cite{Stucki2009}. The high clock speed was reached with a
continuous-wave distributed fiber-Bragg telecom laser diode, a
10\,GHz Lithium Niobate intensity modulator and Peltier cooled
InGaAs avalanche photo diodes in free-running mode for short
distances and SNSPDs operating at sub-4\,K with lower noise for
long distances. Synchronization was achieved by wavelength
division multiplexing of a synchronization channel and a classical
communication channel through a second optical fiber. Using
ultra-low loss fibers and low-noise superconducting detector
operating at 2.5\,K a distance of 250\,km was
reached~\cite{Stucki2009a}. While the previous implementations all
used an asymptotic security proof finite-size effects were taken
into consideration in the implementation described in
2014~\cite{Walenta2014} which reached 21\,kbit per second over
25\,km fiber with gated InGaAs detectors and a key distillation in
FPGAs. Here, the COW QKD system was tested with one single optical
fiber only using dense-wavelength division multiplexing for
quantum and all classical channels.

The distance record of a system implementing the coherent one-way
protocol was reported in 2015~\cite{Korzh2015} reaching 307 km.
Novel free-running InGaAs/InP negative feedback avalanche
detectors operated at 153\,K with low background noise (few dark
counts per second) and low loss optical fibers as well as a novel
composable finite-key size security analysis enabled the result.
The experimental implementation is schematically depicted in
Fig.~\ref{fig:dvqkdexp}d.

\begin{figure*}
  \includegraphics{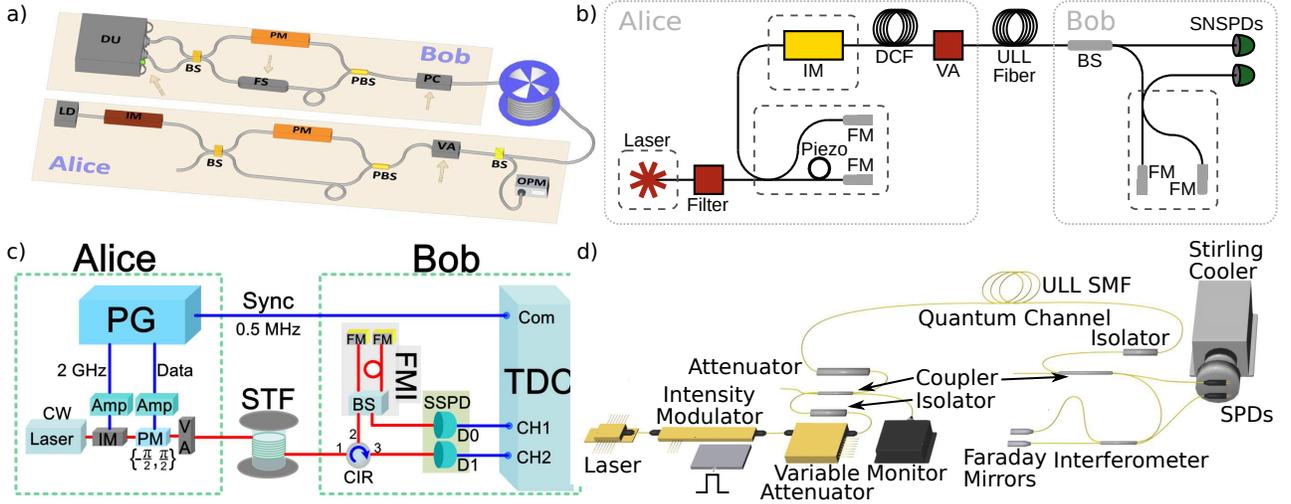}
  \caption{Exemplary experimental implementation of discrete variable QKD.
    a) Two-decoy state BB84 protocol with biased basis choice reported in~\cite{Lucamarini2013a}. A laser diode emitted pulses at 1550 nm which were intensity modulated (IM) to generate the different intensity of the states. An unbalanced Mach-Zehnder interferometer with a phase modulator (PM) in one arm was used to generate the different BB84 states, i.e. 0 and $\pi$ for the Z basis and $\pi/2$ and $3\pi /2$ for the X basis. After attenuation to the single photon level with a variable attenuator (VA), the states were transmitted through a fiber. At Bob's side decoding was performed with an identical Mach-Zehnder interferometer and a PM either set to 0 or $\pi$. A fiber stretcher (FS) matched the two interferometers. The detection unit (DU) consisted of two InGaAs APDs. BS: Beam Splitter, PBS: Polarizing Beam Splitter, OPM: Optical Power Meter, PC: Polarization Controller.
    b) Simplified one-decoy state BB84 protocol with three states implemented over 421 km~\cite{Boaron2018}. Alice uses a phase randomized laser pulse with a repetition rate of 2.5 GHz which is tightly bandpass filtered around 1550 nm. The pulses pass through an unbalanced Michelson interferometer with 200 ps delay made of beam splitter and two Faraday mirrors (FM) and a piezo in one of the arms to control the phase, to enable time bin encoding. Afterwards the pulses are intensity modulated (IM) to generate the different qubit states. After dispersion compensation (DCF) and attenuation to the single photon level (variable attenuator: VA), the pulses are transmitted through an ultra-low-loss (ULL) fiber. To implement the different bases choices at Bob's station the pulses are split with a beam splitter. One of its outputs is directly detected with an SNSPD, measuring the arrival time in Z basis which is used for the raw key. The other is used to measure the X basis by passing the pulses through an unbalanced interferometer identical to Alice's. This measurement is used to estimate the eavesdropper information.
        c) Implementation of the differential phase shift protocol reported in~\cite{Wang2012} over 260 km with a rate of 2 GHz. A continuous wave (CW) laser at 1560 nm is chopped into pulses with an intensity modulator (IM). A phase modulator then randomly applies a $\pi/2$ or $-\pi/2$ phase shift on the pulses before they are attenuated to the single photon level. The pulses are then transmitted through standard telecom fiber (STF). At Bob's side the encoded information is decoded by a Faraday Michelson interferometer (FMI) which interferes a pulse with the one before and after it. The two outputs of the interferometer were detected by superconducting single photon detectors (SSPD). TDC: time to digit converter.
        d) Coherent one-way protocol implementation over 307 km with a repetition rate of 625 MHz reported in~\cite{Korzh2015}. Pulses were carved into a continuous wave laser beam at 1550 nm using two different intensities to encode bits using consecutive time bins. After attenuating to the single photon level the pulses were sent through an ultra-low loss (ULL) single mode fiber (SMF). Bob's receiver is similar to the receiver described in b). }
  \label{fig:dvqkdexp}
\end{figure*}

\subsection{DV MDI-QKD}

DV MDI-QKD was first experimentally demonstrated in 2013 by three
groups. The first group implemented MDI-QKD between three
locations in Calgary with a distance of about 12 km between Alice
and the untrusted relay Charlie and about 6 km between Bob and
Charlie~\cite{Rubenok2013}. Alice's and Bob's transmitter
generated time-bin qubits at a rate of 2 MHz using an attenuated
pulsed laser at 1552 nm and an intensity and phase modulator. The
generated states were chosen by Alice and Bob independently from
the set $\vert \psi_{A,B}\rangle \in \left[\vert 0\rangle, \vert
1\rangle, \vert +\rangle, \vert -\rangle \right]$ where $\vert
\pm\rangle = (\vert 0\rangle \pm \vert 1\rangle)/\sqrt{2}$. By
choosing between three intensity levels, vacuum, a decoy state
level and a signal state level, the decoy state protocol was
implemented. Both transmitters were synchronized by a master clock
located at Charlie which was optically transmitted to the
respective stations through another deployed fiber. After
receiving the photons Charlie performed a Bell state measurement
by superimposing the pulses at a balanced beam splitter and
detecting the outputs with gated InGaAs single photon detectors
with 10 $\mu s$ dead time. If the two detectors coincidentally
clicked within 1.4 ns the states were projected into a Bell state.
Those instances were publicly announced by Charlie.

The second group implemented the protocol over 50 km in the
lab~\cite{Liu2013}. They implemented a similar qubit time-bin
encoding scheme as in the Calgary experiment, but used four decoy
intensity levels with 0, 0.1, 0.2 and 0.5 photons per pulse on
average. A pulsed laser was fed through an unbalanced Mach-Zehnder
interferometer to generate two time-bin pulses. The encoding of
qubits and decoy were implemented with three amplitude and one
phase modulator situated in a thermostatic container for stability
reasons. After traveling through 25 km of fiber the untrusted
relay Charlie performed a Bell state measurement identically to
described above. The employed photo detectors used an upconversion
technique where a nonlinear process in periodically poled lithium
niobate converted the 1550 nm photons to 862 nm detected by
Silicon avalanche photo detectors with a dark count rate of 1 kHz.

The third implementation~\cite{FerreiraDaSilva2013} was a
proof-of-principle demonstration based on polarization qubits
instead and demonstrated MDI-QKD over 8.5 km long  fiber links
between the two trusted parties and the relay. Using a continuous
wave laser pulses were carved with an amplitude modulator. The
decoy state levels were chosen by variable optical attenuators and
the polarization encoding was performed with an automatic
polarization controller. The relay was built from a balanced beam
splitter and two polarization beam splitters. Four gated InGaAs
avalanche single photon detectors with a dark count probability of
15 ppm and 10 $\mu s$ dead time detected their output.

The distance of MDI-QKD was then boosted to 200 km~\cite{Tang2014}
and 404 km~\cite{Yin2016} using ultra-low loss fiber with an
attenuation of 0.16 dB/km. To achieve such a large communication
length of 404 km the MDI-QKD protocol was optimized to improve on
the effects of statistical fluctuations on the estimation of
crucial security parameters. The protocol consisted of four decoy
states with three levels in the X basis and only one in the Z
basis. The probabilities for each was carefully optimized to
obtain largest key rate. Five intensity modulators and one phase
modulator was employed to implement those. The receiver was
implemented in the same way as described above for the first two
experiments. Superconducting single photon detectors improved the
quantum efficiency (about $65 \%$) and dark count rate (30 Hz).
Furthermore to achieve 404 km in the order of $10^{14}$ successful
transmissions were recorded which took with a clock rate of 75 MHz
over 3 months. The achieved secret key rate was $3.2 \times
10^{-4}$ bits per second.

Furthermore at zero transmission distance a secret key rate of 1.6
MBit/s was reached~\cite{Comandar2016} by introducing a pulsed
laser seeding technique to achieve indistinguishable laser pulses
at 1 GHz repetition rate. The new technique where a master laser
pulse is injected into a slave laser as a seed to trigger
stimulated emission at a defined time yielded very low timing
jitter and close-to-transform limited pulses.

To demonstrate MDI-QKD over quantum networks in star topology
extending over 100 km distance, cost-effective and commercially
available hardware was used to build a robust MDI-QKD system based
on time-bin encoding~\cite{Valivarthi2017}. Similar plug and play
systems with time-bin or polarization encoding and different level
of immunity against environmental disturbances have been
implemented as well in other
groups~\cite{Choi2016,Tang2016,Wang2017a,Park2018,Liu2018a}.



\subsection{High-dimensional QKD}

Most discrete variable (DV) QKD schemes encode quantum states in
qubits ($d=2$), such as the polarization states used in the first
QKD experiment~\cite{Bennett1992}. Going back to the early 2000s,
there has been considerable interest in developing large-alphabet
DV QKD schemes that encode photons into qudits: high-dimensional
basis states with $d>2$. Such schemes offer the ability to encode
multiple ($\log_2 d$) bits of information in each photon. This
benefit is not without a drawback; the information density per
mode decreases as $(\log_2 d)/d$. Nevertheless, high-dimensional
QKD (HD QKD) can offer major advantages over their qubit
counterparts.

HD QKD can increase the effective secret key generation rate when
this rate is limited by the bandwidth mismatch between the
transmitter and the receiver. This mismatch happens when either
the transmitter is limited to a flux below the available receiver
bandwidth or the single-photon detector is saturated by the high
photon flux received. While the former does not typically occur
with attenuated laser source, the latter often arises due to
detector dead time. In a superconducting nanowire single photon
detector (SNSPD), the dead time is dominated by the time it takes
to recover its supercurrent (which flows with zero
resistance)---during which the nanowire is insensitive to any
photon~\cite{Nat12}.

Fig.~\ref{fig:qkd_trend} shows a representative plot of
qubit-based DV QKD secret key rate versus distance for currently
achievable parameters. Three distinct regimes are apparent: regime
II denotes normal operation where the secret key rate scales as
the transmissivity in the fiber, which decays exponentially with
distance. At longer distances, we enter regime III where the
received photon rate is comparable to the detectors' background
rate---masking any correlation between the key-generating parties
and abruptly reducing the secret key rate. However, at short
distances with low photon loss (regime I with distances up to
$\sim 100$~km), the secret key rate is limited due to the detector
dead time. The highest QKD key rate is achieved in this regime and
it currently amounts to $13.72$ Mb/s \cite{YPTD+18}. To increase
this key rate further, more detectors could be added so to
distribute the initial intensity among them. Another strategy
would be increasing the dimensionality of the alphabet to reduce
the transmitted photon rate until the detectors are just below
saturation. To date, multiple degrees of freedom have been
investigated for high-dimensional QKD, including
position-momentum~\cite{Zhang2008},
temporal-spectral~\cite{Tittel1999,Nunn2013,Lee2014,Thew2004,Qi2006,Ali-Khan2007},
and orbital angular momentum
(OAM)~\cite{Mafu2013,Mirhosseini2015,Sit2016}.

\begin{figure}[ht]
   \includegraphics[width=\columnwidth]{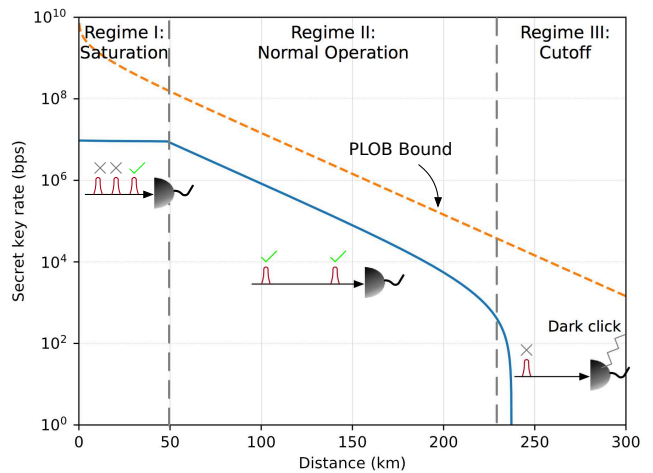}
    \caption{Representative plot of secret key generation rate against channel distance
    for a traditional qubit DV QKD protocol for currently achievable device parameters.
    The plot assumes a 1~GHz clock rate, a 93\% detector efficiency, a 1000~cps dark count rate,
    and a 100~ns detector dead time. We denote three distinct regimes: I. Short metropolitan-scale
    distances, where the secret key rate is limited by detector saturation; II. Longer distances,
    where the secret key rate decays exponentially with distance; III. Extremely long distances,
    where the secret key rate is sharply limited by detector dark count
    rates. The PLOB bound~\cite{PLOB} is plotted for comparison.
    }
    \label{fig:qkd_trend}
\end{figure}

Initial security analysis by Cerf \textit{et al.} for discrete
large-alphabet QKD showed improved resilience against noise and
loss~\cite{Cerf2002a}. HD QKD with discrete quantum states is
capable of tolerating error rates than the 11\% limit for
qubit-based protocols. However, the proposed scheme with its two
early proposals---one using OAM and another using
temporal-spectral encoding---was challenging to demonstrate. The
main difficulty lies in the measurement of discrete
high-dimensional states within at least two mutually unbiased
bases. Efficient implementation of the scheme for the two proposed
degrees of freedom required single-photon detectors that scale
with the dimensionality $d$---prohibiting the use of large $d$.
Therefore, there has been a strong desire in developing HD QKD
schemes with the ability to measure higher-order correlations
using only a few single photon detectors.

One detector-efficient temporal scheme---borrowing techniques from
continuous variable (CV) QKD and applying them to the
temporal-spectral mode---demonstrated QKD operations with an
extremely high alphabet of $d = 1278$, i.e., over 10~bits per
photon~\cite{Ali-Khan2007}. However, no security proof against
collective or coherent attacks was available at the time. The
challenge is that time and energy states are not inherently
discrete, but rather they form a continuous basis. Therefore, the
security proof on discrete dimensional bases do not transfer
directly to these continuous-basis schemes. Considerable effort
was made to extend the proofs for CV QKD to HD QKD by realizing
that the security of temporal-spectral HD QKD can be guaranteed by
measuring the covariance matrices between Alice's and Bob's
information.

Measuring the covariance matrices involves detection in the
frequency basis. Direct spectral detection of the incoming light
can be done using a single-photon-limited spectrometer: a spectral
grating followed by $d$ single photon detectors. However, the
required number of detectors would again prevent reaching a large
dimensionality. To work around these limitations, new techniques
were introduced to convert the spectral information to time
information by using group-velocity dispersion~\cite{Mower2013},
Franson interferometers~\cite{Zhang2014a}, or a time-varying
series of phase shifts~\cite{Nunn2013}.

The development of temporal-spectral encoded HD QKD spurred record
demonstrations of secret key capacity at 7.4~secret bits per
detected photon~\cite{Zhong2015b} and secret key generation rates
of 23~Mbps~\cite{Lee2016} and 26.2~Mbps~\cite{Islam2017} with $d =
16$ at 0.1~dB loss and $d = 4$ at 4~dB induced loss, respectively.
Furthermore, a 43-km (12.7~dB loss) field demonstration between
two different cities show a maximum secret key generation rate of
1.2~Mbps~\cite{Lee2016}. Since HD QKD is vulnerable against photon
number splitting attacks as it relies on transmission of single
photons, these demonstrations make use of decoy state techniques
to close this security loophole~\cite{Bunandar2014}. More
recently, the security of temporal-spectral HD QKD has been
extended to include the composable security framework, which takes
into account statistical fluctuations in estimating parameters
through only a finite number of
measurements~\cite{Lee2013a,Niu2016}.

High-dimensional QKD with OAM has also witnessed rapid development
due as it is directly compatible with free-space QKD
systems~\cite{Erhard2018}. Since OAM modes rely on the preparation
and the measurement of discrete high-dimensional states, the
security proofs extend directly from the work by Cerf \textit{et
al.} Recently, the security proof has also been successfully
extended to include finite-key analysis for composable
security~\cite{Bradler2015}.

A photon carrying an OAM information has a helical or twisted wave
front with an azimuthal phase $\varphi$ which wraps around $\ell$
(helicity) times per wavelength. For the popular Laguerre-Gauss
mode, a photon carrying an $\ell \hbar$ OAM can be described as
$\ket{\Psi^{\ell}_Z} = e^{i \ell \varphi}$. $\ell$ is an unbounded
integer, which allows arbitrarily high encoding dimension, but
practically one limits $\ell \in [-L,L]$ to achieve a
dimensionality $d = 2L+1$. A mutually unbiased basis set can be
constructed using a linear combination of OAM modes
\begin{equation}
    \ket{\Psi^{n}_X} = \frac{1}{\sqrt{d}} \sum_{\ell = -L}^L
    \exp\left( i\frac{2\pi n \ell}{d}\right) \ket{\Psi^{\ell}_Z}.
\end{equation}
Both sets of quantum states can be generated using a spatial light
modulator (SLM)~\cite{Neff1990}, a digital micro-mirror device
(DMD)~\cite{Blanche2013}, or a tunable liquid crystal device known
as $q$-plates~\cite{Slussarenko:11,Zhang:2014aa}.

The first laboratory demonstration of high-dimensional OAM QKD
achieved a secret key generation rate of 2.05~bits per sifted
photon using a seven-dimensional alphabet ($L = 3$ and $d =
7$)~\cite{Mirhosseini2015}. More recently, a 300-m free-space
field demonstration in Ottawa with four-dimensional quantum states
achieved 0.65 bits per detected photon with an error rate of 11\%:
well below the QKD error rate threshold for $d=4$ at
18\%~\cite{Sit2016}. Although moderate turbulence was present
during the experiment, going to longer distances will require
active turbulence monitoring and compensation~\cite{Ren:14}.

The main challenge in high-dimensional OAM QKD towards achieving a
high secret key generation rate is the relatively low switching
speed of the encoding and decoding devices when compared to the
multi-gigahertz-bandwidth electro-optic modulators used in
time-bin encoded high-dimensional QKD. QKD demonstrations
involving SLM, DMD, and $q$-plates so far have required a time in
the order of 1~ms to reconfigure---limiting the QKD clock rate in
the kHz regime. While $q$-plates can potentially be operated at
GHz rates by using electro-optic tuning, these have yet to be
demonstrated~\cite{Karimi2009}. One appealing new direction is the
use of photonic integrated circuits (PICs), which may dramatically
reduce the configuration time. Thermo-optically tuned on-chip ring
resonators have demonstrated a switching time of
$20$~$\mu$s~\cite{Strain:2014aa,Cai2012}. More recently, precise
control of OAM mode generation has been demonstrated using a
$16\times 16$ optical phase array which allows for generation of
higher fidelity OAM states~\cite{Sun:14}. Furthermore, large scale
on-chip MEMS-actuation has also been demonstrated with a switching
time of $2.5$~$\mu$s with the potential of application to OAM
generation and control~\cite{Han:15}.

Demonstrations of HD QKD using a single set of conjugate photonic
degrees of freedom, such as time-energy or OAM, to increase the
secret key generation rate have been successful. Investigation in
new techniques, which include the miniaturized photonic integrated
circuit platform (see Sec.~\ref{sec:PIC}), to manipulate and
detect multiple degrees of freedom simultaneously can dramatically
increase the dimensionality that would improve the secret key rate
even further. Moreover, a more detailed study into the choices of
degrees of freedom and the choice of mutually unbiased bases can
shed light into which means of encoding is most robust for the
different QKD settings. For example, it has been hinted that the
Laguerre-Gauss OAM modes show greater resilience to cross talk in
turbulent environments than the Hermite-Gaussian OAM
modes~\cite{Restuccia:16}. With the potential of high-dimensional
QKD systems generating secret keys at rates commensurate to those
of data communication rates, further study into HD QKD in a
measurement-device-independent configuration is warranted.

\subsection{Photonic integrated circuits}\label{sec:PIC}

QKD devices have more demanding requirements than those offered by
standard off-the-shelf telecommunication equipments. QKD
transmitter needs single photon sources or weak coherent sources
modulated  at an extremely high ($\geq20$~dB) extinction ratio for
low-error QKD operations. Furthermore, quantum-limited detectors
such as single photon detectors or shot-noise limited homodyne
detectors are also required on the receiver side.

Photonic integrated circuits (PICs) provide a compact and stable
platform  for the integration of multiple high-speed quantum
photonic operations into a single compact monolithic circuit. PICs
allow experimentalists to engineer quantum devices in the
different material platforms at lithographic precision to meet the
stringent requirements of QKD devices. The amount of complexity
that can be achieved with PICs has been shown to enable practical
implementation of wavelength multiplexing for higher secret key
rates~\cite{Price:18,Bunandar18b}, multi-protocol operations for
flexibility~\cite{Sibson2017}, and additional monitoring and
compensation capabilities against timing and polarization drifts
in the channel~\cite{Bunandar2018}. Various material platforms
have been explored for building high-performance QKD
devices---each with its own strengths and weaknesses.
(See~\cite{Androvitsaneas2016} for further discussion of the
different material platforms.)

\begin{figure*}[ptbh]
   \includegraphics[width=\textwidth]{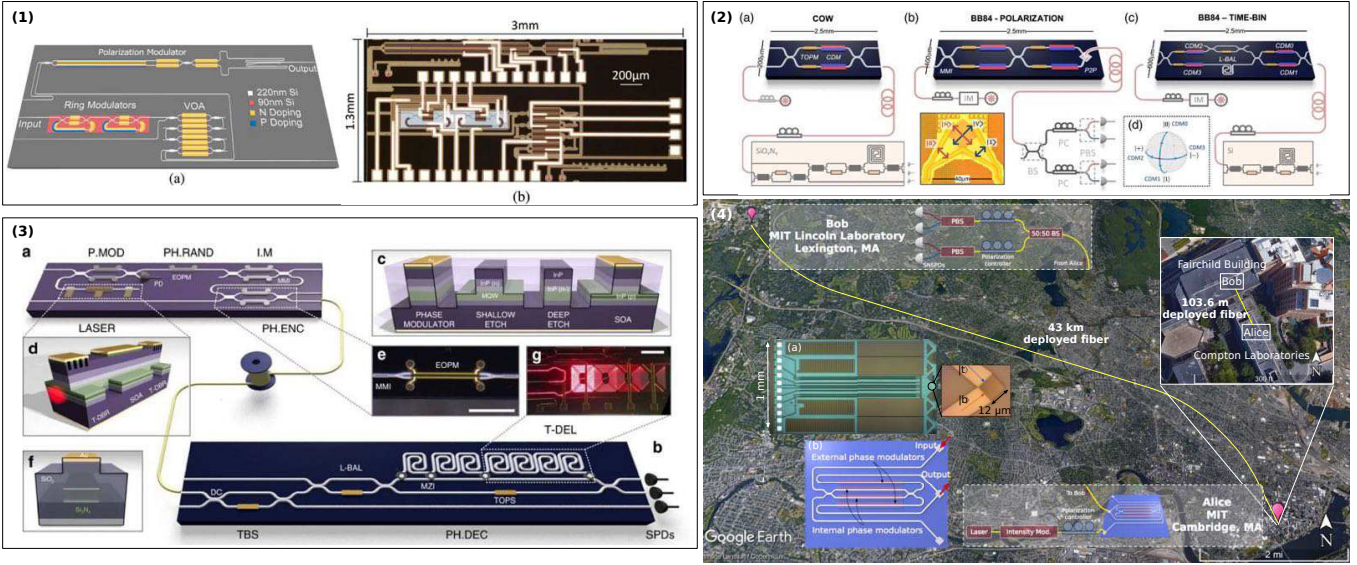}
    \caption{Experimental demonstrations of QKD using PICs.
    (1) (a) Schematic and (b) micrograph of the silicon photonics polarization QKD transmitter. The full transmitter consists of ring pulse generators, a variable optical attenuator (VOA), and  polarization controller~\cite{Ma2016}.
    (2) Schematic of integrated silicon photonics QKD transmitters for (a) coherent-one-way (COW) protocol, (b) polarization BB84 protocol, and (c) time-bin BB84 protocol~\cite{Sibson2016}.
    (3) (a) Schematic of the InP QKD transmitter, which combines a laser, pulse modulator, phase randomization, intensity modulator, and phase encoder. (b) Schematic of the TriPleX QKD time-bin receiver, which either immediately sends the signal for direct detection or interferes the signal before sending it to detectors. Cross-section of (c) InP PIC, (d) laser in InP PIC, and (f) TriPleX PIC. (e) and (g) are micrographs of the PICs~\cite{Sibson2017}.
    (4) Aerial view of the intercity polarization QKD field test between the cities of Cambridge and Lexington and the local field test between two adjacent buildings. Insets: (a) Micrograph and (b) schematic of the polarization silicon photonics QKD transmitter used~\cite{Bunandar2018}. \textcopyright Google. Map data from Google, Landsat/Copernicus.
    All figures are adapted with permission from: Ref.~\cite{Ma2016} \copyright OSA (2016), Ref.~\cite{Sibson2016} \copyright OSA (2017), Ref.~\cite{Sibson2017} \copyright NPG (2017), and Ref.~\cite{Bunandar2018} \copyright APS (2018).}
    \label{fig:pic_qkd}
\end{figure*}

Active III--V laser materials, such as indium phosphide (InP), is
a promising platform for QKD transmitters because of the
availability of gain laser medium for producing weak coherent
light~\cite{Smit2014}. The InP platform also has the advantage of
building quantum well structures using other ternary and
quaternary III--V semiconductors that are lattice-matched to InP,
such as InGaAs, InGaAsP, or InAlAsP~\cite{Roelkens2010}. Within
these quantum wells, carriers---electrons and holes---are confined
within the resulting one-dimensional potential wells. Applying
electric field to the well shifts the energies of the carriers,
which in turn changes its absorption spectrum and its refractive
index shift. This process, named Quantum-confined Stark effect
(QCSE)~\cite{Miller1984}, is the strongest electro-optic
modulation available in the platform---albeit with the undesirable
phase-dependent loss. Intensity and phase modulation with QCSE has
been demonstrated to achieve high extinction ratio beyond 50~dB at
bandwidths $\geq40$~GHz~\cite{Chiu2002}.

The SiO$_2$-Si$_3$N$_4$ TriPleX technology has record low loss
passive components at $\sim10^{-4}$~dB/cm~\cite{Worhoff2015} which
makes it an attractive platform for time-based or phase-based QKD
receiver components in high-speed gigahertz-clocked QKD
operations, where Bob has to interfere weak coherent pulses spaced
by $\sim1$~ns. The combination of low propagation loss and high
interference visibility and stability can enable Bob to maintain
low error-rate QKD operations without sophisticated stabilization
circuitry typically required for fiber- based or bulk optical
interferometers~\cite{Sibson2017}. The TriPleX platform, however,
relies solely on thermo-optic phase modulation which is slow (with
$\sim$MHz bandwidth) for high-speed QKD operations.

Silicon photonics recently has gained traction as the leading
platform for quantum communications with the promise of its high
density integration with the existing complementary
metal-oxide-semiconductor (CMOS) processes that have enabled
monolithic integration of both photonic and electronic components.
With no natural electro-optic nonlinearity, silicon photonics rely
on the slow thermo-optic phase modulation~\cite{Harris:14} to
achieve high-visibility interference~\cite{Wilkes:16}. Carrier
injection and depletion within an intrinsic region between p-doped
and n-doped silicon offer high-speed modulation within silicon
photonics, but with a phase-dependent loss that must be
mitigated~\cite{Reed2005,Soref1987}. Recently, MEMS-based phase
shifters have shown great promise in miniaturizing the device
further, in lowering the power consumption, and in achieving
gigahertz-bandwidth phase shifts without the undesirable phase-
dependent loss~\cite{mi7040069}.

While the development of a fully integrated light source within
the silicon photonics platform is still underway, the platform has
been proven to be highly amenable to heterogeneous bonding of the
active III--V materials mentioned
above~\cite{ma3031782,Keyvaninia:13,Heck2013,BenBakir:11,Liang:2010aa}.
Moreover, superconducting nanowire single photon detectors
(SNSPDs) have been integrated into silicon photonics using a
pick-and-place method, paving the way for a possible monolithic
compact QKD receiver with single photon counting
capabilities~\cite{Najafi2014}. Quantum-limited homodyne detectors
have also been demonstrated with sufficiently large noise
clearance between shot noise and electronic noise that can be
useful for CV QKD applications~\cite{Raffaelli2018,Michel2010}.

A recent demonstration of QKD with PIC uses an InP transmitter to
leverage its on-chip source capability and a TriPleX to leverage
its low-loss performance. The experiment showcased PIC's
flexibility in being able to demonstrate multiple time-bin encoded
protocols using the same chip set at a clock rate of one
GHz~\cite{Sibson2017}. More recently, recent demonstrations of
time-bin and polarization QKD transmitters in silicon photonics
with further miniaturized components hinted at possible
performance advantage over off-the-shelf fiber optical components
with LiNbO$_3$-based modulators~\cite{Sibson2016, Ma2016}. Silicon
photonics recently proved possible QKD operations using
polarization encoding over a 43-km intercity fiber link which was
commonly thought too unstable because of fiber polarization
drifts~\cite{Bunandar2018}. The experiment demonstrated secret-key
rate generation comparable to state-of-the-art time-bin
demonstrations but with polarization stabilization capabilities.
See Fig.\ref{fig:pic_qkd}.

The PIC platform also offers new methods of generating quantum
sources of light: single photons and entangled photon sources.
While weak coherent light is currently the most popular approach
for QKD operations, its Poissonian statistics create side-channel
vulnerability that must be closed with decoy state
approaches~\cite{decoy1,decoyWang,Wang4decoy,decoy,decoyfin}. QKD
with true single photons or entangled photons can circumvent this
problem without needing decoy state protocols~\cite{rev2}, which
consume random bits. In the InP platform, single photons can be
generated from quantum dots that are grown epitaxially to emit
light in the standard telecom 1550~nm window. In silicon
photonics, on-chip entangled pair sources based on spontaneous
four-wave mixing (SFWM) have been demonstrated without the need of
any off-chip filtering~\cite{Harris14,Leuthold2010}.

One important challenge that remains in these novel quantum
sources is in increasing the brightness to be sufficient for
gigahertz-clocked QKD operations. Currently, the amount of output
flux of these quantum sources has been limited at $\sim10$~MHz
even at near unity collection
efficiency~\cite{Wang2016a,Somaschi2016a}. These quantum sources
are typically pumped using a coherent laser. Increasing the pump
power of the quantum dot sources induces multi-photon emissions
which degrade the single photon purity. However, alternative
schemes of excitation have shown great promise of reducing the
probability of multi-photon emissions by several orders of
magnitude~\cite{Hanschke2018}. Increasing the pump power of the
entangled SFWM sources has been shown to induce two-photon
absorption which saturates the source brightness~\cite{Xu:06}.

Integrated photonics is poised to deliver major benefits towards
building QKD networks. The miniaturization of devices coupled with
highly robust manufacturing processes can accelerate the adoption
of QKD for real-world data encryption, especially with the MDI
configuration. In this setting, only several central receiver
nodes need to have cryogenic high-efficiency
SNSPDs~\cite{Marsili:2013aa}, while all the clients can make use
of personal PICs to generate secret keys among each other. The
lithographic precision afforded by the platform also promises the
possibility of identical integrated light sources for MDI
QKD~\cite{Preble2015,Li2017}.

In conclusion, PIC presents a novel opportunity to design new
devices that meet the needs of low-error QKD operations.
Investigations into new device physics through heterogeneous
integration of the multiple platforms can enable the development
of new quantum sources and receivers with superior
performance~\cite{Fathpour2015}. Furthermore, PIC's phase stable
platform also lends itself to highly-dense-multiplexed QKD
operations, which can dramatically increase the secret-key
generation rate.


\section{Satellite quantum communications}

\subsection{Introduction}

The quantum communication protocols on which QKD is based are very
well suited to be applied in space. Space channels, in connection
with ground single-links and networks, may be exploited in a
number of scenarios embracing the entire planet Earth, the
satellite networks around it and novel and more ambitious projects
aimed at more distant links with the Moon or other planets. In the
context of an evolving society that leverage more and more on
secure communications, space is expected to play a crucial role in
quantum communications as it is  now playing for global
communication, navigation and positioning, time distribution,
imaging and sensoring, realized by several generation of
satellites.

\subsection{The satellite opportunity}

The extension of the QKD to secure links to long distance, to
connect nodes of networks spanning large scales, including
national, continental, planetary as well as space missions, was
devised in feasibility studies more than a decade
ago~\cite{Rarity2002, Aspelmeyer2003,Pfennigbauer2005,
BonatoQComm9, TomaelloASR11}. The extension to space of quantum
communications (QC) was initially proposed in combination with
experiments devoted to testing fundamental principles and
resources of quantum information in the novel space context. Some
of these were directed to the development of a payload for the
International Space Station (ISS)~\cite{Ursin2009, Scheidl2013a},
others as standalone satellites~\cite{Aspelmeyer2003}.

These proposals were supported by the early evidence of long
distance free-space QC experiments on the
ground~\cite{Kurtsiefer2002, Hughes2002, UrsinNPhys2007}. In this
way it was proved that significant portion of atmosphere paths
were suitable not only for classical optical communications but
also for the quantum one. Indeed,  the  degrading role of the
atmosphere  on the channel performances was already assessed in
terms of  beam widening and wandering, fading of  signal and
scintillation at the receiver as a function of the turbulence
level, wavelength and link length~\cite{Fante1975a, Phillips}.
However, the single photon discrimination at the correct
wavelength, arrival time and direction as well as the detection
with an effective rejection of the background noise is more
demanding than the classical counterpart.

Starting in 2003 with an experimental campaign at the Matera Laser
Ranging Observatory (Italy), it was possible  to demonstrate that
the exchange of single photons are suitably achievable between a
LEO satellite and the ground~\cite{VilloresiMatera8}. In this
case, even without an active photon source in orbit, the
demonstration was obtained by exploiting  satellites equipped with
optical retroreflectors, and directing to them a train of pulses
with calibrated energy such that the collected portion that is
retroreflected back toward the transmitter on the Earth is a
coherent state with a content of a single photon or less. A
suitable bidirectional telescope on the ground allows for the
transmission of the uplink train of pulses and of the single
photons in downlink. This technique was then extended to
demonstrate QC using different degrees of freedom, as later
discussed~\cite{Vallone2015b, Vallone2016b} and is a candidate for
QKD with a very compact payload~\cite{Vallone2015b}.

The application of space QC for a global QKD was considered since
the beginning as an effective solution to joint separate networks
of fibre-based ground links. Indeed, the key exchange between a
trusted satellite and two ground terminals may then be used to
generate a secure key between the two
terminals via one-time pad.  
Despite these attracting opportunities for the improvement of
secure communications on ground, as well as other that have been
conceived for the use in space (described below), the realization
of a satellite for QKD was kept on hold in Europe and USA and
found at the beginning of this decade a concrete interest in Asia.
More in detail China and Japan put in their roadmaps the
demonstration of the space QKD with ambitious but concrete plans
to develop and launch satellites for QC. The Japanese SOTA
satellite was indeed launched in 2014 and Chinese Micius in 2015,
as will be described below. The perspective of using a very
compact payloads as nanosat o cubesat has recently vamped the
European initiatives, spurring for the development of space
components of great efficiency and small dimension~\cite{Oi2017,
Bedington2017}. Such direction is expected to be beneficial for
the ground QKD as well, for the realization of high performance
small components to be used in  compact and power-saving QKD
terminals on ground networks.

\subsection{Type of orbits and applications}

The  type of key exchange provided by an orbiting terminal changes
significantly with the type of orbit. The altitude has relevant
implications in the losses of the optical link. Although the
possible configuration of a space QKD setup could use the
transmitter in both the space terminal (downlink of the qubits)
and the ground terminal (uplink), the detrimental impact of the
atmosphere is asymmetric. Indeed in the uplink, the propagation of
the wavefront associated to the qubit stream in the turbulent
atmosphere occurs at the beginning of the path. This induce a non
uniform modulation of the wavefront phase. The subsequent
propagation results in the development of an amplitude modulation
at the satellite altitude, with a significant beam diameter
broadening and a scintillation that causes a fluctuation of the
link transmissivity. On the contrary, in the downlink the
propagation of the qubit train occurs in vacuum and get degraded
by the atmosphere only in the final portion, with an exponential
air density increase within the last 10 km. The broadening of the
beam at the receiving terminal is then mainly due to the
diffraction and the scintillation is also reduced. Therefore the
downlink is the common configuration, and the subsequent analysis
will be referred to it.

\subsubsection{Space-link losses}

The evaluation of the QKD rate in a space link is based on the
analysis of the losses and the fluctuations of the corresponding
optical channel. From classical studies in satellite optical
communications~\cite{Phillips}, we know that the geometric losses
(namely the losses due to diffraction) may be modeled considering,
at the transmitter, a Gaussian beam with waist $w_0$ passing
through a telescope aperture of diameter $D$. The far-field
distribution at distance $d\gg D$, can be written in terms of the
coordinates $(x,y)$ of the plane transverse to propagation as
\begin{align}
E(x,y) \propto \frac{D}{2a} \int_{X^2+Y^2\leq 1}
e^{\left[i\frac{2\pi}{\lambda}\frac{D}{2d}(x X+y
Y)-\frac{X^2+Y^2}{a^2}\right]} {\rm d}X{\rm d}Y,
\end{align}
where $a=2w_0/D$ is the ratio between the beam waist and the Tx
aperture radius. As first obtained by Siegman, to optimize the
received power it is necessary to choose $a\simeq 0.89$ for
classical communication. However, for QKD we may use different
values if we consider the single-photon regime after the Tx
aperture. By using $a\simeq2$ we obtain in the far-field (at
distance $d$ from the transmitter) a beam which is
well-approximated by a Gaussian beam with radius $w(d)\simeq
0.9\frac{\lambda}{D}d$. The total losses of the channel is
evaluated in dB as $\simeq-10\log_{10}\frac{D^2}{2w^2(d)}$ by
assuming a receiver aperture with equal diameter $D$.

In Fig.~\ref{F_sat_losses} we show the expected losses with a
selection of significant wavelengths and telescope diameters as a
function of the terminal separation. The range of losses is
radically different according to the orbit altitude, conditioning
the possible applications. In the classification below, we discuss
the roles played by the different satellites and types of orbits
for the purpose of QC.

\begin{figure}[h]
\includegraphics[width=0.47\textwidth]{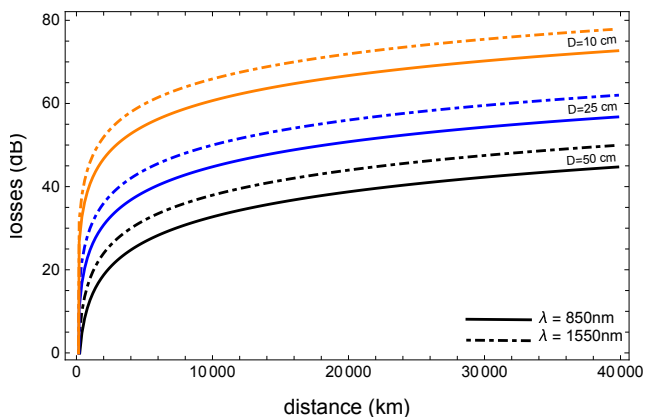}
\caption{Space-link losses for an optimized beam waist $w_0$ at
the transmitter, different telescope diameters $D$, and two
relevant wavelengths for space QKD, i.e., $\lambda=850$nm (solid
line) and $\lambda=1550$nm (dotted-dashed line).}
\label{F_sat_losses}
\end{figure}

\subsubsection{Low-Earth-orbit (LEO)}
This type of orbit, reaching not above the  altitude of 2000 km,
was the first choice to demonstrate QC protocols from space. This
is because of the relative ease to reach the orbit with multiple
launcher options combined with the lower exposition to the
aggressive ionizing radiation affecting higher altitudes. The
rapid round-trip time around Earth of about one to two hours
combined with a wide selection of orbit inclinations,  open
possibilities so as to cover all the planet in hours with a single
sat or to maintain a constant position relative to the Sun. Among
the limitations of LEO there is the fact that the passage over a
ground terminal is limited to just a few minutes of effective
link, whereas the sat is above the 10 degrees of elevation from
the horizon. Moreover, satellite speed relative to ground may
reach 7 km/s for a 400-km orbit, like that of the ISS, which
causes a varying Doppler shift of the order of tens of GHz.

LEO sats for QKD were the first to be considered~\cite{Ursin2009,
BonatoNJP9}, initially as payload to be operated on the ISS for
six months to one year, and then as independent spacecrafts.
Ajisai, a LEO sat devoted to geodynamic studies, was used as the
first source of single photons in orbit using its corner-cube
retroreflectors. These were illuminated by a train of pulses from
the Matera Laser Ranging Observatory (MLRO, Italy) in such a way
that a single photon was reflected on average by the
satellite~\cite{VilloresiMatera8}. This approach was later used
with 4 satellites equipped with polarization preserving
retroreflectors to realize an orbiting source of polarization
qubits, providing the experimental feasibility of the BB84
protocol on a space-link~\cite{Vallone2015b}. See
Fig.~\ref{VilloSAT}. The QBER observed was well within the
applicability of the BB84 protocol, and in line with criteria of
both general or pragmatic security~\cite{BaccoNComms2013}. Later
and still at MLRO, the use of temporal modes, or phase encoding,
was also demonstrated~\cite{Vallone2016b}.

\begin{figure}[ptbh]
\vspace{+0.1cm}
\par
\begin{center}
\includegraphics[width=0.47\textwidth] {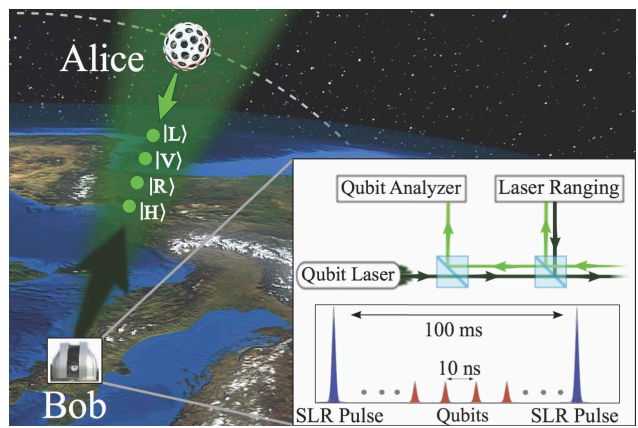} \vspace{-0.4cm}
\end{center}
\caption{Satellite QKD demonstration. A train of qubit pulses is
sent to the satellite with a repetition rate of 100 MHz. These are
reflected back from the satellite at the single photon level,
therefore mimicking a QKD source. In order to achieve
synchronization the experiment also employed a train of bright
satellite laser ranging (SLR) pulses at a repetition rate of 10
Hz. Reprinted figure with permission from Ref.~\cite{Vallone2015b}
\copyright APS (2015).} \label{VilloSAT}
\end{figure}

The Chinese satellite Micius was announced as a major step of the
space program of the Chinese Academy of Science~\cite{Xin2011} and
was launched on the 16th of August 2016. It provided the
experimental verification of various QC protocols in space. Indeed
the spacecraft was equipped as a quantum optics lab capable not
only to generate coherent and entangled states and to transmit
them to the ground but also to measure qubit sent by the ground
terminal. In this way, ground-to-satellite quantum teleportation
was realized by  teleporting  six input states in mutually
unbiased bases with an average fidelity of $0.80 \pm 0.01$ from
the Ngari ground station in Tibet~\cite{Ren2017}.  The decoy-BB84
protocol was realized with a key rate exceeding 10 kbps at about
the central part of the satellite passage. This remarkable result
was possible by very accurate pointing of the downward beam, whose
far field angle was about 10 $\mu$rad at $1 / e^2$ and the
pointing fluctuation was reported to be a factor five lower
\cite{Liao2017}. The wavelength for the qubit was chosen to be
about 850 nm and the observed losses were about 22 dB, in line
with the theoretical modeling based on a 300-mm telescope.

Entangled-based QKD was also demonstrated by Micius, using a high
visibility source onboard. The twin beam downlink was used to
establish a secret key (via violation of Bell inequalities)
between the two stations of Delingha and Lijiang, with a slant
distance of about 1200~km~\cite{Yin2017b}. Due to the composition
of the losses of the two channels, the QKD rate resulted of the
order of half a bps. In  2017, entangled-based QKD (in particular,
the Ekert protocol~\cite{Eke91}) was also realized with one of the
entangled photons measured  at the satellite while the other one
was detected at the receiver in the Delingha ground station. The
link losses ranged from 29 dB at 530 km to 36 dB at 1000 km,
allowing for a max key rate of a few bps~\cite{Yin2017}. Micius
was also used for demonstrating an intercontinental quantum
network, distributing the keys for a text and video exchange
between the ground stations of Xinglong (China), Nanshan (China)
and Graz (Austria)~\cite{Liao2018}.

%
%

In 2017, QKD was also demonstrated in downlink from the Tiangong-2
Space Lab, where a compact transmitter with a 200-mm telescope was
installed. Photons were transmitted down to the 1.2-m telescope at
the Nanshan ground station~\cite{Liao2017b}. The key rate was
assessed to reach beyond 700 bps with about 30 dB of losses. In
the same year, a Japanese team at the NICT  developed the SOTA
lasercom terminal for testing the optical downlinks as well as QC
with a low-cost platform, the microsatellite SOCRATES at an
altitude of 650 km~\cite{Takenaka2017a}. 

Since the beginning of this decade, the cubesats have grown
rapidly in the several areas of space science and technology,
including space QC~\cite{Oi2017, Bedington2017}. Two main tasks
are envisaged for such small sats: the test of novel technology
for QC in the space context and the operation of a apace network
for capillary coverage of low rate QKD. For the first purpose, the
Singapore NUS team developed a prototype of source and detector
that was successfully operated first on a balloon and then in
Space~\cite{Tang2015a, Tang2016a}. Several proposal of cubesat use
have been put forward worldwide (e.g., see~\cite{Bedington2017}).

\subsubsection{Higher Earth orbits (MEO and GEO)}

The medium Earth orbit (MEO) is above LEO and below the
geostationary orbit (GEO), the latter being at 35,786 km above
Earth's equator. The MEO includes the Global Navigation Satellite
Systems (GNSS) while GEO includes weather and communication
satellites. These higher orbits are preferable because they would
extend the link duration (becoming permanent for a GEO). However,
they involve larger losses and the payloads are exposed to much
more aggressive ionizing radiation from the Sun.

The first experimental single-photon exchange with a MEO sat at
7000 km of slant distance was realized in 2016 at
MLRO~\cite{Dequal2016}. The QKD links were modeled in previous
studies~\cite{BonatoQComm9,TomaelloASR11}. A recent result
addressing the photon exchange with two Glonass sats has supported
the future possibility of QKD-enabled secure services for the GNSS
satellites~\cite{Calderaro2018a}. This opportunity is seriously
considered, given the critical service that the navigation system
are playing in several continents. Finally, the feasibility of
quantum-limited measurement of optical signal from an existing GEO
communication satellite has been recently carried
out~\cite{Gunthner2017}.



\subsubsection{Night and day use of the link}

Space QKD was so far investigated experimentally during night-time
only. However, the operation  in daylight is of great interest for
a significative expansion of the satellite usage. The possibility
of a daylight use in inter-satellite communication was supported
by a study on the ground~\cite{Liao2017a}. The key ingredients
were a strong rejection of the background radiation, via a precise
pointing and a narrow field-of-view, together with the reduction
of the temporal integration interval for the arrival qubits,
obtained by means of a very precise temporal synchronization.
Finally, the wavelength of 1550 nm was used thanks to its lower
scattering.

\subsection{Beyond satellite QKD}

Several areas may be found in which quantum communication from and
in space are crucial. Below we review some possible other
protocols (beyond QKD) that can be realized by sending single
photons at large distance in space. We discuss some fundamental
tests that were and can be realized in this context.

\subsubsection{Other protocols} Some possible protocols that can be
realized with long distance quantum communication are quantum
digital signature (QDS) and blind quantum computing (BQC). A QDS
refers to the quantum mechanical equivalent of a digital signature
(see Sec.~\ref{SectionQDSs}). In a QDS protocol, Alice sends a
message with a digital signature to two recipients, Bob and
Charlie. Then QDS guarantees nonrepudiation, unforgeability, and
transferability of a signature with information-theoretical
security. A very recent long-distance ground demonstration
exploiting decoy states has been realized without assuming any
secure channel. A one-bit message was successfully signed through
a 102-km optical fiber~\cite{HLyin2017}.

A second example is BQC where a client sends a quantum state
$|\psi\rangle$ to the server, with such state encoding both the
chosen algorithm and the input (for a review see
Ref.~\cite{fitzsimons2017}). For a cloud computer (an in
particular a cloud quantum computer), the privacy of the users may
be a serious issue. BQC allows a client to execute a quantum
algorithm by using one or more remote quantum computers while at
the same time keeping the results of the computation hidden. By
satellite quantum communication it could be possible to send
quantum states from a satellite to ground servers that may perform
the BQC.

\subsubsection{Tests of quantum mechanics in space}
Quantum communication in free space at large distance is the key
ingredient to perform fundamental tests of quantum mechanics in an
unexplored scenario. Indeed, as for any scientific theory, quantum
mechanics can be considered valid only within the limits in which
it has been experimentally verified. By exploiting quantum
communication in space it is possible to extend such limits, by
observing quantum phenomena in unexplored conditions, such as
moving terminals and/or larger and larger distances. The possibly
interplay of quantum mechanics with general (or special)
relativity can be studied in this
context~\cite{Rideout2012,Agnesi2018}. Bell's inequality with
observers at relative motion and gravitational-induced redshifts
on quantum objects are some very significant experiments that can
be performed in space (for a detailed review of these possible
experiments see~\cite{Rideout2012}). As paradigmatic examples of
the possibilities offered by space quantum communication we may
recall two recent demonstrations: the violation of a Bell's
inequality at a distance of about 1200km~\cite{Yin2017b} and the
Wheeler's delayed-choice experiment along a 3500-km space
channel~\cite{Vedovato2017}.

As we know, Bell's inequalities~\cite{bell1964} demonstrate that a
local hidden variable model cannot reproduce the experimental
results that can be achieved by entangled states. Nowadays, Bell's
inequality are used as a simple and effective tool to {\it
certify} the presence of entanglement between separate observers.
In 2017, the Micius satellite, orbiting at an altitude of about
500km and hosting a source of polarization entangled photons,
allowed the demonstration of the persistence of entanglement at
the record distance of 1200km between the two ground station of
Delingha and Lijiang in China~\cite{Yin2017b}. The experiment
realized the violation of the CHSH inequality, with a value $S =
2.37 \pm 0.09 $ larger than the limit of 2 by four standard
deviations. This result confirmed the nonlocal feature of quantum
mechanics excluding the local models of reality on the thousand km
scale. Previous demonstrations~\cite{inagaki2013,UrsinNPhys2007},
using fiber or ground free-space links were limited to one order
of magnitude less in distance, due to photon loss in the fiber or
the Earth curvature for ground free-space links.

Quantum mechanics predicts that quantum entanglement should be
measured at any distance: however, it is tempting to challenge
such prediction and verify if some unexpected effects (such as
gravitational influence) will put some limits of such distance.
The availability of quantum communication in space now allows to
extend such limit at larger and larger distance. For instance, by
using an entangled source on a GEO satellite that sends the two
photons on ground, it would be possible to increase by one order
of magnitude the distance between two entangled photons.

The second example is Wheeler's delayed-choice
experiment~\cite{wheeler1978}, a wave-particle duality test that
cannot be fully understood using only classical concepts.
Wave-particle duality implies that is not possible to reveal both
the wave- and particle-like properties of a quantum object at the
same time. Wheeler's gedankenexperiment was invented to highlight
the contradictory interpretation given by classical physics on a
single photon measured by Mach-Zehnder interferometer (MZI). In
his idea, a photon emerging from the first beam splitter (BS) of a
MZI may find two alternative configurations: the presence or
absence of a second BS at the output of the interferometer. In the
former/latter case the apparatus reveals the wave/particle-like
character of the photon. In a classical interpretation, one could
argue that the photon decides its nature at the first BS. However,
if the MZI configuration is chosen {\it after} the photon entered
the interferometer (hence the name delayed-choice), a purely
classical interpretation of the process would imply a violation of
causality. Several implementations of Wheeler's Gedankenexperiment
have been realized on the ground~\cite{ma2016}. In the experiment
of Ref.~\cite{jacques2007}, a space-like separation between the
choice of the measurement and the entry of the particle into the
interferometer was achieved with a 48-m-long polarization
interferometer and a fast EOM controlled by a QRNG.

Then, in Ref.~\cite{Agnesi2018}, the delayed-choice paradigm has
been extended to space, by exploiting the temporal degree of
freedom of photons reflected by a rapidly moving satellite in
orbit. The two time bins represents the two distinct paths of the
interferometer. Photon polarization was used as an ancillary
degree of freedom to choose the insertion or removal of the BS at
the measurement apparatus and thus observe interference  or
which-path information. The experiment showed the correctness of
the wave-particle model for a propagation distance of up to 3500
km, namely at a much larger scale than all previous experiments.

\subsection{Concluding remarks} We have reviewed the
opportunities offered by space quantum communications and their
possible applications. In particular, they are expected to have a
great impact in the creation of a secure quantum network around
the globe. The design of a QKD network in space encompass the
realization of the single-link connections, the modeling of their
performances and their further exploitation in networks based on
multiples ground stations. The study of such features needs
further investigations both theoretically and experimentally.


\section{Continuous-variable QKD\label{CV-QKD}}

\subsection{Brief introduction to CV systems}



Recall that CV quantum systems are described by an
infinite-dimensional Hilbert space~\cite{SamRMPm,RMPwee}. Consider
a quantum system made of $n$ bosonic modes of the electromagnetic
field with tensor-product Hilbert space
$\otimes_{k=1}^{n}\mathcal{H}_{k}$ and associated $n$ pairs of
field operators $\hat{a}_{k}^{\dag}$, $\hat{a}_{k}$, with
$k=1,\dots,n$. For each mode $k$ we can define the following field
quadratures
\begin{align} \label{quadDEF}
\hat{q}_{k}  &  :=\hat{a}_{k}+\hat{a}_{k}^{\dag},\\
\hat{p}_{k}  &  :=i\left(  \hat{a}_{k}^{\dag}-\hat{a}_{k}\right).
\end{align}
These operators can be arranged in an $N$-mode vector
$\mathbf{\hat{x}:=}\left(
\hat{q}_{1},\hat{p}_{1},\dots,\hat{q}_{n},\hat{p}_{n}\right)
^{T}$. Using the standard bosonic commutation relation, for
field's creation ($\hat{a}_{k}^{\dag}$) and annihilation
($\hat{a}_{k}$) operators, one can easily verify that the any
pairs of entries of vector $\mathbf{x}$ satisfy the following
commutation relation
\begin{equation}\label{symFORM}
\left[  \hat{x}_{l},\hat{x}_{m}\right]
=2i\Omega_{lm},~~\Omega_{lm}=\left(
\begin{array}
[c]{cc}
0 & 1\\
-1 & 0
\end{array}
\right), \eeq where $\Omega_{lm}$ is the symplectic form
\cite{RMPwee}. From Eqs.~(\ref{quadDEF})-(\ref{symFORM}) we can
see that the vacuum noise is here set to $1$.

An $n$-mode quantum state can be represented either as a density
operator $\hat{\rho}$ acting on $\otimes_{k=1}^{n}\mathcal{H}_{k}$
or as a Wigner function defined over a $2n$-dimensional phase
space (see Ref.~\cite{RMPwee} for more details). In particular, a
state is Gaussian if its Wigner function is Gaussian, so that it
is completely characterized by the first two statistical moments,
i.e., the mean value $ \bar{\mathbf{x}}:=\left\langle
\mathbf{\hat{x}}\right\rangle = \mathrm{Tr} \( \mathbf{\hat{x}}
\hat{\rho} \) $  and covariance matrix (CM) $\mathbf{V}$, whose
arbitrary element is defined by
\begin{equation}
V_{ij}:=\tfrac{1}{2}\left\langle \left\{  \Delta\hat{x}_{i},\Delta\hat{x}%
_{j}\right\}  \right\rangle , \label{CM_definition}%
\end{equation}
where $\Delta\hat{x}_{i}:=\hat{x}_{i}-\langle\hat{x}_{i}\rangle$
and $\{,\}$ is the anti-commutator.

For a single-mode, one can consider different classes of quantum
states, the most known are the coherent states. These are states
with minimum (vacuum) noise uncertainty, symmetrically distributed
in the two quadratures, and characterized by their complex
amplitudes in the phase space. They are denoted as
$|\alpha\rangle$, where $\alpha = \bar{q}+i \bar{p}$, where
$(\bar{q},\bar{p})$ are the components of the mean value. Another
important class is that of squeezed states, where the noise is
less than the vacuum in one of the two quadratures (while greater
than in the other)~\cite{RMPwee}. The reader can consult
Appendix~\ref{sec:GaussAPPENDIX} for more details on the CV
notation (which typically varies between quantum information and
quantum optics) and a number of formulas that are relevant for
calculations with Gaussian states.

The basic one-way CV-QKD protocols can be classified with respect
to the quantum states employed (coherent or squeezed), the type of
encoding adopted (Gaussian modulation or discrete alphabet), and
the type of measurement used (homodyne or heterodyne detection).
In particular, Gaussian protocols based on the Gaussian modulation
of Gaussian states have received an increasing attention in the
latest years, not only because Gaussian states are routinely
produced in quantum optics labs but also because they are
relatively easy to study, due to their description based on mean
value and CM.

\subsection{Historical outline}

As an alternative to DV-QKD protocols, which are ideally based on
a single photon detection, CV-QKD protocols encode keys into CV
observables of light fields~\cite{SamRMPm} that can be measured by
shot-noise limited homodyne detection. In a homodyne detector an
optical signal is coupled to a shot-noise limited strong local
oscillator (LO) beam on a balanced beamsplitter and the light
intensities on the output ports are measured. Depending on the
optical phase difference between the signal and LO, the difference
of photocurrents produced at each of the two detectors will be
proportional to one of the two field quadratures. The LO therefore
carries the phase reference, which allows to switch between the
measurement of $q-$ and $p-$quadrature (or more generally perform
the state tomography by measuring the Wigner function associated
to the state).

The first proposal of using the quadratures of the bosonic field
for implementing QKD dates back to 1999, when
Ralph~\cite{Ralph1999} considered the encoding of key bits by
using four fixed quadrature displacements of bright coherent or
two-mode entangled beams. Later, Ralph discussed the security of
the two-mode entanglement-based scheme in more
detail~\cite{Ralph2000}, considering not only intercept-resend
attacks but also CV teleportation. The latter was identified as an
optimal attack against the protocol, imposing the requirements of
high signal squeezing and low channel loss~\cite{Ralph2000}.
Independently, Hillery~\cite{Hillery2000} suggested a CV-QKD
protocol based on quadrature encoding of a single-mode beam,
randomly squeezed in one of the quadrature directions. Security
against intercept-resend and beam-splitting attacks were assessed
on the basis of the uncertainty principle. Another early CV-QKD
scheme was suggested by Reid~\cite{Reid2000} and based on the
verification of EPR-type correlations to detect an eavesdropper.

In 2000 Cerf et al.~\cite{Cerf2001} proposed the first \textit{all
continuous} QKD protocol, where the quadratures of a squeezed beam
were used to encode a Gaussian-distributed secure key. The
security of the protocol was shown against individual attacks
based on the uncertainty relations and the optimality of a quantum
cloner. Later, reconciliation procedures were introduced for
Gaussian-distributed data, which allowed to implement error
correction and privacy amplification close to the theoretical
bounds~\cite{VanAssche2004}. Another CV-QKD protocol based on the
Gaussian modulation of squeezed beams was suggested by Gottesman
and Preskill~\cite{Gottesman2001}. This protocol was shown to be
secure against arbitrary attacks at feasible levels of squeezing,
by using quantum error-correcting codes.

In 2001 Grosshans and Grangier introduced a seminal coherent-state
protocol with Gaussian quadrature modulation and showed its
security against individual attacks~\cite{Grosshans2002} by
resorting to the CV version of the no-cloning
theorem~\cite{Grosshans2001}. The standard protocol based on the
direct reconciliation (DR), where Alice is the reference side for
the information post-processing, was however limited to 50\%
channel transmittance, i.e., 3dB. As an attempt to beat the 3dB
limit, the use of post-selection in CV-QKD was suggested by
Silberhorn et al.~\cite{Silberhorn2002}. Alternatively, it was
shown that the use of the reverse reconciliation (RR), where the
reference side is Bob, allowed the coherent-state protocol to be
secure against individual attacks up to arbitrarily-low channel
transmittances~\cite{Grosshans2002b}. In 2004, the heterodyne
detection was then suggested for coherent-state protocols by
Weedbrook et al.~\cite{Weedbrook2004}. This \textit{non-switching
protocol} has the advantage that both the quadratures are
measured, therefore increasing the key rate.

The security of CV-QKD against collective Gaussian attacks was
shown independently by Navascu\'es et al.~\cite{Navascues2006} and
by Garc\'ia-Patr\'on and Cerf~\cite{Garcia-Patron2006}. Collective
Gaussian attacks were fully characterized by Pirandola et
al.~\cite{pirs-attacks-PRL08}, who later derived the secret-key
capacities for CV-QKD~\cite{ReverseCAP,PLOB}. Security against
collective attacks was extended to the general attacks by Renner
and Cirac~\cite{renner-cirac} using the quantum de Finetti theorem
applied to infinite-dimensional systems. This concluded the
security proofs for the basic one-way CV-QKD protocols in the
asymptotic limit of infinitely large data
sets~\cite{Garcia-Patron2007,Usenko2016}. Next developments were
the study of finite-size effects and fully composable proofs (e.g.
see Ref~\cite{leverrier-comp}).

\subsection{One-way CV-QKD protocols}\label{CVonewayQKD}

The family of one-way CV-QKD protocols can be divided into four
major ones, depending on the signal states and the type of
measurements applied. It was already mentioned that CV-QKD can be
realized using coherent or squeezed signal states, and the
homodyne measurement is used to obtain quadrature value of an
incoming signal. As an alternative to the homodyne detection, the
heterodyne measurement can be applied. Here the signal mode is
divided on a balanced beamsplitter and $q$- and $p$-quadratures
are simultaneously detected using homodyne detectors at the
outputs. A vacuum noise is then unavoidably being mixed to the
signal.

The \textquotedblleft prepare and measure\textquotedblright\
realization of a generic one-way CV-QKD protocol includes the
following steps:

\begin{itemize}
\item Alice encodes a classical variable $\alpha$ in the amplitudes of
Gaussian states which are randomly displaced in the phase space by
means of a zero-mean Gaussian distribution, whose variance is
typically large. If coherent states are used, the modulation is
symmetric in the phase space. If squeezed states are used instead,
then the displacement is along the direction of the squeezing and
Alice randomly switches between $q$- and $p$- squeezings.

\item Alice then sends the modulated signal states to Bob through the quantum
channel, which is typically a thermal-loss channel with
transmissivity $\eta$ and some thermal noise, quantified by the
mean number of thermal photons in the environment $\bar{n}$ or,
equivalently, by the excess noise $\varepsilon = \eta^{-1}
(1-\eta) \bar{n}$. In some cases, one may have a fading channel
where the channel's transmissivity varies over time (e.g. due
turbulence)~\cite{PanosFADING}.

\item At the output of the quantum channel, Bob performs homodyne or
heterodyne detection on the incoming signals, thus retrieving his
classical variable $\beta$. If homodyne is used, this is randomly
switched between the $q$- and the $p$- quadratures.

\item If Alice and Bob have switched between different quadratures, they will
implement a session of classical communication (CC) to
reconciliate their bases, so as to keep only the choices
corresponding to the same quadratures (sifting).

\item By publicly declaring and comparing part of their sifted data, Alice and
Bob perform parameter estimation. From the knowledge of the
parameters of the quantum channel, they can estimate the maximum
information leaked to Eve, e.g., in a collective Gaussian attack.
If this leakage is above a certain security threshold, they abort
the protocol.

\item Alice and Bob perform error correction and privacy amplification on
their data. This is done in DR if Bob aims to infer Alice's
variable, or RR if Alice aims to infer Bob's one.
\end{itemize}

\subsection{Computation of the key rate}

In a Gaussian CV-QKD protocol, where the Gaussian signal states
are Gaussianly-modulated and the outputs are measured by homodyne
or heterodyne detection, the optimal attack is a collective
Gaussian attack. Here Eve combines each signal state and a vacuum
environmental state via a Gaussian unitary and collects the output
of environment in a quantum memory for an optimized and delayed
joint quantum measurement. The possible collective Gaussian
attacks have been fully classified in Ref.~\cite{ReverseCAP}. A
realistic case is the so-called entangling
cloner~\cite{Grosshans2001} where Eve prepares a TMSV state with
variance $\omega=\bar{n}+1$ and mixes one of its modes with the
signal mode via a beam-splitter with transmissivity $\eta$,
therefore resulting in a thermal-loss channel. In this scenario,
the asymptotic secret key rates in DR ($\blacktriangleright$) or
RR ($\blacktriangleleft$) are respectively given by
\begin{align}
R^{\blacktriangleright}  &  =\xi I(\alpha:\beta)-I(\alpha:E),\label{k1d}\\
R^{\blacktriangleleft}  &  =\xi I(\alpha:\beta)-I(\beta:E), \label{k2d}%
\end{align}
where $\xi\in(0,1)$ is the reconciliation efficiency, defining how
efficient are the steps of error correction and privacy
amplification, $I(\alpha:\beta)$ is Alice and Bob's mutual
information on their (sifted) variables $\alpha$ and $\beta$,
while $I(\alpha:E)$ [$I(\beta:E)$] is Eve's Holevo information on
Alice's (Bob's) variable.

Theoretical evaluation of these rates is performed in the
equivalent entanglement-based representation of the protocol,
where Alice's preparation of signal states on the input mode $a$
is replaced by a TMSV state $\Phi_{aA}^{\mu}$ in modes $a$ and
$A$. A Gaussian measurement performed on mode $A$ is able to
remotely prepare a Gaussian ensemble of Gaussian states on mode
$a$. For instance, if $A$ is subject to heterodyne, then mode $a$
is projected onto a coherent state whose amplitude is one-to-one
with the outcome of the heterodyne and is Gaussianly modulated in
phase space with variance $\mu-1$. In this representation, Alice's
classical variable is equivalently represented by the outcome of
her measurement.

Once mode $a$ is propagated through the channel, it is perturbed
by Eve and received as mode $B$ by Bob. Therefore, Alice and Bob
will share a bipartite state $\rho_{AB}$. In the worst case
scenario, the entire purification of $\rho_{AB}$ is assumed to be
held by Eve. This means that we assume a pure state $\Psi_{ABE}$
involving a number of extra modes $E$\ such that
$\mathrm{Tr}_{E}(\Psi_{ABE})=\rho_{AB}$. For a Gaussian protocol
under a collective Gaussian attack, we have that $\Psi_{ABE}$ is
pure, so that the Eve's reduced output state
$\rho_{E}:=\mathrm{Tr}_{AB}(\Psi_{ABE})$ has the
same entropy of $\rho_{AB}$, i.e.,%
\begin{equation}
S(E):=S(\rho_{E})=S(\rho_{AB}):=S(AB). \label{ent1d}%
\end{equation}

Assuming that Alice and Bob performs rank-1 Gaussian measurements
(like homodyne or heterodyne), then they project on pure states.
In DR, this means that the output $\alpha$ of Alice measurement,
with probability $p(\alpha)$, generates a pure conditional
Gaussian state $\Psi_{BE|\alpha}$ whose CM does not depend on the
actual value of $\alpha$. Then, because the reduced states
$\rho_{B(E)|\alpha}:=\mathrm{Tr}_{E(B)}(\Psi_{BE|\alpha})$ have
the same
entropy, we may write the following equality for the conditional entropies%
\begin{align}
S(E|\alpha)  &  :=\int d\alpha~p(\alpha)S(\rho_{E|\alpha})\nonumber\\
&  =S(\rho_{E|\alpha})=S(\rho_{B|\alpha})\nonumber\\
&  =\int d\alpha~p(\alpha)S(\rho_{B|\alpha}):=S(B|\alpha). \label{ent2d}%
\end{align}
Similarly, in RR, we have Bob's outcome $\beta$ with probability
$p(\beta)$ which generates a pure conditional Gaussian state
$\Psi_{AE|\beta}$ with similar properties as above. In terms of
the reduced states $\rho_{A(E)|\beta
}:=\mathrm{Tr}_{E(A)}(\Psi_{AE|\beta})$ we write the conditional entropies%
\begin{align}
S(E|\beta)  &  :=\int d\beta~p(\beta)S(\rho_{E|\beta})\nonumber\\
&  =S(\rho_{E|\beta})=S(\rho_{A|\beta})\nonumber\\
&  =\int d\beta~p(\beta)S(\rho_{A|\beta}):=S(A|\beta). \label{ent3d}%
\end{align}

By using Eqs.~(\ref{ent1d}), (\ref{ent2d}) and~(\ref{ent3d}) in
the key rates of Eqs.~(\ref{k1d}) and~(\ref{k2d}) we may simplify
the Holevo quantities as
\begin{align}
I(\alpha &  :E):=S(E)-S(E|\alpha)=S(AB)-S(B|\alpha),\label{ii1}\\
I(\beta &  :E):=S(E)-S(E|\beta)=S(AB)-S(A|\beta). \label{ii2}%
\end{align}
This is a remarkable simplification because the two rates are now
entirely computable from the output bipartite state $\rho_{AB}$
and its reduced versions $\rho_{B|\alpha}$ and $\rho_{A|\beta}$.
In particular, because all these state are Gaussian, the von
Neumann entropies in Eqs.~(\ref{ii1}) and~(\ref{ii2}) are very
easy to compute from the CM of $\rho_{AB}$. Similarly, the mutual
information $I(\alpha:\beta)$ can be computed from the CM. Given
the expressions of the rates, one can also compute the security
thresholds by solving $R^{\blacktriangleright}=0$ or
$R^{\blacktriangleleft }=0$.

\subsection{Ideal performances in a thermal-loss channel}

The ideal performances of the main one-way Gaussian protocols can
be studied in a thermal-loss channel, assuming asymptotic
security, perfect reconciliation ($\xi=1$), and infinite Gaussian
modulation. Let us consider the entropic function
\begin{equation}\label{entropicH}
s(x):=\frac{x+1}{2}\log_{2}\frac{x+1}{2}-\frac{x-1}{2}\log_{2}\frac{x-1}{2},
\end{equation}
so that $s(1)=0$ for the vacuum noise. For the protocol with
Gaussian-modulated coherent states and homodyne
detection~\cite{Grosshans2001}, one has
\begin{align}
R_{\text{coh,hom}}^{\blacktriangleright}  &
=\frac{1}{2}\log_{2}\frac {\eta\left(  1-\eta+\eta\omega\right)
}{\left(  1-\eta\right)  \left[
\eta+(1-\eta)\omega\right]  }-s(\omega)\nonumber\\
&  +s\left[
\sqrt{\frac{\eta+(1-\eta)\omega}{1-\eta+\eta\omega}\omega
}\right]  ,\\
R_{\text{coh,hom}}^{\blacktriangleleft}  &
=\frac{1}{2}\log_{2}\frac{\omega }{(1-\eta)\left[
\eta+(1-\eta)\omega\right]  }-s(\omega).
\end{align}
For the non-switching protocol with Gaussian-modulated coherent
states and
heterodyne detection~\cite{Weedbrook2004}, one instead has%
\begin{align}
R_{\text{coh,het}}^{\blacktriangleright}  &
=\log_{2}\frac{2}{e}\frac{\eta
}{(1-\eta)\left[  1+\eta+(1-\eta)\omega\right]  }-s(\omega)\nonumber\\
&  +s\left[  \eta+\omega(1-\eta)\right]  ,\label{coPROT}\\
R_{\text{coh,het}}^{\blacktriangleleft}  &
=\log_{2}\frac{2}{e}\frac{\eta
}{(1-\eta)\left[  1+\eta+(1-\eta)\omega\right]  }-s(\omega)\nonumber\\
&  +s\left[  \frac{1+(1-\eta)\omega}{\eta}\right]  . \label{coPROT2}%
\end{align}

For the protocol with Gaussian-modulated squeezed states (in the
limit of infinite squeezing) and homodyne
detection~\cite{Cerf2001}, here we
analytically compute%
\begin{align}
R_{\text{sq,hom}}^{\blacktriangleright}  &  =\frac{1}{2}\left[  \log_{2}%
\frac{\eta}{1-\eta}-s(\omega)\right]  ,\\
R_{\text{sq,hom}}^{\blacktriangleleft}  &  =\frac{1}{2}\left[  \log_{2}%
\frac{1}{1-\eta}-s(\omega)\right]  . \label{negg}%
\end{align}
Note that, for this specific protocol, a simple bound can be
derived at low $\eta$ and low $\bar{n}$, which is given
by~\cite{Lasota2017}
$R_{\text{sq,hom}}^{\blacktriangleleft}\approx(\eta-\bar{n})\log_{2}{e}%
+\bar{n}\log_{2}\bar{n}$, which provides a security threshold $\bar{n}%
_{max}(\eta)=\exp[1+W_{-1}(-\eta/e)]$ in terms of the Lambert
W-function.

Finally, for the protocol with Gaussian-modulated
infinitely-squeezed states
and heterodyne detection~\cite{Garcia2009}, here we analytically compute%
\begin{align}
R_{\text{sq,het}}^{\blacktriangleright}  &
=\frac{1}{2}\log_{2}\frac{\eta
^{2}\omega}{(1-\eta)\left[  1+(1-\eta)\omega\right]  }-s(\omega),\\
R_{\text{sq,het}}^{\blacktriangleleft}  &
=\frac{1}{2}\log_{2}\frac {1-\eta+\omega}{(1-\eta)\left[
1+(1-\eta)\omega\right]  }-s(\omega
)\nonumber\\
&  +s\left[  \sqrt{\frac{\omega\left[  1+\omega(1-\eta)\right]  }%
{1+\omega-\eta}}\right]  .
\end{align}
Note that this is a particular case of protocol where trusted
noise added at the detection can have beneficial effects on its
security threshold~\cite{Usenko2016}. In CV-QKD this effect was
studied in Refs.~\cite{Garcia2009,RevCohINFO,Madsen2012} and later
in Refs.~\cite{CarloSpie,GanEPJD,TQC} as a tool to increase the
lower bound to the secret key capacity of the thermal-loss and
amplifier channels. In particular, the protocol presented in
Ref.~\cite{TQC} has the highest-known security threshold so far
(see also Sec.~\ref{sec:ultimateQKD}).

Also note that for a pure-loss channel ($\omega=1$), we find
\begin{equation}
R_{\text{sq,het}}^{\blacktriangleleft}=R_{\text{sq,hom}}^{\blacktriangleleft
}=\frac{1}{2}\log_{2}\frac{1}{1-\eta}, \label{sif}%
\end{equation}
which is half of the PLOB bound $-\log_{2}(1-\eta)$. According to
Ref.~\cite{ReverseCAP}, this bound is achievable if one of these
two protocols is implemented in the entanglement-based
representation and with a quantum memory. In particular for the
squeezed-state protocol with homodyne detection, the use of the
memory allows Alice and Bob to always choose the same quadrature,
so that we may remove the sifting factor $1/2$ from
$R_{\text{sq,hom}}^{\blacktriangleleft}$ in Eq.~(\ref{sif}).

\subsection{Finite-size aspects}

The practical security of CV-QKD~\cite{rev3} deals with finite
data points obtained experimentally. In this finite-size regime
the security of CV-QKD was first analyzed against collective
attacks~\cite{Leverrier2010} by including corrections to the key
rate taking into account of the data points used and discarded
during parameter estimation and the convergence of the smooth
min-entropy towards the von Neumann entropy. The channel
estimation in the finite-size regime of CV-QKD was further studied
in Ref.~\cite{Ruppert2014} where it was suggested the use of a
double Gaussian modulation, so that two displacements are applied
and each signal state can be used for both key generation and
channel estimation. See also Ref.~\cite{Thearle2016} for excess
noise estimation using the method of moments.


The finite-size security under general coherent attacks have been
also studied. Ref.~\cite{furrer-comp,furrer-comp2} used entropic
uncertainty relations for the smooth entropies to show this kind
of security for an entanglement-based protocol based on TMSV
states. The analysis was extended to the protocol with squeezed
states and homodyne detection~\cite{Furrer2014}. Finite-size
security for one-way coherent-state protocols against general
attacks was studied in Ref.~\cite{Leverrier2013} by using
post-selection and employing phase-space symmetries. More
recently, it was shown that, for coherent-state protocols, the
finite-size security under general attacks can be reduced to
proving the security against collective Gaussian attacks by using
the de Finetti reduction~\cite{leverrier-comp}.

\subsection{Two-way CV-QKD protocols\label{two-way-QKD}}

In a two-way CV-QKD scheme~\cite{PIRS-2way}, similar to its DV
counterpart~\cite{felbinger,LM05}, Alice and Bob\ use twice the
insecure channel in order to share a raw key. During the first
quantum communication, Bob randomly prepares and sends a reference
state to Alice who, in turn, encodes her information by performing
a unitary transformation on the received state, before sending it
back to Bob for the final measurement. The appeal of this type of
protocol is its increased robustness to the presence of excess
noise in the channel. From an intuitive point of view, this is due
to the fact that Eve needs to attack both the forward and backward
transmissions in order to steal information, resulting in an
increased perturbation of the quantum system. The promise is a
higher security threshold with respect to one-way protocols.

Let us now describe in detail a two-way protocol based on coherent
states and heterodyne detection~\cite{PIRS-2way}. Here Bob
prepares a reference coherent state $|\beta\rangle$ whose
amplitude is Gaussianly-modulated with variance $\mu_{B}$. This is
sent through the quantum channel and received as a mixed state
$\rho(\beta)$ by Alice. At this point, Alice randomly decides to
close (ON) or open (OFF) the \textquotedblleft
circuit\textquotedblright\ of the quantum communication. When the
circuit is in ON, Alice encodes a classical variable $\alpha$ on
the reference state by applying a displacement $\hat {D}(\alpha)$
whose amplitude is Gaussianly-modulated with variance $\mu_{A}$.
This creates the state
$\hat{D}(\alpha)\rho(\beta)\hat{D}(-\alpha)$ which is sent back to
Bob, where heterodyne detection is performed with outcome
$\gamma\simeq\alpha+\beta$. From the knowledge of $\gamma$\ and
$\beta$, Bob can generate a post-processed variable
$\alpha^{\prime}\simeq\alpha$. In DR, the key is generated by Bob
trying to infer Alice's variable $\alpha$. In RR, the situation is
reversed with Alice guessing Bob's variable $\alpha^{\prime}$.

When the circuit is in OFF, Alice first applies heterodyne
detection on the incoming reference state $\rho(\beta)$, obtaining
a variable $\beta^{\prime}$. Then she prepares a new
Gaussian-modulated coherent state $|\alpha\rangle$ to be sent back
to Bob. His heterodyne detection provides an output variable
$\alpha^{\prime}$. In such a case the parties may use the
variables $\{\beta,\beta^{\prime}\}$ and
$\{\alpha,\alpha^{\prime}\}$ to prepare the key. In DR the
variables $\beta$ and $\alpha$ are guessed while, in RR, the
primed variables $\beta^{\prime}$ and $\alpha^{\prime}$ are. For
both the ON and the OFF configuration, Alice and Bob publicly
declare and compare a fraction of their data in order to estimate
the transmissivity and noise present in the round-trip process.

Note that the control of the parties over the ON/OFF setup of the
two-way communication represents an additional \emph{degree of
freedom} evading Eve's control. As a matter of fact, Alice and Bob
may decide which setup to use for the generation of the key. The
safest solution is to use the ON configuration if Eve performs an
attack of the round-trip based on two memoryless channels, while
the OFF\ configuration is used when Eve performs an attack which
has memory, i.e., with correlations between the forward and the
backward paths.

\subsubsection{Asymptotic security of two-way CV-QKD}

The security of two-way CV-QKD\ protocols has been first studied
in the asymptotic limit of infinitely-many uses of the channel and
large Gaussian
modulation~\cite{PIRS-2way,ott-scirep016,ott-pra015}. The most
general coherent attack can be reduced by applying de Finetti
random permutations~\cite{renner-cirac} which allow the parties to
neglect possible correlations established by Eve between different
rounds of the protocol. In this way the attack is reduced to a
two-mode attack which is coherent within a single round-trip
quantum communication. Then, the security analysis can be further
simplified by using the extremality of Gaussian
states~\cite{wolf-cirac}, which allows one to just consider
two-mode Gaussian attacks. The most realistic of these attacks is
implemented by using two beam-splitters of transmissivity $\eta$,
where Eve injects two ancillary modes $E_{1}$ and $E_{2}$, the
first interacting with the forward mode and the second with the
backward mode. Their outputs are then stored in a quantum memory
which is subject to a final collective measurement.

In each round-trip interaction, Eve's ancillae $E_{1}$ and $E_{2}$
may be coupled with to another set of modes so as to define a
global pure state. However, these additional ancillary modes can
be neglected if we consider the asymptotic limit where Eve's
accessible information is bounded by the Holevo
quantity~\cite{pirs-attacks-PRL08}. As a result, we may just
consider a two-mode Gaussian state $\rho_{E_{1}E_{2}}$ for Eve's
input ancillary modes.
In practical cases, its CM can be assumed to have the normal form%
\begin{equation}
\mathbf{V}_{E_{1}E_{2}}=\left(
\begin{array}
[c]{cc}%
\omega\mathbf{I} & \mathbf{G}\\
\mathbf{G} & \omega\mathbf{I}%
\end{array}
\right)  ,\text{ }\mathbf{G:=}\left(
\begin{array}
[c]{cc}%
g & 0\\
0 & g^{\prime}%
\end{array}
\right)  , \label{VE1E2}%
\end{equation}
where $\omega$ is the variance of the thermal noise, $\mathbf{I}=$
diag($1,1$), $\mathbf{Z}=$ diag($1,-1$), and matrix $\mathbf{G}$
describes the two-mode correlations. Here the parameters $\omega,$
$g$ and $g^{\prime}$ must
fulfill the bona fide conditions~\cite{PIRS-NJP}%
\begin{equation}
|g|<\omega,~|g^{\prime}|<\omega,~\omega^{2}+gg^{\prime}-1\geq\omega\left\vert
g+g^{\prime}\right\vert . \label{CMconstraints}%
\end{equation}
We notice that when $g=g^{\prime}=0$, then CM of Eq.~(\ref{VE1E2})
describes the action of two independent entangling cloners, i.e.,
the attack simplifies to one-mode Gaussian attack.

\subsubsection{Asymptotic key rates}

Here we provide the asymptotic secret key rates of the main
two-way protocols
based on coherent states and heterodyne or homodyne detection~\cite{PIRS-2way}%
. For each protocol, we summarize the key rates in DR
($\blacktriangleright$) or RR ($\blacktriangleleft$) for the two
configurations in ON or OFF. In particular, ON is assumed against
one-mode Gaussian attacks, while OFF\ is assumed to be used under
two-mode Gaussian
attacks~\cite{PIRS-2way,ott-pra015,ott-scirep016}. For an ON key
rate under two-mode attacks see Ref.~\cite{ott-pra015}, but we do
not consider this here.

For the two-way protocol based on coherent states and heterodyne
detection we
write the key rates%
\begin{align}
R_{\text{ON}}^{\blacktriangleright}  &  =\log_{2}\frac{2\eta(1+\eta)}%
{e(1-\eta)\Lambda}-s(\omega).\label{R-DR-HET2-ON}\\
R_{\text{ON}}^{\blacktriangleleft}  &  =\log_{2}\frac{\eta(1+\eta)}%
{e(1-\eta)\Lambda}+\sum_{i=1}^{3}s(\bar{\nu}_{i})-2s(\omega),\\
R_{\text{OFF}}^{\blacktriangleright}  &
=\log_{2}\frac{2\eta}{e(1-\eta
)\tilde{\Lambda}}+%
{\displaystyle\sum\limits_{k=\pm}}
\frac{s(\bar{\nu}_{k})-s(\nu_{k})}{2},\label{R-DR-HET2-OFF}\\
R_{\text{OFF}}^{\blacktriangleleft}  &
=\log_{2}\frac{2\eta}{e(1-\eta
)\tilde{\Lambda}}+\sum_{k=\pm}\frac{s(\bar{\nu}_{k}^{\prime})-s(\nu_{k})}{2},
\end{align}
where $\Lambda:=1+\eta^{2}+(1-\eta^{2})\omega$, $\tilde{\Lambda}%
:=1+\eta+(1-\eta)\omega$, the eigenvalues $\bar{\nu}_{i}$ are
computed numerically and
\begin{align}
\nu_{\pm}  &  =\sqrt{(\omega\pm g)(\omega\pm g^{\prime})}%
,\label{spectrum-HET2-TOT-OFF}\\
\bar{\nu}_{\pm}^{\prime}  &  =\frac{\sqrt{[(\omega\pm
g)(1-\eta)+1][(\omega\pm
g^{\prime})(1-\eta)+1]]}}{\eta}. \label{spectrum-HET2-COND-RR-OFF}%
\end{align}
Note that, if we set $g=g^{\prime}=0$ in the OFF\ rates, we
retrieve the two rates of the one-way coherent-state protocol in
Eqs.~(\ref{coPROT}) and~(\ref{coPROT2}).

Consider now the two-way coherent state protocol with homodyne
detection. In this case, the solution is fully analytical. In
fact, we have the following
key rates%
\begin{align}
R_{\text{ON}}^{\blacktriangleright}  &
=\frac{1}{2}\log_{2}\frac{\eta
(1+\eta)\omega}{(1-\eta)[\eta^{2}+(1-\eta^{2})\omega]}-s(\omega),\\
R_{\text{ON}}^{\blacktriangleleft}  &
=\frac{1}{2}\log_{2}\frac{\eta
^{2}+\omega+\eta^{3}(\omega-1)}{(1-\eta)[\eta^{2}+(1-\eta^{2})\omega
]}+s(\tilde{\nu})-s(\omega),\\
R_{\text{OFF}}^{\blacktriangleright}  &
=\frac{1}{2}\log_{2}\frac{\eta
\sqrt{\lbrack1+\eta(\omega-1)]^{2}-\eta^{2}g^{2}}}{(1-\eta)[\eta
+(1-\eta)\omega]}\nonumber\\
&  -{\sum\limits_{k=\pm}}\frac{s(\delta_{k})-s\left(  \nu_{k}\right)  }{2},\\
R_{\text{OFF}}^{\blacktriangleleft}  &  =\frac{1}{2}\log_{2}\frac
{\sqrt[4]{(\omega^{2}-g^{2})(\omega^{2}-g^{\prime2})}}{(1-\eta)[\eta
+(1-\eta)\omega]}\nonumber\\
&  -\sum_{k=\pm}\frac{s(\nu_{k})}{2},
\end{align}
where the $\nu_{k}$ are given in Eq.~(\ref{spectrum-HET2-TOT-OFF}), and%
\begin{align}
\tilde{\nu}  &  :=\sqrt{\frac{\omega\lbrack1+\eta^{2}\omega(1-\eta)+\eta^{3}%
]}{\eta^{2}+\omega+\eta^{3}(\omega-1)}},\\
\delta_{\pm}  &  =\sqrt{\frac{(\omega\pm
g^{\prime})[\eta+(\omega\pm g)(1-\eta)]}{1-\eta+\eta(\omega\pm
g^{\prime})}}.
\end{align}
Other cases with encoding by squeezed states and decoding by
heterodyne/homodyne detection, have been discussed in
Ref.~\cite{ott-scirep016}.

\subsubsection{Further considerations}

As discussed in Refs.~\cite{ott-pra015,ott-scirep016,PIRS-2way}
the security thresholds of the two-way protocols, given by setting
$R=0$ in the expression above, are higher of the corresponding
one-way protocols. This makes two-way CV-QKD a good choice in
communication channels affected by high thermal noise (e.g., at
the THz or microwave regime). The security analysis provided in
Refs.~\cite{ott-pra015,ott-scirep016,PIRS-2way} is limited to the
asymptotic regime. However, recently the composable security has
been also proven in Ref.~\cite{Ghorai-2way}. Other studies on the
security of two-way CV-QKD protocols have been carried out in
Refs.~\cite{SunGuo12,LeoGuo17}, besides proposing the use of
optical amplifiers~\cite{LeoGuo14}. It is also important to note
that, besides the Gaussian two-way protocols, one may also
consider schemes that are based on quantum
illumination~\cite{ZheshenQKD} or may implement floodlight
QKD~\cite{floodlight1,floodlight2,floodlight3,floodlight4}. The
latter is a two-way quantum communication scheme which allows one
to achieve, in principle, Gbit/s secret-key rate at metropolitan
distances. This is done by employing a multiband strategy where
the multiple optical modes are employed in each quantum
communication.

\subsection{Thermal-state QKD\label{thermal-schemes}}

In the protocols treated so far one assumes that the Gaussian
states are pure. This requirement can however be relaxed. The
possibility of using \textquotedblleft noisy\textquotedblright\
coherent states, i.e., thermal states, was first considered in
Ref.~\cite{Filip2008} which showed that thermal states are
suitable for QKD if the parties adopt RR and the signals are
purified before transmission over the channel. This approach was
later reconsidered in Ref.~\cite{Usenko2010a}, which proved its
security in realistic quantum channels.
Refs.~\cite{Weedbrook2010,Weedbrook2012} showed that thermal
states can be directly employed in CV-QKD (without any
purification at the input) if the protocol is run in DR.
Similarly, they can be directly employed in two-way CV-QKD if the
protocol is run in RR~\cite{weedbrook2014_2way}. This possibility
opened the way for extending CV-QKD to longer wavelengths down to
the microwave regime, where the protocols can be implemented for
short-range applications and are sufficiently robust to
finite-size effects~\cite{Papanastasiou2018}. More recently, the
terahertz regime has been also proposed for short-range uses of
CV-QKD~\cite{Ottaviani2018}. Following this idea, another
work~\cite{MalaneyTHZ} has proposed the same regime for satellite
(LEO) communications where the issue of the thermal noise is
mitigated.

\subsubsection{One-way thermal communication}

For simplicity, we focus on the one-way protocol where Bob
homodynes the incoming signals, randomly switching between the
quadratures. An alternative no-switching implementation based on
heterodyne detection can be considered as well. The protocol
starts with Alice randomly displacing thermal states in the phase
space according to a bivariate Gaussian distribution. She then
sends the resulting state to Bob, over the insecure quantum
channel. The generic quadrature $\hat{A}=(q_{A},p_{A})$ of Alice's
input mode $A$ can be written as $\hat{A}=\hat{0}+\alpha$, where
the real number $\alpha$ is the Gaussian encoding variable with
variance $V_{\alpha}$, while operator $\hat{0}$ accounts for the
thermal `preparation noise', with variance $V_{0}\geq1$. The
overall variance of Alice's average state is therefore
$V_{A}=V_{0}+V_{\alpha }$.

The variance $V_{0}$ can be broken down as $V_{0}=1+\mu_{th}$,
where $1$ is the variance of the vacuum shot-noise, and
$\mu_{th}\geq0$ is the variance of an extra trusted noise confined
in Alice's station and uncorrelated to Eve. Bob homodynes the
incoming signals, randomly switching between position and momentum
detections. In this way, Bob collects his output variable $\beta$
which is correlated to Alice's encoding $\alpha$. After using a
public channel to compare a subset of their data, to estimate the
noise in the channel and the maximum information eavesdropped, the
parties may apply classical post-processing procedures of error
correction and privacy amplification in order to extract a shorter
secret-key.

The security analysis of this type of protocol is analogous to
that of the case based on coherent states. Being the protocol
Gaussian, the security is upper-bounded by collective Gaussian
attacks and Eve's accessible information overestimated by the
Holevo bound. Assuming a realistic entangling-cloner attack, in
the typical limit of large variance $V_{\alpha}\gg1$, we obtain
the following expressions for the asymptotic
key-rates~\cite{Weedbrook2010,Weedbrook2012}%
\begin{align}
R_{\text{th}}^{\blacktriangleright}  &
=\frac{1}{2}\log_{2}\frac{\eta
\Lambda(\omega,V_{0})}{(1-\eta)\Lambda(V_{0},\omega)}\nonumber\\
&  +s\left[  \sqrt{\frac{\omega\Lambda(1,\omega V_{0})}{\Lambda(\omega,V_{0}%
)}}\right]  -s(\omega)~,\label{rate1DR-S-TH}\\
R_{\text{th}}^{\blacktriangleleft}  &
=\frac{1}{2}\log_{2}\frac{\omega
}{(1-\eta)\Lambda(V_{0},\omega)}-s(\omega),
\end{align}
where function $\Lambda(x,y):=\tau x+(1-\tau)y$ and
$s\left(x\right) $ is defined in Eq.~(\ref{entropicH}).

\subsubsection{Two-way thermal communication}

The two-way thermal protocol~\cite{weedbrook2014_2way} extends the
one-way thermal protocol~\cite{Weedbrook2010,Weedbrook2012} to
two-way quantum communication. The steps of the protocol are the
same as described in Sec. \ref{two-way-QKD} but with thermal
states replacing coherent ones. Therefore, Bob has an input
mode $B_{1}$, described by the generic quadrature $\hat{B}_{1}=\hat{0}%
+\beta_{1}$, where $\beta_{1}$ is the encoding Gaussian variable
having variance $V_{\beta_{1}}$ while mode $\hat{0}$ has variance
$V_{0}=1+\mu _{th}\geq1$. After the first quantum communication
Alice receives the noisy mode $A_{1}$ and randomly switches
between the two possible
configurations~\cite{weedbrook2014_2way,ott-scirep016}. In case of
ON configuration, Alice encodes a Gaussian variable $\alpha$ with
variance $V_{\alpha}=V_{\beta_{1}}$, randomly displacing the
quadrature of the incoming mode
$\hat{A}_{1}\rightarrow\hat{A}_{2}=\hat{A}_{1}+\alpha$. When the
two-way circuit is set OFF, Alice homodynes the incoming mode
$A_{1}$ with classical output $\alpha_{1}$, and prepares another
Gaussian-modulated thermal state $\hat{A}_{2}=\hat{0}+\alpha_{2}$,
with the same preparation and signal variances as Bob, i.e.,
$V_{0}$ and $V_{a_{2}}=V_{\beta_{1}}$. In both cases, the
processed mode $A_{2}$ is sent back to Bob in the second quantum
communication through the channel. At the output, Bob homodynes
the incoming mode $B_{2}$ with classical output $\beta_{2}.$

At the end of the double quantum communication, Alice publicly
reveals the configuration used in each round of the protocol, and
both the parties declare which quadratures were detected by their
homodyne detectors. After this stage, Alice and Bob possess a set
of correlated variables, which are $\alpha _{1}\approx\beta_{1}$
and $\alpha_{2}\approx\beta_{2}$ in OFF configuration, and
$\alpha\approx\beta$ in ON configuration. By comparing a small
subset of values of these variables, the parties may detect the
presence of memory between the first and the second use of the
quantum channel. If two-mode coherent attacks are present then
they use the OFF\ configuration, extracting a secret-key from
$\alpha_{1}\approx\beta_{1}$ and $\alpha_{2}\approx\beta _{2}$. If
memory is absent, the parties assume one-mode collective attacks
against the ON\ configuration, and they post-process $\alpha$ and
$\beta$. We remark that the switching between the two
configurations can be used as a virtual basis against
Eve~\cite{ott-scirep016}, who has no advantage in using two-mode
correlated attacks against the CV two-way protocol.

Let us assume the realistic Gaussian attack composed by\ two
beam-splitters of transmissivity $\eta$, where Eve injects two
ancillary modes $E_{1}$ and $E_{2}$ in a Gaussian state whose CM
is specified in Eq.~(\ref{VE1E2}. This is a two-mode (one-mode)
attack for $g,g^{\prime}\neq0$ ($=0$). Assuming ideal
reconciliation efficiency, working in the asymptotic limit of many
signals and large Gaussian modulations, one can compute the
following secret\ key rates
for the two-way thermal protocol with homodyne decoding%
\begin{align}
R_{\text{2-th}}^{\blacktriangleright}  &
=\frac{1}{2}\log_{2}\frac {\eta(1+\eta)\omega}{(1-\eta)\left[
\eta^{2}V_{0}+(1-\eta^{2})\omega\right]
}-s(\omega),\\
R_{\text{2-th}}^{\blacktriangleleft}  &
=\frac{1}{2}\log_{2}\frac{\eta ^{2}V_{0}+\omega+\eta^{3}\left(
\omega-V_{0}\right)  }{\left[  V_{0}\eta
^{2}+\left(  1-\eta^{2}\right)  \omega\right]  (1-\eta)}-s(\omega)\nonumber\\
&  +s\left(  \sqrt{\frac{\omega\left[  1+\eta^{2}V_{0}\omega+\eta^{3}%
(1-V_{0}\omega)\right]  }{\eta^{2}V_{0}+\omega+\eta^{3}(\omega-V_{0})}%
}\right)  .
\end{align}

\subsection{Unidimensional protocol}

As an alternative to the standard CV-QKD protocols described
above, where both quadratures have to be modulated and measured
(be it simultaneously or subsequently), one may consider a
unidimensional (UD) protocol~\cite{Usenko2015,Gehring2016}, which
relies on a single quadrature modulation at Alice's side while Bob
performs a randomly-switched homodyne detection. Because it
requires a single modulator, the UD CV-QKD protocols provide a
simple experimental realization with respect to conventional CV
QKD. This also means that the trusted parties are not able to
estimate the channel transmittance in the un-modulated quadrature,
which remains an unknown free parameter in the protocol security
analysis. This parameter however can be limited by considerations
of physicality of the obtained CMs. In other words, Eve's
collective attack should be pessimistically assumed to be
maximally effective, but is still limited by the physicality
bounds related to the positivity of\ the CM and its compliance
with the uncertainty principle~\cite{bonafideCM,RMPwee}.

Therefore, Eve's information can be still upper-bounded and the
lower bound on the key rate can be evaluated. The performance of
the protocol was compared to standard one-way CV-QKD in the
typical condition of a phase-insensitive thermal-loss channel
(with the same transmittance and excess noise for both the
quadratures). While the UD protocol is more fragile to channel
loss and noise than conventional CV-QKD, it still provides the
possibility of long-distance fiber-optical communication. In fact,
in the limit of low transmissivity $\eta$\ and infinitely strong
modulation, the key rate for the UD CV-QKD protocol with
coherent-state and homodyne detection is approximately given by
$(\eta\log_{2}{e})/3$~\cite{Usenko2015}, which is slightly smaller
than the similar limit for the standard one-way protocol with
coherent states and homodyne detection~\cite{Grosshans2005} for
which the rate is approximately given by $(\eta\log_{2}{e})/2$. UD
CV-QKD was recently extended to squeezed
states~\cite{Usenko-UD-2018}, which were shown to be advantageous
only in the DR scenario if the anti-squeezed quadrature is
modulated. The security of coherent-state UD CV-QKD was recently
extended to the finite-size regime~\cite{Wang2017}.

\subsection{CV-QKD with discrete modulation}

In CV-QKD, information is encoded in quantum systems with
infinite-dimensional Hilbert spaces. This allows the sender to use
bright coherent states and highly-efficient homodyne detections,
which naturally boost the communication rate. These features do
not come for free. At the error correction stage, one pays a
penalty in mapping the continuous output data from the physical
Gaussian channel into a binary-input additive white Gaussian-noise
channel. This mapping is more accurate by employing discrete
modulation~\cite{Leverrier2009}. The first discrete-modulated
CV-QKD protocol was based on a binary encoding of coherent
states~\cite{Silberhorn2002} and was designed to overcome the
$3$dB limitation of CV-QKD in DR. Later protocols have consider
three~\cite{Bradler3states} or arbitrary number of phase-encoded
coherent states~\cite{PanosDriscrete}.

The basic idea in Ref.~\cite{Silberhorn2002} is to perform a
binary encoding which assigns the bit-value $0$ ($1$) to a
coherent state with positive (negative) displacement. Then, the
receiver switches the homodyne detection setup, measuring
quadrature $q$ or $p$. After the quantum communication, the
parties discard \emph{unfavorable} data by applying an advantage
distillation routine~\cite{maurer93,maurer97}, which is a
post-selection procedure which extracts a key by using two-way
classical communication. The asymptotic security of this protocol
was first studied under individual attacks~\cite{Silberhorn2002}
and later against collective attacks, with also a proof-of-concept
experiment~\cite{symul07}. In general, the security of CV-QKD with
non-Gaussian modulation remains an open question. In the
asymptotic limit, its security has been proven against Gaussian
attacks~\cite{Leverrier2011} and, more recently, general
attacks~\cite{GhoraiDIS}.

\subsection{CV MDI-QKD\label{CVMDISection}}

\subsubsection{Basic concepts and protocol}

As we know, MDI-QKD~\cite{BP,Lo} has been introduced to overcome a
crucial vulnerability of QKD systems, i.e., the side-channel
attacks on the measurement devices of the parties. The basic
advantage of MDI scheme is that Alice and Bob do not need to
perform any measurement in order to share a secret key. The
measurements are in fact performed by an intermediate relay, which
is generally untrusted, i.e., controlled by Eve. This idea can
also be realized in the setting of CV-QKD with the promise of
sensibly higher rates at metropolitan distances. The protocol was
first introduced on the arXiv at the end of 2013 by
Ref.~\cite{RELAY} and independently re-proposed in
Ref.~\cite{CVMDIQKDleo}.

The protocol proceeds as follows: Alice and Bob possess two modes,
$A$ and $B$ respectively, which are prepared in coherent states
$|\alpha\rangle$ and $|\beta\rangle$. The amplitude of these
coherent states is randomly-modulated, according to a bi-variate
Gaussian distribution with large variance. Each one of the parties
send the coherent states to the intermediate relay using the
insecure channel. The modes arriving at the relay, say
$A^{\prime}$ and $B^{\prime}$, are measured by the relay by means
a CV Bell detection~\cite{BellFORMULA}. This means that
$A^{\prime}$ and $B^{\prime}$ are first mixed on a balanced beam
splitter, and the output ports conjugately homodyned: on one port
it is applied a homodyne detection on quadrature $\hat{q}$, which
returns the outcome $q_{-}$, while the other port is homodyned in
the $\hat{p}$-quadrature, obtaining an outcome $p_{+}$. The
outcomes from the CV-Bell measurement are combined to form a new
complex outcome $\gamma:=(q_{-}+ip_{+})/\sqrt{2}$ which is
broadcast over a public channel by the relay.

For the sake of simplicity, let us consider lossless links to the
relay. Then we can write $\gamma\simeq\alpha-\beta^{\ast}$, so
that the public broadcast of $\gamma$ creates \textit{a
posteriori} correlations between Alice's and Bob's variables. In
this way, each of the honest parties may infer the variable of the
other. For instance, Bob may use the knowledge of $\beta$\ and
$\gamma$ to compute $\beta^{\ast}-\gamma\simeq\alpha$ recovering
Alice's variable up to detection noise~\cite{RELAY}. Eve's
knowledge of the variable $\gamma$ does not help her to extract
information on the individual variables $\alpha$ and $\beta$. This
means that Eve needs to attack the two communication links with
the relay in order to steal information, which results in the
introduction of loss and noise to be quantified by the parties. In
terms of mutual information this situation can be described
writing that $I(\alpha:\gamma)=I(\beta:\gamma)=0$, while as a
consequence of the broadcast of variable $\gamma$ Alice-Bob
conditional mutual information is non-zero, i.e.,
$I(\alpha:\beta|\gamma)>I(\alpha:\beta)=0$.


As discussed in Ref.~\cite{RELAY}, the best decoding strategy is
to guess the variable of the party who is closer to the relay
(i.e., whose link has the highest transmissivity). Also note that,
as proven in Ref.~\cite{lupo-comp}, the whole raw data can be used
to perform both secret key extraction and parameter estimation.
This is because the protocol allows the parties to recover each
other variable from the knowledge of $\gamma$, so that they can
locally reconstruct the entire CM of the shared data without
disclosing any information.

\subsubsection{Security and key rates}

The security of CV MDI-QKD has been first studied in the
asymptotic limit~\cite{RELAY,RELAY-SYMM} (including fading
channels~\cite{PanosFADING}) and recently extended to finite-size
security~\cite{papanastasiou018,guo018} and then composable
security~\cite{lupo-comp,lupo-PE}. The asymptotic security
analysis starts by considering the general scenario of a global
unitary operation correlating all the uses of the protocol.
However, using random permutations~\cite{renner2,renner-cirac},
Alice and Bob can reduce this scenario to an attack which is
coherent within the single use of the protocol. After de Finetti
reduction, this is a joint attack of both the links and the relay.
In particular, since the protocol is based on the Gaussian
modulation and Gaussian detection of Gaussian states, the optimal
attack will be
Gaussian~\cite{RMPwee,Garcia-Patron2006,Navascues2006}. More
details can be found in Ref.~\cite{RELAY}.

In analogy with the two-way CV-QKD protocol, a realistic two-mode
Gaussian attack consists of Eve attacking the two links by using
two beam-splitters of transmissivity $\eta_{A}$ and $\eta_{B}$
that are used to inject to modes $E_{1}$ and $E_{2}$ in a Gaussian
state with CM given in Eq.~(\ref{VE1E2}). Detailed analysis of the
possible two-mode attacks showed that, in the asymptotic regime,
the optimal attack is given by the \emph{negative EPR attack},
which corresponds to the case where $g=-g^{\prime}$ with
$g^{\prime }=-\sqrt{\omega^{2}-1}$ in Eq.~(\ref{VE1E2}). In such a
case Eve injects maximally entangled states with correlations
contrasting those established by the CV-Bell detection, resulting
in a reduction of the key rate.

Indeed, assuming the asymptotic limit of many uses, large variance
of the signal modulation, and ideal reconciliation efficiency, it
is possible to obtain a closed formula describing the secret key
rate for CV MDI-QKD. In particular, we can distinguish between two
setups: the symmetric
configuration, where the relay lies exactly midway the parties ($\eta_{A}%
=\eta_{B}$), and the asymmetric configuration
($\eta_{A}\neq\eta_{B}$). Assuming that Alice is the encoding
party and Bob is the decoding party (inferring Alice's variable),
the general expression of the asymmetric configuration takes the
form
\begin{align}
R_{\text{asy}}  &  =\log_{2}\frac{2\left( \eta_{A}+\eta_{B}\right)
}{e\left\vert \eta_{A}-\eta_{B}\right\vert \bar{\chi}}+s\left[
\frac{\eta
_{A}\bar{\chi}}{\eta_{A}+\eta_{B}}-1\right] \nonumber\\
&  -s\left[  \frac{\eta_{A}\eta_{B}\bar{\chi}-\left( \eta_{A}+\eta
_{B}\right)  ^{2}}{\left\vert \eta_{A}-\eta_{B}\right\vert \left(
\eta
_{A}+\eta_{B}\right)  }\right]  , \label{key-CVMDI-A}%
\end{align}
where $\bar{\chi}:=2\left(  \eta_{A}+\eta_{B}\right)
/(\eta_{A}\eta _{B})+\varepsilon$, $\varepsilon$ is the excess
noise, and $s\left( x\right) $ is defined in
Eq.~(\ref{entropicH}). For pure-loss links ($\varepsilon=0$) the
rate of
Eq.~(\ref{key-CVMDI-A}) reduces to%
\begin{align}
R_{\text{asy}}  &  =\log_{2}\frac{\eta_{A}\eta_{B}}{e|\eta_{A}-\eta_{B}%
|}+s\left(  \frac{2-\eta_{B}}{\eta_{B}}\right) \nonumber\\
&  -s\left( \frac{2-\eta_{A}-\eta_{B}}{|\eta_{A}-\eta_{B}|}\right)
.
\label{key-CVMDI-A-PL}%
\end{align}

The asymmetric configuration, under ideal conditions, allows to
achieve long-distance secure communication. In particular, for
$\eta_{A}=1$ and arbitrary $\eta_{B}$ the maximum achievable
distance can be of $170$~km (in standard optical fibers with
attenuation $0.2$ dB/Km) and key rate of $2\times10^{-4}$
bit/use~\cite{ideal-relay}. Under such conditions,
the rate of Eq.~(\ref{key-CVMDI-A-PL}) becomes%
\begin{equation}
R_{\text{asy}}=\log_{2}\frac{\eta_{B}}{e(1-\eta_{B})}+s\left(
\frac {2-\eta_{B}}{\eta_{B}}\right)  ,
\end{equation}
which coincides with the RR rate of the one-way protocol with
coherent states and heterodyne detection. The performance degrades
moving the relay in symmetric position with respect to Alice and
Bob. In such a case, we set $\bar{\chi}=2/\eta+\varepsilon$ where
$\eta:=\eta_{A}=\eta_{B}$, and we write
the rate~\cite{RELAY,RELAY-SYMM}%
\begin{equation}
R_{\text{sym}}=\log_{2}\frac{16}{e^{2}\bar{\chi}\left(
\bar{\chi}-4\right) }+s\left(  \frac{\bar{\chi}}{2}-1\right)  .
\end{equation}
For pure-loss links, this simplifies to%
\begin{equation}
R_{\text{sym}}=\log_{2}\frac{\eta^{2}}{e^{2}(1-\eta)}+s\left(
\frac{2-\eta }{\eta}\right)  ,
\end{equation}
and the maximum achievable distance is about $3.8$~km of standard
optical fiber from the relay.

Finite-size analysis and composable security have been developed
for CV MDI-QKD. In Refs.~\cite{papanastasiou018,guo018}
finite-size corrections have been studied assuming Gaussian
attacks. The estimation of the channel parameters is provided
within confidence intervals which are used to identify the
worst-case scenario, corresponding to assuming the lowest
transmissivity and the highest excess noise compatible with the
limited data. The analysis showed that using signal block-size in
the range of $10^{6}\div10^{9}$ data points is sufficient to
obtain a positive secret key rate of about $10^{-2}$ bits/use.

The composable security proof of CV MDI-QKD has been developed in
Ref.~\cite{lupo-comp} using the lower bound provided by the
smooth-min entropy, and designing a novel parameter estimation
procedure~\cite{lupo-PE} which allows to simplify the analysis.
The security has been proved against general attacks using the
optimality of Gaussian attacks for Gaussian protocols, and the de
Finetti reduction of general attacks to collective ones.
The lower bound to the key rate is given by the following expression%
\begin{align}
R_{n}^{\varepsilon^{\prime\prime}}  &  \geq\frac{n-k}{n}\left(
\xi I_{AB}-I_{E}\right)
-\frac{\sqrt{n-k}}{n}\Delta_{\mathrm{AEP}}\left(
\frac{2p\epsilon_{s}}{3},d\right) \nonumber\\
&  +\frac{1}{n}\log_{2}\left(  p-\frac{2p\epsilon_{s}}{3}\right)  +\frac{2}%
{n}\log_{2}2\epsilon\nonumber\\
&  -\frac{2}{n}\log_{2}\left(
\begin{array}
[c]{c}%
K+4\\
4
\end{array}
\right)  , \label{kEY-COMPOSABLE}%
\end{align}
where $\xi$ accounts for all sources of non-ideality in the
protocol, $I_{AB}$ is Alice-Bob mutual information and $I_{E}$
Eve's accessible information. The parameter $k$ is the number of
signals used for the energy test, $n$ is the
total number of signal exchanged,\textbf{ }and%
\begin{equation}
K=\max\left\{  1,n(d_{A}+d_{B})\frac{1+2\sqrt{\frac{\ln(8/\epsilon^{\prime}%
)}{2n}}+\frac{\ln(8/\epsilon^{\prime})}{n}}{1-2\sqrt{\frac{\ln(8/\epsilon
^{\prime})}{2k}}}\right\}  ,
\end{equation}
The quantity $\epsilon^{\prime\prime}=k^{4}\epsilon^{\prime}/50$
is the overall security parameter with
$\epsilon^{\prime}:=\epsilon+\epsilon
_{s}+\epsilon_{\mathrm{EC}}+\epsilon_{\mathrm{PE}}$. Here
$\epsilon$ comes from the leftover hash lemma, $\epsilon_{s}$ is
the smoothing parameter, $\epsilon_{\mathrm{EC}}$ is the error
probability of the error correction routine, and
$\epsilon_{\mathrm{PE}}$ that of the parameter estimation. The
functional $\Delta_{\mathrm{AEP}}\left(
\frac{2p\epsilon_{s}}{3},d\right)  $ quantifies the error
committed bounding the smooth-min entropy using the asymptotic
equipartition property (AEP)~\cite{lupo-comp,MTomamichel-PhD}. The
dimensionality parameter $d$ describes the number of bits used in
the analog-to-digital conversion (ADC) sampling, by which the
unbounded continuous variables used in the protocols $(q,p)$ are
mapped into discrete variables, described as a set of $2^{2d}$
elements (cardinality). The two parameters, $d_{A}$ and $d_{B}$,
describe the dimensions of the sampling of Alice and Bob
variables, set to perform the energy test. Parameter $p$ gives the
probability of success of the error correction. The result
obtained by Ref.~\cite{lupo-comp} confirmed that CV MDI-QKD is
composably secure against general attacks and the use of
block-size of $10^{7}\div10^{9}$ data points is sufficient to
generate a positive key rate against general coherent attacks.

\subsubsection{Variants of CV MDI-QKD}

Several schemes have been introduced to modify the original design
of\ the CV MDI-QKD scheme. One approach has been based on the used
of non-Gaussian operations, like noiseless linear amplifiers (NLA)
and photon subtraction/addition (whose importance is well-known in
entanglement distillation~\cite{eisertPRL2002}). The use of NLAs
in MDI-QKD setups was investigated in Ref.~\cite{GuoNLA} while
photon subtraction has been explored in
Refs.~\cite{zhaoPRA2018,MAPRA2018}. Among other approaches, CV
MDI-QKD has been studied with squeezed
states~\cite{SqueezedCVMDIQKD} (with composable
security~\cite{ChenGuoPRA18}), discrete modulation (alphabet of
four coherent states)~\cite{CVMDIQKD4states}, phase
self-alignment~\cite{CVMDIQKDphase}, imperfect phase reference
calibration~\cite{CVMDIcalibration}, and unidimensional
encoding~\cite{XXMA1}. A multi-party version of the CV MDI-QKD\
protocol~\cite{OttavianiMultiKey2017} has been also introduced and
it is discussed below.

\subsubsection{Multipartite CV MDI-QKD\label{Multipartite}}

An interesting feature to achieve in quantum cryptography is the
ability to reliably connect many trusted users for running a
secure quantum conference or quantum secret-sharing protocols
\cite{Hillery,Cleve,Sanders,Keet}. The MDI architecture,
restricted to two~\cite{RELAY} or three users~\cite{WuPRA2016},
has been recently generalized in this direction. In the
\textit{scalable} MDI network of
Ref.~\cite{OttavianiMultiKey2017}, an arbitrary number $N$ of
remote users send Gaussian-modulated coherent states
$|\alpha_{k}\rangle$ to an untrusted relay where a generalized
multipartite Bell detection is performed. This detection consists
of a suitable cascade of beam-splitters with increasing
transmissivities $T_{k}=1-k^{-1}$, followed by $N-1$ homodyne
detection in the $\hat{q}$-quadrature, and a final homodyne
detection in the $\hat{p}$-quadrature. The result can be denoted
as a single variable $\gamma:=(q_{2},\ldots,q_{N},p)$ which is
broadcast to all parties. This measurement is responsible for
creating bosonic GHZ-type correlations among the parties. Ideally,
it projects on an asymptotic bosonic state with
Einstein-Podolsky-Rosen (EPR) conditions $%
{\textstyle\sum\nolimits_{k=1}^{N}}
\hat{p}_{k}=0$ and $\hat{q}_{k}-\hat{q}_{k^{\prime}}=0$ for any
$k,k^{\prime }=1,\ldots,N$.\

After the measurement is broadcast, the individual variables
$\alpha_{k}$ of the parties share correlations which can be
post-processed to obtain a \ common secret key. To implement
quantum conferencing, the parties choose the $i$th user as the one
encoding the key, with all the others decoding it in DR. To
realize quantum secret sharing, the parties split in two ensembles
which locally cooperate to extract a single secret key across the
bipartition. The scheme is studied in a symmetric configuration,
where the users are equidistant from the relay and the links are
modeled by memoryless thermal channels, with same transmissivity
and thermal noise. In this scenario, high rates are achievable at
relatively short distances. The security of the quantum
conferencing has been proved in both the asymptotic limit of many
signals and the composable setting that incorporates finite-size
effects. The analysis shows that, in principle, $50$ parties can
privately communicate at more than $0.1$ bit/use within a radius
of $40$m. With a clock of $25$MHz this corresponds to a key rate
of the order of $2.5$Mbits per second for all the users.

\section{Experimental CV-QKD}

\subsection{Introduction}

As discussed in the previous section, the various kinds of CV-QKD
protocols basically differ by the choice of input states (coherent
or squeezed states), input alphabets (Gaussian or discrete) and
detection strategy (homodyne or heterodyne detection). Most of
these schemes have been tested in proof-of-concept experiments in
a laboratory setting while a few have been going through different
stages of developments towards real-life implementations.
Specifically, the scheme based on Gaussian modulation of coherent
states and homodyne detection has matured over the last 15 years
from a simple laboratory demonstration based on bulk optical
components creating keys with very low
bandwidth~\cite{Grosshans2003,Lorenz2004,Lance2005,Lodewyck2005}
to a robust telecom-based system that generates keys with
relatively high
bandwidth~\cite{Lodewyck2007,Qi2007,Xuan2009,Jouguet2011,Jouguet2013,Wang2013,Shen2014,Huang2015a,Huang2015b,Huang2016,Laudenbach2017,Hirano2017,Brunner2017}
and allows for in-field
demonstrations~\cite{Fossier2009,Jouguet2012} and network
integration~\cite{Karinou2018,Eriksson2018,LeoJoseph2019}. In the
following sections, we will first describe the experimental
details of the standard point-to-point coherent state protocol
with emphasis on the most recent developments followed by a
discussion of some proof-of-concept experiments demonstrating more
advanced CV-QKD protocols such as squeezed state QKD and
measurement-device-independent QKD.

\subsection{Point-to-point CV-QKD}

The very first implementation of CV-QKD was based on coherent
state modulation and homodyne detection~\cite{Grosshans2003}. The
optical setup comprised bulk optical components and the operating
wavelength was 780~nm. This seminal work together with some
follow-up experiments~\cite{Lorenz2004,Lance2005,Lodewyck2005}
constituted the first important generation of CV-QKD systems.
Despite its successful demonstration of the concept of CV-QKD, it
was however unsuitable for realizing robust long-distance and
high-speed QKD in optical fibers because of the use of
telecom-incompatible wavelengths, the relatively low mechanical
stability of the systems and the low efficiency of the employed
error-correction protocols.

To overcome these impediments, a new generation of CV-QKD systems
was developed. This new generation made use of telecom wavelength,
was mainly based on telecom components, combined optimized
error-correction schemes and comprised several active feedback
control systems to enhance the mechanical
stability~\cite{Lodewyck2007,Jouguet2013,Huang2015a,Huang2016}.
With these new innovations, key rates of up to 1Mbps for a
distance of 25km~\cite{Huang2015a} and key rates of around 300bps
for a distance of 100km~\cite{Huang2016} have been obtained.
Recently, two different field tests of CV-QKD through commercial
fiber networks were performed over distances up to $\simeq50$km
with rates $>6$kpbs~\cite{LeoJoseph2019}, which are the longest
CV-QKD field tests so far, achieving two orders-of-magnitude
higher secret key rates than previous tests.

A third generation of CV-QKD systems are now under development.
They are based on the generation of power for a phase reference
(or local oscillator (LO)) at the receiver station in contrast to
previous generations where the power of the LO was generated at
the transmitter station and thus co-propagating with the signal in
the fiber. These systems have also evolved from simple
proof-of-concept demonstrations~\cite{Qi2015,Soh2015} to
technically more advanced demonstrations using telecom
components~\cite{Huang2015b,Kleis2017,Laudenbach2017,Hirano2017,Wang2018,Karinou2018}.

\begin{figure*}[pth]
\vspace{0.2cm}
\includegraphics[width=0.45\textwidth]{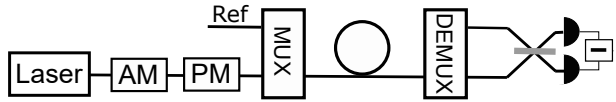}
\vspace{-1.5cm} \caption{Schematic setup of a generic experimental
CV-QKD system. A laser beam is modulated with an amplitude and a
phase modulator, and subsequently multiplexed with a reference
beam, sent through the optical fiber and de-multiplexed at the
receiver site. The quadrature variables of the signal is then
measured with a homodyne or heterodyne detector. A subset of the
measurement outcomes are used for the estimation of channel
parameters while the rest are used for
secret key generation using error-reconciliation and privacy amplification.}%
\label{CVQKDExp_fig1}%
\end{figure*}

The basic optical configuration for realizing CV-QKD is shown in
Fig.~\ref{CVQKDExp_fig1}. The signal is modulated in amplitude and
phase according to a certain distribution (often a continuous
Gaussian or a discrete QPSK distribution). It is then multiplexed
in time, polarization and/or frequency with a phase reference (a
strong local oscillator or a weak pilot tone) and subsequently
injected into the fiber channel. At the receiver side, the signal
and reference are de-multiplexed and made to interfere on a
balanced homodyne (or heterodyne) detector. A subset of the
measurement data are used for sifting and parameter estimation,
while the rest are used for the generation of a secret key via
error correction and privacy amplification.

In Fig.~\ref{CVQKDExp_fig2} we show the main layouts of three
different types of point-to-point CV-QKD experiments based on
coherent state encoding with a Gaussian distribution. The three
experiments represent important steps in the development of a
telecom compatible QKD system, and they illustrate different
techniques for encoding and detection. The experiment in
Fig.~\ref{CVQKDExp_fig2}a~\cite{Jouguet2013} applies a
time-multiplexed LO propagating along the fiber with the signal
while the experiments in
Fig.~\ref{CVQKDExp_fig2}b~\cite{Huang2015b} and
Fig.~\ref{CVQKDExp_fig2}c~\cite{Brunner2017} use a locally
generated LO. The two latter experiments deviate by the signal
encoding strategy (centered or up-converted base-band), the
detection method (homodyne or heterodyne) and the phase and
frequency difference determination. The experimental details are
described in the figure caption and discussed in the following
sections.

\begin{figure*}[pth]
\includegraphics[width=0.7\textwidth]{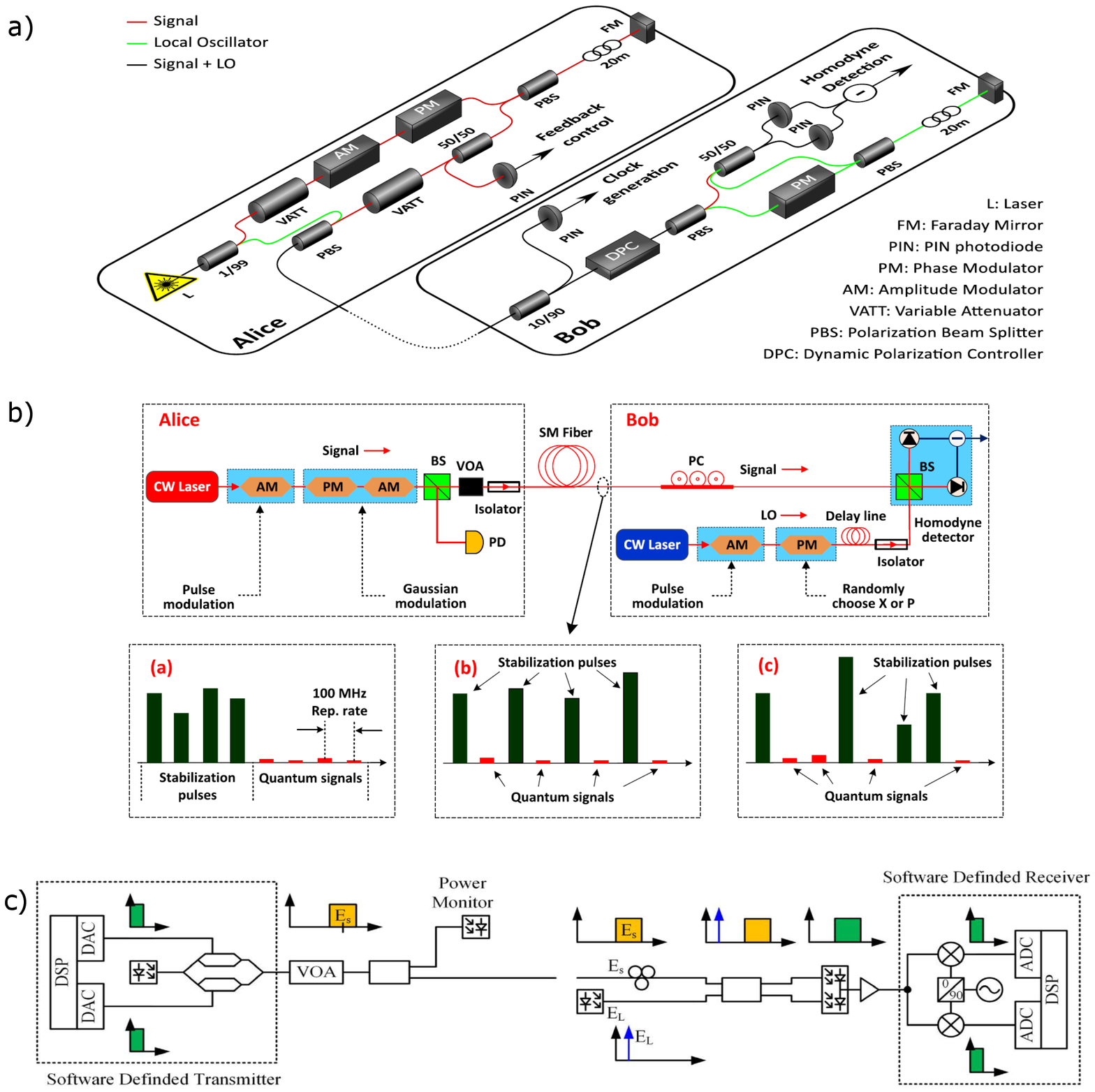}
\vspace{+0.5cm} \caption{Details of three experimental setups for
CV-QKD based on a Gaussian coherent state alphabet. a) Coherent
state pulses of 100ns with repetition rate 1MHz generated by a
1550nm telecom laser diode are split by an asymmetric beam
splitter (99/1) into a signal beam (which is attenuated by
variable attenuators (VATTs)) and a bright local oscillator. The
signal is modulated by an amplitude (AM) and phase modulator (PM)
and subsequently delayed by 200ns using a fiber delay line and a
Faraday mirror. The LO and signal are then multiplexed in time and
further in polarization (with a polarizing beam splitter (PBS))
before being injected into the fiber channel. At Bob's site the LO
and signal are polarization de-multiplexed with a PBS and a
dynamical polarization controller (DPC) and
 time-multiplexed with a delay line of the LO. The pulses are finally interfered in a homodyne detector.
 The PM in the LO enables random $\pi$ phase shift and thus random quadrature measurement. Taken
 from Ref.~\cite{Jouguet2013}. b) A CW telecom laser at 1550nm is transformed into 1nsec pulses
 with a repetition rate of 100MHz using an amplitude modulator (AM). The Gaussian distributed
 signal is produced with a pair of modulators (AM and PM) and its brightness is controlled with
 a variable optical attenuator (VOA). Phase synchronization signals are also produced in the
 modulators time-multiplexed with the quantum signal, either regularly as in (a) and (b) or
 randomly as in (c). The signals are injected into the channel and measured with a locally
 generated local oscillator at Bob. An AM produces local oscillator pulses while a PM randomly
 switches their phases by $\pi$ to allow for a random quadrature measurement. Phase and frequency
 synchronization between LO and signal is attained through DSP of the data produced by the
 interference between the LO and the reference pulses. Taken from Ref. ~\cite{Huang2015b}.
 c) Transmitter and receiver are to a large extent controlled by software, thereby reducing
 the complexity of the hardware. At the transmitter, software defines the pulse shape and the
 modulation pattern (here QPSK), and it ensures that the single side-banded signal - concentrated
 in a 10MHz band - is upconverted and combined with a pilot-tone which is used for controlling
 the frequency carrier-frequency offset and the phase noise. At the receiver, the signal is
 measured with a heterodyne detector using a locally generated, and frequency shifted LO.
 The output of the detector is demodulated into amplitude and phase quadrature components with
 a sampling rate of 200MS/s. Carrier-frequency offset, phase noise, clock skew and quadrature
 imbalance are subsequently corrected in digital-signal-processing, and the output is fed to
 the post-processing steps for key generation. Taken from Ref.~\cite{Brunner2017}. Figures adapted with permissions from:
 a) Ref.~\cite{Jouguet2013} \copyright NPG (2013), b) Ref.~\cite{Huang2015b} \copyright OSA (2015), and c) Ref.~\cite{Brunner2017} \copyright IEEE (2017).}
\label{CVQKDExp_fig2}%
\end{figure*}

The overarching aim for all QKD systems is to generate secret keys
with as high speed as possible and over as long distance as
possible. These two quantifying parameters for QKD are strongly
connected and they crucially depend on the system's clock rate,
the excess noise produced by the system and the efficiency,
quality and speed of the post-processing algorithms. With these
critical parameters in mind, in the following we describe the
technical details associated with transmitter, the receiver and
the post-processing schemes.

\subsubsection{Coherent state encoding}

At the transmitter station, a telecom laser is often transformed
into a train of pulses using an amplitude modulator with a certain
clock rate (e.g. 1MHz~\cite{Jouguet2013} or
50MHz~\cite{Huang2015a}). It is also possible to use a CW signal
where the clock rate is determined by the measurement bandwidth.
The clock rate should be large as it dictates the upper bound for
the final rate of the secret key and the accuracy in estimating
the parameters of the channel. However, using high clock rates
also places extra demands on the detection system and the
post-processing schemes as discussed later. After pulse
generation, a pair of modulators encode information using
different strategies. The traditional approach is to create a base
band signal around the carrier frequency but in more recent
implementations, the base band signal is up-converted to the GHz
range to limit the amount of photons scattered from the
carrier~\cite{Kleis2017}. As the base band signal is separated
from the carrier in frequency, it is possible to significantly
suppress the carrier through interference and thus reduce the
amount of excess noise resulting from scattered carrier
photons~\cite{Brunner2017,Laudenbach2017}.

\subsubsection{Detection}

At the receiver side, the signal is detected using either
homodyne, dual-homodyning or heterodyne detection. A homodyne
detector (sometimes called single-quadrature intradyne detection)
basically consists of the interference of two light beams with
identical frequencies - the signal and the local oscillator - on a
balanced beam splitter, and two PIN diodes combined in subtraction
and followed by a transimpedance amplification stage. In
dual-homodyne detection (also known a phase diverse detection),
the LO and the signal are split in balanced couplers and sent to
two individual homodyne detector to measured orthogonal
phase-space components simultaneously. It enhances the phase
estimation capabilities as two quadratures are measured but it
also increases the complexity of the setup. The heterodyne
detection scheme combines the advantages of the homodyne and
dual-homodyne schemes: It mimics the low-complexity of homdyning
and outputs the enhanced phase information of dual-homodyning. In
heterodyne detection, the LO frequency is offset with respect to
the carrier frequency of the signal, and therefore is
down-converting the signal band to an intermediate frequency.
Electronic downconversion is subsequently able to extract
information about orthogonal phase space quadratures similar to
dual-homodyning. It is worth noting that in the quantum community
dual-homodyne detection is often referred to as a heterodyne
detection while in the classical community heterodyne detection is
reserved for the scheme described above. The gained signal
information and noise penalties are basically similar for the two
approaches but the hardware implementations are very different.

The figure of merits associated with the detectors are the quantum
efficiency, the bandwidth and the electronic noise power relative
to the shot noise power. Naturally, a large clock rate requires a
large homodyne detector bandwidth. E.g. in Ref.~\cite{Huang2015a}
a detector with GHz bandwidth was used to resolve a 50MHz clock
rate while in Ref.~\cite{Jouguet2013} a much lower bandwidth
suffices to detect 1MHz clocked pulses. Homodyne detectors with
large bandwidths are commercially available but they have
relatively poor electronic noise performance and low quantum
efficiency. Although these detectors have been widely used for
CV-QKD, a new generation of improved homodyne detectors are under
development~\cite{Kumar2012,Huang2013} which in turn will improve
the performance of future QKD systems.

Homodyne and heterodyne detectors require a stable phase and
frequency reference also known as a local oscillator (LO). Two
strategies for realizing such a reference have been studied:

{\it Transmission of LO.} The traditional strategy is to use a LO
from the same laser as the signal and let it co-propagate with the
signal through the optical fiber. Different techniques for
combining the LO and signal have been tested including time
multiplexing~\cite{Lodewyck2007} and time-polarization
multiplexing~\cite{Jouguet2013}. This method however entails some
significant problems. First, due to channel loss, the power of the
LO at the receiver is strongly reduced and thus in some cases
insufficient for proper homodyne detection, second, the large
power of the LO in the fiber scatters photons and thus disturbs
the other quantum or classical fiber
channels~\cite{Qi2007,Xuan2009,Huang2015a}, and third, the
co-propagating LO is vulnerable to side-channel
attacks~\cite{Ma2013,Jouguet2013}. The approach is thus
incompatible with the existing telecom infra-structure and it
opens some security loopholes.

{\it Receiver generation of LO.} The alternative strategy which is
now gaining increasing interest and which is compatible with
classical coherent communication is to use a LO that is generated
at the receiver station, thereby avoiding the transmission of the
large powered LO through the optical fiber. To enable coherent
detection between the LO and the signal, strong synchronization of
the frequencies and phases is required. This can be performed in
post-processing similar to carrier-phase recovery schemes applied
in classical communication: The phase and frequency
synchronization of the LO do not have to be carried out prior to
signal measurements but can be corrected a posteriori in digital
signal processing (DSP). This is done by measuring the phase and
frequency differences and subsequently counter-rotate the
reference axes to correct for the drifts. Phase and frequency
estimation cannot be performed by using the quantum signal as a
reference since its power is too weak. Therefore, a small
reference beam or pilot tone must be sent along with the signal in
the fiber channel to establish the phase and frequency at the
receiver. However, it is important to note that this pilot tone is
very dim compared to a LO and thus do not result in the
complications associated with the transmission of a LO. CV-QKD
with a locally generated LO have been demonstrated by transmitting
the reference beam with the signal using time
multiplexing~\cite{Huang2015b,Wang2018}, frequency
multiplexing~\cite{Kleis2017} and frequency-polarization
multiplexing~\cite{Laudenbach2017}. In all these works, advanced
DSP was used to correct for phase and frequency mismatch in
post-measurements. The quality of the DSP algorithm is of utmost
importance as inaccuracies in correcting for drifts directly lead
to excess noise and thus a reduction of the resulting key rate and
distance. Recovery of the clock has been achieved either by using
or wavelength multiplexed clock laser~\cite{Huang2015a}, known
patterns as a header to the quantum signal~\cite{Laudenbach2017}
or a second frequency multiplexed pilot tone~\cite{Kleis2017}.

\subsubsection{Post-processing}

A critical part for the successful completion of CV-QKD is the
remaining post-processing protocols which include error
reconciliation, parameter estimation and privacy amplification.
The latter scheme is standard for any communication system (and
thus will not be discussed further in this review) while the two
former schemes are more complicated as they require sophisticated
computational algorithms specifically tailored for CV-QKD.
Furthermore, the performance of CV-QKD, that is key rate and
distance, critically depends on the efficiency of the error
correction algorithm and the quality of parameter estimation
scheme.

{\it Error reconciliation.~} In long-distance communication, the
signal-to-noise ratio (SNR) of the acquired data is usually very
low and thus the reconciliation of the Gaussian correlated
variables is computationally very hard. Earlier versions of
high-efficiency error correction protocols for CV-QKD could only
handle data with an SNR larger than around 1, thereby inherently
limiting the secure communication distance to about 25km
corresponding to a small metropolitan sized network. However, in
recent years there have been numerous new developments in
constructing high-efficiency error-correcting codes operating at
very low SNR. New code developments have resulted in a significant
improvement of the performance of CV-QKD and was the key stepping
stone for realizing long-distance communication. E.g. the work in
ref.~\cite{Jouguet2013} performed error correction with an
efficiency of 96\% at an SNR=$0.08$ as obtained with an 80km link
while in ref.~\cite{Huang2016} the efficiency was 95.6\% at
SNR=$0.002$ corresponding to a 150km fiber link.

The key innovation leading to these improved codes is to combine
multi-edge low density parity check (LDPC) codes with
multidimensional reconciliation
techniques~\cite{Leverrier2008,Jouguet2011,Lin2015,Milicevic2017,LeoGuoCodes17,LeoGuoCodes18}.
A rate-adaptive reconciliation protocol was proposed in
Ref.~\cite{LeoGuoCodes17} to keep high efficiency within a certain
range of SNR, which can effectively reduce the SNR fluctuation of
the quantum channel on the system performance. Throughputs of
25Mbit/s~\cite{Lin2015} and 30Mbit/s~\cite{LeoGuoCodes18} have
been achieved, which are not yet compatible with high speed system
using clock speeds of more than 100 MHz~\cite{Huang2015b}.
However, the speed of the LDPC decoder do not currently
representing the key bottleneck in extending the distance of QKD.
One of the main limiting factors in extending the distance is the
efficiency and quality in estimating the parameters of the system.

{\it Parameter Estimation.~} The quality of practical parameter
estimation is crucial for the reliable extraction of secret keys
for long-distance communication. In addition to the estimation of
the phase and frequency differences for LO adjustments as
discussed above, it is important also to estimate the
transmitter's modulation variance $\mu$, the channel transmittance
$\eta$ and the variance of the excess noise $\eps$. The two
variances are expressed in shot noise units and thus a careful
calibration of the shot noise level is also required.
Once all parameters are estimated, they are used to compute the
bounds on the Eve's information. However, the reliability in
warranting security strongly depends on the precision in
estimating the parameters. Very large data block sizes are often
required to reduce the finite-size effects to a level that is
sufficiently low to claim security (see also \ref{sectionMB_MT}).
As an example, in Ref.~\cite{Jouguet2013}, a secret key was
generated by using blocks of size $10^9$ for a 80km channel but by
reducing the block size to $10^8$, no key could be extracted as a
result of the finite-size effect. Likewise in
Ref.~\cite{Huang2016} a block size of $10^{10}$ was used to enable
secret key generation over 100km. To extend the distance further,
even larger data blocks are required. This places more stringent
demands on the stability of the system to allow for longer
measurement series, and it calls for an increase in the
measurement rate realized by a larger clock rate and higher
detector bandwidth.

{\it Practical key rate:} Within a certain data block, half of the
data are used for sifting (if homodyne detection is used), a
subset is used for parameter estimation and another subset is used
for phase synchronization of the local oscillator. The remaining
data are then used for secret key generation and thus undergo
error correction with efficiency $\xi$ and privacy amplification.
In a stable system it is also possible to perform error correction
first, using the SNR estimate from the previous round, and then
use all the data for parameter estimation. If the clock rate is
given by $C$ and the fraction of data used for key extraction is
$f$, the final practical key rate in RR is given by
\begin{equation}
R_{\mathrm{prac}}=f C (\xi I_{AB}-\chi_{BE})
\end{equation}
where $I_{AB}$ is Alice-Bob mutual information and $\chi_{BE}$ is
Eve's Holevo information on Bob's variable. The fastest system of
today produces secret keys of 1Mbps (for 25km) while the longest
distance attained is 150km with a key rate of 30kbps. Both
realizations are secure against collective attacks and include
finite-size effects.



\subsection{Implementation of advanced CV-QKD}

The point-to-point coherent state protocol discussed above is by
far the most mature CV-QKD scheme developed, and furthermore,
since it is reminiscent of a coherent classical communication
system, it is to some extend compatible with the existing telecom
networks. However, the point-to-point scheme might in some cases
be vulnerable to quantum hacking attacks and it possess some
limitations in speed and distance. To circumvent some of these
vulnerabilities and limitations, more advanced CV-QKD protocols
have been proposed and experimentally tested in proof-of-principle
type experiments using bulk optical components and often without
post-processing. In the following we briefly discuss these
demonstrations.

\subsubsection{Squeezed-state protocols} There has been two
implementations of QKD based on squeezed states. The schemes were
discussed in Section~\ref{CVonewayQKD}, and were shown to be
capable of extending the distance of QKD~\cite{Madsen2012} or to
enable composable security~\cite{Gehring2015}. In the following we
briefly address the experimental details of these realizations.

In both experiments, two-mode squeezed states were generated by
interfering two single-mode squeezed states on a balanced beam
splitter. The squeezed states were produced in cavity-enhanced
parametric down-conversion using non-linear crystals (PPKTP)
inside high-quality optical cavities~\cite{Andersen2016}. One mode
of the two-mode squeezed state was measured with high-efficiency
homodyne detection at the transmitter station while the other mode
was transmitted in free space to Bob who performed
homodyne/heterodyne measurements. The homodyne measurements at
Alice's station steered the state at Bob's station into a Gaussian
distribution of amplitude or phase quadrature squeezed states and
thus the scheme is effectively similar to a single-mode squeezed
state protocol. The noise suppression below shot noise of the
two-mode squeezed states were measured to be 3.5dB at
1064nm~\cite{Madsen2012} and 10.5dB at 1550nm~\cite{Gehring2015}.
The clock rate of the experiments was in principle limited by the
bandwidth of the cavity-enhanced parametric amplifiers (21MHz in
Ref.~\cite{Madsen2012} and 63MHz in Ref.~\cite{Gehring2015}) but
in the actual implementations, the bandwidth was set to a few kHz
given by filters in the homodyne detectors.

In such squeezed state systems, the size of the alphabet is often
limited to the degree of anti-squeezing and thus it is not
possible in practice to maximize the key rate with respect to the
modulation depth. However, in Ref.~\cite{Madsen2012} the Gaussian
alphabet was further extended by modulating the mode sent to Bob
with a pair of modulators. With this approach, the system becomes
more robust against excess noise and thus it is possible to extend
the distance over which secure communication can be realized. The
squeezed state experiment in Ref.~\cite{Gehring2015} was used to
demonstrate CV-QKD with composable security. E.g. for a channel
loss of 0.76~dB (corresponding to 2.7km fiber), they achieved
composable secure key generation with a bit rate of 0.1 bit/sample
using a reconciliation efficiency of 94.3\%. Furthermore, the
system was also used to demonstrate one-sided-device-independent
security against coherent attacks.

In a recent work~\cite{Scheffman2018}, it was experimentally
demonstrated that by using squeezed states encoding with a
uni-dimensional and small alphabet, it is possible to completely
eliminate the information of Eve in a purely lossy channel. This
reduces the computational complexity of the post-processing
protocols and therefore might be of interest in future CV-QKD
schemes despite the limited size of the alphabet.

\subsubsection{CV MDI-QKD} As discussed in Section~\ref{MDISection}, MDI-QKD schemes circumvent
quantum hacking attacks on the measurement system. A CV version of
the MDI-QKD scheme (see Section~\ref{CVMDISection}) has been
realized in a proof-of-concept experiment~\cite{RELAY}. Here
amplitude and phase modulation were applied to CW beams both at
Alice's and at Bob's site. The modulated beams underwent free
space propagation before being jointly measured in a CV Bell state
analyzer. Such an analyzer consists normally of a balanced beam
splitter - in which the two incoming beams interfere - followed by
two homodyne detectors measuring orthogonal quadratures. In the
current experiment, however, the Bell analyzer was significantly
simplified by using the carriers of the signals as local
oscillators: Through a proper phase space rotation of the carriers
enabled by the interference, information about the sum of the
amplitude quadratures and the difference of the phase quadratures
were extracted by simple direct measurements of the beam splitter
output modes followed by an electronic subtraction and summation.
Losses of the channels were simulated in the experiments by
varying the modulation depths of the modulators, and for an
attenuation of respectively 2\% and 60\% for the two channels, a
bit rate of 0.1 secret bits per use was deduced from a measured
excess noise of 0.01SNU and an assumed post-processing efficiency
of 97\%.



\section{Theoretical security aspects}\label{sectionMB_MT}

\subsection{Finite-size analysis in QKD}

Here, we describe how to lift the security analysis of QKD
protocols from more physical considerations to the same level of
rigor as found for classical primitives in theoretical
cryptography. This is crucial to assure that QKD can be safely
used as a cryptographic routine in any type of applications. We
have already seen the general structure of QKD protocols and
discussed the precise composable security criterion in
Section~\ref{BasicQKDSection}. Here we describe in detail how to
give mathematically precise finite key security analyzes. We
emphasize that such finite key analyzes are crucial to understand
the security properties of any practical quantum hardware, just as
every distributed key will have finite length and will only be
approximately secure.

We start by reviewing the different known methods for finite key
analyzes in Section~\ref{sec:finite-size} and then describe in
more detail the state-of-the-art approach based on entropic
uncertainty relations with side quantum information
(Section~\ref{sec:finite/eur}). As we will see, the security
intuitively follows from the two competing basic principles of
quantum physics: the uncertainty principle and entanglement. We
then discuss CV protocols in Section~\ref{sec:finite/contvar} and
end with an outlook Section~\ref{sec:finite/outlook}, commenting
on extensions to device-independent QKD.

\subsection{Finite-size statistical
analysis}\label{sec:finite-size}

It is natural to split the finite-size statistical analysis into
two steps, and in fact most security proofs respect that
structure. In the first step, discussed in
Subsection~\ref{sec:pa}, called \emph{privacy amplification}, we
explain how the criterion for composable security (introduced in
Subsection~\ref{sec:compos}) can be satisfied as long as we can
guarantee a sufficiently strong lower bound on a quantity called
the \emph{smooth min-entropy} $H^{\varepsilon}_{\mathrm{min}}$ of
Alice and Bob's corrected raw key conditioned on Eve's side
information. The second step, discussed in Subsection~\ref{sec:lb}
is then to find such lower bounds.

\subsubsection{Privacy amplification}\label{sec:pa}

Privacy amplification is a procedure that allows Alice and Bob,
who are assumed to share a random bit string (called the raw key)
about which the eavesdropper has only partial information, to
extract a shorter random bit string (the secret key) that is
guaranteed to be uncorrelated with the eavesdropper's information.
To make this more precise, we first need to define what
\emph{partial information} about the raw key means. Since we are
interested in finite-size effects, it turns out that the proper
way to measure the eavesdropper's information is by assessing its
probability of guessing the random bit string. This is done using
the \emph{smooth min-entropy}~\cite{renner1}. The higher the
smooth min-entropy of the eavesdropper on the raw key, the more
secret key can be extracted using the privacy amplification
scheme.

A simple way of extracting Alice's secret key is by applying a
random hash function to her raw bit strings (technically, the
random hash function must form a two-universal family). This was
shown to work even when the eavesdropper has a quantum memory by
Renner~\cite{renner1} who provided a quantum generalization of the
so-called \emph{Leftover Hashing Lemma}. This method has the
advantage that the random seed used to decide on the hash function
is independent of the resulting random bit string (i.e.\ the
extractor is strong) and hence the seed can be published over the
public channel to Bob. This way both Alice and Bob can apply the
same hash function, and since their initial bit strings agreed
their final strings will too. Thus, we end up with Alice and Bob
both holding bit strings that are independent of the
eavesdropper's information\,---\,and thus a secret key between
them has been successfully established.

Therefore, as long as we can guarantee some lower bound on the
min-entropy of the raw key, privacy amplification can be invoked
to extract a secret key.

\subsubsection{Guaranteeing large smooth min-entropy}\label{sec:lb}

There are a plethora of techniques available to analytically show
that the smooth min-entropy of the raw key is indeed large given
the observed data but it is worth summarizing the most important
ones. We give a more detailed exposition of one of the most
powerful techniques, based on entropic uncertainty relations, in
Section~\ref{sec:finite/eur}. We restrict our attention here to
finite-dimensional quantum systems and discuss CV protocols in
Section~\ref{sec:finite/contvar}.

\paragraph{Asymptotic equipartition and exponential de
Finetti~\cite{renner1}.} Under the  independent and identically
distributed (i.i.d.) assumption where the total state of the
system after the quantum phase has product form, the smooth
min-entropy can be bounded by the von Neumann entropy of a single
measurement, using the so-called quantum asymptotic equipartition
property~\cite{tomamichel08}. This von Neumann entropy can in turn
be estimated using state tomography, performed on the quantum
state shared between Alice and Bob. 
This approach often works directly when we only consider
individual or collective attacks, as the state after the
distribution phase then usually admits an i.i.d.\ structure.
However, with general coherent attacks such a structure can no
longer be guaranteed. Nonetheless, Renner~\cite{renner1}, in his
seminal work establishing the security of BB84 for finite length
keys, uses an exponential de Finetti theorem to argue that, in a
suitable sense, the general case can be reduced to the i.i.d.\
setting as well. This comes at a significant cost in extractable
key, however.

\paragraph{Asymptotic equipartition and
post-selection~\cite{christandl09}.} In particular the latter
reduction, using the exponential de Finetti theorem, leads to
large correction terms that make the security proof impractical.
It can be replaced by a significantly tighter method (based on
similar representation-theoretic arguments), the so-called
post-selection technique~\cite{christandl09}.

\paragraph{Virtual entanglement
distillation~\cite{SP00,hayashi11}.} This technique can be traced
back to one of the early security proofs by Shor and
Preskill~\cite{SP00} and has been adapted to deal with finite key
lengths by Hayashi and Tsurumaru~\cite{hayashi11}. The basic idea
is to interpret part of the raw key as a syndrome measurement of a
Calderbank-Shor-Steane (CSS) code and use it to virtually distill
entanglement on the remaining qubits. The crucial point is that
this can be done after measurement (hence virtual) so that no
multi-qubit quantum operations are required. The correctness and
secrecy of Alice and Bob's key then follows directly from the fact
that they effectively measured a close to maximally entangled
state.

\paragraph{Entropic uncertainty relations with side quantum
information~\cite{berta10,tomamichellim11}.} This approach is
based on an entropic uncertainty relation~\cite{tomamichel11} that
allows us to directly reduce the problem of lower bounding the
smooth min-entropy of the raw key conditioned on the
eavesdropper's information with a different, more tractable
problem. Namely, instead of bounding the min-entropy we need only
ensure that the correlations between Alice's and Bob's raw keys
are strong enough in an appropriate sense, which can be done by a
comparably simple statistical tests. This reduction naturally
deals with general attacks and gives tight bounds for finite keys.

The intuition and some of the details behind this approach will be
discussed in detail in the next section. The  proof technique has
recently been reviewed in detail and in a self-contained way
in~\cite{tomamichel17b}.

\paragraph{Entropy accumulation~\cite{arnon18}.}
A recent proof technique uses entropy accumulation~\cite{DFR} to
argue that the smooth min-entropy accumulates in each round of the
protocol. This security proof naturally deals with the so-called
device independent setting (see Section~\ref{sec:DIQKD}). The
bounds, while still not as strong as the ones that can be achieved
using the last two methods discussed above, have recently been
improved in~\cite{dupuis18}.

\subsection{Uncertainty principle versus entanglement: an intuitive
approach to QKD security}\label{sec:finite/eur}

One of the basic principles of quantum physics that is intuitively
linked to privacy is Heisenberg's uncertainty
principle~\cite{heisenberg27}. In its modern information-theoretic
form due to Maassen-Uffink~\cite{maassen88} it states that for any
two measurements $X,Z$ with eigenvectors $|x\rangle,|z\rangle$,
respectively, we have
\begin{align}\label{eq:shannonUR}
H(X)+H(Z)\geq-\log_2\max_{x,z}|\langle x|z\rangle|^2,
\end{align}
where $H(X)=-\sum_xp_x\log_2 p_x$ denotes the entropy of the
post-measurement probability distribution. Importantly, the bound
on the right-hand side is independent of the initial state and the
first ideas of directly making use of this uncertainty principle
for security proofs can be traced back
to~\cite{grosshans04,koashi05}.

It turns out, however, that when taking into account the most
general coherent attacks, the adversary might have access to a
quantum memory and with that to the purification of the state held
by the honest parties. Now, when starting with a maximally
entangled bipartite state $\Phi_{AB}$ and applying measurement $X$
or $Z$ on the $A$-system, then it is easily checked that there
always exists a measurement on the $B$-system that reproduces the
measurement statistics on $A$, independent if $X$ or $Z$ was
measured! This phenomenon was first discussed in the famous EPR
paper~\cite{epr35} and in terms of entropies it implies that
\begin{align}\label{eq:mes}
H(X|B)+H(Z|B)=0,
\end{align}
where $H(X|B):=H(XB)-H(B)$ denotes the conditional von Neumann
entropy of the post-measurement classical-quantum state. We
emphasize that this is in contrast to classical memory systems
$B$, for which the left-hand side of~\eqref{eq:mes} would always
respect the lower bound from~\eqref{eq:shannonUR}. 

Luckily, entanglement turns out to be monogamous in the sense that
for tripartite quantum states $ABC$ the more $A$ is entangled with
$B$ the less $A$ can be entangled with $C$ (and vice versa).
Moreover, it is now possible to make this monogamy principle of
entanglement quantitatively precise by showing that the
Maassen-Uffink bound is recovered in the tripartite
setting~\cite{berta10}
\begin{align}\label{eq:nature-vN}
H(X|B)+H(Z|C)\geq-\log_2\max_{x,z}|\langle x|z\rangle|^2.
\end{align}
It is this type of entropic uncertainty relation with quantum side
information that is employed for deducing the security of QKD.
(Entropic uncertainty relations with and without quantum side
information as well as their applications in quantum cryptography
are also reviewed in~\cite{coles17}.) Note that there is now no
need to distinguish between individual, collective, and coherent
attacks but rather~\eqref{eq:nature-vN} directly treats the most
general attacks. This is crucial for not ending up with too
pessimistic estimates for finite secure key rates.

To continue, we actually need an entropic uncertainty relation
suitable for the finite key analysis. This is in terms of smooth
conditional min- and max-entropies, taking the
form~\cite{tomamichel11}
\begin{align}
H_{\min}^{\varepsilon}(Y^n|E\Theta^n)+H_{\max}^{\varepsilon}(Y^n|\hat{Y}^n)\geq
n,
\end{align}
where $Y^n$ is Alice's raw key (of length $n$ bits), $\hat{Y}^n$
is Bob's raw key, $E$ denotes Eve's information and $\Theta^n$
labels the basis choice of the measurements made in the n rounds
by the honest parties (e.g., for BB84 if X or Z was chosen in each
round). An information reconciliation protocol can then be used to
make sure that Alice and Bob hold the same raw key. Taking into
account the maximum amount of information that gets leaked to the
eavesdropper in this process, denoted
$\textrm{leak}_{\textrm{ec}}$, this yields the bound
\begin{align}
H_{\min}^{\varepsilon}(Y^n|E\Theta^n) \geq n -
H_{\max}^{\varepsilon}(Y^n|\hat{Y}^n) -
\textrm{leak}_{\textrm{ec}}
\end{align}
Therefore, it only remains to statistically estimate
$H_{\max}^{\varepsilon}(Y^n|\hat{Y}^n)$, which can be done by
calculating the number of bit discrepancies between Alice and Bob
on a random sample of the raw keys~\cite{tomamichellim11}.

\subsection{CV protocols}\label{sec:finite/contvar}

For infinite-dimensional systems, finite-key approaches based on
exponential de Finetti or post-selection unfortunately
fail~\cite{Christandl2007}, unless additional assumptions are made
that lead to rather pessimistic finite-key rate
estimates~\cite{renner-cirac}. Fortunately, for protocols based on
the transmission of TMSV states measured via homodyne detection
(therefore squeezed-state protocols), the proof principle via
entropic uncertainty applies as well, leading to a tight
characterisation~\cite{furrer-comp,furrer-comp2}. This follows the
intuition from the early work~\cite{grosshans04} and is based on
the entropic uncertainty relations with side quantum information
derived in~\cite{furrer13,furrer11}.

Importantly, the analysis does not directly work with continuous
position and momentum measurements but rather with a binning
argument leading to discretized position and momentum
measurements. Namely, a finite resolution measurement device gives
the position $Q$ by indicating in which interval
$\mathcal{I}_{k;\delta}:=\big(k\delta,(k+1)\delta\big]$ of size
$\delta > 0$ the value $q$ falls $(k\in\mathbb{Z})$. If the
initial state is described by a pure state wave function
$|\psi(q)\rangle_{Q}$ we get
$\{\Gamma_{Q_{\delta}}(k)\}_{k\in\mathbb{Z}}$,
\begin{align}\label{eq:discrete_distribution}
& \Gamma_{Q_{\delta}}(k)=\int_{k\delta}^{(k+1)\delta}
\big|\psi(q)\big|^2\,\mathrm{d}q\; \\ & \text{with entropy
$H(Q_{\delta}):=-\sum_{k=-\infty}^{\infty}\Gamma_{Q_{\delta}}(k)\log_2\Gamma_{Q_{\delta}}(k)$.}
\end{align}

For these definitions we then recover a discretized version of the
of Everett-Hirschman~\cite{everett57,hirschman57} entropic
uncertainty relation
\begin{align}
H(Q_\delta)+H(P_\delta) & \geq\log_2(2\pi) \label{eq:finite_pq}
\\ & -\log_2\left[\delta_{q}\delta_{p}~S_{0}^{(1)}\left(1,\frac{\delta_q\delta_p}{4}\right)^{2}\right],\nonumber
\end{align}
where $S_{0}^{(1)}(\cdot,\cdot)$ denotes the $0$th radial prolate
spheroidal wave function of the first kind~\cite{rudnicki12}.
Extending this to quantum side information we find similarly as in
the finite-dimensional case that~\cite{furrer13}
\begin{align}
H(Q_{\delta q}|B)+H(P_{\delta p}|C) &
\geq\log_2(2\pi)\label{eq:discrete_qm}
\\ &
-\log_2\left[\delta_{q}\delta_{p}
~S_{0}^{(1)}\left(1,\frac{\delta_q\delta_p}{4}\right)^{2}\right].\nonumber
\end{align}
Extending this to the smooth min-entropy then allows for the same
security analysis as in~\eqref{eq:nature-vN}.

For other classes of CV protocols where the entropic uncertainty
approach is not known to work, finite key security is analyzed via
a recently discovered Gaussian de Finetti reduction that exploits
the invariance under the action of the unitary group $U(n)$
(instead of the symmetric group $S(n)$ as in usual de Finetti
theorems)~\cite{Leverrier15,Leverrier16,leverrier-comp,Ghorai-2way}.
This then recovers the widely held belief that Gaussian attacks
are indeed optimal in the finite key regime as well.

\subsection{Extensions and Outlook}\label{sec:finite/outlook}

Going forward from a theoretical viewpoint some of the main
challenges in the security analysis of quantum key distribution
schemes are:
\begin{itemize}
\item To refine the mathematical models on which the security proofs
are based to more accurately match the quantum hardware used in the
actual implementations. This is of crucial importance to decrease the
vulnerability to quantum hacking, which is typically based on side
channel attacks exploiting weaknesses of the quantum hardware~\cite{lydersen10}.
Intensified collaborations of theorists and experimentalists should help to
close this gap between realistic implementations and provable security.

\item Device independent QKD makes fewer assumptions on the devices used
and hence naturally takes care of issues with imperfect hardware.
However, it still remains to determine the ultimate finite key
rates possible in device-independent QKD. The state-of-the-art
works are based on entropy accumulation~\cite{arnon18,DFR} and
have recently been improved~\cite{dupuis18}. In contrast to the
tightest device dependent approach based on entropic uncertainty
relations with side quantum information (as presented in
Section~\ref{sec:finite/eur}), the lower bounds on the smooth
min-entropy are achieved in a device independent way by ensuring
that there is enough entanglement present. The details are beyond
the scope of this review but the open question is then to
determine if the same experimentally feasible trade-off between
security and protocol parameters is available as in the device
dependent case.

\item From a more business oriented perspective it is crucial to argue that
QKD schemes not only offer security in an abstract information-theoretic
sense but are actually more secure in practice compared to widely used classical
encryption schemes. That is, it is important to realize that in typical every
day use cases no cryptographic scheme is absolutely secure but instead the
relevant question is how much security one can obtain for how much money. Given
the ongoing development around post-quantum or quantum-secure-cryptography~\cite{NIST18}
we believe that there is still significant territory to conquer for QKD.
\end{itemize}

\section{Quantum hacking}

A practical implementation of QKD protocols is never perfect and
the performance of protocols depends on the applicability of the
security proofs and assumptions to the real devices, as well as on
numerous parameters, including post-processing efficiency and the
level of noise added to the signal at each stage (including the
noise added due to attenuation). Broadly speaking, quantum hacking
encompasses all attacks that allow an eavesdropper to gain more
information about messages sent between the trusted parties than
these parties assume to be the case, based on their security
proofs. Since security proofs are constructed on physical
principles, this can only be the case if one or more of the
assumptions required by the security proof does not
hold~\cite{scarani_black_2009}. If this is the case, the proof
will no longer be valid, and Eve may be able to gain more
information about the message than Alice and Bob believe her to
have. These assumptions include the existence of an authenticated
channel between Alice and Bob, the isolation of the trusted
devices (i.e. that Eve cannot access Alice and Bob's devices) and
that the devices perform in the way that they are expected to.
Certain forms of quantum hacking have already been mentioned in
this review, such as the photon-number splitting (PNS) attack
against DV-QKD protocols.

Exploitable imperfections in the trusted parties' devices that
allow quantum hacking are called side-channels. These could take
the form of losses within the trusted devices that could
potentially contribute to Eve's information about the signal, or
added noise within the devices that could be partially controlled
by Eve, in order to influence the key data. Such partially
controllable losses and noises constitute threats to the security
of QKD protocols, if overlooked. Quantum hacking often serves one
of two purposes: to directly gain information about the secret key
or to disguise other types of attack on a protocol, by altering
the trusted parties' estimation of the channel properties. To
restore security, the trusted parties can either incorporate the
side-channels into their security analysis, in order to not
underestimate Eve's information, or can modify their protocol to
include countermeasures. In this section, we will discuss some
common side-channel attacks, and how their effects can be
mitigated. See also Ref.~\cite{hackrev} for a very recent review
on the topic. It is clear that the study of quantum hacking is an
important aspect for the real-life security of QKD
implementations; it is central to the ongoing effort for the
standardization of QKD by the European Telecommunications
Standards Institute~\cite{LSA+18}.

\subsection{Hacking DV-QKD protocols}

The security proofs for many DV protocols, such as BB84
\cite{bennett_quantum_2014} and B92~\cite{B92}, assume the use of
single-photon sources. However, real QKD implementations often use
strongly attenuated laser pulses, rather than true single-photon
sources,
which will send some pulses with multiple photons~\cite%
{gaidash_revealing_2016}. The existence of such pulses allows the
use of the (previously mentioned) PNS attack. This is where Eve
beamsplits off all but one of the photons from the main quantum
channel. Since Bob is expecting to receive a single-photon pulse,
and since this pulse will be undisturbed (if Eve does not carry
out any other attack), the trusted parties will not detect any
additional error on the line. Eve can then store the photons she
receives in a quantum memory until after all classical
communication has been completed. She can then perform a
collective measurement on her stored qubits, based on the
classical communication, to gain information about the secret key,
without revealing her presence to the trusted parties. For
instance, in BB84, she will know all of the preparation bases used
by Alice, after the classical communication is complete, and so
will be able to gain perfect information about all of the key bits
that were generated by multi-photon pulses.

\subsubsection{PNS and intensity-based attacks}

A method used to counter the PNS attack is the use of decoy
states. For instance, the BB84 protocol can be modified into BB84
with decoy
states~\cite{decoy1,decoyWang,Wang4decoy,decoy,decoyfin}. In this
protocol, Alice randomly replaces some of her signal states with
multi-photon pulses from a decoy source. Eve will not be able to
distinguish between decoy pulses, from the decoy source, and
signal states, and so will act identically on both types of pulse.
In the post-processing steps, Alice will publicly announce which
pulses were decoy pulses. Using the yields of these decoy pulses,
the trusted parties can then characterize the action of the
channel on multi-photon pulses, and so can detect the presence of
a PNS attack. They can then adjust their key-rate accordingly, or
abort the protocol if secret key distribution is not possible.

Imperfections in Alice's source can give rise to exploitable
side-channels, which can allow Eve to carry out the PNS attack
undetected. Huang et al~\cite{huang_decoy_2017} tested a source
that modulates the intensity of the generated pulses by using
different laser pump-currents, and found that different
pump-currents cause the pulse to be sent at different times, on
average. This means that the choice of intensity setting
determines the probability that the pulse will be sent at a given
time, and hence it is possible for Eve to distinguish between
decoy states and signal states, based on the time of sending. Eve
can then enact the PNS attack on states that she determines to be
more likely to be signal states, whilst not acting on states that
she determines to be likely to be decoy states. This would allow
her PNS attack to go undetected by the trusted parties. They then
bounded the key rate for BB84 with decoy states, using an
imperfect source (for which the different intensity settings are
in some way distinguishable).

Huang et al.~\cite{huang_decoy_2017} also tested a source that
uses an external intensity modulator to determine the intensity
settings (meaning that the intensity is modulated after the laser
pulse is generated). They found that such sources do not give a
correlation between intensity setting and sending time, giving a
possible countermeasure to attacks based on this side-channel.
Another option is for Alice to change the time at which the
pump-current is applied depending on the intensity setting, in
order to compensate for this effect. Eve may be able to circumvent
this countermeasure, however, by heating Alice's source using
intense illumination. Fei et al.~\cite{fei_strong_2018} found that
if gain-switched semiconductor lasers are heated, the pulse
timings of different intensity settings shift relative to each
other, so Alice will no longer be able to compensate for the
timing differences unless she knows that they have been changed.
They also found that heating the gain medium can cause the time
taken for the carrier density to fall to its default level between
pulses to increase. This could lead to unwanted (by Alice)
correlations between pulses, which could compromise the security
of the protocol.

\subsubsection{Trojan horse attacks}

Another form of hacking that can be used against DV-QKD protocols
is the Trojan horse attack (THA)
\cite{vakhitov_large_2001,gisin_trojan-horse_2006,jain_risk_2015}.
This encompasses a variety of different types of attack that
involve sending quantum systems into one or both of the trusted
parties' devices in order to gain information. For instance,
Vakhitov et al.~\cite{vakhitov_large_2001} considered the use of
large pulses of photons to gain information about Alice's choice
of basis and about Bob's choice of measurement basis, in BB84 and
B92. The information is gained by sending a photon pulse into the
trusted device via the main channel and performing measurements on
the reflections. Considering the case in which the qubit is
encoded via a phase shift, if Eve is able to pass her pulse
through Alice's phase modulator, measuring the resulting pulse
will give some information about the signal state. This is
possible because Alice's phase modulator operates for a finite
amount of time (rather than only being operational for exactly
long enough to modulate the signal state), giving Eve a window in
which to send her own pulse through, to be modulated similarly to
the signal pulse. The process of Eve gaining information about the
basis choice via reflectometry is described in some detail by
Gisin et al.~\cite{gisin_trojan-horse_2006}.

The information may be partial, giving only the basis used, or may
directly give the key bit. Even in the case in which only the
basis can be obtained, the security of the protocol is still
compromised, as Eve is now able to always choose the same
measurement basis as Alice, for an intercept and resend attack,
gaining complete information about the key without introducing any
error. Alternatively, Eve may be able to target Bob's device. For
B92 or SARG04, it is sufficient to know Bob's measurement basis in
order to gain complete information about the key. In BB84, if Eve
is able to ascertain the measurement basis that Bob will use prior
to the signal state arriving at his device, she can carry out an
undetectable intercept and resend attack by choosing the same
basis as him. Vakhitov et al.~\cite{vakhitov_large_2001} also note
that even if Eve only gains information about Bob's basis after
the signal state has been measured, this can help with a practical
PNS attack, since it reduces the need for a quantum memory (Eve
can carry out a measurement on the photons she receives
immediately, rather than having to wait until after all classical
communication is completed). This does not aid an eavesdropper who
is limited only by the laws of physics, but it could help one who
is using current technology.

A number of methods of protecting against THAs have been proposed.
Vakhitov et al.~\cite{vakhitov_large_2001} suggested placing an
attenuator between the quantum channel and Alice's setup, whilst
actively monitoring the incoming photon number for Bob's setup.
Gisin et al.~\cite{gisin_trojan-horse_2006} calculated the
information leakage due to a THA, assuming heavy attenuation of
the incoming state, and suggested applying phase randomization to
any outgoing leaked photons. Lucamarini et
al.~\cite{lucamarini_practical_2015} calculated the key rates for
BB84, with and without decoy states (in the without case, assuming
an ideal single-photon source), in the presence of a Trojan horse
side-channel, parameterized by the outgoing photon number of the
Trojan horse state. They then proposed an architecture to
passively limit the potential information leakage via the Trojan
horse system (without using active monitoring). This was done by
finding the maximum incoming photon number in terms of the Laser
Induced Damage Threshold (LIDT) of the optical fibre, which can be
treated as an optical fuse. The LIDT is the power threshold over
which the optical fibre will be damaged. The minimum energy per
photon, which depends only on the frequency of the photons, is
lower-bounded using an optical fibre loop, which selects for
frequency (photons of too low a frequency will not totally
internally reflect, and so will leave the loop). The maximum time
for Eve's pulse is also known and bounded by the time it takes
Alice's encoding device to reset between signals. Hence, the
maximum number of photons per pulse can be upper-bounded. By
correctly setting the attenuation of incoming photons, Alice can
then upper-bound and reduce the information leakage due to any
THA.

Jain et al.~\cite{jain_risk_2015} considered using THAs at
wavelengths lower than the signal pulse in order to reduce the
risk of detection by the trusted parties. This could reduce the
efficacy of active monitoring of the incoming average photon
number, as detectors often have a frequency band at which they are
most sensitive, and so may not detect photons outside of this
band. It could also lead to reduced attenuation of the Trojan
state by passive attenuators, as the transmittance of a material
is frequency-dependent. Sajeed et al.~\cite{sajeed_security_2015}
carried out experimental measurements on equipment used in
existing QKD implementations at a potential THA wavelength as well
as at the wavelength used by the signal state, and found that the
new wavelength (1924 nm) reduced the probability of afterpulses in
Bob's detectors, which could alert the trusted users to the
attack. This came at the cost of increased attenuation and a lower
distinguishability, due to the phase modulator being less
efficient at the new wavelength. Jain et al.~\cite{jain_risk_2015}
suggest the use of a spectral filter to prevent attacks of this
type. Further, the optical fibre loop in the setup proposed by
Lucamarini et al.~\cite{lucamarini_practical_2015} would help
prevent attacks at low wavelengths.

Eve could also use a THA to target the intensity modulator, used
by Alice to generate decoy states. If Eve can distinguish between
decoy pulses and signal states, she can carry out a PNS attack
without being detected, by ignoring decoy pulses and only
attacking multi-photon signal states. Tamaki et
al.~\cite{tamaki_decoy-state_2016} created a formalism for
bounding the information leakage from the intensity modulator in
terms of the operation of the modulator (i.e. the unitaries
enacted by the modulator for each intensity level). They also
develop the formalism for calculating the information leakage due
to a THA against the phase modulator. Using these, they then
calculated the key rate for BB84 with various types of THA
(different assumptions about the details of the attack, and hence
about the outgoing Trojan state), for fixed intensity of the
outgoing Trojan horse mode.

Vinay et al.~\cite{vinay_burning_2018} expanded on the work by
Lucamarini et al. in~\cite{lucamarini_practical_2015}, and showed
that coherent states (displaced vacuum states) are the optimal
Trojan horse state, amongst the Gaussian states, for Eve to use in
an attack on BB84, assuming attenuation of the Trojan mode and a
limited outgoing photon number. Based on a calculation of the
distinguishability (a measure of the information leakage of a
THA), they showed that adding thermal noise to both the signal
state and the Trojan horse mode can provide an effective defence
against a THA, greatly increasing the key rate for a given
outgoing photon number. They then upper-bounded the
distinguishability for THA attacks using different photon number
statistics, expanding from the case of Gaussian states to more
general separable states (allowing correlations between different
Fock states, but assuming no entanglement between the Trojan horse
mode and some idler state held by Eve). They found that this upper
bound on the distinguishability was higher than the bound found in
Ref.~\cite{lucamarini_practical_2015} but that it could be reduced
to below the Lucamarini bound by applying their thermal noise
defence. They also proposed the use of a shutter between Alice's
device and the main channel, in conjunction with a time delay
between the encoding apparatus and the shutter, as a defence that
could be used in place of an attenuator. This would work by
forcing the Trojan pulse to make several journeys through Alice's
encoding apparatus, making it more difficult for Eve to accurately
determine the encoded phase.

\subsubsection{Backflash attacks}

A different type of side-channel attack is the detector backflash
attack, introduced by Kurtsiefer et
al.~\cite{kurtsiefer_breakdown_2001}. Detectors based on avalanche
photodiodes (APDs) sometimes emit light (referred to as backflash
light) upon detecting a pulse. This backflash light can give
information to Eve about Bob's measurement outcome in a variety of
ways. The polarization of the backflash light can give an
indication of which components of Bob's system it has passed
through, which could tell Eve which detector it originated
from~\cite{pinheiro_eavesdropping_2018}. Alternatively, the travel
time of the backflash photons (after entering Bob's detector) or
path-dependent alterations to the profile of the outgoing light
could also give Eve this information. This could tell Eve which
measurement basis was chosen by Bob, and for certain detector
setups could even reveal Bob's measurement outcome.

Meda et al.~\cite{meda_backflash_2017} characterized two
commercially used InGaAs/InP APDs and found that backflash light
could be detected for a significant proportion of avalanche
events. Pinheiro et al.~\cite{pinheiro_eavesdropping_2018} then
built on this work by characterizing a commercial Si APDs; they
also found that the backflash probability was significant, with a
backflash probability greater than or equal to 0.065. Both papers
found that the backflash light was broadband, and so could be
reduced using a spectral filter. Pinheiro et
al.~\cite{pinheiro_eavesdropping_2018}  also characterized a
photomultiplier tube, and found that it had a much lower backflash
probability. They therefore suggested using photomultiplier tubes
in place of APDs in Bob's detectors.

\subsubsection{Faked states and detector efficiency mismatch}

The security of BB84 (and most DV protocols) is based on Eve and
Bob's basis choices being independent. If Eve is able to exploit
some imperfection in Bob's device that lets her influence Bob's
basis or even choose it for him, this independence would no longer
hold, and the security of the protocol may be breached. Makarov et
al.~\cite{makarov_*_faked_2005} proposed a number of schemes that
could allow an eavesdropper to control or influence Bob's detector
basis or measurement results. These schemes use faked states; this
is where Eve does not attempt to gain information without
disturbing the signal state, but instead sends a state designed to
take advantage of flaws in Bob's detection device to give him the
results that she wants him to receive.

Two of the proposed schemes take advantage of Bob's passive basis
selection. Receiver implementations can select the measurement
basis using a beamsplitter: this will randomly allow a photon
through or reflect it onto another path. We call this passive
basis selection, and differentiate it from active basis selection,
in which Bob explicitly uses a random number generator to select
the basis and changes the measurement basis accordingly
\cite{stipcevic_preventing_2014}. If the beamsplitter used is
polarization-dependent, Eve can tune the polarization of the
pulses she sends such that they go to the basis of her choosing,
allowing her to intercept and resend signals whilst ensuring that
she and Bob always choose the same basis.

The second scheme proposes a way around the countermeasure of
placing a polarizer in front of the beamsplitter. Makarov et
al.~\cite{makarov_*_faked_2005} suggest that imperfections in the
polarizer will allow Eve to still choose Bob's basis if she uses
sufficiently large pulses (albeit with heavy attenuation due to
the polarizer). They also suggest the use of a polarization
scrambler as a defence against these two schemes. Li et
al.~\cite{li_attacking_2011} proposed a similar type of attack, in
which they exploit the frequency-dependence of the beamsplitter to
allow them to choose Bob's basis with high probability.

Makarov et al.~\cite{makarov_*_faked_2005} proposed two further
schemes using faked states. One takes advantage of unaccounted for
reflections in Bob's device, which could allow Eve to choose Bob's
basis by precisely timing her pulses such that part of them
reflects into the detectors in the correct time window for the
chosen basis. This would require Eve to have a very good
characterisation of Bob's device and carries the risk of detection
due to side-effects caused by the part of the pulse that is not
reflected. The second attack, briefly introduced in
\cite{makarov_*_faked_2005} and then expanded on in
\cite{makarov_effects_2006}, exploits detector efficiency mismatch
(DEM). This is where the detector corresponding to the outcome 0
has a different efficiency to the detector corresponding to the
outcome 1 (here we are considering setups in which both bases are
measured using the same pair of detectors, with a phase modulator
beforehand to determine the basis used; it is also possible to
have DEM in a setup in which four detectors are used) for some
values of a parameter. This parameter could be time or some other
parameter such as polarization, space or frequency.

If Bob's detectors have DEM, a photon with a certain value of the
parameter (e.g. a certain polarization) would be more likely to be
detected by the detector giving the value 0, if it were to hit it,
than to be detected by the detector giving the value 1, if it were
to hit that detector (or vice versa). Note, however, that in
reality the photon will only hit one of the detectors, depending
on what value it encodes. In the attack, Eve intercepts Alice's
states and measures them in some basis. She then sends a state to
Bob in the opposite basis, with the timing chosen such that Bob is
likely to receive no result rather than an error, reducing Bob's
error rate at the cost of increasing loss (with the likelihood of
this depending on how mismatched the detector efficiencies are).
We say that one of Bob's detectors has been blinded, since it is
less able to pick up signals.

Lydersen et al.~\cite{LWWESM} demonstrated the possibility of
using faked states to attack commercial QKD systems. They blinded
the detectors using a continuous wave laser, exploiting the fact
that APDs can be made to operate in linear mode; in this mode, the
detectors do not register single photons. Eve sends in pulses to
trigger Bob's detectors, and they will only give a result when he
chooses the same basis as Eve (with the result in this case being
the same as Eve's value). Lydersen et
al.~\cite{lydersen_thermal_2010} showed that the APDs could also
be blinded by heating them using bright light. Yuan et
al.~\cite{yuan_avoiding_2010,yuan_resilience_2011} argued that a
properly operated APD would be difficult to keep in linear mode
and that faked states attacks
of this type could be identified by monitoring the photocurrent. Stip\v{c}evi%
\'{c} et al.~\cite{stipcevic_preventing_2014} showed that detector
blinding attacks can be treated as an attack on Bob's random
number generator, and suggested that an actively selected basis
with a four detector configuration (rather than using the same two
detectors to measure both bases) could mitigate the effects of
many types of blinding attack.

Qi et al.~\cite{qi_time-shift_2005} suggested a different attack
based on DEM with the time parameter, which they called the
time-shift attack. In this attack, Eve does not attempt to measure
the signal state, but simply shifts the time at which it enters
Bob's device, such that, if Bob has a detection event, it is more
likely to be one value than another. For instance, if the detector
corresponding to the outcome 0 has a higher efficiency at some
given time than the detector corresponding to the outcome 1, Eve
can know that if the time of arrival of the pulse is shifted such
that it arrives at the detector at that particular time, and the
pulse is detected by one of the detectors, it is more likely that
the outcome was 0 than that it was 1. The greater the DEM, the
more information Eve can gain about the key. This attack does not
introduce any error in the resulting key bits, and was shown to be
practical using current technology by Zhao et
al.~\cite{zhao_quantum_2008}, who carried out the attack on a
modified commercial system.

In fact, there is no requirement that the DEM be parameterized by
time. The detectors could be mismatched over polarization,
frequency or even spatially~\cite{fung_security_2008}. Sajeed et
al.~\cite{sajeed_security_2015} and Rau et
al.~\cite{rau_spatial_2015} both showed that by altering the angle
at which pulses enter Bob's device, it is possible to alter the
relative sensitivity of the detectors, since the angle of entry
will determine the angle at which the light hits the detectors,
but small changes in the configuration of the setup could lead to
different angles of incidence on each detector. The angle at which
the pulse hits the detector determines the surface area of the
detector that is hit by the pulse, and hence the sensitivity of
the detector.

One proposed countermeasure against attacks exploiting DEM is for
Bob to use a four detector configuration, with the mapping of
outcomes to detectors randomly assigned each time. However, this
configuration is still vulnerable to a time-shift (or other DEM)
attack, if the attack is coupled with a THA on Bob's device (which
learns the detector configuration after the measurement has been
carried out)~\cite{lydersen_security_2008}. A THA of this type is
hard for Bob to defend against, and making Bob's device more
complicated by adding hardware safeguards risks opening up more
vulnerabilities for Eve to exploit, and so it is desirable to find
a software solution (i.e. to calculate the key rate accounting for
the possibility of DEM attacks).

Fung et al.~\cite{fung_security_2008} found a formula for the key
rate in the presence of a DEM for a very broad class of attacks,
using a proof devised by Koashi~\cite{koashi_unconditional_2006}.
Lydersen et al.~\cite{lydersen_security_2008} generalized this
proof slightly. Both formulae require a thorough characterisation
of the detector efficiencies over possible values of the DEM
parameter. This may be difficult for the trusted parties,
especially since they may not always know which parameters of the
detector give rise to DEM. Fei et al.~\cite{fei_practical_2018}
calculated the key rate of BB84 with decoy states in the presence
of DEM, using a new technique based on treating the detection
process as a combination of the case in which there is no DEM and
the case in which there is complete DEM (i.e. for some value of
the parameter, there is 0\% detection efficiency for one detector,
but not for the other, and vice versa for some other value of the
parameter). They then numerically simulated QKD in this case, and
found that DEM decreases the secure key rate.

\subsection{Hacking CV-QKD protocols}

The differences between the types of protocols and devices used in
DV and CV protocols mean that not all hacking attacks on DV
systems are directly applicable to CV systems. Some
vulnerabilities, such as attacks on the local oscillator (LO), are
specific to CV protocols, whilst some, such as THAs, are analogous
to the attacks used on DV protocols. An important practical issue
in the implementation of CV-QKD is calibration of the equipment
used. The shot-noise must be determined, since it affects the
parameter estimation. If this is not carried out accurately, the
security of CV QKD may be undermined~\cite{Jouguet2012b}. During
calibration, the phase noise introduced during modulation should
be estimated. By taking it into account in the security analysis,
the key rate can be increased, since the phase noise added by the
modulator can then be treated as trusted noise.

\subsubsection{Attacks on the local oscillator}

To carry out measurements of Alice's signal states, Bob interferes
them with an LO. Due to the difficulty of maintaining coherence
between Alice's source and Bob's LO, implementations of CV-QKD
often send the LO through the quantum channel. Since security
proofs of CV-QKD do not account for this (as it is not
theoretically necessary to send the LO through the channel), this
leaves open some side-channels, which Eve can exploit. H\"{a}seler
et al.~\cite{haseler_testing_2008} showed that the intensity of
the LO must be monitored, in order to prevent Eve from replacing
the signal state and LO with squeezed states, in such a way as to
disguise an intercept and resend attack, by reducing the error
relative to what the trusted parties would expect for such an
attack. Huang et al.~\cite{huang_quantum_2013} and Ma et
al.~\cite{ma_wavelength_2013} proposed an attack on the LO based
on the wavelength-dependency of beamsplitters. They found that by
exploiting the wavelength-dependence of the beamsplitters in Bob's
device, they could engineer Bob's outcomes, whilst preventing Bob
from accurately determining the LO intensity. Huang et al.
proposed a countermeasure in which a wavelength filter is applied
at random, and any difference in channel properties between the
cases in which it is applied and not applied is monitored.

Jouguet et al.~\cite{jouguet_preventing_2013} devised another
attack on the LO. This attack uses the fact that Bob's clock is
triggered by the LO pulse. By changing the shape of the LO pulse,
Eve can delay the time at which the clock is triggered. This can
lead to Bob incorrectly calculating the shot-noise, and hence can
allow Eve to carry out an intercept and resend attack undetected.
As a countermeasure, Jouguet et al. suggest that Bob should
measure the shot-noise in real-time by randomly applying strong
attenuation to the signal. Huang et al.~\cite{huang_quantum_2014}
built on this by showing that an attack exploiting the
wavelength-dependence of beamsplitters could be used to defeat
Bob's attempt to measure the shot-noise in real-time. However,
they found that by adding a third attenuation value (rather than
just on or off) to the strong attenuation could prevent their
attack. Xie et al.~\cite{xie_practical_2018} also found that a
jitter effect in the clock signal can lead to an incorrect
calculation of the shot noise, and Zhao et
al.~\cite{ZhaoGuoPola19} identified a polarization attack where
the eavesdropper attacks unmeasured LO pulses to control and
tamper with the shot-noise unit of the protocol.

In order to prevent LO attacks altogether, Qi et al.~\cite{Qi2015}
and Soh et al.~\cite{Soh2015} proposed and analyzed a way in which
Bob could generate the local oscillator locally (LLO). Alice
regularly sends phase reference pulses, and Bob applies a phase
rotation to his results during post-processing, to ensure that
they are in phase with Alice's source. Marie et
al.~\cite{marie_self-coherent_2017} improved on this scheme, in
order to reduce the phase noise. Ren et
al.~\cite{ren_reference_2017} proposed that even an LLO could be
vulnerable to a hacking attack if the trusted parties assume that
the phase noise is trusted, and cannot be used by Eve. In this
case, Eve can lower the phase noise, by increasing the intensity
of the phase reference pulses, and compensate for the reduced
phase noise by increasing her attack on the signal states, so that
the total noise on Bob's measurements remains the same.

\subsubsection{Saturation attacks on detectors}

Qin et al.~\cite{qin_quantum_2016} considered saturation attacks
on Bob's homodyne detectors. Such attacks exploit the fact that
CV-QKD security proofs assume a linear relationship between the
incoming photon quadratures and the measurement results (that the
quadrature value linearly corresponds to the measurement result),
but in reality, homodyne detectors have a finite range of
linearity. Above a certain quadrature value, homodyne detectors
will saturate, meaning that the measurement result will be the
same whether the quadrature value is at the threshold level or
above it. For instance, a quadrature value of 100 shot-noises
could give the same measurement result as a quadrature value of
200 shot-noises. Qin et al. considered exploiting this by using an
intercept and resend attack and then rescaling and displacing the
measured states (multiplying them by some factor and then adding a
constant displacement to them). By causing Bob's measurement
results to partially overlap with the saturation region, Eve can
alter the distribution of measurement results, and so reduce the
trusted parties' error estimation.

Qin et al.~\cite{qin_quantum_2016} also proposed some
countermeasures, including the use of a Gaussian post-selection
filter~\cite{Fiurasek2012,Walk2013} to try and ensure that the
measurement results used for key generation fall within the linear
range of the detector and the use of random attenuations of Bob's
signal, to test whether the measurement results are linearly
related to the inputs. Qin et al.~\cite{qin_homodyne_2018}
expanded on their previous work, considering a slightly different
attack in which an incoherent laser is used to displace Bob's
measurement results into the saturated range. They also
demonstrated saturation of a homodyne detector experimentally and
numerically simulated their attack to show feasibility.

\subsubsection{Trojan horse attacks}

CV protocols are also vulnerable to THAs~\cite{Stiller2015}. By
sending Trojan states into Alice's encoding device, Eve can try to
learn how the signal states have been modulated, without
disturbing the signal state itself. Derkach et
al.~\cite{Derkach2016} considered a leakage mode side-channel in
Alice's device. They modeled this side-channel as a beamsplitter
in Alice's device, coupling the signal state to a vacuum state
after modulation. They also considered a side-channel allowing Eve
to couple an untrusted noise to the signal state prior to
detection. They then calculated the resulting key rates for both
coherent state and squeezed state protocols, using reverse
reconciliation. Derkach et al. suggest some countermeasures to the
sender side-channel based on manipulation of the input vacuum
state to the beamsplitter, and to the receiver side-channel based
on measuring the output of the coupled noise. They then expanded
on their earlier work~\cite{derkach_continuous-variable_2017},
considering two types of side-channel leakage: leakage after
modulation of the signal state and leakage prior to modulation,
but after squeezing of the signal state (in the squeezed state
protocol). They allowed multiple leakage modes from each
side-channel. They calculated the key rates for both direct and
reverse reconciliation, for side-channels of this type, and
optimized the squeezing for post-modulation leakage.

Pereira et al.~\cite{pereira_hacking_2018} considered a THA on
Alice, in the coherent state protocol, in which Eve is able to
send a Trojan state with a bounded average photon number into
Alice's box. This state is then modulated in a similar way to the
signal state and returned to Eve. The key rate and security
threshold are calculated, for reverse reconciliation. Active
monitoring of incoming light is suggested as a countermeasure. Ma
et al.~\cite{ma_quantum_2016} considered a THA on the two-way
protocol, in which Eve sends a state into Alice's device,
following Bob's signal state, and then measures this state to gain
information about the modulation applied by Alice. They suggest
the use of active monitoring to remove the Trojan state.

Part of the noise originating from the trusted parties' devices
can be assumed to be trusted and therefore not under the control
of an eavesdropper. Such trusted noise could be the noise of the
signal states (e.g. thermal noise in thermal-state protocols),
noise added by the modulators or the noise of the detection.
Trusted noise can have different impacts on CV QKD depending on
the reconciliation direction. Trusted noise on the reference side
of the protocol can even be helpful due to decoupling Eve's
systems from the information shared by the trusted parties. On the
other hand, noise on the remote side of reconciliation protocols
can be harmful for the protocols, despite being
trusted~\cite{Usenko2016}.

\subsection{General considerations}

A number of more general attacks exist, which can be used against
both DV or CV protocols (although much of the current research has
been focused on DV protocols), based on altering the properties of
the devices used. These type of attacks can be used to create
vulnerabilities even in well-characterized devices.

Jain et al.~\cite{jain_device_2011} suggested and experimentally
tested an attack that Eve could carry out during the calibration
phase of a QKD protocol. The attack targets the system whilst Bob
is calibrating his detectors (for a DV protocol) using a line
length measurement (LLM). Attacks of this type are implementation
dependent; in the system under consideration, Jain et al. found
that by changing the phase of the calibration pulses sent by Bob
during the LLM, they could induce a DEM (in the time parameter).
This would open up the system to other types of attack (such as
those previously mentioned), and would be especially problematic,
since the trusted parties would not realise that Bob's device was
miscalibrated. Building on this, Fei et
al.~\cite{fei_quantum_2018} found that by sending faked
calibration pulses during the LLM process, they could induce DEM
or basis-dependent DEM with high probability. Fei et al. suggest
adding a system to allow Bob to test his own device for
calibration errors after the calibration process.

Even if an implementation is perfect, it could be possible for Eve
to create vulnerabilities, by damaging components of the trusted
parties' devices using a laser. Bugge et
al.~\cite{bugge_laser_2014} suggested that Eve could use a laser
to damage to components such as the detectors or any active
monitoring devices, allowing other attacks to be enacted. They
showed that APDs could be damaged by intense laser light, reducing
their detection efficiency and hence permanently blinding them.
This creates loopholes for Eve, without requiring her to
continuously ensure that the detectors are kept blinded. Higher
laser powers rendered APDs completely non-functional; this could
be exploited by Eve if an APD were being used as a monitoring
device (e.g. for Trojan pulses). Makarov et
al.~\cite{makarov_creation_2016} demonstrated this on a commercial
system and then showed that they were able to melt a hole in a
spatial filter, meant to protect against spatial DEM.

Sun et al.~\cite{sun_effect_2015} considered an attack on Alice's
source. By shining a continuous wave laser onto Alice's gain
medium, Eve is able to control the phase of Alice's pulses. This
could open up loopholes for other attacks in both DV and CV
systems. Sun et al. suggest monitoring the light leaving Alice's
source and the use of active phase randomization.

\subsection{Device-independence as a solution?}

A conceptually different approach to dealing with side-channels is
the development of device-independent protocols
(DI-QKD)~\cite{Eke91}, which can prevent a lot of side-channel
attacks. As discussed in Section~\ref{sec:DIQKD}, this is a type
of QKD that allows for untrusted devices, which could even have
been produced by Eve. Schemes for implementing DI-QKD have been
designed for both the DV~\cite{PABGMS} and the
CV~\cite{marshall_device-independent_2014} cases. Where sources
can be trusted, measurement device-independent QKD (MDI-QKD)
schemes can be used instead.  These have also been designed for
both the DV~\cite{BP,Lo} and the CV~\cite{RELAY} cases.  DI- and
MDI-QKD protocols are harder to implement and so in general give
lower key rates than device-dependent protocols.  In spite of
improving security, neither are immune to attack.  In all
protocols there is a requirement that Alice and Bob's devices be
isolated from the outside world (in particular from Eve). If there
is a hidden channel that allows Eve to gain access to measurement
outcomes, then the key will not be secure. MDI-QKD is also
vulnerable to source imperfections, such as the
previously-mentioned attack by Sun et al.~\cite{sun_effect_2015}.
Therefore, device-independence cannot be seen as a panacea for
side-channels.



\section{Limits of point-to-point QKD}\label{sec:ultimateQKD}

\subsection{Overview}

One of the crucial problems in QKD is to achieve long distances at
reasonably-high rates. However, since the proposal of the BB84
protocol~\cite{BB84}, it was understood that this is a daunting
task because even an ideal implementation of this protocol (based
on perfect single-photon sources, ideal detectors and perfect
error correction) shows a linear decay of the secret key rate $R$
in terms of the loss $\eta$ in the channel, i.e., $R=\eta/2$. One
possible way to overcome the rate problem was to introduce CV QKD\
protocols. Their ideal implementation can in fact beat any DV QKD
protocol at any distance, even though current practical
demonstrations can achieve this task only for limited distances
due to some implementation problems connected with finite
reconciliation efficiency and other technical issues.

One of the breakthrough in CV QKD\ was the introduction of the
reverse reconciliation (RR)~\cite{Grosshans2002}, where it is
Alice to infer Bob's outcomes $\beta$, rather than Bob guessing
Alice's encodings $\alpha$, known as direct reconciliation (DR).
This led the CV\ QKD community to considering a modified
Devetak-Winter rate~\cite{winter} in RR. This takes the form of
$I(\alpha:\beta)-\chi(E:\beta)$, where the latter is Eve's Holevo
information on Bob's outcomes. In a CV QKD\ setup, where both the
energy and the entropy may hugely vary at the two ends of a lossy
communication channel, there may be a non-trivial difference
between the two reconciliation methods. Most importantly, it was
soon realized that RR allowed one to achieve much longer
distances, well beyond the $3$dB limit of the previous CV\
approaches. At long distances (i.e., small transmissivity $\eta$),
an ideal implementation of the CV QKD\ protocols proposed in
Refs.~\cite{Grosshans2003,Weedbrook2004} has rate
$R\simeq\eta/(2\ln2)\simeq0.72\eta$. An open question was
therefore raised:

\begin{itemize}
\item What is the maximum key rate (secret key capacity)
achievable at the ends of a pure-loss channel?
\end{itemize}

With the aim of answering this question, a 2009
paper~\cite{ReverseCAP} introduced the notion of reverse coherent
information (RCI) of a bosonic channel. This was quantity was
previously defined in the setting of
DVs~\cite{RevCohINFO,DeveRuskai}. It was called \textquotedblleft
negative cb-entropy of a channel\textquotedblright~in
Ref.~\cite{DeveRuskai} and ``pseudocoherent information'' in
Ref.~\cite{Hayashi}; Ref.~\cite{RevCohINFO} introduced the
terminology of RCI and, most importantly, it showed its
fundamental use as lower bound for entanglement distribution over
a quantum channel (thus extending the hashing
inequality~\cite{winter} from states to channels).
Ref.~\cite{ReverseCAP} extended the notion to CVs where it has its
more natural application.

Given a bosonic channel $\mathcal{E}$, consider its asymptotic
Choi matrix
$\sigma_{\mathcal{E}}:=\lim_{\mu}\sigma_{\mathcal{E}}^{\mu}$. This
is defined
over a sequence of Choi-approximating states of the form $\sigma_{\mathcal{E}%
}^{\mu}:=\mathcal{I}_{A}\otimes\mathcal{E}_{B}(\Phi_{AB}^{\mu})$,
where $\Phi_{AB}^{\mu}$ is a two-mode squeezed vacuum (TMSV)
state~\cite{RMPwee} with $\bar{n}=\mu-1/2$ mean thermal photons in
each mode. Then, we define its RCI as~\cite{ReverseCAP}
\begin{align}
I_{\text{RCI}}(\mathcal{E})  &  :=\lim_{\mu}I(A\langle B)_{\sigma
_{\mathcal{E}}^{\mu}},\\
I(A\langle B)_{\sigma_{\mathcal{E}}^{\mu}}  &  :=S[\mathrm{Tr}_{B}%
(\sigma_{\mathcal{E}}^{\mu})]-S(\sigma_{\mathcal{E}}^{\mu}), \label{eqRCI}%
\end{align}
with $S(\sigma):=-\mathrm{Tr}(\sigma\log_{2}\sigma)$ is the von
Neumann entropy of $\sigma$. Here first note that, by changing
$\mathrm{Tr}_{B}$ with $\mathrm{Tr}_{A}$ in Eq.~(\ref{eqRCI}), one
defines the coherent information of a bosonic
channel~\cite{ReverseCAP}, therefore extending the definition of
Refs.~\cite{Schu96,Lloyd97}\ to CV\ systems. Also note that $I_{\text{RCI}%
}(\mathcal{E})$ is easily computable for a bosonic Gaussian
channel, because $\sigma_{\mathcal{E}}^{\mu}$ would be a two-mode
Gaussian state.



Operationally, the RCI of a bosonic channel represents a lower
bound for its secret key capacity and, more weakly, its
entanglement distribution capacity~\cite{ReverseCAP}. A powerful
CV QKD protocol reaching the RCI of a bosonic channel consists of
the following steps:

\begin{itemize}
\item Alice sends to Bob the $B$-modes of TMSV states $\Phi_{AB}^{\mu}$ with
variance $\mu$.

\item Bob performs heterodyne detections of the output modes sending back a
classical variable to assist Alice.

\item Alice performs an optimal and conditional joint detection of all the $A$-modes.
\end{itemize}

The achievable rate can be computed as a difference between the
Alice Holevo information $\chi(A:\beta)$ and Eve's Holevo
information $\chi(E:\beta)$ on Bob's outcomes. Note that this is
not a Devetak-Winter rate (in RR) but rather a generalization,
where the parties' mutual information is replaced by the Holevo
bound. Because Eve holds the entire purification of $\sigma
_{\mathcal{E}}^{\mu}$, her reduced state $\rho_{E}$ has entropy
$S(\rho _{E})=S(\sigma_{\mathcal{E}}^{\mu})$. Then, because Bob's
detections are rank-1 measurements (projecting onto pure states),
Alice and Eve's global state $\rho_{AE|\beta}$ conditioned to
Bob's outcome $\beta$ is pure. This means that
$S(\rho_{E|\beta})=S(\rho_{A|\beta})$. As a result, Eve's Holevo
information becomes%
\begin{equation}
\chi(E:\beta):=S(\rho_{E})-S(\rho_{E|\beta})=S(\sigma_{\mathcal{E}}^{\mu
})-S(\rho_{A|\beta}).
\end{equation}
On the other hand, we also write%
\begin{equation}
\chi(A:\beta):=S(\rho_{A})-S(\rho_{A|\beta}),
\end{equation}
where $\rho_{A}:=\mathrm{Tr}_{B}(\sigma_{\mathcal{E}}^{\mu})$ and
$\rho_{A|\beta}$ is conditioned to Bob's outcome. As a result we
get the following achievable rate
\begin{equation}
R^{\mu}(\mathcal{E}):=\chi(A:\beta)-\chi(E:\beta)=I(A\langle
B)_{\sigma _{\mathcal{E}}^{\mu}}.
\end{equation}

By taking the limit for large $\mu$, this provides the key rate
$R(\mathcal{E}):=\mathrm{lim}_{\mu} R^{\mu}(\mathcal{E})
=I_{\text{RCI}}(\mathcal{E})$, so that the secret key capacity of
the channel can be bounded as
\begin{equation}
\mathcal{K}(\mathcal{E})\geq I_{\text{RCI}}(\mathcal{E})~.
\end{equation}
In particular, for a pure-loss channel $\mathcal{E}_{\eta}$ with
transmissivity $\eta$, we may write~\cite{ReverseCAP}
\begin{equation}
\mathcal{K}(\mathcal{E}_{\eta})\geq I_{\text{RCI}}(\mathcal{E}_{\eta}%
)=-\log_{2}(1-\eta). \label{LB2009}%
\end{equation}
At long distances $\eta \simeq 0$, this achievable key rate decays
as $\simeq\eta/\ln 2\simeq1.44\eta$ bits per channel use.

\begin{figure*}[ptbh]
\begin{center}
\vspace{0.1cm}
\includegraphics[width=0.8\textwidth]{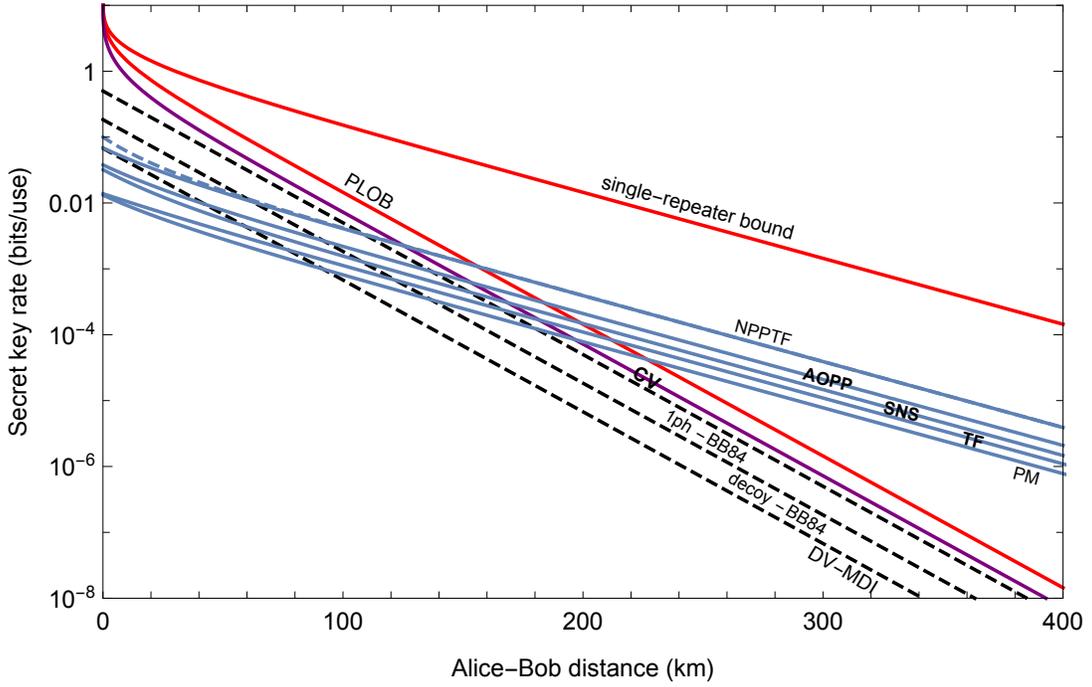}
\end{center}
\par
\vspace{-0.2cm}\caption{State of the art in high-rate QKD. We plot
the ideal key rates of several point-to-point and relay-assisted
protocols with respect to the PLOB bound~\cite{PLOB} of
Eq.~(\ref{PLOBeqn}) and the single-repeater
bound~\cite{net2006,net2006p} of Eq.~(\ref{singleREPbound}). The
key rates are expressed in terms of bits per channel use and
plotted versus distance (km) at the standard fiber-loss rate of
$0.2$~dB per km. In particular, below the PLOB bound we consider:
(\textbf{CV}) One-way coherent-state protocol with heterodyne
detection~\cite{Weedbrook2004}, which coincides with the most
asymmetric protocol for CV-MDI-QKD~\cite{RELAY}. For long
distances, the rate scales as $1/2$ of the PLOB bound. The same
asymptotic scaling is found for the coherent-state protocol with
homodyne detection~\cite{Grosshans2003}; (\textbf{1ph-BB84}) The
ideal BB84 protocol, implemented with single-photon
sources~\cite{BB84}. This achieves the ideal rate of $\eta/2$;
(\textbf{decoy-BB84}) BB84 protocol implemented with weak coherent
pulses and decoy states achieving the ideal rate of $\eta/(2e)$;
(\textbf{DV-MDI}) Ideal implementation of a passive MDI-QKD
node~\cite{BP,Lo}. In particular, we plot the ideal performance of
decoy state DV-MDI-QKD~\cite{Lo} with a rate of $\eta/(2e^{2})$.
Then, we consider relay-assisted protocols able to beat the PLOB
bound. In particular: (\textbf{TF}) Twin-field QKD
protocol~\cite{TFQKD}; (\textbf{PM}) phase-matching QKD
protocol~\cite{PMQKD}; (\textbf{SNS}) sending or not sending
version of TF-QKD~\cite{SNSwang}; (\textbf{AOPP}) Active
odd-parity pair protocol~\cite{SNSAOPP}, which is an improved
formulation of the SNS protocol; (\textbf{NPPTF})
No-phase-postselected TF-QKD protocol~\cite{Cuo3}, including the
variant of Ref.~\cite{NPPTF3} with improved rate at short
distances (blue dashed line).} \label{poi}
\end{figure*}

With the aim of providing an upper bound to the key rate of CV QKD
protocols, in 2014 Ref.~\cite{TGW}\ introduced the
Takeoka-Guha-Wilde (TGW) bound by employing the notion of squashed
entanglement~\cite{Squashed} for a bosonic channel. This led to
the upper bound
\begin{equation}
K(\mathcal{E}_{\eta})\leq\log_{2}\left(
\frac{1+\eta}{1-\eta}\right)  ,
\label{TGW}%
\end{equation}
which is $\simeq2.88\eta$ bits per use at long distances. By
comparing the lower bound in Eq.~(\ref{LB2009}) and the upper
bound in Eq.~(\ref{TGW}), we see the presence of a gap.

This gap was finally closed in 2015 by Ref.~\cite{PLOB} which
derived the Pirandola-Laurenza-Ottaviani-Banchi (PLOB) upper bound
for the pure-loss channel
\begin{equation}
K(\mathcal{E}_{\eta})\leq -\log _{2}(1-\eta).\label{PLOBintrosec}
\end{equation}
This was done by employing the relative entropy of entanglement
(REE)~\cite{RMPrelent,VedFORMm,Pleniom}, suitably extended to
quantum channels, combined with an adaptive-to-block reduction of
quantum protocols. As a result, Ref.~\cite{PLOB} established the
secret key capacity of the pure-loss channel to be
\begin{equation}
\mathcal{K}(\mathcal{E}_{\eta})=-\log_{2}(1-\eta)\simeq1.44\eta~~\text{(at
high loss).} \label{PLOBeqn}%
\end{equation}

This capacity cannot be beaten by any point-to-point QKD\ protocol
at the two ends of the lossy channel. It can only be outperformed
if Alice and Bob pre-share some secret randomness or if there is a
quantum repeater splitting the quantum communication channel and
assisting the remote parties. For this reason, the PLOB bound not
only completely characterizes the fundamental rate-loss scaling of
point-to-point QKD but also provides the exact benchmark for
testing the quality of quantum repeaters. Note that a weaker
version of the PLOB may also be written by explicitly accounting
for the overall detector efficiency $\eta_{\mathrm{det}}$ of a
protocol. This corresponds to Alice and Bob having a composite
channel of transmissivity $\eta_{\mathrm{det}} \eta$, so that the
PLOB bound weakens to $-\log_{2}(1-\eta_{\mathrm{det}} \eta)$.

Soon after the introduction of the PLOB bound, in early 2016
Ref.~\cite{net2006} (later published as Ref.~\cite{net2006p})
established the secret key capacities achievable in chains of
quantum repeaters and, more generally, quantum networks connected
by pure-loss channels. In particular, in the presence of a single
repeater, in the middle between the remote parties and splitting
the overall pure-loss channel $\mathcal{E}_{\eta}$ of
transmissivity $\eta$, one finds the following single-repeater
secret key capacity

\begin{equation}
\mathcal{K_{\mathrm{1rep}}}(\mathcal{E}_{\eta})=-\log_{2}(1-\sqrt{\eta}). \label{singleREPbound}%
\end{equation}

In Fig.~\ref{poi} we show the ideal key rates of state-of-the-art
point-to-point and relay-assisted QKD protocols, compared with the
PLOB bound of Eq.~(\ref{PLOBeqn}) and the single-repeater bound of
Eq.~(\ref{singleREPbound}). By ideal rates we mean the optimal
ones that can be computed assuming zero dark counts, perfect
detector efficiency, zero misalignment error, as well as perfect
error correction and reconciliation efficiency. Point-to-point
protocols cannot beat the PLOB bound and asymptotically scales as
$\simeq \eta$ bits per channel use. This is the case for the BB84
protocol (both with single-photon sources and decoy-state
implementation) and one-way CV-QKD protocols. Even though MDI-QKD
is relay assisted, its relay is not efficient (i.e., it does not
repeat), which is why DV-MDI-QKD is below the PLOB bound. After
TF-QKD was introduced, a number of TF-inspired relay-assisted
protocols were developed, all able to beat the PLOB bound. The
middle relays of these protocols are efficient (i.e., they
repeat). Their key rates cannot overcome the single repeater
bound, but clearly follow its asymptotic rate-loss scaling of
$\simeq \sqrt{\eta}$ bits per channel use.

In the following subsections, we provide the main mathematical
definitions, tools, and formulas related to the study of the
ultimate limits of point-to-point QKD protocols over an arbitrary
quantum channel. Then, in subsequent
Sec.~\ref{repeater-assisted-SEC} we discuss the extension of these
results to repeater-assisted quantum communications.

\subsection{Adaptive protocols and two-way assisted capacities}

Let us start by defining an adaptive point-to-point protocol
$\mathcal{P}$ through a quantum channel $\mathcal{E}$. Assume that
Alice has register $\mathbf{a}$ and Bob has register $\mathbf{b}$.
These registers are (countable) sets of quantum systems which are
prepared in some state $\rho_{\mathbf{ab}}^{0}$ by an adaptive
LOCC $\Lambda_{0}$ applied to some
fundamental separable state $\rho_{\mathbf{a}}^{0}\otimes\rho_{\mathbf{b}}%
^{0}$. Then, for the first transmission, Alice picks a system $a_{1}%
\in\mathbf{a}$ and sends it through channel $\mathcal{E}$; at the
output, Bob
receives a system $b_{1}$ which is included in his register $b_{1}%
\mathbf{b}\rightarrow\mathbf{b}$. Another adaptive LOCC
$\Lambda_{1}$ is applied to the registers. Then, there is the
second transmission $\mathbf{a}\ni a_{2}\rightarrow b_{2}$ through
$\mathcal{E}$, followed by another LOCC $\Lambda_{2}$ and so on
(see Fig.~\ref{longPIC}). After $n$ uses, Alice and Bob share an
output state $\rho_{\mathbf{ab}}^{n}$ which is epsilon-close to
some target state $\phi^{n}$ with $nR_{n}^{\varepsilon}$ bits.
This means that, for any $\varepsilon>0$, one has $\left\Vert
\rho_{\mathbf{ab}}^{n}-\phi^{n}\right\Vert \leq\varepsilon$ in
trace norm. This is also called an
($n,R_{n}^{\varepsilon},\varepsilon$)-protocol. Operationally, the
protocol $\mathcal{P}$ is completely characterized by the sequence
of adaptive LOCCs $\mathcal{L}=\{\Lambda_{0},\Lambda_{1}\ldots\}$.
\begin{figure}[pth]
\vspace{-2.3cm}
\par
\begin{center}
\includegraphics[width=0.50\textwidth]{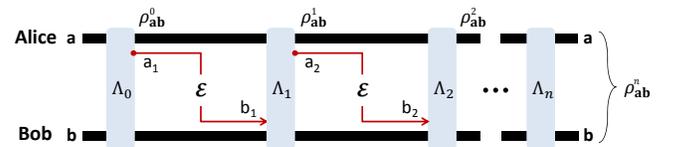} \vspace{-2.9cm}
\end{center}
\caption{Point-to-point adaptive protocol. Each transmission
$a_{i}\rightarrow b_{i}$ through the quantum channel $\mathcal{E}$
is interleaved by two adaptive LOCCs, $\Lambda_{i-1}$ and
$\Lambda_{i}$, applied to Alice's and Bob's local registers
$\mathbf{a}$ and $\mathbf{b}$. After $n$ transmissions, Alice and
Bob share an output state $\rho_{\mathbf{ab}}^{n}$ close to some
target state $\phi^{n}$. Adapted with permission from
Ref.~\cite{TQC} \copyright IOPP (2018).}
\label{longPIC}%
\end{figure}

The (generic) two-way assisted capacity of the quantum channel is
defined by taking the limit of the asymptotic weak-converse rate
$\lim_{\varepsilon ,n}R_{n}^{\varepsilon}$ and maximizing over all
adaptive protocols
$\mathcal{P}$, i.e.,%
\begin{equation}
\mathcal{C}(\mathcal{E}):=\sup_{\mathcal{P}}\lim_{\varepsilon}\lim_{n}%
R_{n}^{\varepsilon}.
\end{equation}
The specification of the target state $\phi^{n}$\ identifies a
corresponding type of two-way capacity. If $\phi^{n}$ is a
maximally-entangled state, then we have the two-way
entanglement-distribution capacity $D_{2}(\mathcal{E})$.
The latter is in turn equal to the two-way quantum capacity $Q_{2}%
(\mathcal{E})$, because transmitting qubits is equivalent to
distributing ebits under two-way CCs. If $\phi^{n}$ is a private
state~\cite{KD1}, then we have the secret key capacity
$K(\mathcal{E})$ and we have $K(\mathcal{E})\geq
D_{2}(\mathcal{E})$, because a maximally-entangled state is a
particular type of private state. Also note that
$K(\mathcal{E})=P_{2}(\mathcal{E})$, where $P_{2}$ is the two-way
private capacity, i.e., the maximum rate at which Alice may
\textit{deterministically} transmit secret bits~\cite{Devetak}.
Thus, we may write the chain of (in)equalities
\begin{equation}
D_{2}(\mathcal{E})=Q_{2}(\mathcal{E})\leq
K(\mathcal{E})=P_{2}(\mathcal{E}).
\label{hierarchy}%
\end{equation}

\subsection{General weak-converse upper bound}

The two-way capacity $\mathcal{C}(\mathcal{E})$ [i.e., any of the
capacities in Eq.~(\ref{hierarchy})] can be bounded by a general
expression in terms of the REE~\cite{RMPrelent,VedFORMm,Pleniom}.
First of all, recall that the REE of a quantum state $\sigma$ is
given by
\begin{equation}
E_{\mathrm{R}}(\sigma)=\inf_{\gamma\in\text{\textrm{SEP}}}S(\sigma||\gamma),
\label{REEbona}%
\end{equation}
where $\gamma$ is a separable state and $S$ is the quantum
relative entropy, defined by~\cite{RMPrelent}
\begin{equation}
S(\sigma||\gamma):=\mathrm{Tr}\left[  \sigma(\log_{2}\sigma-\log_{2}%
\gamma)\right]  .
\end{equation}

The notion of REE can be extended to an asymptotic state
$\sigma:=\lim_{\mu }\sigma^{\mu}$, which is defined as a limit of
a sequence of states $\sigma^{\mu}$ (e.g., this is the case for
energy unbounded states of CV\ systems). In this case, we may
modify Eq.~(\ref{REEbona}) into the following expression
\begin{equation}
E_{\mathrm{R}}(\sigma):=\inf_{\gamma^{\mu}}\underset{\mu\rightarrow+\infty
}{\lim\inf}S(\sigma^{\mu}||\gamma^{\mu}), \label{REE_weaker}%
\end{equation}
where $\gamma^{\mu}$ is sequence of separable states that
converges in trace-norm, i.e., such that
$||\gamma^{\mu}-\gamma||\overset{\mu}{\rightarrow }0$ for some
separable $\gamma$, and the inferior limit comes from the lower
semi-continuity of the quantum relative entropy (valid at any
dimension, including for CV\ systems~\cite{HolevoBOOK}).

With these notions in hand, we may write a general upper bound. In
fact, for any quantum channel $\mathcal{E}$ (at any dimension,
finite or infinite), we have~\cite{PLOB}
\begin{equation}
\mathcal{C}(\mathcal{E})\leq E_{\mathrm{R}}^{\bigstar}(\mathcal{E}%
):=\sup_{\mathcal{P}}\underset{n}{\lim}\frac{E_{R}(\rho_{\mathbf{ab}}^{n})}%
{n}~,\label{mainweak}%
\end{equation}
where $E_{\mathrm{R}}^{\bigstar}(\mathcal{E})$\ is defined by
computing the REE of the output state $\rho_{\mathbf{ab}}^{n}$,
taking the limit for many channels uses, and optimizing over all
the adaptive protocols $\mathcal{P}$.


To simplify the REE bound $E_{\mathrm{R}}^{\bigstar}(\mathcal{E})$
into a single-letter quantity, we adopt a technique of
adaptive-to-block reduction or
protocol \textquotedblleft stretching\textquotedblright%
~\cite{PLOB,TQC,FiniteSTRET}. A preliminary step consists in using
a suitable simulation of the quantum channel, where the channel is
replaced by a corresponding resource state. Then, this simulation
argument can be exploited to stretch the adaptive protocol into a
much simpler block-type protocol, where the output is decomposed
into a tensor product of resource states up to a trace-preserving
LOCC.

\subsection{LOCC simulation of quantum channels}

Given an arbitrary quantum channel $\mathcal{E}$, we may consider
a corresponding simulation $S(\mathcal{E})=(\mathcal{T},\sigma)$
based on some LOCC $\mathcal{T}$ and resource state $\sigma$. This
simulation is such that, for any input state $\rho$, the output of
the channel can be expressed as
\begin{equation}
\mathcal{E}(\rho)=\mathcal{T}(\rho\otimes\sigma). \label{sigma}%
\end{equation}
See also Fig.~\ref{simulationPIC}. A channel $\mathcal{E}$ which
is simulable as in Eq.~(\ref{sigma}) can also be called
\textquotedblleft$\sigma $-stretchable\textquotedblright. Note
that there are different simulations for the same channel. One is
trivial because it just corresponds to choosing $\sigma$ as a
maximally-entangled state and $\mathcal{T}$ as teleportation
followed by $\mathcal{E}$ completely pushed in Bob's LO.
Therefore, it is implicitly understood that one has to carry out
an optimization over these simulations, which also depend on the
specific functional under study.

More generally, the simulation can be asymptotic, i.e., we may
consider sequences of LOCCs $\mathcal{T}^{\mu}$ and resource
states $\sigma^{\mu}$ such that~\cite{PLOB}
\begin{equation}
\mathcal{E}(\rho)=\lim_{\mu}\mathcal{E}^{\mu}(\rho),~~\mathcal{E}^{\mu}%
(\rho):=\mathcal{T}^{\mu}(\rho\otimes\sigma^{\mu}). \label{asymptotic}%
\end{equation}
In other words a quantum channel $\mathcal{E}$ may be defined as a
point-wise limit of a sequence of approximating channels
$\mathcal{E}^{\mu}$ that are simulable as in
Eq.~(\ref{asymptotic}). We call $(\mathcal{T},\sigma
):=\lim_{\mu}(\mathcal{T}^{\mu},\sigma^{\mu})$ the asymptotic
simulation of $\mathcal{E}$. This generalization is crucial for
bosonic channels and some classes of DV channels. Furthermore, it
may reproduce the simpler case of Eq.~(\ref{sigma}). Note that
both Eq.~(\ref{sigma}) and~(\ref{asymptotic}) play an important
role in quantum resource theories (e.g., see also Eq.~(54) of
Ref.~\cite{Gour}).

Given an asymptotic simulation of a quantum channel, the
associated simulation error is correctly quantified in terms of
the energy-constrained diamond (ECD) norm. Consider the compact
set of energy-constrained states
\begin{equation}
\mathcal{D}_{\bar{N}}:=\{\rho_{\mathbf{ab}}~|~\mathrm{Tr}(\hat{N}\rho
_{AB})\leq\bar{N}\},
\end{equation}
where $\bar{N}$ is the total multi-mode number operator. For two
bosonic channels, $\mathcal{E}$ and $\mathcal{E}^{\prime}$, and
$\bar{N}$ mean number of photons, we define the ECD distance
as~\cite[Eq.~(98)]{PLOB}
\begin{equation}
\left\Vert \mathcal{E}-\mathcal{E}^{\prime}\right\Vert _{\diamond\bar{N}%
}:=\sup_{\rho_{AB}\in\mathcal{D}_{\bar{N}}}\left\Vert \mathcal{I}_{A}%
\otimes\mathcal{E}(\rho_{AB})-\mathcal{I}_{A}\otimes\mathcal{E}^{\prime}%
(\rho_{AB})\right\Vert _{1}.
\end{equation}
(See also Ref.~\cite{Shirokov} for a slightly different
definition, where the constraint is only enforced on the $B$
part.) The condition in Eq.~(\ref{asymptotic}) means that, for any
finite $\bar{N}$, we may write the bounded-uniform convergence
\begin{equation}
\left\Vert \mathcal{E}-\mathcal{E}^{\mu}\right\Vert _{\diamond\bar{N}}%
\overset{\mu}{\rightarrow}0. \label{deltaN}%
\end{equation}
\begin{figure}[pth]
\vspace{-0.0cm}
\par
\begin{center}
\includegraphics[width=0.28\textwidth]{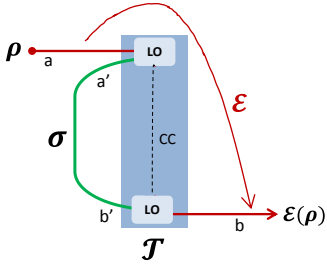} \vspace{-0.6cm}
\end{center}
\caption{LOCC simulation of an arbitrary quantum channel
$\mathcal{E}$ by means of an LOCC $\mathcal{T}$ applied to the
input state $\rho$ and a resource state $\sigma$, according to
Eq.~(\ref{sigma}). For asymptotic simulation, we have the
approximate channel $\mathcal{E}^{\mu}$ which is simulated by
$(\mathcal{T}^{\mu},\sigma^{\mu})$. We then take the point-wise
limit for infinite $\mu$, which defines the asymptotic channel
$\mathcal{E}$
as in Eq.~(\ref{asymptotic}).}%
\label{simulationPIC}%
\end{figure}

Since the first teleportation-based simulation of Pauli channels
introduced in Ref.~\cite[Section~V]{BDSW_96}, the tool of channel
simulation has been progressively developed over the years thanks
to the contributions of several
authors~\cite{Niset,MHthesis,Leung,Cirac,HoroTEL,SougatoBowen} and
it is still today a topic of improvements and generalizations
(e.g., see Table~I in Ref.~\cite{TQC}). Here we have presented the
most general formulation for the LOCC simulation of a quantum
channel at any dimension (finite or infinite) as it has been
formalized in Ref.~\cite{PLOB}. This formulation enables one to
deterministically simulate the amplitude damping (AD) channel.

As a matter of fact, today we only know asymptotic simulations for
the AD channel which either involves a limit in the dimension of
the Hilbert space or a limit in the number of systems forming the
resource state (e.g., implementing port-based
teleportation~\cite{PBT,PBT1,PBT2,PBT2bis} over an infinite number
of Choi matrices~\cite{PBTsimulator}). It is an open problem to
find a deterministic and non-asymptotic simulation for this
channel, which would provide a better estimate of its secret key
capacity, today still unknown. Also note that the tool of
conditional channel simulation~\cite{CCS} seems to fail to
simulate the AD channel, while it can easily simulate a diagonal
amplitude damping (\textquotedblleft DAD\textquotedblright)\
channel or a \textquotedblleft dephrasure\textquotedblright\
channel~\cite{dephrasure}.

Finally, note that the LOCC simulation is also useful to simplify
adaptive protocols of quantum metrology and quantum channel
discrimination~\cite{Metro}. See Ref.~\cite{RevMetro} for a review
on adaptive quantum metrology and Ref.~\cite{SensingREV} for a
recent review on quantum channel discrimination with applications
to quantum illumination~\cite{QIll1,QIll2}, quantum
reading~\cite{QR} and optical
resolution~\cite{OptRES1,OptRES2,OptRES3}.

\subsection{Teleportation covariance and simulability}

For some channels, the LOCC simulation takes a very convenient
form. This is
the case for the \textquotedblleft teleportation covariant\textquotedblright%
\ channels, that are those channels commuting with the random
unitaries of quantum
teleportation~\cite{teleBENNETT,teleCV,Samtele2,telereview}, i.e.,
Pauli operators in DVs~\cite{NielsenChuang}, phase-space
displacements in CVs~\cite{RMPwee,SamRMPm}. More precisely, a
quantum channel $\mathcal{E}$ is called teleportation covariant
if, for any teleportation unitary $U$,\ we may
write%
\begin{equation}
\mathcal{E}(U\rho U^{\dagger})=V\mathcal{E}(\rho)V^{\dagger}~,
\label{stretchability}%
\end{equation}
for another (generally-different) unitary $V$. This property was
discussed in Ref.~\cite{MHthesis,Leung} for DV\ systems and then
in Ref.~\cite{PLOB} for systems of any dimension.

Note that this is a wide family, which includes Pauli channels
(e.g., depolarizing or dephasing), erasure channels and bosonic
Gaussian channels. Thanks to the property in
Eq.~(\ref{stretchability}), the random corrections of the
teleportation protocol can be pushed at the output of these
channels. For this reason, they may be simulated by teleportation.
In fact, a teleportation-covariant channel $\mathcal{E}$ can be
simulated as
\begin{equation}
\mathcal{E}(\rho)=\mathcal{T}_{\text{tele}}(\rho\otimes\sigma_{\mathcal{E}}),
\label{kkkll}%
\end{equation}
where $\mathcal{T}_{\text{tele}}$\ is a teleportation\ LOCC (based
on Bell detection and conditional unitaries) and
$\sigma_{\mathcal{E}}$ is the Choi
matrix of the channel, defined as $\sigma_{\mathcal{E}}:=\mathcal{I}%
\otimes\mathcal{E}(\Phi)$, with $\Phi$ being a maximally entangled
state.

For single-mode bosonic channels (Gaussian or non-Gaussian), we
may write the asymptotic simulation~\cite{PLOB}
\begin{equation}
\mathcal{E}(\rho)=\lim_{\mu}\mathcal{T}_{\text{tele}}^{\mu}(\rho\otimes
\sigma_{\mathcal{E}}^{\mu}), \label{asymptotics}%
\end{equation}
where $\mathcal{T}_{\text{tele}}^{\mu}$ is a sequence of
teleportation-LOCCs (based on finite-energy versions of the ideal
CV\ Bell detection) and
$\sigma_{\mathcal{E}}^{\mu}:=\mathcal{I}\otimes\mathcal{E}(\Phi^{\mu})$
is a sequence of Choi-approximating states (recall that
$\Phi^{\mu}$ is a TMSV state with $\bar{n}=\mu-1/2$ mean thermal
photons in each mode).

When a quantum channel can be simulated as in Eq.~(\ref{kkkll}) or
(\ref{asymptotics}) may be called \textquotedblleft
Choi-stretchable\textquotedblright\ or \textquotedblleft
teleportation simulable\textquotedblright. Let us also mention
that, recently, non-asymptotic types of teleportation simulations
have been considered for bosonic Gaussian
channels~\cite{GerLimited,FiniteSTRET,WildeKaur,Spyros}. These
simulations remove the limit in the resource state (while the
infinite-energy limit is still assumed in the CV Bell detection).
It has been found that these simulations cannot provide tight
results for quantum and private capacities as the asymptotic one
in Eq.~(\ref{asymptotics}), as long as the energy of the resource
state remains finite~\cite{Spyros2}.

\subsection{Strong and uniform convergence}

An important point here is to evaluate the quality of the
simulation of Eq.~(\ref{asymptotics}) for bosonic channels. It is
known that the Braunstein-Kimble (BK) protocol for
continuous-variable
teleportation~\cite{teleCV,Samtele2,telereview} strongly converges
to the identity channel in the limit of infinite squeezing (both
in the resource state and in the Bell detection). In other words,
for any input state $\rho$, we may write the point-wise limit
\begin{equation}
\lim_{\mu}\mathcal{T}_{\text{tele}}^{\mu}(\rho\otimes\Phi^{\mu})=\mathcal{I}%
(\rho).
\end{equation}
Because of this, the channel simulation of any
teleportation-covariant bosonic channel \textit{strongly}
converges to the channel. This condition can be expressed in terms
of the ECD norm, so that for any finite $\bar{N}$ we may
write%
\begin{equation}
\left\Vert \mathcal{E}-\mathcal{T}_{\text{tele}}^{\mu}(\rho\otimes
\sigma_{\mathcal{E}}^{\mu})\right\Vert
_{\diamond\bar{N}}\overset{\mu }{\rightarrow}0.
\end{equation}

However, it is also known that the BK\ protocol does not converge
\textit{uniformly} to the identity channel~\cite{TQC}. In other
words, if we consider the standard diamond norm which is defined
over the entire set $\mathcal{D}$ of bipartite states, then we
have
\begin{equation}
\left\Vert
\mathcal{I}-\mathcal{T}_{\text{tele}}^{\mu}(\rho\otimes\Phi^{\mu
})\right\Vert _{\diamond}\overset{\mu}{\rightarrow}2.
\end{equation}
For this reason, the uniform convergence in the teleportation
simulation of bosonic channels is not guaranteed. However, it can
be explicitly proven to hold for specific types of bosonic
Gaussian channels.

Recall that a single-mode Gaussian channel transforms the
characteristic function as follows~\cite{HolevoCanonical}
\begin{equation}
\mathcal{G}:\chi(\boldsymbol{\xi})\rightarrow\chi(\mathbf{T}\boldsymbol{\xi
})\exp\left(
-\tfrac{1}{2}\boldsymbol{\xi}^{T}\mathbf{N}\boldsymbol{\xi
}+i\mathbf{d}^{T}\boldsymbol{\xi}\right)  ~, \label{Gaussian_Map}%
\end{equation}
where $\mathbf{d}\in\mathbb{R}^{2}$ is a displacement, while the
transmission matrix $\mathbf{T}$ and the noise matrix $\mathbf{N}$
are $2\times2$ real, with $\mathbf{N}^{T}=\mathbf{N}\geq0$ and
$\det\mathbf{N}\geq\left( \det\mathbf{T}-1\right)  ^{2}$. In terms
of mean value $\mathbf{\bar{x}}$ and covariance matrix
$\mathbf{V}$, Eq.~(\ref{Gaussian_Map}) corresponds
to~\cite{HolevoCanonical,Caruso,HolevoVittorio,RMPwee}
\begin{equation}
\mathbf{\bar{x}\rightarrow\mathbf{T}\bar{x}+d},~~\mathbf{V}\rightarrow
\mathbf{TVT}^{T}+\mathbf{N.} \label{mapGG}%
\end{equation}
Then, the asymptotic simulation of a single-mode Gaussian channel
uniformly converges to the channel if and only if the channel's
noise matrix $\mathbf{N}$ has full rank. As a specific case, this
is true for phase-insensitive channels having diagonal
$\mathbf{T}$ and $\mathbf{N}$. See~\cite{EPJD,WildePRA} for other
details.

\subsection{Stretching of an adaptive protocol\label{BosonSTTT}}

By exploiting the LOCC simulation
$S(\mathcal{E})=(\mathcal{T},\sigma)$ of a quantum channel
$\mathcal{E}$, we may completely simplify an adaptive protocol. In
fact, the output state $\rho_{\mathbf{ab}}^{n}$ can be decomposed
into a tensor-product of resources states $\sigma^{\otimes n}$ up
to a trace-preserving LOCC $\bar{\Lambda}$. In other words, we may
write~\cite[Lemma~3]{PLOB}
\begin{equation}
\rho_{\mathbf{ab}}^{n}=\bar{\Lambda}\left(  \sigma^{\otimes
n}\right)  .
\label{StretchingMAIN}%
\end{equation}
For non-asymptotic simulations the proof goes as follows. As shown
in Fig.~\ref{pppPIC}, for the generic $i$th transmission, we
replace the original
quantum channel $\mathcal{E}$ with a simulation $S(\mathcal{E})=(\mathcal{T}%
,\sigma)$. Then, we collapse the LOCC $\mathcal{T}$ into the
adaptive LOCC $\Lambda_{i}$ to form the composite LOCC
$\Delta_{i}$. As a result, the
pre-transmission state $\rho_{\mathbf{ab}}^{i-1}:=\rho_{\mathbf{a}%
a_{i}\mathbf{b}}$ is transformed into the following
post-transmission state
\begin{equation}
\rho_{\mathbf{ab}}^{i}=\Delta_{i}\left(
\rho_{\mathbf{ab}}^{i-1}\otimes
\sigma\right)  . \label{toite}%
\end{equation}
\begin{figure*}[pth]
\vspace{-4.5cm}
\par
\begin{center}
\includegraphics[width=0.98\textwidth]{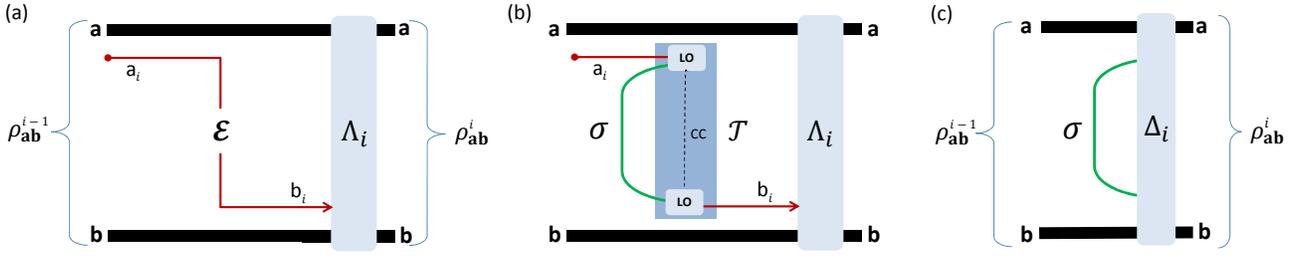} \vspace{-4.8cm}
\end{center}
\caption{Stretching of the $i$th transmission of an adaptive
protocol. (a)~We depict the original transmission through the
channel $\mathcal{E}$ which
transforms the register state $\rho_{\mathbf{ab}}^{i-1}:=\rho_{\mathbf{a}%
a_{i}\mathbf{b}}$ into the output $\rho_{\mathbf{ab}}^{i}$. (b)~We
simulate the channel by means of an LOCC $\mathcal{T}$ and a
resource state $\sigma$, as in previous Fig.~\ref{simulationPIC}.
(c)~We collapse $\mathcal{T}$ and the adaptive LOCC $\Lambda_{i}$
into a single LOCC $\Delta_{i}$ applied to the tensor product
$\rho_{\mathbf{ab}}^{i-1}\otimes\sigma$, as in Eq.~(\ref{toite} ).
Adapted with permission from Ref.~\cite{PLOB} \copyright NPG
(2017).}
\label{pppPIC}%
\end{figure*}
The next step is to iterate Eq.~(\ref{toite}). One finds%
\begin{equation}
\rho_{\mathbf{ab}}^{n}=(\Delta_{n}\circ\cdots\circ\Delta_{1})(\rho
_{\mathbf{ab}}^{0}\otimes\sigma^{\otimes n}).
\end{equation}
Because $\rho_{\mathbf{ab}}^{0}$ is separable, its preparation may
be included in the LOCCs and we get Eq.~(\ref{StretchingMAIN}) for
a complicated but single trace-preserving LOCC $\bar{\Lambda}$.

For a bosonic channel with asymptotic simulation as in
Eq.~(\ref{asymptotic}), the procedure is more involved. One first
considers an imperfect channel simulation
$\mathcal{E}^{\mu}(\rho):=\mathcal{T}^{\mu}(\rho\otimes\sigma^{\mu
})$ in each transmission. By adopting this simulation, we realize
an imperfect stretching of the protocol, with output state
$\rho_{\mathbf{ab}}^{\mu ,n}:=\bar{\Lambda}_{\mu}\left(
\sigma^{\mu\otimes n}\right)  $\ for a trace-preserving LOCC
$\bar{\Lambda}_{\mu}$. This is done similarly to the steps in
Fig.~\ref{pppPIC}, but considering $\mathcal{E}^{\mu}$ in the
place of the original channel $\mathcal{E}$. A crucial point is
now the estimation of the error in the channel simulation, which
must be controlled and propagated to the output state. Assume
that, during the $n$ transmissions of the protocol, the total mean
number of photons in the registers is bounded by some large but
finite value $\bar{N}$. By using a \textquotedblleft peeling
argument\textquotedblright\ over the trace distance, which
exploits the triangle inequality and the monotonicity under
completely-positive maps, we may write the output simulation error
in terms of the channel simulation
error, i.e.,~\cite{PLOB,Metro,TQC}%
\begin{equation}
\left\Vert
\rho_{\mathbf{ab}}^{n}-\rho_{\mathbf{ab}}^{n,\mu}\right\Vert \leq
n\left\Vert \mathcal{E}-\mathcal{E}^{\mu}\right\Vert
_{\diamond\bar{N}}~.
\end{equation}
Therefore, we may write the trace-norm limit
\begin{equation}
\left\Vert \rho_{\mathbf{ab}}^{n}-\bar{\Lambda}_{\mu}\left(
\sigma ^{\mu\otimes n}\right)  \right\Vert
\overset{\mu}{\rightarrow}0,
\label{stretch2}%
\end{equation}
i.e., the asymptotic stretching
$\rho_{\mathbf{ab}}^{n}=\lim_{\mu}\bar
{\Lambda}_{\mu}(\sigma^{\mu\otimes n})$. This is true for any
finite energy bound $\bar{N}$, an assumption that can be removed
at the very end of the calculations.

Let us note that protocol stretching simplifies an
\textit{arbitrary} adaptive protocol over an \textit{arbitrary}
channel at \textit{any} dimension, finite or infinite.\ In
particular, it works by maintaining the original communication
task. This means that\ an adaptive protocol of quantum
communication (QC), entanglement distribution (ED) or key
generation (KG), is reduced to a corresponding block protocol with
exactly the same original task (QC, ED, or KG), but with the
output state being decomposed in the form of
Eq.~(\ref{StretchingMAIN}) or Eq.~(\ref{stretch2}). In the
literature, there were precursory arguments, as those in
Refs.~\cite{BDSW_96,Niset,MHthesis,Leung,Cirac,HoroTEL}, which
were about the transformation of a protocol of QC into a protocol
of ED, over restricted classes of quantum channels. Most
importantly, no control of the simulation error was considered in
previous literature.

\subsection{Single-letter upper bound for two-way assisted capacities}

A\ crucial insight from Ref.~\cite{PLOB} has been the combination
of protocol stretching with the REE, so that its properties of
monotonicity and sub-additivity can be powerfully exploited. This
is the key observation that leads to a single-letter upper bound
for all the two-way capacities of a quantum channel. In fact, let
us compute the REE\ of the output state
decomposed as in Eq.~(\ref{StretchingMAIN}). We derive%
\begin{equation}
E_{\mathrm{R}}(\rho_{\mathbf{ab}}^{n})\overset{(1)}{\leq}E_{\mathrm{R}}%
(\sigma^{\otimes n})\overset{(2)}{\leq}nE_{\mathrm{R}}(\sigma)~, \label{toREP}%
\end{equation}
using (1) the monotonicity of the REE under trace-preserving LOCCs
and (2) its subadditive over tensor products. By replacing
Eq.~(\ref{toREP}) in Eq.~(\ref{mainweak}), we then find the
single-letter upper bound
\begin{equation}
\mathcal{C}(\mathcal{E})\leq E_{\mathrm{R}}(\sigma)~. \label{UB1}%
\end{equation}
In particular, if the channel $\mathcal{E}$ is
teleportation-covariant, it is Choi-stretchable, and we may write
\begin{equation}
\mathcal{C}(\mathcal{E})\leq E_{\mathrm{R}}(\sigma_{\mathcal{E}}%
)=E_{\mathrm{R}}(\mathcal{E}):=\sup_{\rho}E_{\mathrm{R}}\left[  \mathcal{I}%
\otimes\mathcal{E}(\rho)\right]  , \label{UB2}%
\end{equation}
where $E_{\mathrm{R}}(\mathcal{E})$ is the REE\ of the channel $\mathcal{E}%
$~\cite[Theorem~5]{PLOB}.

These results are suitable extended to asymptotic simulations.
Using the weaker definition of REE in Eq.~(\ref{REE_weaker}), the
bounds in Eqs.~(\ref{UB1}) and~(\ref{UB2}) are also valid for
bosonic channels with asymptotic simulations. For a bosonic
Gaussian channel $\mathcal{E}$, the upper bound in Eq.~(\ref{UB2})
is expressed in terms of its asymptotic Choi matrix
$\sigma_{\mathcal{E}}:=\lim_{\mu}\sigma_{\mathcal{E}}^{\mu}$. By
inserting Eq.~(\ref{REE_weaker}) in Eq.~(\ref{UB2}), we derive%
\begin{equation}
\mathcal{C}(\mathcal{E})\leq\Phi(\mathcal{E})\leq\underset{\mu\rightarrow
+\infty}{\lim\inf}S(\sigma_{\mathcal{E}}^{\mu}||\tilde{\gamma}^{\mu})~,
\label{liminf}%
\end{equation}
for a suitable converging sequence of separable states
$\tilde{\gamma}^{\mu}$. Here
$\sigma_{\mathcal{E}}^{\mu}:=\mathcal{I}\otimes\mathcal{E}(\Phi^{\mu})$
is Gaussian and also $\tilde{\gamma}^{\mu}$ can be chosen to be
Gaussian, so that we are left with a simple computation of
relative entropy between Gaussian states.

In related investigations, Ref.~\cite{WTB}\ found that the weak
converse upper bound $E_{\mathrm{R}}(\mathcal{E})$ in
Eq.~(\ref{UB2}) is also a strong converse rate, while
Ref.~\cite{CMH} found that replacing the REE with the max-relative
entropy of entanglement the strong converse bound can be written
for any quantum channel (but the resulting bound is generally
larger).

\subsection{Bounds for teleportation-covariant channels\label{teleBOUNDS}}

Because the upper bounds in Eqs.~(\ref{UB2}) and~(\ref{liminf})
are valid for any teleportation-covariant channel, they may be
applied to Pauli channels and bosonic Gaussian channels. Consider
a qubit Pauli channel
\begin{equation}
\mathcal{E}_{\text{Pauli}}(\rho)=p_{0}\rho+p_{1}X\rho X+p_{2}Y\rho
Y+p_{3}Z\rho Z,
\end{equation}
where $\{p_{k}\}$ is a probability distribution and $X$, $Y$, and
$Z$ are Pauli operators~\cite{NielsenChuang}. Let us call $H_{2}$
the binary Shannon
entropy and $p_{\max}:=\max\{p_{k}\}$. Then, we may write~\cite[Eq.~(33)]%
{PLOB}
\begin{equation}
\mathcal{C}(\mathcal{E}_{\text{Pauli}})\leq\left\{
\begin{array}
[c]{c}%
1-H_{2}(p_{\max}),~~\mathrm{if~~}p_{\max}\geq1/2,\\
\\
0,\text{~~}\mathrm{if~~}p_{\max}<1/2,\text{~~~~~~~~~~~~~~~}%
\end{array}
\right.
\end{equation}
which can be easily generalized to arbitrary finite dimension
(qudits).

Consider now phase-insensitive Gaussian channels. The most
important is the thermal-loss channel $\mathcal{E}_{\eta,\bar{n}}$
which transforms input
quadratures $\mathbf{\hat{x}}=(\hat{q},\hat{p})^{T}$ as $\mathbf{\hat{x}%
}\rightarrow\sqrt{\eta}\mathbf{\hat{x}}+\sqrt{1-\eta}\mathbf{\hat{x}}_{E}$,
where $\eta\in(0,1)$ is the transmissivity and $E$ is the thermal
environment with $\bar{n}$ mean photons. For this channel, we may
derive~\cite[Eq.~(23)]{PLOB}
\begin{equation}
\mathcal{C}(\mathcal{E}_{\eta,\bar{n}})\leq\left\{
\begin{array}
[c]{c}%
-\log_{2}\left[  (1-\eta)\eta^{\bar{n}}\right]  -h(\bar{n}),~~\mathrm{if~}%
\bar{n}<\frac{\eta}{1-\eta},\\
\\
0,\text{~~}\mathrm{if~~}\bar{n}\geq\frac{\eta}{1-\eta}%
,\text{~~~~~~~~~~~~~~~~~~~~~~~~~}%
\end{array}
\right.\label{thermalBB}
\end{equation}
where we have set
\begin{equation}
h(x):=(x+1)\log_{2}(x+1)-x\log_{2}x. \label{hEntropyMAIN}%
\end{equation}

\bigskip

The thermal-loss channel is particularly important for QKD. From
the variance parameter $\omega=\bar{n}+1/2$, we define the
so-called \textquotedblleft excess noise\textquotedblright\
$\varepsilon$ of the channel $\omega
=(1-\eta)^{-1}\eta\varepsilon+1/2,$which leads to%
\begin{equation}
\varepsilon=\eta^{-1}(1-\eta)\bar{n}. \label{excessFORLL}%
\end{equation}
Then, for a generic QKD protocol, we may write its rate as
$R=R(\eta ,\varepsilon)$. The security threshold of the protocol
is therefore obtained for $R=0$ and expressed as
$\varepsilon=\varepsilon(\eta)$, providing the maximum tolerable
excess noise as a function of the transmissivity. An open question
is to find the optimal security threshold in CV-QKD. From
Eq.~(\ref{thermalBB}), it is easy to see that this must be lower
than the entanglement-breaking value $\varepsilon_{\text{UB}}=1$.
In terms of lower bounds, we may consider the RCI which is however
beaten by QKD protocols with trusted noise or two-way quantum
communication. The highest security thresholds known in CV-QKD are
plotted in Fig.~\ref{excessPIC}, where we may also note the big
gap between the best-known achievable thresholds and the upper
bound.

\begin{figure}[pth] \vspace{0.1cm}
\par
\begin{center}
\includegraphics[width=0.4\textwidth]{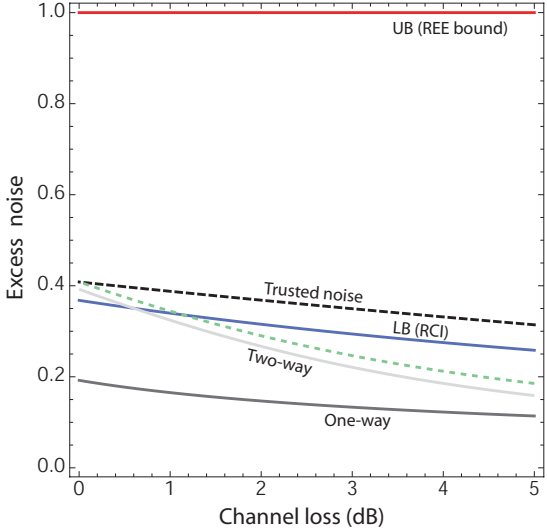} \vspace{-0.2cm}
\end{center}
\caption{Best-known security thresholds in CV-QKD expressed as
maximum tolerable excess noise $\varepsilon$ versus channel loss
(dB). Each protocol is secure only below its threshold. The red
line corresponds to the upper bound $\varepsilon_{\text{UB}}=1$.
The blue line is the lower bound $\varepsilon_{\text{LB}}$
computed from the RCI of the thermal-loss
channel~\cite{ReverseCAP}. The black dashed line is the best-known
security threshold, which is achieved by the one-way trusted noise
protocol described in Ref.~\cite[Sec.~VII]{TQC}. Then, we show the
thresholds for the one-way coherent-state protocol with heterodyne
detection~\cite{Weedbrook2004} and the two-way protocols with
coherent states~\cite{PIRS-2way} (solid line) and largely-thermal
states~\cite{weedbrook2014_2way} (green dashed line).
Reproduced with permission from Ref.~\cite{TQC} \copyright IOPP (2018).}%
\label{excessPIC}%
\end{figure}

For a noisy quantum amplifier $\mathcal{E}_{g,\bar{n}}$ we have
the transformation
$\mathbf{\hat{x}}\rightarrow\sqrt{g}\mathbf{\hat{x}}+\sqrt
{g-1}\mathbf{\hat{x}}_{E}$, where $g>1$ is the gain and $E$ is the
thermal environment with $\bar{n}$ mean photons. In this case, we
may
compute~\cite[Eq.~(26)]{PLOB}%
\begin{equation}
\mathcal{C}(\mathcal{E}_{\eta,\bar{n}})\leq\left\{
\begin{array}
[c]{c}%
\log_{2}\left(  \dfrac{g^{\bar{n}+1}}{g-1}\right)  -h(\bar{n}),~~\mathrm{if~~}%
\bar{n}<(g-1)^{-1},\\
\\
0,\text{~~}\mathrm{if~~}\bar{n}\geq(g-1)^{-1}.\text{~~~~~~~~~~~~~~~~~~~~~~~~~}%
\end{array}
\right.
\end{equation}
Finally, for an additive-noise Gaussian channel
$\mathcal{E}_{\xi}$, we have
$\mathbf{\hat{x}}\rightarrow\mathbf{\hat{x}}+(z,z)^{T}$ where $z$
is a classical Gaussian variable with zero mean and variance
$\xi\geq0$. In this
case, we have the bound~\cite[Eq. (29)]{PLOB}%
\begin{equation}
\mathcal{C}(\mathcal{E}_{\xi})\leq\left\{
\begin{array}
[c]{c}%
\frac{\xi-1}{\ln2}-\log_{2}\xi,~~\mathrm{if~~}\xi<1,\\
\\
0,\text{~~}\mathrm{if~~}\xi\geq1\text{~~~~~~~~~~~~~~}%
\end{array}
\right.
\end{equation}

\subsection{Capacities for distillable channels\label{teleDISTILLABLE}%
}

Within the class of teleportation-covariant channels, there is a
sub-class for
which the upper bound $E_{\mathrm{R}}(\sigma_{\mathcal{E}})$ in Eq.~(\ref{UB2}%
) coincides with an achievable rate for one-way entanglement
distillation. These \textquotedblleft distillable
channels\textquotedblright\ are those for
which we may write%
\begin{equation}
E_{\mathrm{R}}(\sigma_{\mathcal{E}})=D_{1}(\sigma_{\mathcal{E}}),
\label{coinCCC}%
\end{equation}
where $D_{1}(\sigma_{\mathcal{E}})$ is the distillable
entanglement of the Choi matrix $\sigma_{\mathcal{E}}$ via one-way
(forward or backward) CC. This quantity is also suitably extended
to asymptotic Choi matrices in the case of bosonic channels.

The equality in Eq.~(\ref{coinCCC}) is a remarkable coincidence
for three reasons:

\begin{enumerate}
\item Since $D_{1}(\sigma_{\mathcal{E}})$ is a lower bound to $D_{2}%
(\mathcal{E})$, all the two-way capacities of these channels
coincide
($D_{2}=Q_{2}=K=P_{2}$) and are fully established as%
\begin{equation}
\mathcal{C}(\mathcal{E})=E_{\mathrm{R}}(\sigma_{\mathcal{E}})=D_{1}%
(\sigma_{\mathcal{E}}). \label{coincidence}%
\end{equation}

\item The two-way capacities are achieved by means of rounds of one-way CC, so
that adaptiveness is not needed and the amount of CC is limited.

\item Because of the hashing inequality, we have
\begin{equation}
D_{1}(\sigma_{\mathcal{E}})\geq\max\{I_{\mathrm{C}}(\sigma_{\mathcal{E}%
}),I_{\mathrm{RC}}(\sigma_{\mathcal{E}})\},
\end{equation}
where $I_{\mathrm{C}}$ and $I_{\mathrm{RC}}$ are the
coherent~\cite{Schu96,Lloyd97} and reverse
coherent~\cite{RevCohINFO,ReverseCAP} information of the Choi
matrix. Such quantities (and their asymptotic versions) are easily
computable and may be used to show the coincidence in
Eq.~(\ref{coincidence}).
\end{enumerate}

\noindent In this way we can write simple formulas for the two-way
capacities of fundamental quantum channels, such as the pure-loss
channel, the quantum-limited amplifier, the dephasing and erasure
channels (all distillable channels).

In particular, for a bosonic pure-loss channel
$\mathcal{E}_{\eta}$\ with transmissivity $\eta$, one has the PLOB
bound~\cite[Eq. (19)]{PLOB}
\begin{equation}
\mathcal{C}(\eta)=-\log_{2}(1-\eta)~. \label{formCloss}%
\end{equation}
The secret-key capacity $K(\eta)$ determines the maximum rate
achievable by any QKD protocol in the presence of a lossy
communication line (see also Fig.~\ref{poi}). Note that the PLOB
bound can be extended to a multiband lossy channel, for which we
write $\mathcal{C}=-\sum_{i}\log_{2}(1-\eta_{i})$, where
$\eta_{i}$ are the transmissivities of the various bands or
frequency components. For instance, for a multimode telecom fibre
with constant
transmissivity $\eta$ and bandwidth $W$, we have%

\begin{equation}
\mathcal{C}=-W\log_{2}(1-\eta).
\end{equation}

Now consider the other distillable channels. For a quantum-limited
amplifier $\mathcal{E}_{g}$ with gain $g>1$ (and zero thermal
noise $\bar{n}=0$), one
finds~\cite[Eq.~(28)]{PLOB}%
\begin{equation}
\mathcal{C}(g)=-\log_{2}\left(  1-g^{-1}\right)  ~. \label{Campli}%
\end{equation}
In particular, this proves that $Q_{2}(g)$ coincides with the
unassisted quantum capacity $Q(g)$~\cite{HolevoWerner,Wolf}. For a
qubit dephasing channel $\mathcal{E}_{p}^{\text{deph}}$ with
dephasing probability $p$, one
finds~\cite[Eq.~(39)]{PLOB}%
\begin{equation}
\mathcal{C}(p)=1-H_{2}(p)~, \label{dep2}%
\end{equation}
where $H_{2}$ is the binary Shannon entropy. Note that this also
proves
$Q_{2}(\mathcal{E}_{p}^{\text{deph}})=Q(\mathcal{E}_{p}^{\text{deph}})$,
where the latter was derived in ref.~\cite{degradable}.
Eq.~(\ref{dep2}) can be extended to arbitrary dimension $d$, so
that~\cite[Eq.~(41)]{PLOB}
\begin{equation}
\mathcal{C}(p,d)=\log_{2}d-H(\{P_{i}\})~, \label{depDgen}%
\end{equation}
where $H$ is the Shannon entropy and $P_{i}$ is the probability of
$i$ phase
flips. Finally, for the qudit erasure channel $\mathcal{E}_{p,d}%
^{\text{erase}}$ with erasure probability $p$, one finds~\cite[Eq.~(44)]%
{PLOB}
\begin{equation}
\mathcal{C}(p)=(1-p)\log_{2}d~. \label{erase2}%
\end{equation}
For this channel, only $Q_{2}$ was previously
known~\cite{ErasureChannel}, while~\cite{PLOB,GEW} co-established
$K$.

\subsection{Open problems}

There are a number of open questions that are currently subject of
theoretical investigation. While the secret key capacity has been
established for a number of important channels, there are others
for which the gap between best-known lower bound and best-known
upper bound is still open. This is the case for the thermal-loss
channel, the noisy quantum amplifier, the additive-noise Gaussian
channel, the depolarizing channel, and the amplitude damping
channel. For most of these channels, an improvement may come from
refined calculations (e.g., including non-Gaussian states in the
optimization of the REE, or by employing the regularized REE). As
we already mentioned before, for the amplitude damping channel the
problem is also its LOCC simulation, which is not good enough to
provide a tight upper bound. Recently this simulation has been
improved in the setting of DVs by resorting to the convex
optimization of programmable quantum processors~\cite{PQC1,PQC2},
e.g., based on the port-based teleportation
protocol~\cite{PBT,PBTsimulator}.

For the thermal-loss channel, we also know that the lower bound to
the secret key capacity given by the RCI is not tight. There are
in fact QKD protocols with trusted-noise in the detectors whose
rates may beat the RCI, as shown in
Refs.~\cite{ReverseCAP,CarloSpie}. Similarly, for the noisy
amplifier, we know~\cite{GanEPJD} trusted-noise protocols that are
able to beat the CI of the channel, which is therefore not tight.
The non-tightness of the CI (and RCI) is also a feature in the
computation of energy-constrained quantum capacities of bosonic
Gaussian channels~\cite{Noh1}. An interesting approach to bound
the quantum capacities of these channels has been recently pursued
in Refs.~\cite{Noh2,Noh3,Noh4} by using the
Gottesman-Preskill-Kitaev (GKP) states~\cite{GKP1}, realizable
with various technologies~\cite{GKP2,GKP3,GKP4,GKP5,GKP6}.

\section{Repeater chains and quantum
networks}\label{repeater-assisted-SEC}

\subsection{Overview}

In order to overcome the fundamental rate-loss scaling of QKD, one
needs to design a multi-hop network which exploits the assistance
of quantum repeaters. In an information-theoretic sense, a quantum
repeater or quantum relay is \textit{any type} of middle node
between Alice and Bob which helps their quantum communication by
breaking down their original quantum channel into sub-channels. It
does not matter what technology the node is employing, e.g., it
may or may not have quantum memories. A repeater actually
\textit{repeats} only when it is able to beat the performance of
any point-to-point QKD protocol, i.e., the PLOB bound. We may call
these \textquotedblleft effective repeaters\textquotedblright.
Example of non-effective repeaters are MDI-nodes simply based on
Bell detections~\cite{BP,Lo,RELAY}\ while examples of effective
repeaters are MDI-nodes exploiting phase-randomization, such as
TF-QKD~\cite{TFQKD,TFQKD2} and the related protocols of
PM-QKD~\cite{PMQKD},
SNS-QKD~\cite{SNSwang,SNSpractical,SNSpractical2,SNSAOPP}, and
NPPTF-QKD~\cite{Cuo1,Cuo2,Cuo3,NPPTF2,NPPTF3}. According to our
definitions above, the proof-of-concept experiment in
Ref.~\cite{TREL19} represents the first effective repeater ever
implemented. See Fig.~\ref{poi} for an overview of the ideal
performances of these relay-assisted protocols.

While the violation of the PLOB bound provides a simple criterion
for benchmarking the quality of quantum repeaters, one also needs
to understand what optimal performance an ideal repeater may
achieve. In other words, how much key rate can be generated by a
chain of quantum repeaters or by a more general multi-hop quantum
network? In the information-theoretic setting and considering the
general case of adaptive protocols, upper bounds on this key rate
have been studied in Refs.~\cite{net2006,net2006p} and later in a
series of other papers~\cite{netO1,netO2,netO3,netO4,netO5,netO6}.
Here we report the tightest-known bounds that establish the
end-to-end capacities in chains and networks connected by
fundamental channels (subsections~\ref{sec:idealREP}
and~\ref{sec:idealNET}). This has been developed in
Refs.~\cite{net2006,net2006p} which extended the channel
simulation methods from point-to-point to repeater-assisted
quantum communications. In the second part of the section
(subsection~\ref{sec:repeaterPRACT}), we will then discuss
practical aspects and designs of quantum repeaters.

\subsection{Ideal chains of quantum repeaters\label{sec:idealREP}}

Consider a linear chain of $N$ quantum repeaters, labeled by $\mathbf{r}%
_{1},\ldots,\mathbf{r}_{N}$. This is characterized by an ensemble
of $N+1$ quantum channels $\{\mathcal{E}_{i}\}$ describing the
sequence of
transmissions $i=0,\ldots,N$ between the two end-points Alice $\mathbf{a}%
:=\mathbf{r}_{0}$ and Bob $\mathbf{b}:=\mathbf{r}_{N+1}$ (see
Fig.~\ref{intropic}). Assume the most general adaptive protocol
$\mathcal{P}$, where the generation of the secret key between
Alice and Bob is assisted by adaptive LOCCs involving all the
parties in the chain. After $n$ uses of the chain, Alice and Bob
will share an output state $\rho_{\mathbf{ab}}^{n}$ which depends
on $\mathcal{P}$. By taking the limit of large $n$ and
optimization over the protocols $\mathcal{P}$, we may define the
repeater-assisted secret
key capacity $K(\{\mathcal{E}_{i}\})$. This quantity satisfies the bound%
\begin{equation}
K(\{\mathcal{E}_{i}\})\leq
E_{R}^{\bigstar}(\{\mathcal{E}_{i}\}):=\sup
_{\mathcal{P}}\lim_{n}E_{R}(\rho_{\mathbf{ab}}^{n}). \label{REEbound}%
\end{equation}
where the REE $E_{R}$ is defined in Eq.~(\ref{REEbona}) and, more
weakly, in Eq.~(\ref{REE_weaker}) for asymptotic states.
\begin{figure}[ptbh] \vspace{-2.5cm}
\par
\begin{center}
\includegraphics[width=0.48\textwidth] {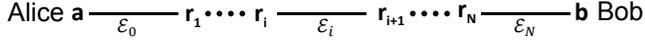} \vspace{-1.9cm}
\vspace{-1.5cm}
\end{center}
\caption{Chain of $N$ quantum repeaters
$\mathbf{r}_{1},\ldots,\mathbf{r}_{N}$
between Alice $\mathbf{a}:=\mathbf{r}_{0}$ and Bob $\mathbf{b}:=\mathbf{r}%
_{N+1}$. The chain is connected by $N+1$ quantum channels $\{\mathcal{E}%
_{i}\}$.}%
\label{intropic}%
\end{figure}

In order to bound this capacity, let us perform a cut \textquotedblleft%
$i$\textquotedblright\ which disconnects channel $\mathcal{E}_{i}$
between $\mathbf{r}_{i}$ and $\mathbf{r}_{i+1}$. We may then
simulate channel $\mathcal{E}_{i}$ with a resource state
$\sigma_{i}$, either exactly as in Eq.~(\ref{sigma}) or
asymptotically as in Eq.~(\ref{asymptotic}). By stretching the
protocol with respect to $\mathcal{E}_{i}$, we may decompose
Alice and Bob's output state as $\rho_{\mathbf{ab}}^{n}=\bar{\Lambda}%
_{i}\left(  \sigma_{i}^{\otimes n}\right)  $ for a
trace-preserving LOCC $\bar{\Lambda}_{i}$, which is local between
\textquotedblleft super-Alice\textquotedblright\ (i.e., all the
repeaters with $\leq i$) and the \textquotedblleft
super-Bob\textquotedblright\ (i.e., all the others with $\geq
i+1$). This decomposition may be asymptotic for bosonic channels,
as specified in Eq.~(\ref{stretch2}).

If we now compute the REE\ on the output state, we find
$E_{R}(\rho _{\mathbf{ab}}^{n})\leq nE_{R}(\sigma_{i})$ for any
$i$ and protocol $\mathcal{P}$. By replacing this inequality in
Eq.~(\ref{REEbound}), we establish the single-letter
bound~\cite{net2006,net2006p}
\begin{equation}
K(\{\mathcal{E}_{i}\})\leq\min_{i}E_{R}(\sigma_{i})~. \label{chain1}%
\end{equation}
Consider now a chain of teleportation-covariant channels $\{\mathcal{E}_{i}%
\}$, so that each quantum channel satisfies the condition in
Eq.~(\ref{stretchability}). These channels $\{\mathcal{E}_{i}\}$
can all be simulated by their (possibly-asymptotic) Choi matrices
$\{\sigma _{\mathcal{E}_{i}}\}$. Therefore, Eq.~(\ref{chain1})
takes the form
\begin{equation}
K(\{\mathcal{E}_{i}\})\leq\min_{i}E_{R}(\sigma_{\mathcal{E}_{i}})~.
\label{ub22}%
\end{equation}
Furthermore, if the quantum channels are distillable, i.e., $E_{R}%
(\sigma_{\mathcal{E}_{i}})=D_{1}(\sigma_{\mathcal{E}_{i}})$ as in
Eq.~(\ref{coinCCC}), then we have $E_{R}(\sigma_{\mathcal{E}_{i}%
})=K(\mathcal{E}_{i})$ and Eq.~(\ref{ub22}) becomes $K(\{\mathcal{E}%
_{i}\})\leq\min_{i}K(\mathcal{E}_{i})$. Remarkably, this upper
bound coincides with a simple lower bound, where each pair of
neighbor repeaters, $\mathbf{r}_{i}$ and $\mathbf{r}_{i+1}$,
exchange a key at their channel capacity $K(\mathcal{E}_{i})$ and
one-time pad is applied to all the keys to generate an end-to-end
key at the minimum rate $\min_{i}K(\mathcal{E}_{i})$. As a result,
for distillable chains, we have an exact result for their secret
key capacity~\cite{net2006,net2006p}%
\begin{equation}
K(\{\mathcal{E}_{i}\})=\min_{i}K(\mathcal{E}_{i})~. \label{resultKEY}%
\end{equation}

Thanks to the result in Eq.~(\ref{resultKEY}), we know the
repeater-assisted secret key capacities of chains composed of
fundamental channels, such as bosonic pure-loss channels,
quantum-limited amplifiers, dephasing and erasure channels. In
particular, for a chain of repeaters connected by pure-loss
channels with transmissivities $\{\eta_{i}\}$, we may write the
secret key capacity~\cite{net2006,net2006p}
\begin{equation}
K_{\text{loss}}=-\log_{2}\left[  1-\min_{i}\eta_{i}\right]  ~,
\label{EqLOSSY2}%
\end{equation}
which is fully determined by the minimum transmissivity in the
chain. In particular, consider an optical fiber with
transmissivity $\eta$ which is split into $N+1$ parts by inserting
$N$ equidistant repeaters, so that each part has transmissivity
$\sqrt[N+1]{\eta}$. Then, we write the capacity
\begin{equation}
K_{\text{loss}}(\eta,N)=-\log_{2}\left(  1-\sqrt[N+1]{\eta}\right)
~.
\label{optLOSScap}%
\end{equation}

In a chain connected by quantum-limited amplifiers with gains
$\{g_{i}\}$, we may write~\cite{net2006,net2006p}
\begin{equation}
K_{\text{amp}}=-\log_{2}\left[  1-\left(  \max_{i}g_{i}\right)
^{-1}\right] ~.
\end{equation}
For a chain connected by dephasing channel $\mathcal{E}_{i}$ with
probability $p_{i}\leq1/2$, we find~\cite{net2006,net2006p}
\begin{equation}
K_{\text{deph}}=1-H_{2}(\max_{i}p_{i})~,
\end{equation}
where $H_{2}$ is the binary Shannon entropy. Finally, for a chain
of erasure channels
$K_{\text{erase}}=1-\max_{i}p_{i}$~\cite{net2006,net2006p}.

\subsection{Quantum communication networks\label{sec:idealNET}}

The results for repeater chains can be generalized to arbitrary
quantum networks by combining methods of channel simulation with
powerful results from the classical network theory. Here we do not
present the details of this methodology but only an introduction
to the main notions and the specific results for pure-loss
channels. The reader interested in further details may consult
Refs.~\cite{net2006,net2006p} where they can find a comprehensive
treatment and general results for arbitrary quantum channels.

We may represent a quantum communication network as an undirected
finite graph $\mathcal{N}=(P,E)$, where $P$ is the set of points
and $E$ is the set of edges. Each point $\mathbf{p}$ has a local
set of quantum systems and two points, $\mathbf{p}_{i}$\textbf{
}and\textbf{ }$\mathbf{p}_{j}$, are logically connected by an edge
$(\mathbf{p}_{i},\mathbf{p}_{j})\in E$ if and only if
they are physically connected by a quantum channel $\mathcal{E}_{ij}%
:=\mathcal{E}_{\mathbf{p}_{i}\mathbf{p}_{j}}$. Between the two
end-points, Alice $\mathbf{a}$ and Bob $\mathbf{b}$, there is an
ensemble of possible routes $\Omega=\{1,\ldots,\omega,\ldots\}$.
Here the generic route $\omega$ is an undirected path between
$\mathbf{a}$ and $\mathbf{b}$, and is associated to a sequence of
quantum channels $\{\mathcal{E}_{0}^{\omega},\ldots
,\mathcal{E}_{k}^{\omega}\ldots\}$. Then, a cut $C$ is a
bipartition $(A,B)$ of the points $P$ such that $\mathbf{a}\in A$
and $\mathbf{b}\in B$. Correspondingly, the cut-set $\tilde{C}$ of
$C$ is the set of edges with one
end-point in each subset of the bipartition, i.e., $\tilde{C}=\{(\mathbf{x}%
,\mathbf{y})\in E:\mathbf{x}\in A,\mathbf{y}\in B\}$. Given these
notions we may define two type of network protocols, which are
based either on sequential or parallel routing of the quantum
systems.

In a sequential protocol $\mathcal{P}_{\operatorname{seq}}$, the
network $\mathcal{N}$ is initialized by a network LOCCs, where
each point classically communicates with all the others (via
unlimited two-way CCs) and each point adaptively performs LOs on
its local quantum systems on the basis of the information
exchanged. Then, Alice connects to some point $\mathbf{p}_{i}$ by
exchanging a quantum system (with forward or backward transmission
depending on the physical direction of the quantum channel). This
is followed by a second network LOCC. Then, point $\mathbf{p}_{i}$
connects to another point $\mathbf{p}_{j}$ by exchanging another
quantum system, which is followed by a third network LOCC and so
on. Finally, Bob is reached via some route $\omega$, which
completes the first sequential use of $\mathcal{N}$. For the
second use, a different route may be chosen. After $n$ uses of
$\mathcal{N}$, Alice and Bob's output state
$\rho_{\mathbf{ab}}^{n}$ is $\varepsilon$-close to a private state
with $nR_{n}^{\varepsilon}$ secret bits. Optimizing their rate
over $\mathcal{P}_{\operatorname{seq}}$ and taking the limit for
large $n$ and small $\varepsilon$ (weak converse), one may define
the single-path (secret key capacity) capacity of the network
$K(\mathcal{N})$.

Following Refs.~\cite{net2006,net2006p}, a general upper bound may
be written for $K(\mathcal{N})$, which is particularly simple for
networks of teleportation-covariant channels. For networks of
distillable channels, the formula for $K(\mathcal{N})$ can exactly
be found. In particular, consider an optical network, so that two
arbitrary points $\mathbf{x}$ and $\mathbf{y}$ are either
disconnected or connected by a pure-loss channel with
transmissivity $\eta_{\mathbf{xy}}$. The single-path capacity of
the lossy network $\mathcal{N}_{\text{loss}}$ is determined
by~\cite{net2006,net2006p}
\begin{equation}
K(\mathcal{N}_{\text{loss}})=-\log_{2}(1-\tilde{\eta}),~~\tilde{\eta}=\min
_{C}\max_{(\mathbf{x},\mathbf{y})\in\tilde{C}}\eta_{\mathbf{xy}},
\end{equation}
where $\tilde{\eta}$ is found by computing the maximum
transmissivity along a cut, and then minimizing over the cuts.

This result can be formulated in an equivalent way. In fact, for
any route $\omega$ of pure-loss channels with transmissivities
$\{\eta_{i}^{\omega}\}$, we may compute the end-to-end
transmissivity of the route as $\eta_{\omega
}:=\min_{i}\eta_{i}^{\omega}$. Then the single-path capacity is
determined by the route with maximum
transmissivity~\cite{net2006,net2006p}
\begin{equation}
K(\mathcal{N}_{\text{loss}})=-\log_{2}(1-\tilde{\eta}),~~\tilde{\eta}%
:=\max_{\omega\in\Omega}\eta_{\omega}.\label{lossSINGLE}%
\end{equation}
Finding the optimal route $\tilde{\omega}$ corresponds to solving
the widest path problem~\cite{Pollack}. Adopting the modified
Dijkstra's shortest path algorithm~\cite{MITp}, this is possible
in time $O(\left\vert E\right\vert \log_{2}\left\vert P\right\vert
)$, where $\left\vert E\right\vert $ is the number of edges and
$\left\vert P\right\vert $ is the number of points.

Consider now a parallel protocol, where multiple routes between
the end-points are used simultaneously. More precisely, after the
initialization of the network $\mathcal{N}$, Alice exchanges
quantum systems with all her neighbor points (i.e., all points she
share a quantum channel with). This multipoint communication is
followed by a network LOCC. Then, each receiving point exchanges
quantum systems with other neighbor points and so on. This is done
in such a way that these subsequent multipoint communications are
interleaved by network LOCCs and they do not overlap with each
other, so that no edge of the network is used twice. The latter
condition is achieved by imposing that receiving points only
choose unused edges for their subsequent transmissions.
This routing strategy is known as \textquotedblleft flooding\textquotedblright%
~\cite{flooding}. Eventually, Bob is reached as an end-point,
which completes the first parallel use of $\mathcal{N}$. The next
parallel uses of $\mathcal{N}$\ may involve different choices by
the intermediate nodes. After
$n$ uses, Alice and Bob share a private state with $nR_{n}^{\varepsilon}%
$\ secret bits. By optimizing over the flooding protocols and
taking the limit for many uses ($n\rightarrow\infty$) and the weak
converse limit ($\varepsilon\rightarrow0$), one defines the
multi-path capacity of the network $K^{\text{m}}(\mathcal{N})$.

Again, a general upper bound may be written for
$K^{\text{m}}(\mathcal{N})$, which simplifies for network of
teleportation-covariant channels and even more for distillable
channels for which $K^{\text{m}}(\mathcal{N})$ is completely
established. As an example, consider again a network $\mathcal{N}%
_{\text{loss}}$ composed of pure-loss channels, so that each edge
$(\mathbf{x},\mathbf{y})$ has transmissivity $\eta_{\mathbf{xy}}$.
For any cut
$C$, define its total loss as $l(C):=%
{\textstyle\prod\nolimits_{(\mathbf{x},\mathbf{y})\in\tilde{C}}}
(1-\eta_{\mathbf{xy}})$. By maximizing $l(C)$ over all cuts we
define the total loss of the network, i.e.,
$l(\mathcal{N}_{\text{loss}}):=\max_{C}l(C)$. The multi-path
(secret key capacity) capacity of $\mathcal{N}_{\text{loss}}$ is
therefore given by~\cite{net2006,net2006p}
\begin{equation}
K^{\text{m}}(\mathcal{N}_{\text{loss}})=-\log_{2}l(\mathcal{N}_{\text{loss}}).
\end{equation}
The optimal multi-path routing of is provided by the solution of
the maximum flow problem, which can be found in $O(|P|\times|E|)$
time by using Orlin's algorithm~\cite{Orlin}. As one may expect,
the multi-path capacity $K^{\text{m}}(\mathcal{N}_{\text{loss}})$
generally outperforms the single-path version
$K(\mathcal{N}_{\text{loss}})$.



\subsection{Practical designs for quantum
repeaters}\label{sec:repeaterPRACT}

One of the key features of future quantum communications networks
is the ability to transfer quantum states reliably from one point
to another. As basic as it sounds, this is one of the most
challenging implementation tasks that quantum technologies face.
There are two main approaches to solving this problem. One relies
on using teleportation techniques, which themselves rely on one of
the pillars of quantum information science, i.e., entanglement. In
this scenario, the state transfer problem reduces to how we can
efficiently distribute entanglement across a
network~\cite{Briegel, MIT-NU, DLCZ_01, ProbReps:RevModPhys.2011}.
The second solution relies on using quantum error correction
techniques to overcome loss and operation
errors~\cite{Loss-Tolerant-Code, Munro:NatPhot:2012, Azuma2015,
Liang:NoMemRep_PRL2014, Syndrome-QEC, QPolyCode_Rep}. This is
similar to what we have in data communications networks where by
adding some redundancy to our message we can correct some of the
errors that might be added by the channel. In the quantum case,
not only we have to correct the bit flip errors, but also phase
flip and erasure errors in possibly a fault-tolerant way. This
approach will then require advanced quantum computing modules.
Both above solutions are considered to be part of an underlying
platform that enables quantum networks to operate at any distance,
i.e., the platform of quantum repeaters.

Original proposals on quantum repeaters rely on the use of
entanglement swapping in a nested way. Suppose you have
distributed and stored a Bell state between nodes $A$ and $B$ at
distance $L_0$ in a network. Suppose node $B$ also shares a Bell
state with node $C$ farther apart by $L_0$. Then, by performing a
Bell-state measurement (BSM) on the two subsystems in node $B$, we
can entangle the systems in node $A$ and $C$. That means that if
we can distribute entanglement over distance $L_0$, by using
entanglement swapping, we can extend it to $2L_0$. By using this
technique $n$ times, in a nested way, we can then in principle
cover entanglement over a distance $L = 2^n L_0$, where $n$ would
be the nesting level for our repeater system; see
Fig.~\ref{FigCh4:Repeater}. Looking at this from a different
angle, what we have basically done is that in order to distribute
entanglement over distance $L$, we have divided the entire
distance into $2^n$ segments, distributed and stored entanglement
over elementary links with distance $L_0$, and then have applied
BSMs on the middle nodes to extend the entanglement over distance
$L$.

The key advantage of the above nested way for entanglement
distribution is its efficiency. Distributing entanglement over an
elementary $L_0$-long link often requires the transmission and
detection of single-photon-level light across the channel, which
would only succeed with a probability on the order of
$\exp(-\alpha L_0)$, where $\alpha$ is proportional to the channel
attenuation parameter. Note that using a similar technique over
the total distance $L$ would be exponentially worse as now the
success rate scales with $\exp(-\alpha L)$. That implies that
entanglement distribution over elementary links must be repeated
until it succeeds. The distributed entanglement should also be
stored somewhere until entanglement is also distributed over
neighboring links. That is why quantum storage will be needed for
such probabilistic entanglement distribution techniques. But,
entanglement distribution over elementary links can be attempted
in parallel, and, although they do not necessarily succeed all at
the same time, sooner or later we are in a position to perform
BSMs on the middle nodes. The entanglement distribution rate over
distance $L$ would then scale with $\exp(-\alpha L_0)$.

The above discussion makes some idealistic assumptions on the
system components. In practice, we should also account for the
imperfections in the setup. For instance, the distributed state
over elementary links may not be a maximally entangled states, in
which case, by every BSM, we deviate further from the ideal state.
The measurement operations may also not be error free or
deterministic to begin with. These issues require us to apply
certain entanglement distillation techniques to improve the
quality of distributed entangled states
\cite{Purification_Bennett, BDSW_96}. But, that would add to the
quantum computational cost of the system and makes its
implementation even more challenging. Depending on the state of
the art on quantum computing, we can then envisage several
different stages of development for quantum repeaters
\cite{Muralidharan2016, Razavi_book}. In the following, we review
three such classes, or generations~\cite{Repeater_review}, of
quantum repeaters.

\subsubsection{Probabilistic quantum repeaters} \label{Sec:Prep}
\begin{figure}[!t]
\centering
\includegraphics[width=.75\columnwidth]{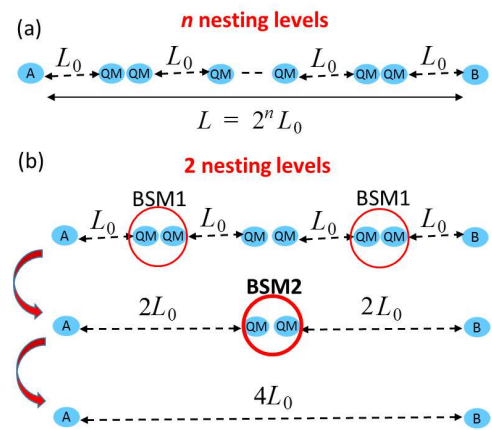}
\caption{(a) A quantum repeater link with nesting level $n$. (b)
An example quantum repeater link with nesting level 2. BSM1
operations extend the entanglement over $2L_0$. BSM2 would then
extend it further to $4L_0$. Note that if BSM operations are
probabilistic, BSM2 should not be done until the middle node
learns about the success of BSM1 operations. This requirement
slows down the process and makes the coherence time requirements
on the memories more demanding. Reproduced from
Ref.~\cite{Razavi_book} under permission of the IOPP.
\label{FigCh4:Repeater}}
\end{figure}
Since the introduction of quantum repeaters in 1998
\cite{Briegel}, experimentalists have been looking for practical
ways to implement the underlying ideas. The original repeater
protocol requires distribution, storage, swapping, and
distillation of entanglement. A lot of research has therefore been
directed into devising quantum memory units that can interact
efficiently with light and can store quantum states for a
sufficiently long time. The interaction with light is necessary
for such devices as it would allow us to use photonic systems for
both distribution and swapping of entanglement. Photon-based
systems are, however, fragile against loss and that could result
in probabilistic operations, which, in turn, require us to repeat
a certain procedure until it succeeds. Probabilistic quantum
repeaters are those that rely on probabilistic techniques for
entanglement distribution and entanglement swapping. This class of
repeaters has been at the center of experimental attention in the
past 20 years.

There are different ways of distributing entangled states between
quantum memories of an elementary link. In some proposals
\cite{MIT-NU}, entangled photons are generated at the middle of
the link and sent toward quantum memories located at the two end
of the elementary link. If these photons survive the path loss and
can be stored in the memories in a {\em heralding} way we can then
assume that the two memories are entangled. This technique
requires us to have a verification technique by which we can tell
if the storing procedure has been successful. Alternatively, in
some other proposals, we start with entangling a photon with the
memory and either send it to the other side for a similar
operation or swap entanglement in the middle between two such
memories. The most famous proposal of this type is that of Duan,
Lukin, Cirac, and Zoller~\cite{DLCZ_01}, known as DLCZ, whose many
variants~\cite{ProbReps:RevModPhys.2011} have been proposed and
partly demonstrated in practice~\cite{EntgDist3m,
Zhao:Robust:2007}.

The BSM operation in probabilistic quantum repeaters is typically
done by first converting the state of quantum memories back into
photonic states and then use linear optics modules to perform the
BSM. Such linear optics modules can, however, be inefficient and
face certain limitations in offering a full
BSM~\cite{Lutkenhaus:BSMLinear_2001}. There are certain
tricks~\cite{Grice_BSM, vanLoock_BSM, RUS_PRA} by which their
performance can be improved, but, in the end, the chance of
success in most practical settings would remain below one. An
implication of a probabilistic BSM is that we cannot perform BSMs
in a certain nesting level until we have learned about the results
of the BSMs in the previous nesting level. That requires
exchanging data between intermediate nodes, which can not be done
faster than the transmission delay between such nodes. This would
result in requiring long coherence time and a low entanglement
generation rate for probabilistic repeaters.

\begin{figure}[!t]
\centering
\includegraphics[width=\columnwidth]{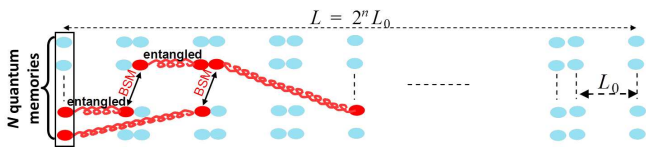}
\caption{A probabilistic quantum repeater with multiple memories
per node. In each round, entanglement distribution is attempted on
all available elementary links. BSMs, at different nesting levels,
will also be performed by matching as many entangled pairs as
possible~\cite{Razavi.Lutkenhaus.09} Reproduced from
Ref.~\cite{Razavi_book} under permission of the IOPP.
\label{FigCh4:MultMem}}
\end{figure}

One remedy to the above problems is the multiple-memory
configuration in Fig.~\ref{FigCh4:MultMem}. In this setup, instead
of one QM in each site, we use a bank of $N$ memories. In each
round of duration $T_0=L_0/c$, with $c$ being the speed of light
in the classical channel, we attempt to entangle as many
elementary links as possible. We then mix and match entangled
pairs across two neighboring links and perform as many BSM
operations as possible. As we continue doing informed BSMs and, at
the same time, refilling available elementary links with fresh
entangled states, we get to a steady state that in every period we
roughly generate $N P_S P_M^n$ entangled states, for $N \gg 1$,
between the far two ends of the network
\cite{Razavi.Lutkenhaus.09}. Here, $P_S$ is the success
probability for the employed entanglement distribution scheme for
elementary links and $P_M$ is the success probability for the BSM
operation. Moreover, it roughly takes $T = L/c$ to generate an
entangled pair, which minimizes the requirements on the memory
coherence time. Multi-mode structures for quantum memories have
also been proposed and implemented to improve the performance in
repeater setups~\cite{Tittel_Nature_AFC_2011,
Gisin_Nature_AFC_2011, AFC_multi}.

Probabilistic repeaters are perhaps the simplest repeater
technology to be implemented in practice. In fact, even the
simplest setup, when there is only one repeater node in the
middle, can offer certain advantages for QKD applications. These
setups, known as memory-assisted QKD~\cite{Brus:MDIQKD-QM_2013,
Panayi_NJP2014}, can soon offer better rate-versus-distance
scaling than most conventional QKD systems in operation by using
existing quantum memory technologies~\cite{piparo2015measurement,
MA_NV_centers, MAQKD_SPS, Nicolo_single_NV,
Filip-MAQKD,Filip-MAQKD2}. Over long distances, however,
probabilistic repeater would suffer from a low rate, or require a
large number of memories to perform well~\cite{Razavi.Amirloo.10,
LoPiparo:2013, Nicolo_paper3}. Part of the reason for such low key
rates is the use of probabilistic BSM modules. We next see what
can be achieved if we have a deterministic BSM unit.

\subsubsection{Deterministic quantum repeaters} \label{Sec:Drep}
Unlike our previous scheme, the original proposal for quantum
repeaters relies on {\em deterministic}, but possibly {\em
erroneous}, gates for BSM, or similar, operations \cite{Briegel}.
Proposed in 1998, the authors assume that the initial entanglement
distribution and storage have already taken place, and now we need
to manipulate a bunch of entangled quantum memories in such a way
that we end up with a high quality entangled states between nodes
A and B. In Sec.~\ref{Sec:Prep}, although we assumed that the BSM
operation may succeed probabilistically, we did not account for
possible errors that may be caused by BSM modules. Further, we
assumed that the initial entanglement over the elementary links
was of the ideal form of a maximally entangled state, e.g., Bell
states. The proposed work in~\cite{Briegel} looks at the latter
two issues by assuming that BSMs can directly be applied to QMs in
a deterministic way. The immediate advantage is that $P_M$ would
become 1, which increases the rate and also reduces the waiting
time caused by the probabilistic events and their corresponding
transmission delays. Once one accounts for errors, however, other
problems arise.

In reality, it is very challenging to generate and distribute
truly maximally entangled states. In practice, we should often
allow for deviations from this ideal case, which can be measured
in different ways. For instance, for two pairs of entangled states
in a Werner state with parameter $p$, a BSM on the middle memories
would leave the remote memories in a Werner state with parameter
$2p$~\cite{Briegel}; that is, the error, or the deviation from the
ideal state, has been doubled. In a quantum repeater link as in
Fig.~\ref{FigCh4:Repeater}(a), such an error will get doubled,
even if we have perfect deterministic BSM gates, for every nesting
level. The danger is that after a certain number of nesting levels
the quality of the resulting entangled states is so low that it
may not be of any use for quantum applications. If we include the
possible errors in the gates, the situation would become even
worse.

The solution suggested in~\cite{Briegel} is based on the use of
purification, or, entanglement distillation, techniques. In short,
the idea is that if we are given $M$ pairs of non-ideal entangled
states, we can use some LOCCs to come up with $N<M$ entangled
states of higher quality, for instance, higher fidelity. Depending
on the type of distillation techniques used, we may end up with
different rate behaviors. Original distillation schemes relied on
performing CNOT gates on pairs of memories~\cite{DEJMPS_96} and
then measuring one of them. The result should have been compared
with that of a similar measurement at the other end of the link,
and success was dependent on whether the two results, for
instance, matched or not. In essence, you could not tell without
classical communication between the nodes, if the distillation
succeeds or not, despite the fact that deterministic gates are
being used throughout. This, in effect, would have turned a
deterministic repeater setup to a probabilistic one, suffering to
some extent from the same impediments we mentioned in the previous
section.

The alternative solution is to use quantum error correction
schemes to distil entanglement~\cite{Aschauer_thesis,QRep_Enc}. In
essence, we can think of the $M$ non-ideal entangled states that
we wish to distil to have been obtained by hypothetically starting
with $N$ ideal entangled states, adding $M-N$ redundant states to
this batch, and sending all $M$ pairs through an error-prone
channel. In such a setting, one can, in principle, use error
correction techniques to get the original $N$ pairs back. This can
be done with a high probability if the ratio $N/M$ is chosen
properly with respect to the expected amount of error in our
system. It turns out that if we want to do this in an error
resilient way, we need quantum gates with error rates on the order
of 0.001--0.01 or below. Such error correction can also alleviate
some of the errors caused by the memory decoherence once we are
waiting to learn about the success of the initial entanglement
distribution.

Using the above techniques, we can design quantum repeaters with a
modestly high key rate. The limitation is mainly from the original
requirement for entangling elementary links, which is still
probabilistic, and the trade-off between having more nesting
levels, and, therefore, higher $P_S$, versus fewer nesting levels,
hence less accumulated error and distillation. Next we show that
if we allow for sophisticated quantum operations to be used in
quantum repeaters, we can further improve the rate by relaxing the
requirement on entangling elementary links.

\subsubsection{Memory-less quantum repeaters}\label{Sec:MLrep}

The most advanced protocols for quantum repeaters leave as little
as possible to probabilistic schemes. In such schemes,
loss-resilient error correction techniques are used to make sure
that the {\em quantum} information carried by photonic systems can
be retrieved at each intermediate node. This can be achieved in
different ways. The common feature of all these schemes is that we
no longer need quantum memories for storage purposes, although we
may still need them for quantum processing. That is why they are
sometime called memory-less quantum repeaters.

Here, we explain one example that relies on quantum error
correction for loss resilience. This idea was first proposed in
\cite{Munro:NatPhot:2012} and then further work followed up to
also account for not only the loss in the channel, but also
possible errors in the system
\cite{Liang:NoMemRep_PRL2014,Syndrome-QEC}. In
\cite{Munro:NatPhot:2012}, a quantum state $\alpha |0 \rangle +
\beta |1 \rangle$ is encoded as
\begin{equation}
|\Psi\rangle^{(m,n)} = \alpha |+ \rangle_1^{(m)} \cdots |+
\rangle_n^{(m)}   + \beta |- \rangle_1^{(m)} \cdots |-
\rangle_n^{(m)},
\end{equation}
where $n$ is the number of logical qubits and $m$ is the number of
physical qubits in each logical qubit. Here, $|\pm\rangle^{(m)} =
|0\rangle^{\otimes m} \pm |1\rangle^{\otimes m}$. This encoding
has the property that the original quantum state can be recovered
provided that at least (1) one photon survives in each logical
qubit, and (2) one logical qubit, with all its $m$ constituent
photons, is fully received. The authors show that for sufficiently
short channels one can find appropriate values of $m$ and $n$ such
that a high key rate on the order of tens of MHz can, in
principle, be achieved. The requirements are, however, beyond the
reach of current technologies.

Such memory-less quantum repeaters, while offering a substantial
improvement in the key rate, require a set of demanding properties
for their required elements. In particular, we need operation
errors as low as $10^{-4}-10^{-3}$, large cluster states of
photons, whose generation may require a series of other advanced
technologies (e.g. high-rate efficient single-photon sources), and
a large number of intermediate nodes. The latter may cause
compatibility problems with existing optical communications
infrastructure, in which, at the core of the network, nodes are
rather sparsely located. That said, such advanced technologies for
quantum repeaters would perhaps be one of the latest generations
of such systems, by which time sufficient improvement in our
quantum computing capabilities as well as other required devices
and technologies may have already happened. For such an era
memory-less repeaters offer a solution that is of an appropriate
quality for the technologies that rely on the {\em quantum
internet}~\cite{QInternet_Kimble, SamInternet}.


\section{QKD against a bounded quantum memory}

\subsection{Introduction}

QKD is commonly defined under the assumption that a potential
eavesdropper have access to unlimited technology. For example, the
eavesdropper (Eve) may have a universal quantum computer with
unlimited computational power, as well as a perfect quantum memory
of unbounded capacity and ideal detectors. While these strong
assumptions put QKD on a solid theoretical ground, they may be
considered unrealistic given the present stage of development of
quantum technologies. Such strong assumptions create a
disproportion between the technology that will be deployed in a
foreseeable future and what is assumed that is already available
to Eve.

One different security scenario is defined by assuming that a
potential eavesdropper has only access to limited quantum
technologies. Here we consider QKD under two similar though
different assumptions about the technological capability of Eve.
First we review the Bounded Quantum Storage Model (BQSM), in which
Eve is assumed to be able to store only a limited number of
qubits. Furthermore, this number is assumed to grow only
sub-linearly with the number of bits exchanged between Alice and
Bob. Second we consider the effect of Quantum Data Locking (QDL)
and its application to QKD and secure communication under the
assumption that Eve can store unlimited number of qubits but only
for a finite time. In both cases, equivalent constraints are
imposed on the trusted users Alice and Bob. Furthermore we assume
that Eve, unlike Alice and Bob, has access to a universal quantum
computer with unbounded computational power and can perform ideal
quantum measurements.
Entropic uncertainty relations, which are briefly reviewed in the
next section, play a major role in security proofs against an
eavesdropper with constrained quantum memory.

It is of utmost importance to remark that QDL yields composable
security given the above assumption that Eve is forced to make a
measurement (for example because she has no quantum memory or a
quantum memory with finite and known storage time). Under this
assumption one consistently obtain composable security
\cite{Entropy,Portmann}. Otherwise, QDL does not guarantee
composable security against an eavesdropper with access to perfect
and unbounded quantum resources~\cite{KRBM}.

\subsection{Entropic uncertainty relations}

For the sake of self-consistency, we briefly recall the entropic
uncertainty relations (EURs)~\cite{WW,coles17}, already discussed
in Section~\ref{sectionMB_MT} but here adapted to the notation of
the present problem. Consider a collection of $k$ measurements
$\mathcal{M} = \{ \mathcal{M}_j \}_{j=1,\dots k}$. On a given
state $\rho$, the $j$-th measurement produces a random variable
$\mathcal{M}_j(\rho)$ with output $x_j$ and associated probability
$p_{\mathcal{M}_j(\rho)}(x_j)$. An EUR is expressed by an
inequality of the form
\begin{equation}\label{EqEUR}
\inf_\rho \frac{1}{k} \sum_{j=1}^k H[\mathcal{M}_j(\rho)] \geq
c_{\mathcal{M}} \, ,
\end{equation}
where $H[\mathcal{M}_j(\rho)] = - \sum_{x_j}
p_{\mathcal{M}_j(\rho)}(x_j) \log_2{p_{\mathcal{M}_j(\rho)}(x_j)}$
is the Shannon entropy of $\mathcal{M}_j(\rho)$ and
$c_{\mathcal{M}}$ is a constant that only depends on the set of
measurements $\mathcal{M}$. By convexity, it is sufficient to
restrict on pure states.

As an example, consider the case of a $d$-dimensional Hilbert
space and a pair of projective measurements $\mathcal{A}$ and
$\mathcal{B}$. Each measurement is defined by a corresponding
collection of $d$ orthonormal vectors, $\mathcal{A} \equiv \{
|a_1\rangle , \dots , |a_d\rangle \}$ and $\mathcal{B} \equiv \{
|b_1\rangle , \dots , |b_d\rangle \}$. Then the Maassen-Uffink
EUR~\cite{maassen88} states that
\begin{equation}
\inf_\rho \frac{ H[\mathcal{A}(\rho)] + H[\mathcal{B}(\rho)] }{2}
\geq  c_{\mathcal{A},\mathcal{B}} \, ,
\end{equation}
where $c_{\mathcal{A},\mathcal{B}} = \log_2{ \max_{k,h} |\langle
a_k | b_h \rangle| }$ and
\begin{align}
H[\mathcal{A}(\rho)] & = - \sum_{k=1}^d \langle a_k | \rho | a_k \rangle \log_2{\langle a_k | \rho | a_k \rangle} \, ,\\
H[\mathcal{B}(\rho)] & = - \sum_{k=1}^d \langle b_k | \rho | b_k
\rangle \log_2{\langle b_k | \rho | b_k \rangle} \, .
\end{align}
In particular, if the two observables are mutually unbiased, then
$\max_{k,h} |\langle a_k | b_h \rangle| = d^{-1/2}$ and we obtain
\begin{equation}\label{MU2}
\inf_\rho \frac{ H[\mathcal{A}(\rho)] + H[\mathcal{B}(\rho)] }{2}
\geq  \frac{1}{2} \log_2{d} \, .
\end{equation}

Given a collection of $k$ observables, one can always find a state
$\rho$ such that $H[\mathcal{M}_j(\rho)] = 0$ for a given $j$.
Therefore the constant $c_{\mathcal{M}}$ in Eq.\ (\ref{EqEUR}) is
at least larger than $\left( 1 - \frac{1}{k} \right) \log_2{d}$.
An EUR that saturates this bound is said to be maximally strong.
An almost maximally strong EUR is obtained for a maximal choice of
$k = d+1$ mutually unbiased observables, in which case the
constant $c_{\mathcal{M}}$ in Eq.~(\ref{EqEUR}) equals
$\log_2{\frac{d+1}{2}}$~\cite{Sanchez}.

Maximally strong EURs can be obtained for multiple measurements in
a high dimensional space. Ref.~\cite{QDL2} showed that a random
choice of $k$ random observables (distributed according to the
unitary invariant measure) satisfies a maximally strong EUR with
probability arbitrary close to $1$, provided that $d$ is large
enough and $k$ grows at least logarithmically in $d$ (see also
Refs.~\cite{Buhrman,Fawzi}). Recently, Ref.~\cite{Adamczak2016}
showed that this property holds for large $d$ at any fixed value
of $k$.

Uncertainty relations can be expressed not only in terms of the
Shannon entropy. For example, fidelity uncertainty relations have
been defined in Ref.~\cite{Adamczak2017}, and metric uncertainty
relations have been introduced in Ref.~\cite{Fawzi}. These are
stronger forms of uncertainty relations, in the sense that they
always imply an EUR, while the contrary does no necessarily hold.

\subsection{Bounded quantum storage model}

In this section we briefly review some basic notions regarding the
BQSM, Our presentation will mostly follow Ref.~\cite{BQSM} (see
also Ref.~\cite{ChThesis}). To make things more concrete, we
consider a one-way protocol in which the sender Alice encodes a
variable $X$ into a $d$-dimensional Hilbert space, with $|X| = d$.
The protocol is specified by a collection of $k$ different
orthogonal bases. Alice randomly selects one of the bases and then
encodes the classical random variable $X$ by using the $d$
mutually orthogonal vectors in the chosen basis. On the receiver
side, Bob independently selects one of the $k$ bases at random and
applies the corresponding projective measurement. The protocol is
indeed analogous to a $d$-dimensional version of BB84 with $k$
different bases. After the quantum part of the protocol, in which
$n$ states are prepared, transmitted, and measured, the users
proceed with the sifting phase, in which they select only the
signal transmissions for which they have made the same choice of
bases. The protocol then concludes with error reconciliation and
privacy amplification. The difference with standard QKD is that in
the BQSM the eavesdropper Eve is assumed to be only capable of
storing a finite amount of quantum information. More specifically,
it is assumed that Eve has kept no more than $q$ qubits in her
quantum memory after $n$ quantum signal transmissions and before
sifting. Therefore, all remaining quantum states intercepted by
Eve have been already measured before the sifting phase takes
place.

A fundamental estimate of the number of secret bits (excluding
sifting) that can be extracted from such a protocol is given by
(for direct reconciliation):
\begin{equation}\label{bitlen}
\ell^\epsilon \simeq H_\mathrm{min}^\epsilon(X^n|Z E) -
H_\mathrm{max}(C) \, ,
\end{equation}
where $\epsilon$ is a security parameter,
$H_\mathrm{min}^\epsilon(X^n|Z E)$ is the smooth
min-entropy~\cite{RennerPhD,MTomamichel-PhD} conditioned on Eve's
side information for $n$ signal transmissions, and
$H_\mathrm{max}(C)$ is the number of bits publicly exchanged for
error reconciliation. Under the assumptions of the BQSM, here
Eve's side information comprises a quantum part $E$ and a
classical part $Z$. Furthermore, since Eve's quantum memory has
capacity below $q$ qubits, we have
\begin{equation}
\ell^\epsilon \gtrsim H_\mathrm{min}^\epsilon(X^n|Z) - q -
H_\mathrm{max}(C) \, .
\end{equation}

It remains to bound the (classical) conditional smooth min-entropy
$H_\mathrm{min}^\epsilon(X^n|Z)$. It has been shown in Ref.\
\cite{BQSM} that if the set of $k$ bases employed in the protocol
satisfies an EUR as in Eq.~(\ref{EqEUR}), then for any $\lambda
\in (0,1/2)$
\begin{equation}
H_\mathrm{min}^\epsilon(X^n|Z) \geq ( c_\mathcal{M} - 2\lambda) n
\, ,
\end{equation}
with
\begin{equation}
\epsilon = \exp{\left[ - \frac{ \lambda^2 n }{32(\log_2(k d
/\lambda))^2} \right]} \, .
\end{equation}

For example, using two mutually unbiased bases we can apply the
Maassen-Uffink EUR in Eq.\ (\ref{MU2}) and obtain, for
sufficiently small $\lambda$
\begin{equation}
H_\mathrm{min}^\epsilon(X^n|Z) \gtrsim \frac{n}{2} \log_2{d} \, .
\end{equation}
In general, the assumptions of BQSM allow us to increase the
resilience to noise of a QKD protocol, but the rate is not
expected to improve dramatically compared to an unbounded
quantum-capable eavesdropper. We conclude by noting that the
number of secret bits in Eq.~(\ref{bitlen}) must be multiplied by
a factor $1/k$ to account for the probability that Alice and Bob
chose the same basis.

\subsection{Quantum data locking}

The phenomenon of QDL can be exploited to obtain efficient
high-dimensional QKD protocols within the assumption that Eve has
access to a quantum memory of unlimited capacity but that can
store quantum information only for finite time. This assumption
implies that she is forced to measure her share of the quantum
system within a given time after having obtained it. When the
memory time goes to zero, we obtain as a limiting case the setting
of individual attacks, where Eve is forced to measure the signals
as soon as she receives them.

The first QDL protocol was discussed in Ref.~\cite{QDL1}. Such a
protocol is analogous to BB84, with the fundamental difference
that now Alice and Bob share $1$ secret bit at the beginning of
the protocol~\cite{QDL1}. While in BB84 Alice and Bob randomly
select their local basis, and only later reconcile their choice in
the sifting phase, in QDL they use the $1$ bit of information they
secretly share to agree on the choice of the basis in which encode
(and decode) information.
Therefore, according to this secret bit, Alice encodes $n$ bits
into $n$ qubits, using either the computational or the diagonal
basis, and Bob measures the received qubits in the same basis. We
follow the original presentation of Ref.~\cite{QDL1} and assume a
noiseless channel from Alice and Bob. Suppose that Eve intercepts
the $n$ signal qubits. As she is forced to measure them (either
instantaneously or after a given time) the amount of information
that she can obtain about the message can be quantified by the
accessible information.

Consider the joint state representing the classical $n$-bits sent
by Alice together with the quantum state intercepted by Eve. Such
a classical-quantum state reads
\begin{equation}
\rho_{XE} = \sum_{x^n=0}^{2^n-1} 2^{-n} |x^n\rangle \langle x^n |
\otimes \frac{1}{2} \sum_{j=0,1} U_j^n | x^n \rangle \langle x^n |
{U_j^n}^\dag \, ,
\end{equation}
where $X$ denote the classical variable sent by Alice, $U_0$ is
the identity transformation and $U_1$ is the unitary that maps the
computational basis into the diagonal one. Notice that this
expression reflects the fact that Eve does not know which basis
has been used for the encoding.
For such a state the accessible information is
\begin{align}
I_\mathrm{acc}( X : E )_\rho & = \max_{M_{E \to Z}} I( X : Z ) \\
& = \max_{M_{E \to Z}} H(X) - H(X|Z) \\
& = n - \min_{M_{E \to Z}} H(X|Z) \, ,
\end{align}
where the maximum is over all possible measurements $M_{E \to Z}$
performed by Eve on $n$ qubits. A straightforward calculation
yields~\cite{QDL1}
\begin{align}
& I_\mathrm{acc}( X : E )_\rho = n - \min_{M_{E \to Z}} H(X|Z) \\
& \leq n + \max_\phi \frac{1}{2} \sum_{j,x^n} |\langle \phi | U_j
|x^n \rangle|^2 \log_2{|\langle \phi | U_j |x^n \rangle|^2} \, .
\label{ub1}
\end{align}

Notice that the last term on the right hand side is bounded by an
EUR. In particular, here we can apply Maassen-Uffink
EUR~\cite{maassen88} and obtain
\begin{align}
I_\mathrm{acc}( X : E )_\rho & \leq \frac{n}{2} \, .
\end{align}
In summary, being ignorant of one single bit of information, Eve
is able to access only $n/2$ bits of information about the n bits
of information communicated from Alice to Bob. This holds for all
values of $n$.

We remark that QDL violates a fundamental property, know as total
proportionality, that is natural property of any well-behaved
information quantifier~\cite{QDL1}. In fact given a mutual
information $I(X:Z)$ we expect that
\begin{align}\label{TotProp}
I(X:ZK) < I(X:Z) + H(K) \, ,
\end{align}
that is, the knowledge of the variable $K$ can increase the mutual
information by no more than its entropy. As a matter of fact this
property is fulfilled by the classical mutual information as well
as by the quantum mutual information. Interestingly enough, QDL
shows that this is not the case for the accessible information.
After the seminal work by DiVincenzo et al.~\cite{QDL1}, other
works have introduced QDL protocols that presents an even stronger
violation of total proportionality.

From a broader perspective, a QDL protocol is defined by a set of
$k \ll d$ different bases in a Hilbert space of dimensions $d$.
For an eavesdropper that does not have which-basis information
(i.e., $\log_2{k}$ bits) the accessible information is smaller
than $\delta$. Therefore, EURs for $k$ bases in a $d$-dimensional
space can be applied to obtain a corresponding QDL protocol.
%
Ref.~\cite{QDL2} has shown that a random choice of the $k =
(\log_2{d})^3 $ bases (sampling according to the distribution
induced by the Haar measure) in a $d$-dimensional Hilbert space
will yield a QDL protocol with $\delta = \epsilon \log_2{d} +
O(1)$, as long as $d$ is large enough.
The probability that a random choice of bases yield a QDL protocol
with these feature is bounded away from $1$ if $\log_2{d} >
\frac{16}{C'' \epsilon} \ln{\frac{20}{\epsilon}}$, with $C'' =
(1760 \ln{2})^{-1}$. As $\log_2{d}$ grows faster than linearly in
$1/\epsilon$, this implies that QDL is obtained only for
asymptotically large values of $d$. For example, putting $\epsilon
= 10^{-1}$ one gets the condition $\log_2{d} \gtrsim 10^6$. A
typicality argument shows that as long as $k$ is sufficiently
smaller than $d$, these bases are with high probability
approximate mutually unbiased~\cite{Fawzi}. Interestingly enough,
a collection of (exact) mutually unbiased bases does not
necessarily yield QDL~\cite{WW}.

A major advance in QDL was provided by the work of Fawzi et
al.~\cite{Fawzi}, which has introduced the notion of metric
uncertainty relations.
Exploiting this powerful tool they have been able to obtain strong
QDL protocols with $\delta = \epsilon \log_2{d}$ and $\log_2{k} =
4 \log_2{(1/\epsilon)} + O(\log_2{\log_2{(1/\epsilon)}})$, for any
$\epsilon >0$ and for $d$ large enough. While these protocols are
for random unitaries (which cannot be simulated efficiently), they
also demonstrated QDL with a set of unitaries that can be
simulated efficiently on a quantum computer. We remark that these
results still require asymptotically large values of $d$.
The QDL protocols of Ref.~\cite{Fawzi} were the first to allow for
an arbitrary small accessible information. As for
Ref.~\cite{QDL2}, the protocols succeed only for asymptotically
large values of $d$.

Whereas QDL was historically introduced in terms of the accessible
information, it can also be expressed in terms of stronger
security quantifiers, e.g., the total variation distance via
Pinsker inequality~\cite{Pinsker}. As additional examples, the
metric uncertainty relations of Ref.~\cite{Fawzi} and the fidelity
uncertainty relations of Ref.~\cite{Adamczak2017} also yield QDL
with a stronger security quantifier.

\subsection{Quantum data locking for communication: the quantum
enigma machine}

QDL was considered for the first time in a communication scenario
in Refs.~\cite{QEM,PRX}. The authors of Ref.~\cite{PRX} considered
a noisy communication channel from Alice to Bob (notice that
previous works only considered a noiseless channel). Two scenarios
were analyzed: in strong QDL Eve is able to access the input of
the channel; in weak QDL she has access to the output of the
complementary channel from Alice to Bob. Notice that weak QDL is
analogous to the familiar wiretap channel model. Strong QDL is
instead closer to the original formulation of QDL. The notion of
weak and strong QDL capacities were introduced and in part
characterized. In analogy to the notion of private capacity of
quantum channel, the (weak and strong) QDL capacities are defined
as the maximum asymptotic rate at which Alice and Bob can
communicate through the quantum channel with the guarantee that
Eve has no information about the exchanged messages. The
difference with the notion of private capacity is that to achieve
the QDL capacities we assume that Eve is forced to make a
measurement as soon as she obtains a train of $n$ signals (then
$n$ is made arbitrary large to obtain an asymptotic rate).

Since it is defined accordingly to a weaker security definition,
the weak QDL capacity is never smaller than the private capacity.
A consequence of the results of Ref.~\cite{Fawzi} is that the
identity qubit channel has unit strong QDL capacity. Entanglement
breaking channels and Hadamard channels are instead shown having
vanishing weak QDL capacity~\cite{PRX}. Ref.~\cite{Winter}
provided explicit examples of quantum channels with a large gap
between the private capacity and the weak QDL capacity.

A quantum optics device that exploits QDL for secure communication
was dubbed a quantum enigma machine (QEM) by Lloyd~\cite{QEM}. In
fact, the protocol of QDL can be seen as a quantum generalization
of poly-alphabetic ciphers, among which one of the most famous
examples was the Enigma machine. Ref.~\cite{QEM} put forward two
architectures for a QEM, using either unary encoding of a single
photon over $n$ modes (this would be a direct application of the
QDL protocols in Refs.~\cite{QDL2,Fawzi}) or using encoding in
coherent states. Ref.~\cite{PRX} showed that a weak QDL protocol
with coherent state cannot surpass the private capacity by more
than $\log_2{e} \simeq 1.44$ bits per bosonic mode, and an almost
matching lower bound was obtained in Ref.~\cite{CVQDL}.

\subsection{Practical quantum data locking}

The QDL protocols of Refs.~\cite{QDL2,Fawzi} require coherent
control over large (actually asymptotically large) Hilbert space.
For this reason there is little hope that these protocols may be
ever realized experimentally, not even as a proof-of-principle
demonstration.
In order to make an experimental realization of QDL feasible, one
needs to solve two problems: 1) to design QDL protocols that
require control over Hilbert space of reasonably small dimensions;
2) to design protocols that are robust in the presence of a noisy
channel from Alice to Bob.
Step forwards towards the solution of these two problems were made
in Ref.~\cite{PRL}. The authors of this work considered a
collection of $n$ $d$-dimensional systems, where $d$ is supposed
to be a small integer and $n$ is asymptotically large. Instead of
considering random unitaries in a large Hilbert space, they
considered local random unitaries in the small $d$-dimensional
systems. This model can be physically realized by a train of $n$
photons, each living in the space defined by a discrete collection
of $d$ bosonic modes (spanning, for example, spatial, temporal,
frequency, or angular momentum degrees of freedom).

Unlike other QDL protocols that exploit EURs, QDL with local
unitaries is obtained from a different upper bound on the
accessible information, i.e.,
\begin{align}
I_\mathrm{acc}(X:E) \leq n \log_2{d} - \min_\phi H[Q(\Phi)] \, ,
\end{align}
where
\begin{align}
H[Q(\Phi)] = - \sum_{x^n} Q_{x^n}(\phi) \log_2{Q_{x^n}(\phi)} \, ,
\end{align}
and
\begin{align}
Q_{x^n}(\phi) = \frac{1}{k} \sum_{j=1}^k | \langle \phi | U^n |
x_n \rangle |^2 \, .
\end{align}
The quantity $\min_\phi H[Q(\Phi)]$ is then bounded by exploiting
the fact that $Q_{x^n}(\phi)$ typically concentrates around
$1/d^n$ (a similar approach was used in Ref.~\cite{Robust} to
obtain QDL with a set of commuting unitaries).
Exploiting this approach, Ref.~\cite{PRL} demonstrated strong QDL
protocols for QKD through generic memoryless qudit channels, and
Ref.~\cite{NJP} obtained weak QDL protocols for direct secret
communication. The price to pay to deploy QDL with local unitaries
is that the amount of pre-shared secret key bits is no-longer
exponentially smaller than the message but grows linearly with the
number of channel uses, with an asymptotic rate of $1$ bit per use
of the channel. This implies that non-zero rates can only be
obtained for $d>2$, yet any value of $d$ equal or larger than $3$
can yield a non-zero rate of QKD or direct communication.

\subsection{Experimental demonstrations}

The first experimental demonstrations of QDL appeared in 2016.
Ref.~\cite{JWPan} realized the original QDL protocol~\cite{QDL1}
with encoding in heralded single photon polarization. They also
implemented error correction to counteract loss and verified a
violation of the total proportionality inequality of
Eq.~(\ref{TotProp}). Ref.~\cite{Lum} realized the QDL protocol of
Ref.~\cite{PRL} using pulse-position modulation (PPM). In
Ref.~\cite{Lum} a lens was used to implement a Fourier transform
and an array of $128 \times 128$ spatial light modulators (SLM)
was applied to generate random phase shifts. This transformation
provides QDL given that at the receiver end a trusted user applies
the inverse phase shift and inverse Fourier transform to
decode~\cite{PRL}. Finally, Ref.~\cite{Jelena} presented an
on-chip array of programmable ring resonators that can be
naturally applied to QDL with encoding in time of arrival degree
of freedom.


\section{Quantum random number generation}

\subsection{Introduction}\label{sec:qrng_intro}
Generating random numbers is an important task: most cryptographic
protocols rely on them, they are used in simulations, in
lotteries, in games and numerous other places. However, in spite
of their usefulness, random number generators (RNGs) are difficult
to construct and the use of poor-quality random numbers can be
detrimental in applications. For instance, in Ref.~\cite{Lenstra&}
public RSA keys were collected from the web and a significant
number were found to share a prime factor, posing problems for the
security of those running the algorithm.  In general, problems can
arise whenever something that is assumed to be chosen randomly is
in fact not~\cite{HDWH}.

A typical way to make random numbers is to use a pseudo random
number generator, in which a short random seed is expanded into a
longer string.  The idea is that this string is sufficiently
random for the application it will be used in.  However, a pseudo
random number generator is a deterministic algorithm, so, in spite
of its length, the output contains no more randomness than the
input.  It must therefore contain subtle correlations that in
principle could be detected and exploited.  Given a powerful
enough computer, a long enough output sequence could be used to
find the seed and hence all of the remaining purportedly random
numbers.

Since classical physics is deterministic, RNGs based on classical
effects can never be fundamentally random.  Instead classical RNGs
rely on a lack of knowledge making the numbers appear random.
Whether this is good enough for a particular application is a
matter of faith, and an undesirable property of such RNGs is that
it can be difficult to detect if they are functioning badly.
Indeed, while statistical tests are able to attest (beyond
reasonable doubt) to particular shortcomings of a candidate RNG,
there is no set of tests that can take the output from a candidate
RNG and eliminate all shortcomings.

To understand this, it is helpful to define what we mean when we
say that a particular string is random.  Note that, although the
string $S$ will always be classical, we want it to appear random
even to an adversary holding quantum information and hence the
definition is phrased in terms of quantum states.  This definition
is related to the definition of a secure key (cf.\
Section~\ref{sec:compos}). If a random number generation protocol
outputs an $n$ bit string $S$, we would like it to be uniform and
unknown to any other party, i.e., independent of any side
information $E$ held.  Mathematically, for $S$ to be a high
quality random string we would like that
$$D(\rho_{SE},\frac{1}{n}\id_n\ot\rho_E)$$
is small and we say that a protocol is secure if
$$p(\bar{\perp})D(\rho_{SE},\frac{1}{n}\id_n\ot\rho_E)$$
is small, where $p(\bar{\perp})$ is the probability that the
protocol does not abort (note the similarity with the secrecy
error, $\eps_{\secr}$, from Section~\ref{sec:compos}).  As before,
this means that whenever there is a high probability of not
aborting, the output is close to perfect randomness, i.e.,
$\rho_{SE}\approx\frac{1}{n}\id_n\ot\rho_E$.  Unlike in key
distribution, there is no second string that the first one needs
to be perfectly correlated with, so there is no analogue of the
correctness error.

From this definition, it is evident that no amount of statistical
testing on the output can verify that $S$ is a high quality random
string: statistical tests on $S$ can only increase confidence that
$\rho_S\approx\frac{1}{n}\id_n$, but cannot say anything about
whether $\rho_{SE}\approx\frac{1}{n}\id_n\ot\rho_E$, i.e., whether
another party could already know the string $S$. (For some
  applications, it may not be a problem for another party to know the
  string, provided that it is statistically random; here we focus on
  the stronger form of randomness.)  Whether a string is random or not
is ultimately not a property of the string itself, but on how it
is generated.

Like in the case of QKD, we can divide quantum random number
generators (QRNGs) into two types depending on whether or not the
users trust the apparatus they use (there are also hybrids, not
discussed here, in which certain features are trusted and others
not, e.g., semi-device-independent QRNGs~\cite{Lunghi15}). Both
types work by exploiting the fundamental randomness of certain
quantum processes, but with trusted devices, it is more
straightforward to do so.  We briefly mention one example here. A
simple trusted-device QRNG can be based on a 50:50 beamsplitter
and two detectors, one for the reflected arm and the other for the
transmitted arm.  If a single photon is incident on the
beamsplitter, then with probability half it will go to one
detector and with probability half the other. In principle this is
a source of quantum random numbers.

However, building such a QRNG is not as straightforward as it
sounds. Generation and detection of single photons is challenging,
and it is difficult to ensure that the beamsplitter is perfect.
Furthermore, correlations may be brought into the string by other
factors such as fluctuations in the power supply, asymmetries in
the detector responses and dead times.  The standard way to
account for such difficulties is to try to quantify these effects,
estimate the min-entropy of the outputs and then use a classical
extractor to compress the imperfect raw string into arbitrarily
good randomness.

One issue that needs to be considered when doing this is that
extraction of randomness typically requires a seed, i.e., an
independent random string. Fortunately, this seed can act
catalytically if a strong extractor is used, i.e., the seed
randomness remains random and virtually independent of the output
randomness so is not consumed in the process.  Nevertheless, the
need for this seed means that QRNG protocols should more
accurately be described as \emph{quantum randomness expansion}
(QRE) protocols.  In order to have a good rate of expansion,
randomness extractors requiring a short seed should be used. Note
also that to have full security guarantees, \emph{quantum-proof}
randomness extractors should be used.

For the type of QRNG mentioned above the security relies on the
accuracy of the model used to describe it. Like in the case of
QKD, various additional advantages can be gained by moving to
device-independent protocols (see Section~\ref{sec:di_intro}).
These shift reliance away from the model: that the output string
is random is checked on-the-fly and relies on the correctness and
completeness of physical laws (note that correctness and
completeness of quantum theory are
related~\cite{CR_ext,CR_book1}).  The idea has been described
earlier in the review where DI-QKD was introduced (see
Section~\ref{sec:Bell_unpred}).  In essence, if we have some
number of separate systems whose correlations violate a Bell
inequality, then their outcomes must contain some min-entropy,
even conditioning on an adversary holding arbitrary side
information (for instance a quantum system entangled with those
being measured).  This min-entropy can be lower bounded and an
extractor applied leading to arbitrarily good randomness output.

To use this idea, some initial randomness is required, so we need
to ensure that the protocol outputs more randomness than it
requires giving genuine expansion.  This can be achieved using a
protocol analogous to the spot-checking CHSH QKD protocol from
Section~\ref{sec:spotCHSHQKD}.

\subsection{Protocols for DI-QRE}
\subsubsection{The setup for DI-QRE}
The setup is different from that for DI-QKD because there is only
one honest party (Alice) in this protocol, and, because the
protocol is for randomness expansion, we do not give an unlimited
supply of random numbers to Alice. These are the assumptions:
\begin{enumerate}
\item\label{ass_re:1} Alice has a secure laboratory and control over
  all channels connecting her laboratory with the outside world. For
  any devices in her labs, Alice can prevent unwanted information flow
  between it and any other devices.
\item Alice has a reliable way to perform classical information
  processing.
\item Alice has an initial seed of perfectly random (and private)
  bits, known only to her.
\end{enumerate}

Like for DI-QKD, security is proven in a composable way (cf.\
Section~\ref{sec:compos}) allowing the protocol's output to be
used in an arbitrary application.  The remarks made in the last
paragraph of Section~\ref{sec:DIQKDsetup} all apply to QRE as
well.  However, mitigating the device-reuse problem is easier for
QRE than in QKD because QRE does not involve public communication
during the protocol~\cite{bckone}.

\subsubsection{The spot-checking CHSH QRE protocol}
There are many possible types of protocol; we will describe a
specific protocol here, based on the CHSH game with spot-checking.
The protocol has parameters $\alpha\in(0,1)$, $n\in\mathbb{N}$,
$\beta\in(2,2\sqrt{2}]$, $\delta\in(0,2(\sqrt{2}-1))$, which are
to be chosen by the users before it commences.
\begin{enumerate}
\item Alice uses her initial random string to generate an $n$-bit
  string of random bits $T_i$, where $T_i=0$ with probability
  $1-\alpha$ and $T_i=1$ with probability $\alpha$.
\item \label{st_re:1} Alice uses a preparation device to generate an
  entangled pair.  She sends one half to one measurement device and
  the other half to another such device. (As in the case of
    DI-QKD, although this step refers to the generation of an
    entangled state, security does not rely on this taking place
    correctly.)
\item \label{st_re:4} If $T_i=0$ (corresponding to no test) then Alice
  makes fixed inputs into each measurement device, $A_i=0$ and $B_i=0$
  and records the outcomes, $X_i$ and $Y_i$. These inputs are made at
  spacelike separation and each device only learns its own input.\\
  If $T_i=1$ (corresponding to a test) then Alice uses her initial
  random string to independently pick uniformly random inputs
  $A_i\in\{0,1\}$ and $B_i\in\{0,1\}$ to her devices and records the
  outcomes, $X_i$ and $Y_i$.
\item Steps \ref{st_re:1} and \ref{st_re:4} are repeated $n$ times,
  increasing $i$ each time.
\item For all the rounds with $T_i=1$, Alice computes the average CHSH
  value (assigning $+1$ if $A.B=X\oplus Y$ and $-1$ otherwise). If
  this value is below $\beta-\delta$, she aborts the protocol.
\item If the protocol does not abort, for the rounds with $T_i=0$ the
  outputs $X_i$ are fed into a randomness extractor whose seed is
  chosen using Alice's initial random string.  The EAT can be used to
  compute how much randomness can be extracted, depending on the value
  of $\beta$.
\end{enumerate}

The ideal implementation of this protocol is as for the CHSH QKD
protocol in Section~\ref{sec:spotCHSHQKD} (except that $B_i=2$ is
not needed) and the intuition for its operation is the same.  The
completeness error is again exponentially small in $n$.  By taking
$n$ sufficiently large, this protocol can output at a rate
arbitrarily close to $H(X|E)$ from~\eqref{eq:chsh_ent} (see
Section~\ref{sec:quant}). This rate is the amount of randomness
output per entangled pair shared. We make a few remarks about the
protocol.

\begin{enumerate}
\item The protocol aims for randomness expansion, so it is important
  to use as little randomness as possible to implement it.  Since each
  test round consumes two bits of randomness, we would like $\alpha$
  to be chosen to be small.  This also helps reduce the amount of
  randomness required to choose the test rounds, since generating a
  string of $n$ bits with bias $\alpha$ requires roughly
  $nh_2(\alpha)$ bits of uniform randomness from Alice's initial
  random string, where $h_2$ is the binary entropy which drops away
  steeply for small $\alpha$.  The value of $\alpha$ can be chosen
  such that in the large $n$ limit, the randomness required to choose
  it is negligible.
\item If a strong extractor is used in the last step then randomness
  is not consumed for this.  Nevertheless, it is helpful to use an
  extractor with a small seed, e.g., Trevisan's extractor, so as to
  reduce the randomness required to initiate the expansion.
\item In the case of the CHSH QKD protocol the aim is to generate an
  identical key shared by Alice and Bob. Here the aim is to generate
  randomness, so there is no need for the ideal implementation to lead
  to the same outcomes for both devices in the case of no test.  This
  allows randomness expansion rates that go beyond that of the QRE
  protocol given above, while still using maximally entangled qubit
  pairs (see~\cite{BRC} for a robust protocol giving up to two bits of
  randomness per entangled pair). Like in the case of DI-QKD, finding tight bounds on the min-entropy
  in terms of the observed correlations for general protocols is an
  open problem.
\item As the number of rounds, $n$, increases the classical
  computation required by the protocol (e.g., to perform the
  randomness extraction) may become prohibitively slow.
\end{enumerate}

\subsection{Historical remarks and further reading}
The use of non-local correlations for expanding randomness without
trusting the devices used goes back to Ref.~\cite{ColbeckThesis}
and the ideas there were developed in Ref.~\cite{CK2}. The idea
was developed experimentally in Ref.~\cite{PAMBMMOHLMM} and
security proofs against classical adversaries were presented in
Refs.~\cite{FGS,PM}. The first work covering quantum adversaries
was Ref.~\cite{VV}, although this lacked tolerance to noise.
Quantum security with error tolerance was proven in
Ref.~\cite{MS1} and improved in Ref.~\cite{MS2}, where it was
shown that any Bell violation can be used to generate randomness.

Most recently, using the EAT~\cite{DFR} the expansion rate was
improved~\cite{ARV} so as to be asymptotically optimal and a
recent experiment has been performed based on these recent
techniques~\cite{QRNGpan}.

Note that several review articles devoted to the topic of
(quantum) random number generation have appeared in the last few
years~\cite{MYCQZ,AM,HG}. These go beyond the scope of the present
review and provide a useful resource for further reading on the
topic.

\subsection{Implementations}
DI-QRE suffers from some of the same drawbacks as DI-QKD, the most
significant being the difficulty of performing a Bell experiment
while closing the detection loophole.  For DI-QRE this is slightly
easier to do because there is no need for the two measurement
devices to be distant from one another. Instead, they only need to
be far enough apart to enable sufficiently shielding to ensure
they cannot communicate during the protocol. (In particular, each
device should
  make its output independently of the input of the other device on
  each round of the protocol.)  Nevertheless, it remains challenging
to do this.  While the first DI-QRE experiment ran at a very low
rate~\cite{PAMBMMOHLMM}, recent state-of-the-art experiments
achieve reasonable rates~\cite{QRNGpan} and even close the
locality loophole as well as the detection loophole. Such a
demonstration could be turned into a future randomness beacon, but
is still far from being built reliably into a small scale device
that could reasonably be included in a desktop computer or mobile
phone.

Because of this, for the next few years or so DI-QRE is unlikely
to be used in commercial products.  One possibility is to use RNGs
that rely on a detailed model of how the device operates (see,
e.g., Ref.~\cite{FRT}).  To ensure such RNGs work as intended it
will be important to make increasingly sophisticated models of
them and to diagnose and patch any weaknesses as and when they are
identified. Furthermore, the performance of a RNG may change with
time and if and when it degrades, it is important that this is
noticed before the purportedly random outputs are used.  A problem
such as this can be mitigated by combining the outputs of several
random number generators (in an appropriate way) to give the
random string that will be used.

\subsection{Randomness amplification}
As we saw in the last section, in order to generate randomness in
a device-independent way we require some seed randomness to start
the process.  This is necessary: to constrain a device based only
on its input-output behavior we use the violation of a Bell
inequality, and random numbers are needed to choose the inputs
when verifying such a violation.

However, while random numbers are required for this task, the
protocol given above assumes these are perfectly random.  The task
of randomness amplification concerns whether a source of imperfect
randomness can be used to generate perfect randomness. (This
  should not be confused with the related task of randomness
  extraction, where an additional perfect seed is available.)  Like
randomness expansion, this task is impossible classically in the
following sense: given a particular type of imperfect source of
randomness, a Santha-Vazirani source~\cite{SanthaVazirani}, and no
other source of randomness, there is no classical protocol can
generate perfectly random bits~\cite{SanthaVazirani}.

A Santha-Vazirani source is a way of modeling a source of bad
randomness. It has the property that each bit given out can be
biased towards either 0 or 1 within some limits which are
specified by a parameter $\eps\in[0,1/2]$.  More precisely, call
the outputs $S_i$, and let $W_i$ be a random variable representing
arbitrary additional information available that could not be
caused by $S_i$.  The sequence of bits $S_i$ in a Santha-Vazirani
source with parameter $\eps$ if
$$\left|P_{S_i|S_{i-1}=s_{i-1},\ldots,S_1=s_1,W_i=w}(0)-\frac{1}{2}\right|\leq\eps\ \
\forall s_{i-1},\ldots,s_1,w\,.$$ In other words, even given the
entire prior sequence and any other information that could not be
caused by $S_i$, the probability of $0$ and $1$ each lie between
$1/2-\eps$ and $1/2+\eps$.

It turns out that such a source of randomness can be amplified
with a quantum protocol.  The first proof of this appeared in
Ref.~\cite{CR_free} where the task was introduced.  There it was
shown that for $\eps\leq0.058$ a single source of $\eps$-free bits
can be used to generate bits that are arbitrarily close to
uniform. Subsequently it was shown that this bound on $\eps$ could
be extended to cover all $\eps<1/2$, i.e., any source of partially
random bits can be amplified, no matter how small the
randomness~\cite{GMTDAA}.  These initial protocols gave important
proofs of principle, but were impractical due to low noise
tolerance or the need for large numbers of devices, a problem
addressed in~\cite{BRGHH}.  The current state of the art can be
found in~\cite{KA}, which includes a protocol with two devices
that tolerates noise and works for all $\eps<1/2$.

Further works considered other types of imperfect randomness, in
particular, min-entropy sources which take as input a single
string with an assumed lower bound on its min-entropy conditioned
on arbitrary side information, and no further assumptions about
the structure of the randomness.  A security proof in this
scenario is given in~\cite{CSW}.

It is worth noting that while all of the above works prove
security of randomness amplification against quantum adversaries,
several also show security against a post-quantum adversary whose
power is only limited by the impossibility of signalling.
Protocols that work against arbitrary no-signalling adversaries
tend to lack efficiency. One reason for this is the difficulty of
extracting randomness against a no-signalling
adversary~\cite{HRW2,AHT}.

Another noteworthy property of many of the above protocols (all
except the protocol of~\cite{BRGHH}) is that they can work using a
public source of randomness as a seed. This is relevant in the
context of randomness beacons (for example that of NIST). If a
user suspects that the output of the beacon is imperfect in some
specified way, they may be able to use a randomness amplification
protocol to increase their trust in the output randomness.


\section{Quantum Digital Signatures}\label{SectionQDSs}

\subsection{Introduction}\label{9.Sec:intro}

Digital signature is a cryptographic primitive that ensures that a
digital message was (i) created by the claimed sender
(authenticity), (ii) that the message was not altered (integrity)
and (iii) that the sender cannot deny having sent this message
(non-repudiation). It is the digital analogue of handwritten
signatures but comes with a higher level of security guaranteed by
cryptographic means. Digital signatures play a very different role
than encryption in modern communications, but this role is of no
less importance.
Ronald Rivest, one of the inventors of public-key cryptography,
stated in 1990 that ``the notion of digital signature may prove to
be one of the most fundamental and useful inventions of modern
cryptography''. This prediction has been fulfilled, since nowadays
it is a necessary tool for a huge range of applications, from
software distribution, financial transactions, emails to
cryptocurrencies and e-voting.

Here we review the research on quantum digital signatures (QDS)
that demonstrate how using simple quantum communications we can
achieve digital signature schemes that are more secure than most
of the commonly used digital signatures algorithms. We start in
Section~\ref{defsQDS} with definitions and security properties of
digital signatures and motivate the use of quantum means in
Sections~\ref{QDSsecc} and~\ref{QDSsecd}. We present the seminal
Gottesman-Chuang scheme, and identify the practical limitations it
has in Section~\ref{9.Sec:GC-QDS}. In
Section~\ref{9.Sec:practical} we describe how one-by-one these
restrictions were lifted, making QDS a currently realizable
quantum technology. In Section~\ref{9.Sec:generic_QDS} we describe
a generic practical QDS protocol. A reader interested in quickly
catching-up with the current state-of-the-art for QDS, could read
this section directly after the introduction. In
Section~\ref{9.Sec:theory_improvements} we give theoretical and in
Section~\ref{9.Sec:experiments} experimental recent developments.
Finally, in Section~\ref{9.Sec:classical_uss} we give a (fully
classical) alternative to QDS that requires point-to-point secret
keys (potentially obtained via QKD) and then we conclude in
Section~\ref{9.Sec:conclusion}.

\subsection{Definitions and security properties}\label{defsQDS}

A QDS scheme involves multiple parties: one sender and
(potentially many) receivers. It consists of three phases each
described by a corresponding algorithm $Gen, Sign, Ver$.

\begin{itemize}
\item[$(Gen)$]Key generation algorithm. This sets and distributes the ``keys'' to be used in the subsequent interactions. (It is
also known as the ``distribution phase''.) A private key $(sk)$
that is given to the sender, and (possible multiple) public key(s)
$(pk)$ given to the receivers are selected. In protocols where the
public key of different receivers is not the same, a subscript
will indicate which receiver refers to e.g. $pk_i$.

\item[$(Sign)$]Signing algorithm. The sender chooses a message $m$ and uses her private key $sk$ to generate a signature $\sigma=Sign(m)$ and then send the pair $(m,Sign(m))$ to the desired receiver.

\item[$(Ver)$]Verifying algorithm. A receiver has as input a
message-signature pair $(m,\sigma)$ and the public key $pk$ and
checks whether to accept the message as originating from the
claimed sender or not. In certain types of signatures (including
the QDS), there are multiple levels of ``accepting'' a message,
depending on what confidence the receiver has that this message
would also be accepted if forwarded to other receivers.
\end{itemize}

An important property of digital signatures schemes is that after
the $Gen$ phase, the actions of the parties are determined without
further (classical or quantum) communication, i.e. they sign and
decide to accept or reject a message-signature pair, based solely
on the keys $sk$ and $pk$ respectively, that were distributed
during the $Gen$ phase. This is precisely how hand-written
signatures are used, where one signs and accepts/reject a
signature ``locally''. Let us define the correctness and security
notions for a digital signature scheme:

\begin{itemize}
\item A digital signature scheme is correct if a message-signature pair signed with $Sign$ algorithm using the correct private key $sk$
is accepted by the $Ver$ algorithm with unit probability.
\item A digital signature scheme is secure if no adversary without access to the
private key $sk$ can generate a signature that is accepted by the
$Ver$ algorithm with non-negligible probability.
\end{itemize}

These definitions, along with the guarantee that the private key
$sk$ is not leaked and that all parties share the same (correct)
public key $pk$, lead to three important properties:
unforgeability, non-repudiation and transferability. We will see
that ensuring parties received the same and correct public key
becomes a non-trivial task when the keys are quantum. Instead of
using the above security definitions, for analyzing QDS schemes,
we will instead aim to ensure that the following three properties
are satisfied:

\begin{enumerate}
\item Unforgeability: A dishonest party cannot send a message pretending to be someone else.
\item Non-repudiation: A sender cannot deny that she signed a message.
\item Transferability: If a receiver accepts a signature, he should be confident that any other receiver (or judge) would also accept the signature.
\end{enumerate}

Firstly, we need to clarify how the words ``cannot'' and
``confident'' are used. The meaning is that, for any adversary
allowed (which, depending on the setting, may or may not have
restrictions in his computational power), the probability of the
protocol failing can be made arbitrarily small with suitable
choices of parameters. The exact magnitude of how small is
determined by the level of security requested by the use of the
given scheme, and is characterized by a small positive number
$\epsilon$. In other words, formally we should write
$\epsilon$-unforgeability, etc.  Secondly, we note that
non-repudiation and transferability are very closely related. Not
being able to deny a signature typically depends on the way one
resolves a dispute, i.e., if Alice refuses that she signed a
contract, who will decide whether the contract had her signature
or not. In most cases this is the same as asking if a signature
accepted by one receiver would also be accepted by a judge or
other receivers, and this is exactly the transferability property.
Here we will identify non-repudiation with transferability, while
keeping in mind that this may not be the most general treatment,
if one chooses a different (less natural) ``dispute-resolution''
mechanism.

For simplicity, in the following we will refer to the sender as
Alice, the received as Bob and when a second receiver is required
(e.g. for transferability of messages), he will be referred to as
Charlie.

\subsection{What is a {\it quantum} digital signature scheme and
why it is useful?}\label{QDSsecc}

There are various things that one could call QDS, but in this
review we present the research that started with Gottesman's and
Chuang's seminal work~\cite{GC01} and deals with: signing a
classical message and using quantum communication (and
computation) in order to provide information-theoretic security
(ITS), so that the signatures generated are ``one-time'' in the
sense that when a message is signed the corresponding
private/public keys cannot be reused for signing other messages.
Other uses of the term include: signing a {\it quantum} message,
``blind'' quantum signatures, arbitrated quantum signatures,
classical signatures secure against quantum computers~\cite{BZ13},
quantum tokens for signatures~\cite{DS16}, etc.

The most common digital signature schemes are RSA-based, DSA and
ECDSA and ElGamal. The security of all these schemes is based on
the assumption that the adversaries have limited computational
power and that, in particular, it is hard for them to solve the
discrete logarithm or factoring problems. While these problems are
still believed to be hard for classical computers, since Shor's
algorithm~\cite{Shor} we know that an efficient quantum algorithm
exists. In other words, if a large quantum computer is built, then
it could solve these problems efficiently and break the security
of all these signature schemes. This provides a compelling
argument in favor of solutions that provide ITS, which is the
strongest type of security, holding irrespectively of the
computational resources that an adversary has.

Here it is important to stress that while the most commonly used
classical digital signatures schemes (mentioned above) would
break, this is not the case for all classical digital signature
schemes. There exist many (less practical) classical signature
schemes, that appear to remain secure against quantum computers
(post-quantum secure), possibly after small modifications in the
security parameters (e.g. by doubling the key-lengths). Examples
of such schemes are the Lamport~\cite{Lamport79}, Merkle
\cite{Merkle89}, Ring-Learning-With-Errors~\cite{RLWE}, CFS
\cite{CFS01}, etc.

Having said that, there is another strong argument for ITS (and
thus QDS). The research in quantum algorithms is not as mature as
in classical algorithms, therefore the confidence we have on the
hardness of problems still changes. For example, in a recent
result~\cite{BK+17} one of the best candidates for  post-quantum
cryptography, the learning-with-errors (LWE) problem, was proven
to be equivalent to the dihedral-coset problem, for which there is
a sub-exponential quantum algorithm. While this algorithm still
would not fully break the security of LWE, it certainly weakens
its security and one may wonder whether we should base the
security of our communications on such computational assumptions.

\subsection{The Lamport one-time signature scheme}\label{QDSsecd}

QDS schemes were inspired by Lamport's one-time
signatures~\cite{Lamport79} and for this reason we present here a
high-level description of this scheme. Assume that we have a
(classical) one-way function $f$. For such a function, it is
simple to evaluate $f(x)$ given $x$. However, given $f(x)=y$ it is
hard to find the pre-image, i.e. invert the function to get the
value $x$. 
Of course, we can already see that this definition (hardness of
inverting) assumes (i) limited computational resources (otherwise
one could ``brute-force'' by trying all $x$'s until he finds the
pre-image or a collision) and (ii) the function is such that it is
not efficiently invertible. In other words any scheme based solely
on the above cannot offer ITS.

Alice chooses two random inputs $x_0,x_1$ and evaluates
$f(x_0),f(x_1)$. She then publicly broadcasts the pairs
$(0,f(x_0))$ and $(1,f(x_1))$ which will be the public keys $pk$,
while she keeps the values $x_0,x_1$ secretly stored (private key
$sk$). This completes the $Gen$ algorithm. Then to $Sign$, Alice
simply sends the message $b$ along with her stored corresponding
secret key $x_b$. The receivers, to accept/reject ($Ver$
algorithm) they evaluate $f(x_b)$ and check if it agrees with
their public key in order to accept.

The intuition why this is secure comes from the fact that the
function is hard to invert. Therefore an adversary with access
only to the public keys (images) cannot find the secret key
(pre-image) for any message, in order to provide a valid (forged)
signature. At the same time, if anyone receives a valid signature
(with respect to the publicly available public key), they are
convinced that it came from Alice (non-repudiation), even if she
claims it does not, because nobody else could have generated such
signature.

Finally, at the end of such scheme, all used and unused keys are
discarded (thus one-time signatures). Such protocol has been
modified using Merkle trees~\cite{Merkle89} to allow the signing
multiple messages.

\subsection{The Gottesman-Chuang QDS}\label{9.Sec:GC-QDS}

In 2001, Gottesman and Chuang~\cite{GC01} proposed the first QDS
protocol, that we may briefly call GC-QDS. The central idea was to
use the fact that non-orthogonal states cannot be distinguished
perfectly so as to realize a ``quantum one-way function'', where
the inability to invert is not based on computational assumptions
but guaranteed by the laws of quantum mechanics. The basic idea is
that if we have a quantum state $\ket{f(x)}$, where $f(x)$
represents the classical description of the state, and the set of
possible states are non-orthogonal, no-one should be able to
determine the classical description of the state, with high
probability, unless they already know it (otherwise they could
also copy/clone the state). Moreover, the amount of information
obtained is bounded by the Holevo theorem~\cite{Holevo}. In other
words, we have the classical description of the state to play the
role of the secret key $sk$, while the quantum state itself is the
public key $\ket{pk}$. Such ``one-way function'', by construction,
cannot be broken even if one has unlimited computational power.

\subsubsection{The protocol}

A  function $f$ is chosen and is made public. This function takes
input $x$ and returns $f(x)$ that is the classical description of
a quantum state. For example, $x$ can be a two-bit string and
$f(x)$ denotes one of the four BB84 states. There is no need for
this function to be one-way, since what replaces the one-wayness
of classical protocols is that one cannot obtain the classical
description of a quantum state with certainty. In GC-QDS some
choices of functions were made, but this is not crucial for the
general description. 
Let us analyze the various steps.

\paragraph{Key generation.} For the private key, Alice chooses pairs
of bit-string $\{x^i_0,x^i_1\}$, where $1\leq i\leq L$. The
$x_0$'s will be used to sign the message $0$ and the $x_1$'s to
sign the message $1$. The number of pairs $L$ is determined by the
security level requested.

For the public key Alice generates multiple copies of the state
$\{\ket{f(x_0^i)},\ket{f(x_1^i)}\}$. Since only Alice knows the
secret keys, and unknown quantum states cannot be copied, she
generates all the copies. Then she distributes to each potential
receiver the corresponding states, along with the label for which
message they correspond.

In a digital signature scheme, {\it all} parties may be dishonest
(not simultaneously though). 
When the public key is classical, parties could easily check that
they have the same public keys. This is far from trivial in our
case. Gottesman and Chuang proposed to use multiple copies (of
each public key for each party) and they interact by performing
SWAP tests (see Fig.~\ref{fig:9.1}). This is a test that gives
always affirmative answer without disturbing the state, when two
states are identical, while fails probabilistically otherwise.
This comes with considerable cost, since each copy of the public
key (quantum state) circulated makes easier the task for an
adversary to recover the classical description (secret key) and
therefore to forge a message. Finally, all receivers store the
public key into a quantum memory until they receive a signed
message.

\begin{figure}[pth]
\includegraphics[width=0.35\textwidth]{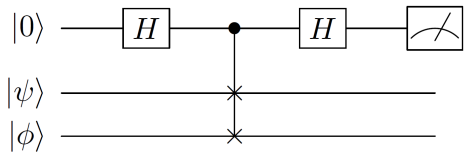}
\caption{The SWAP test. If $\ket{\psi}=\ket{\phi}$ then it gives
always $0$, otherwise the result $1$ is obtained with probability
that depends on the overlap $|\langle \psi\ket{\phi}|$.}
\label{fig:9.1}
\end{figure}

\paragraph{Signing.} To sign a message, Alice simply
chooses the message value $b\in\{0,1\}$ and sends $(b,x_b^i)$ to
the desired receiver. This is a completely classical phase.

\paragraph{Verification.} In order to confirm whether the
message-signature pair is valid, the receiver uses the classical
description: for each $x^i_b$ he generates the corresponding
quantum state $\ket{f(x_b^i)}$ and checks if it is consistent with
his stored public key. Then he counts the fraction of incorrect
keys $s_t$ (out of the $L$ keys). At this point QDS deviates from
standard digital signature schemes. The verification algorithm
takes three answers (rather than the usual two accept/reject). The
receiver can return $REJ$ when convinced that the
message-signature pair is not valid, can return 0-$ACC$ if he is
certain that is valid but is uncertain if this signature would
also be accepted by other receivers when forwarded (or by a
judge), and return 1-$ACC$ if he is certain that is valid and will
also be accepted by other receivers or judges. The reason for this
modification is subtle and is explained below in the section on
the intuition of the security. Depending on the details of the
protocol, there are two parameters $0<s_a < s_v < 1$ that
determine what $Ver$ outputs. If $s_t>s_v$ the receiver $REJ$. If
$s_v>s_t>s_a$ the receiver 0-$ACC$ and if $s_t<s_a$ the receiver
1-$ACC$.

\subsubsection{Security intuition}

Unforgeability is guaranteed by the fact that, given an unknown
quantum state, one cannot guess its classical description with
certainty, even if the state is from within a known
(non-orthogonal) set. A potential forger has in his disposal all
the copies of the quantum public keys, and, if colluding with
other receiver, may even have extra copies. We assume that the
forger performs a minimum-error (or minimum-cost in general)
quantum measurement to obtain his best guess, with an associated
probability $p_f$ of failure. Provided that this probability is
higher than an accepting threshold $p_f>s_v$, a forger cannot
mimic a valid signature, at least not with probability higher than
a decreasing exponential $e^{-c(p_f-s_v)^2L}$ for some constant
$c$. This argument is similar in all QDS protocols, where
calculating $p_f$ and $c$ varies on the details (number of copies
circulated, form of the quantum states sent, method to
measure/identify errors in the key, etc).

To prove non-repudiation is even more subtle. Alice is not forced
to send identical quantum public keys to Bob and Charlie. As
outlined above, they communicate quantumly and perform a number of
SWAP tests on copies of the public keys. The result of these tests
succeeds probabilistically. If there was a single verification
threshold $s_v$, then Alice could tune the public key she sends to
Bob and Charlie to have (expected) $s_vL$ errors. Since the number
of errors is determined by a normal probability distribution with
mean at $s_vL$, the probability that one of them finds more than
$s_vL$ errors is exactly $1/2$. This means that with probability
$1/4$ one of them will practically detect more than $s_vL$ errors
while the other less than $s_vL$ errors, and therefore they would
disagree on whether this signature is valid or not.

This is why Gottesman and Chuang introduced a second threshold
$s_a$ and used both $s_a$ and $s_v$. Now to repudiate, Alice needs
to generate a signature that the first receiver accepts as message
that can be forwarded while the second receiver completely
rejects. In other words we need the errors of Bob to be below
$s_aL$ while those of Charlie to be above $s_vL$. We can see that,
similarly with the forging case, this probability decays
exponentially if $s_a<s_v$ with a rate $e^{-c'(s_a-s_v)^2L}$.

Finally, in any realistic setting, even an honest run would result
to certain errors due to the noise and imperfections in the
quantum communication and quantum memory. This could lead to
honest signatures being rejected, which again is undesirable.
(This is known as correctness, soundness or robustness in
different places in the literature.) We denote the fraction of
those honest errors as $p_e$ and once again we see that the
protocol does not reject honest signatures if $p_e<s_a$, so that
the probability of honest rejection decays as
$e^{-c''(p_e-s_a)^2L}$.

To summarize, we have $0<p_e<s_a<s_v<p_f$. The parameter $p_e$ is
determined by the system, noise, losses, etc, while $p_f$ is
theoretically computed as the best guess/attack. The two
parameters $s_a,s_v$ should be suitably chosen within the gap
$g=p_f-p_e$, and approximately in equal distances so that we have
minimum probability that something undesirable (forging,
repudiation, honest-reject) happens.

\subsubsection{Remarks}

Let us note that exact calculation of the above parameters for
GC-QDS was not done since there were many practical limitations to
actually implement such protocol. It served more as an inspiration
for later works.

Then note that this protocol involves multiple parties (at least
when considering transferability/non-repudiation) and any one of
them could be malicious. This is one of the most crucial
differences compared to QKD. The adversaries in QDS are legitimate
parties in the (honest) protocol (while Eve in QKD is an external
party). This means that even when we are guaranteed that all
quantum communications are done as the sender wishes, there are
still potential attacks. It is exactly this type of attacks that
we have so-far considered. Receivers using their legitimate
quantum public key to make a guess of the private key and forge;
or a sender sending different quantum public key to each receiver
in order to repudiate.

In~\cite{GC01} and in the first few works on QDS, to simplify the
security analysis, it was assumed that these are essentially the
only possible attacks. This formally was described as having an
authenticated quantum channel between the parties. However, to
actually have such a channel (or to have
quantum-message-authentication-codes~\cite{BC+02}), there is a
considerable cost. Subsequent works lifted this assumption.

Finally, while we have described QDS as a public-key cryptosystem,
strictly speaking this is not quite precise. The ``public key'' is
a quantum state, thus it cannot be copied or broadcasted as
classical keys. Therefore to properly ensure that the public key
is the same, point-to-point (quantum) communication is required,
while parties that have not participated in the $Gen$ phase,
cannot enter later (i.e., it lacks the ``universal verifiability''
of classical public key cryptosystems). Whether these issues are
crucial or not, it depends on the use/application that the digital
signatures are required (e.g., how important is the security) and
the efficiency of the QDS protocol after having taken into account
the above issues.

\subsubsection{Practical limitations of GC-QDS}

The GC-QDS scheme highlighted the possibility of a beyond-QKD
quantum cryptographic primitive, but it did not trigger a wide
research burst immediately. The reason that it took more than 10
years to have the wider research community following these steps
was, mainly, because the original protocol was highly impractical
to be actually implemented and used. The three major practical
restrictions were:

\begin{enumerate}
\item The quantum public key is a quantum state received during the $Gen$
phase and then later used again during the $Ver$ phase. However,
in normal practise, the acts of establishing the possibility of
digital signatures and actually signing and even later verifying
(or forwarding) a signature can be separated by long periods of
time (days or even months). Storing quantum information for even
seconds is hard and is one of the major restrictions in building
scalable quantum computers.

\item In order to test that receivers
obtained the same quantum public key, they need to communicate and
have multiple copies of the key, then test whether they are the
same using a comparing mechanism such as the SWAP test, send
copies to other parties and re-use the SWAP test between their
public key and the one received from other parties. This involves
extra quantum communication and, more importantly, using ancillae
and controlled SWAP gates on each qubit. These operations are
operations sufficient for a universal quantum computation and
adding the quantum memory requirement, it appears that all parties
should have a full universal quantum computer to participate in
GC-QDS. This is in sharp contrast with QKD, that requires minimal
quantum technologies, for example preparing and measuring single
qubits.

\item In the analysis of GC-QDS we have assumed that the
quantum states that parties want to send, arrive to the desired
party correctly, i.e., they have an authenticated quantum channel.
While quantum-message-authentication-codes do exist~\cite{BC+02},
they bring extra cost. Alternatively, one could modify a QDS
protocol to be secure even when there is no guarantee about the
quantum channel(s) used.
\end{enumerate}

\subsection{Practical QDS: Lifting the
limitations}\label{9.Sec:practical} Since the appearance of the
GC-QDS protocol there were four major developments, which we
outline here. These lifted all the aforementioned limitations
transforming QDS from a theoretical idea to a practical quantum
communication primitive, technologically as mature as QKD. Further
improvements (in the security proofs/guarantees, performance and
realizations) will be summarized in the subsequent sections.

\subsubsection{Simplifying state comparison} 
Andersson et al.~\cite{ACJ06} introduced a practical quantum
comparison for coherent states (here expressed in the photon
number basis) \EQ{\ket{\alpha}=e^{-\frac{|\alpha|^2}{2}}
\sum_{n=0}^{\infty}\frac{\alpha^n}{\sqrt{n!}}\ket{n}.} Provided
that the quantum public key of a QDS protocol consists of coherent
states, this practical test can replace the SWAP-test in both its
uses. First to ensure that the quantum public keys are identical
and thus Alice is not attempting to repudiate. Second, to check
the validity of a signature by checking that a quantum public key
matches its classical description (given when a signed message is
received).

In Ref.~\cite{ACJ06}, the quantum public key consists of two
states ($b=0$ for signing message $0$ and $b=1$ for message $1$)
whose form $\ket{\psi^b_{pk}}$ corresponds to a string of coherent
states $\ket{|\alpha|e^{i\theta}}$ with the same (known) amplitude
$|\alpha|$ and phase chosen randomly from the angles
$\theta\in\{2\pi\frac1N,2\pi\frac2N,\ldots,2\pi\frac{N-1}N\}$, for
suitable $N$. The classical description of this string (the choice
of phase for each coherent state in the string) is the private key
$sk$.

The main idea of the coherent-state quantum comparison is depicted
in Fig.~\ref{fig:9.2}, where we can see that the ``null-ports''
measure the phase difference between the two incoming coherent
states (and is the vacuum when they are identical). Note, that
this comparison is very simple technologically, since all that is
needed is beam splitters, mirrors and photon detectors. This is
why this set-up is a considerable advancement compared to the
SWAP-test used in GC-QDS.

\begin{figure}[pth]
\includegraphics[width=0.30\textwidth]{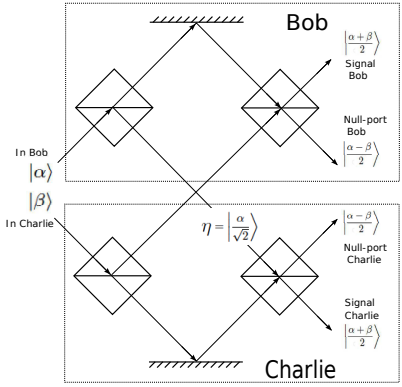}
\caption{Coherent-state quantum comparison introduced in
Ref.~\cite{ACJ06}. If $\ket{\alpha}=\ket{\beta}$ then only the
signal port of Bob and Charlie detects photons. Otherwise, the
null-port detects photons too. Reprinted figure with permission
from Ref.~\cite{DWA14} \copyright APS (2014).} \label{fig:9.2}
\end{figure}

Since this comparison should be performed for the quantum public
keys that receivers (Bob and Charlie) have, the multiport of
Fig.~\ref{fig:9.2} is placed between different locations (Bob's
and Charlie's labs). This has two consequences, one practical and
one theoretical. The practical is that being in a distance,
ensuring path lengths are identical is not trivial, while extra
losses occur, since not only the quantum public key needs to go
from Alice to Bob/Charlie, but then it needs to go through this
comparison process. The theoretical issue is that Alice, in
principle, could also tamper the quantum states communicated
between Bob and Charlie, something that complicates the (full)
security proof. Such proof was only completed for some modified
protocols much later.

To sign a message Alice sends the corresponding message $b$ along
with the string of phases corresponding to $\psi_{pk}^b$. Bob
(Charlie) to accept, reconstruct the state $\ket{\psi_{pk}^b}$ and
use state-comparison with their stored state. They count the
number of positions in the string that the null-port clicks and
compare the fraction with the thresholds ($s_a,s_v$) and decide to
accept as original (1-$ACC$), as forwarded (0-$ACC$) or to reject
the message.

\subsubsection{No quantum memory requirement}

As we have already stressed, in GC-QDS the public key is a quantum
state and one needs to store it until the $Ver$ phase, something
that makes impractical such scheme. Dunjko et al.~\cite{DWA14}
constructed a QDS protocol that does not require a quantum memory.
The central idea is to replace the quantum public key (which needs
to be stored coherently) with a classical ``verification key'',
which is no longer the same for all receivers.

The protocol builds on Refs.~\cite{ACJ06,CC+12} and starts with
Alice distributing a ``quantum public key'' being two strings of
coherent states, where for simplicity, the possible phases of each
state are two, i.e., the strings are of the type
$\ket{\alpha}\ket{-\alpha}\ket{\alpha}\cdots\ket{-\alpha}$ (if
Alice is honest). Bob and Charlie use the multiport scheme of
Fig.~\ref{fig:9.2} to ensure that Alice was (mostly) honest. Then
they directly measure each coherent state pulse, using unambiguous
state discrimination (USD)~\cite{USD1,USD2,USD3}. With this
measurement, Bob and Charlie know with certainty the classical
description of some positions in the string(s) of coherent states
that Alice sent, while they have no information about other
positions.

This information (position in string and value measured) are
stored by Bob and Charlie and will be their verification key.
(Note that the verification key of Bob and Charlie is different,
even in the honest case. This is because for which positions a USD
gives conclusive outcome is probabilistic and happens
independently for the copy that Bob and Charlie have.) When Alice
sends a signature, she needs to return the classical description
for {\it all} the string corresponding to the message $b$ she
wants to sign. Bob and Charlie count the fraction of mismatches
that the string that Alice has with their stored verification
key(s) $s_t$ and they reject, 0-$ACC$ or 1-$ACC$ comparing these
mismatches with the thresholds $s_v$ an $s_a$.

The important detail that makes such scheme secure, is that Alice
does {\it not} know for which positions Bob and Charlie know the
state and for which they do not. It is therefore impossible for
her to send a classical description that agrees with all the
possible verification keys unless she sends the honest private
key.

Note that, to achieve ITS encryption, one uses quantum
communication (QKD) to achieve a shared secret key and then uses
the fully classical one-time-pad protocol to encrypt a message.
Similarly, here, we use quantum communication to achieve
correlations between the classical information of the parties
involved in the signature scheme (their private/verification
keys). Then after establishing these correlations, a fully
classical signing and verifying algorithm follows and achieves the
digital signature functionality.

\subsubsection{QDS from QKD technology}

After removing the quantum memory requirement, the only remaining
difficulty making QDS harder than QKD is the mechanism to ensure
that the same quantum public key was sent to different receivers
(ensuring non-repudiation), whether this mechanism is the
SWAP-test or the much simpler coherent-state comparison (using a
spatially separated multiport).

A crucial observation is that both the SWAP-test and the
coherent-state comparison accept symmetric states. In other words,
if the states compared are in the global state
$\frac1{\sqrt{2}}(\ket{\psi}\ket{\phi}+\ket{\phi}\ket{\psi})$,
both tests would always accept. 
While this may appear as a problem, it turns out that Alice is
unable to repudiate by sending such symmetric states. In fact,
since the state is symmetric, Alice is unable to make Bob accept
(with the lower error threshold $s_a$) and in the same time make
Charlie rejects (with the higher error threshold $s_v$ used for
forwarded messages). It is the symmetry of the state and the gap
$g=(s_v-s_a)$ that ensures non-repudiation.

Starting from this observation, Wallden et al.~\cite{WDKA15}
replaced the comparison test with a ``symmetrizing'' step and
proposed three protocols. This extra step ensures that, even if
Alice did not distribute identical quantum public keys, the
classical verification keys that Bob and Charlie store will be
symmetric and thus Alice is unable to make them disagree. All
protocols given in~\cite{WDKA15} can be performed with BB84
states. Here we outline one of these protocols to demonstrate
these ideas.

Alice selects two strings of BB84 states (one for each future
message $b$), and she generates two copies of these strings and
send them to Bob and Charlie. For each qubit, Bob (Charlie) either
forwards it to Charlie (Bob) or keeps it and measures it in either
$\{\ket{0},\ket{1}\}$ or $\{\ket{+},\ket{-}\}$ basis. Similarly,
he measures the forwarded qubit that he received from Charlie
(Bob). Depending on the result, he rules-out one of the possible
states. For example if for the $n$th qubit he obtains the outcome
``$+$'', Bob stores that the $n$th qubit is {\it not} $\ket{-}$
(something he can know with certainty). Bob (Charlie) stores the
sequence of eliminated states, the position in the string, and
whether he received it directly from Alice or as forwarded from
Charlie (Bob). This classical information will be Bob's
(Charlie's) verification key.

As usual, Alice to sign sends the message and the classical
description of the corresponding string of qubits. Bob checks for
positions that the declaration of Alice contradicts his stored
verification key (i.e. places that Alice sends the state that Bob
has ruled-out). Then the fraction of this mismatches is compared
to the two thresholds $s_a,s_v$ and Bob rejects or 0-$ACC$ or
1-$ACC$.

\subsubsection{Insecure quantum channels}

One major assumption made so far was that the quantum channels
used were authenticated, i.e., the quantum states sent during the
quantum communication part of the protocol were the same as the
one received. While there are general (costly) methods to achieve
this, an intuition why this may not be necessary was already
given. One can imagine ``sacrificing'' part of the communicated
qubits to test (and bound) the tampering that third parties
performed. This is exactly what parameter estimation in QKD
achieves. In  Refs.~\cite{AWKA16,YFC16} the authors made this
intuition precise. As part of the interaction that leads to a
private key for Alice and (partial information) for Bob (Charlie),
they included a parameter estimation phase. As far as the
experiment is concerned, it is now a normal QKD protocol, that
stops before error correction and privacy amplification. The way
to bound forging and repudiation is very different and depends on
the specific protocol. 

In Ref.~\cite{AWKA16} there was another change. The quantum states
(quantum public key) that Alice sends to Bob and Charlie were no
longer the same. Since the only condition that secures Bob and
Charlie against repudiation is the symmetrization procedure,
whether Alice sends initially the same or different states makes
no essential difference. The only practical difference is that
Alice's private key is now composed from the classical description
of both the different strings sent to Bob and Charlie. On the
other hand, by sending different quantum states, Alice limits the
potential forging attacks, since forgers have no longer copies of
the full legitimate quantum public key.

Finally, a very interesting observation is that the (channel)
error rates for which QDS was possible in~\cite{AWKA16} were
higher than those for QKD. This means that by using this QDS
scheme, one may be able to perform QDS in a setting where QKD is
not experimentally feasible.

\subsection{A generic modern QDS protocol}\label{9.Sec:generic_QDS}

\subsubsection{Description}

We can now give a description of a generic modern QDS protocol,
i.e., one that does not require quantum memory, that can be
realized with the same technology as QKD and that makes no
assumption on the quantum channels used. The description below is
restricted to three parties, but can be generalized to more
parties.

\paragraph{Key generation.} We start with (any) QKD system as
basis. Alice performs the first part of a QKD protocol
(separately) twice with Bob and twice with Charlie. The QKD
protocol is completed up to obtaining the raw key (i.e., before
error correction and privacy amplification). As result, Alice has
four bit strings $A_0^B, A_1^B, A_0^C, A_1^C$, Bob has two strings
$K_0^B, K_1^B$ and Charlie has $K_0^C, K_1^C$. By the properties
of a (non-aborted) QKD protocol, the correlation between, say,
$A_0^B$ and $K_0^B$ is greater than the correlation of $A_0^B$ and
a string that any other party can produce. The private key $sk$
that Alice uses to sign a message in the future is the
concatenation of the two corresponding strings
$sk=(A_0^B||A_0^C,A_1^B||A_1^C)$. During this process, the error
rates of the channels are estimated, and values for $s_a,s_v$ are
chosen such that $0<p_e<s_a<s_v<p_f<1$, and $s_a,s_v$ are
``placed'' equally separated within the gap $p_f-p_e$. Here, $p_e$
is the estimated (``honest'') error rate between Alice-Bob using
their quantum channel, while $p_f$ is the minimum error rate that
Eve makes trying to guess Alice's string.

Bob and Charlie perform a symmetrization by exchanging secretly
half of their strings (e.g., using another full QKD link). The new
strings for Bob $S_0^B,S_1^B$ (and similarly for Charlie
$S_0^C,S_1^C$) are each composed from half of the string initially
sent to Bob and half of that to Charlie, but which part from Bob's
initial string and which part from Charlie's is unknown to Alice
(since the symmetrization was performed secretly). The
verification keys for Bob and Charlie are $pk_B=(S_0^B,S_1^B)$ and
$pk_C=(S_0^C,S_1^C)$. (Note that we no longer call them public
keys, being different for Bob and Charlie.)

\paragraph{Signing.} In order to sign a message $m$, Alice sends $(m,A_m^B||A_m^C)$ to Bob.

\paragraph{Verification.} To accept a message coming directly from
Alice, Bob checks the mismatches rate $p^B_t$ between the
signature received $A_m^B||A_m^C$ and his stored verification key
$S_m^B$, for each part of the signature separately (i.e.
mismatches in the part he obtained directly from Alice and
mismatches in the part he obtained from Charlie during the
symmetrization). If $p^B_t<s_a$ he accepts for both cases, i.e.
1-$ACC$. Charlie receives a message with Alice's signature, but
from Bob. He checks the mismatches rate $p^C_t$ similarly to Bob,
and if $p_t^C<s_v$ for both parts, he accepts the message as
coming originally from Alice.

\subsubsection{Security intuition and performance}

Forging is not possible because any potential forger, even a
legitimate party (e.g. Charlie), cannot guess the part of the
Alice's private key that was not directly send to him, at least
not with any probability significantly better than $p_f$. His
forging probability actually scales at best as
$e^{-c(p_f-s_a)^2L}$ which vanishes for sufficiently large length
of string $L$. Similarly, non-repudiation is guaranteed by the
fact that Alice is ignorant on which part of $K_m^B,K_m^C$ is in
$S_m^B$ and which is in $S_m^C$, she is therefore unable to make
Bob accept and Charlie to reject, and her probability of
succeeding in this scales as $e^{-c'(s_v-s_a)^2L}$. Finally, an
honest abort is unlikely, since we chose $s_a>p_e$ which leads to
the honest abort occurring with probability at most
$e^{-c''(p_e-s_a)^2L}$.

A QDS protocol performance is judged by the time taken to
distribute the verification key(s) among the parties, but also the
distance that the parties could be separated. (The signing
algorithm and verification algorithm are both assumed to be much
quicker and thus we judge the protocols, essentially, on the time
required for the key generation.) In most cases, we consider
single-bit message and assume linear scaling, however there may
exist more efficient ways to sign longer messages (e.g.
\cite{AAWA18} for a classical ITS scheme). The time taken to
distribute the verification key(s) depends on the clock-rates (how
many pulses are sent per second) and on how long strings $L$ are
required to achieve a desired level of security. In other words,
what $L$ and other choices should be made, so that the probability
of something going wrong (forging, repudiation, honest-abort) is
below $\epsilon$ -- the desired level of security.

To jointly minimize the probabilities of forging, repudiation and
honest-abort, we first need to determine the values $p_e,p_f$. The
estimated honest-error rate $p_e$ is obtained from the specific
channel/experimental set-up used, and can be thought as a
practical constraint. It is easy to see that $p_e$ increases with
the distance between parties, therefore there is a trade-off
between speed of distributing verification key(s) and distance. We
should keep this in mind when comparing different implementations.
The best forging error attempt $p_f$ is theoretically evaluated
for example by considering the minimum-cost quantum measurement
that adversaries can perform. Once these two are given, optimal
choices for $s_a,s_v$ are calculated to jointly minimize the
probabilities of forging, repudiation and honest abort. Typically
we require equal separation between the intervals
$(s_a-p_e),(s_v-s_a),(p_f-s_v)$, since they all appear in similar
form in the exponential decay of the expressions of honest-abort,
repudiation and forging, respectively.

\subsection{Extending QDS: Multiple parties, longer messages, and MDI}\label{9.Sec:theory_improvements}

The QDS schemes we presented considered the case of three parties,
the smallest number sufficient to illustrate the transferability
property. In that setting, only one party at a time can be an
adversary. In real practise however, multiple parties would be
involved as potential receivers.

A potentially-important disadvantage of QDS compared with
classical schemes is the way the communication required in the
$Gen$ phase scales as a function of the number of parties
involved. For most QDS protocols, a quadratic number of
communication channels is required. Moreover, when multiple
parties are involved, the issue of colluding parties (including
sender colluding with some receivers) should be considered, while
also the issue of multiple transfers of a signed message (and the
corresponding honest parties fraction thresholds) need to be
considered. In Ref.~\cite{AWA16} the general framework for
multiple-party QDS, certain generic properties, and the concept of
multiple levels of transferability (and verification) were
introduced, along with a multi-party generalization of one
protocol. In Ref.~\cite{SY16} the three-party protocol of
\cite{AWKA16} was also extended to multiple parties.

Most of the research on QDS is focused on signing single-bit
messages, and it is usually stated that a simple iteration can be
used for longer messages. While this is mostly true, there are two
issues that require attention. First, as analyzed in
Ref~\cite{WC+15}, there are attacks on longer messages impossible
to be addressed from single-bit signatures, e.g. tampering with
the order of the bits. The second issue is that of efficiency. In
classical schemes, using hash functions one can reduce the extra
cost from being linear in the size of the message (as in simple
iterations) to being logarithmic~\cite{AAWA18}. It is worth
exploring QDS schemes that could improve the scaling with the
message size.

As with QKD, many QDS protocols are vulnerable to side-channel
attacks~\cite{HB+18}, with the best known side-channel attacks
exploiting measurement-device/detector vulnerabilities (e.g. the
``blinding attack''). For this reason, MDI protocols for QDS where
first introduced in Ref.~\cite{PA+16}. The analysis follow closely
that of QKD and of Ref.~\cite{AWKA16} and we omit further details.
One interesting thing to note is that the extra security guarantee
(against some side-channel attacks) comes at no (or low) cost in
terms of practicality, unlike the fully device-independent
protocols. Moreover, the MDI setting that contains untrusted
mediating parties is suitable for QDS (where there are multiple
parties and each party can be adversarial). It allows us to
consider optimization of routing of quantum information in quantum
networks, i.e. consider different (or even flexible) connectivity
of parties to optimize the multiparty versions of QDS schemes.

\subsection{Experimental QDS realizations}\label{9.Sec:experiments}

Since 2012, a number of experiments implementing QDS protocols has
been performed, and from the first proof-of-principle experiments
we now have fully secure, long-distance QDS implementations on
existing quantum networks, suitable for real life applications. As
mentioned in Section \ref{9.Sec:generic_QDS} and similarly to QKD,
there is a trade-off between the distance that parties can be
separated and the speed that verification keys for fixed length
messages are distributed. The distances mentioned below are the
maximum distances that QDS could run, while it is understood that
for smaller distances the ``rate''/performance would improve.

\subsubsection{Proof-of-principle} The first QDS experiment by
Clarke et al.~\cite{CC+12} was based on the QDS protocol outlined
in Ref.~\cite{ACJ06}, where the coherent-state comparison (see
Fig.~\ref{fig:9.2}) was introduced in order to replace the
SWAP-test~\cite{GC01}. The simplest case of three-parties was
implemented, where each coherent-state pulse
$\ket{\alpha}=\ket{|\alpha|\exp (2\pi i \phi)}$ had its phase
randomly chosen from eight possible choices ($\phi=k/8$ for
$k\in\{0,7\}$). Different mean-photon numbers $|\alpha|^2$ were
examined. As explained in the end of Section
\ref{9.Sec:generic_QDS}, one needs to jointly minimize the
forging, repudiation and honest-abort probabilities. Too high
$|\alpha|$ makes $p_f$ small (and forging simple since the states
approach classical states and can be copied), while too low
$|\alpha|$ makes $p_e$ large (dark counts are a larger fraction of
detections, making honest-abort more likely) so an optimal value
for $|\alpha|$ should be sought.

The experiment was meant to be a proof of principle. Firstly, the
parties were all located within small distance (same lab).
Secondly, the signing and verifying happened immediately (no
quantum state stored). In particular, instead of Bob regenerating
the quantum state of the signature from Alice's signature
(classical description), and then compare it with his stored
states, Bob obtained directly the sequence of the qubits from
Alice (that used a beam-splitter before sending the quantum public
key) and compared it with the hypothetically stored quantum public
key.

A second QDS experiment was performed by Collins et
al.~\cite{CD+14}, based on the QDS protocol of Ref.~\cite{DWA14}
that does not require quantum memory. Because of this property,
the only unrealistic assumption was the separation of the parties
(still within the same lab), while the signing and verifying
happened in arbitrary later time. The protocol used a
generalization of the unambiguous discrimination measurement,
namely unambiguous elimination measurement. Again it involved
three parties, sharing strings of phase-encoded coherent states,
where this time the possible phases were four $N=4$.

\subsubsection{Kilometer-range and fully-secure QDS}

Subsequently, based on the idea that one can replace the state
comparison with symmetrization~\cite{WDKA15}, two
experiments~\cite{DC+16,CP+16} were performed that had parties
able to be separated by a distance of the order of kilometer.
Callum et al.~\cite{CP+16} was also the first QDS protocol that
used continuous variables (heterodyne detection measurements) and
the first experiment to be performed through a free-space noisy
1.6 km channel (in Erlangen).

Following~\cite{AWKA16,YFC16}, the last unrealistic assumption was
removed, i.e., authenticated quantum channels. The use of decoy
states, and other theoretical but also technical improvements,
resulted in protocols with far superior performance, having the
parties separated by tens to hundred
kilometers~\cite{CA+16,CA+17}. This brings QDS in par with QKD in
terms of practicality. Finally, MDI-QDS protocols, addressing
measurement-device side-channel attacks, have been implemented
over a metropolitan network~\cite{YW+17} and at high rates by
using a laser seeding technique together with a novel treatment of
the finite-size effects~\cite{RL+17}.

\subsection{Classical unconditional secure
signatures}\label{9.Sec:classical_uss}

The type of digital signatures that are achieved by QDS offer ITS,
but are one-time (cannot be reused) and require a fixed number of
parties all participating during the key generation phase. Only
those parties can sign and verify in the future messages and, if
one wanted to extend the participating parties, new interactions
would be required between (many) parties. In contrast, classical
public-key signatures can be verified by anyone with access to the
public key (that can be obtained later than the Key Generation
phase).

This specific type of signatures that QDS achieve, was actually
first considered by Chaum and Roijakkers~\cite{CR90} and were
termed unconditionally secure digital signatures (USS). In order
to achieve USS, all parties needed to share (long) secret key
pairwise, while another assumption was also necessary (an
authenticated broadcast channel or anonymous channels). Only a few
papers followed this work~\cite{Hanaoka, Shikata, Hanaoka04,
Swanson11}. The main reason for the limited interest was probably
because such protocols were seen as impractical, specifically
because they require point-to-point shared secret keys. Then, the
extra security offered (information theoretic) was not viewed as
necessary. Both of these issues have been revisited with the
recent advances in quantum technologies since: (i) sharing long
secret keys can be achieved with QKD, and (ii) advances in quantum
computers make realistic the prospect of large scale quantum
computers that could break existing cryptosystems in the medium
term. Therefore, it appears likely that interest for this type of
protocols could increase.

In a parallel direction, inspired by QDS, a classical USS protocol
was proposed in Ref.~\cite{WDKA15}, where only pairwise secret
keys were required. This scheme was generalized to multiple
parties in Ref.~\cite{AWA16}. Subsequently, in Ref.~\cite{AAWA18},
a USS protocol that scales much better for longer messages was
obtained using universal hashing (it requires key-sizes that scale
logarithmically with the message length). All these protocols
require only point-to-point secret keys and neither authenticated
broadcast channel nor anonymous channels or trusted third parties
were assumed. Because no further assumptions were made, these
protocols prove a ``reduction'' of the task of USS to that of
point-to-point secret keys and thus to standard QKD.

\subsection{Summary and outlook}\label{9.Sec:conclusion}

QDS is a type of digital signatures that offers information
theoretic security, a very attractive feature, that the progress
in building quantum computers has made even more timely. The
``trade-off'' is that this type of digital signatures that QDS
achieve is missing some of the elements that made digital
signatures such an important functionality (e.g. the universal
verifiability).

In this review we described the research that transformed QDS from
a theoretical interesting observation to a practical possibility.
In Sections \ref{9.Sec:GC-QDS} and \ref{9.Sec:practical} we
presented the developments and choices made in a historical order,
while in Section \ref{9.Sec:generic_QDS} we gave a description and
brief analysis of a generic modern QDS protocol. Latest
state-of-the-art developments were subsequently mentioned briefly,
referring the reader to the original works for further details.

Possibly the biggest challenge for QDS is how do they compare with
the classical digital signature schemes that offer ITS. In Section
\ref{9.Sec:classical_uss} we presented those classical protocols
and noted that all of them require point-to-point (long) secret
keys between the participants. Indeed, it appears that the
classical scheme given in Ref.~\cite{AAWA18} (inspired by QDS)
offers similar guarantees and cost with QDS while being more
efficient for long messages, i.e., exponentially better with
respect to the size of the message signed.

However, there are at least three directions (and reasons) that
further research in QDS is still very promising.
\begin{enumerate}
\item Firstly, it is likely that a QDS protocol with better scaling
for long messages can be developed. So far, the majority of
research in QDS focused on the single-bit message case and the
possibility of better scaling for longer messages has not been
sufficiently examined.

\item Secondly, classical protocols require communication between
{\it all} parties, i.e., quadratic in the number of participants
and number of communication channels. In contrast, with QDS it is
possible to achieve linear scaling with respect to the quantum
channels. The QDS scheme given in Ref.~\cite{YFC16} is an example
that offers such feature. This particular protocol would not scale
so well with more parties for different reasons (sensitive in
forging probability), but it demonstrates the possibility of using
quantum resources to reduce the communication channels.

\item Thirdly, in Ref.~\cite{AWKA16} a QDS protocol was given that could
be secure even when the noise in the channels was too high for
QKD, again demonstrating the possibility that fundamentally
quantum protocols are possible when the ``classical'' ones (that
in any case require QKD) are impossible.

\end{enumerate}

Finally, the so-called ``classical'' ITS protocols, such as the
one given in Ref.~\cite{AAWA18}, require point-to-point secret
keys, and those keys can only be practically achieved using QKD.
In this sense we can view even these schemes as {\it quantum}
digital signature schemes. While the theory of classical ITS
protocols involves little or no new quantum research, their
development makes stronger the case for building a quantum
communication infrastructure and thus increases the impact of
quantum cryptography by offering further functionalities.

In particular, digital signatures are useful when they involve
many (potential) parties, which means that QDS could be useful for
real applications only if the corresponding infrastructure is in
place, i.e., a large quantum network. While this infrastructure is
not currently available, the possibility of QDS (including the USS
protocols given above) offers greater value to quantum networks
and thus makes the argument for developing such infrastructure
more compelling.


\section{Conclusions}

In this review we have presented basic notions and recent advances
in the field of quantum cryptography. We have focused most of the
discussion on QKD, but also presented some developments which goes
beyond the standard setting of key distribution. As a matter of
fact, quantum cryptography is today a big umbrella name which
includes various areas, some of which have not been treated in
this review. For instance, other topics of interest are
cryptographic primitives such as oblivious transfer and bit
commitment, or topics of secure computing, including blind,
verifiable quantum computing, and secure function evaluation.
Post-quantum cryptography (e.g., lattice-based cryptography) is an
interesting area that may offer a temporary solution to security.
Then, a number of protocols have not been treated and worth
mentioning such as quantum fingerprinting, quantum secret sharing,
quantum Byzantine agreement, and quantum e-voting.

While quantum cryptography is certainly the most mature quantum
technology so far, a number of challenges and open questions are
facing both theoretical and experimental work. There is still the
need to develop and implement more robust QKD protocols, which are
able to achieve long distances at reasonably high rates. This
seems to be matter of developing a QKD network based on practical
quantum repeaters. An even more ideal task would be to realize the
end-to-end principle in such a QKD network, so that the middle
nodes may generally be unreliable and untrusted. Today, this idea
seems only to be connected to the exploitation of EPR-like
correlations, either in a direct way (i.e., assuming that middle
nodes produce maximally-entangled states) or in a reverse fashion
(i.e., assuming that the nodes apply projections onto
maximally-entangled states, i.e., Bell detections).

Theoretically, there are efforts directed at establishing the
fully-composable finite-size security of a number of QKD
protocols, both in DV and CV settings. It is then an open question
to determine the secret key capacity of several fundamental
quantum channels, such as the thermal-loss channel and the
amplitude damping channel. While the recently-developed simulation
techniques have been successful in many cases, the two-way
assisted capacities of these channels may need the development of
a completely new and different approach.

Experimentally, the current efforts are going towards many
directions, from photonic integrated circuits to satellite quantum
communications, from more robust point-to-point protocols to
implementations in trusted-node quantum networks, from qubit-based
approaches to higher dimensions and continuous variable systems.
While optical and telecom frequencies are by far the more natural
for quantum communications, longer wavelengths such as THz and
microwaves may have non-trivial short-range applications which are
currently under-developed.

A number of loopholes need to be carefully considered before QKD
can be considered to have become an fully-secure quantum
technology. Practical threats are coming from side-channel
attacks, for which countermeasures are currently being studied and
developed for some of the most dangerous quantum hacks. Weakness
may come from things like imperfections in detectors or the random
number generators. Quantum hacking and countermeasures is
therefore an important and growing area.

In general, for a technological deployment of quantum cryptography
and QKD, we will need to consider its integration with the current
classical infrastructure and develop layers of security, depending
on the degree of confidentiality to be reached which, in turn,
depends on the stakeholder and the type of business involved.
Protocols based on bounded-memories and quantum data locking
provide a temporary low-level of quantum security that may be
suitable for private personal communications. Standard QKD
protocols provide higher levels of security that may be suitable
for financial transactions. Within QKD, different secret key rates
might be considered, for instance, with respect to individual,
collective or fully-coherent attacks. The choice of these rates
may also be associated with a specific sub-level of security to be
reached. Higher level of security, for applications such as
political or strategic decisions, may involve the use of DI-QKD,
which is more robust to both conventional and side-channel
attacks. These aspects will become clearer and clearer as quantum
cryptography will progressively become a wider technological
product.

\bigskip


\begin{acknowledgments}
This work has been supported by the EPSRC via the `UK Quantum
Communications Hub' (EP/M013472/1) and the First Grant
EP/P016588/1; the European Union via the project `Continuous
Variable Quantum Communications' (CiViQ, no 820466), the project
UNIQORN (no 820474), and the H2020 Marie Sklodowska Curie Project
675662 (QCALL); the Danish National Research Foundation bigQ
(DNRF142); the Czech Science Foundation (project No 19-23739S);
the Czech Ministry of Education via the INTER-COST grant No
LTC17086; the Air Force Office of Scientific Research program
FA9550-16-1-0391 (supervised by Gernot Pomrenke) and the Office of
Naval Research CONQUEST program. Authors are indebted to Marco
Lucamarini for his help, comments and suggestions, and for
providing the file for the rate of the TF-QKD protocol, plotted in
Figs.~3 and~11. A special thank also goes to Xiongfeng Ma, Pei
Zeng, Xiang-Bing Wang, Zhen-Qiang Yin, and Federico Grasselli for
having provided files and data points for the rates plotted in
Fig.~11. Authors also thank Wolfgang D\"{u}r, Zeng-Bing Chen,
Hua-Lei Yin, and Yichen Zhang for feedback.
\end{acknowledgments}

\appendix


\section{Formulas for Gaussian states}\label{sec:GaussAPPENDIX}


Consider $n$ bosonic modes described by the creation and
annihilation operators $\hat{a}_{j}^{\dagger}$, $\hat{a}_{j}$ with
$j=1,\dots,n$ and define the quadrature operators as
\begin{align} \hat{q}_{j}=(\hat{a}_{j}+\hat{a}_{j}^{\dagger
})/\sqrt{2\kappa},~~\hat{p}_{j}=-i(\hat{a}_{j}-\hat{a}_{j}^{\dagger
})/\sqrt{2\kappa},
\end{align}
where the factor $\kappa$ is introduced to consistently describe
different notations used in the literature (see also
Ref.~\cite{FerraroNotes}). The canonical choice is $\kappa=1$
(vacuum noise equal to $1/2$) where we recover the canonical
commutation relations $[\hat{q}_{k},\hat{p}_{j} ]=i\delta_{kj}$,
while a popular alternative in quantum information is $\kappa=1/2$
(vacuum noise equal to $1$). For any general $\kappa$, the
quadrature operator can be grouped into a vector
$\mathbf{\hat{x}}$ with $2n$ components that satisfies the
following commutation relation
\begin{equation}
\lbrack\mathbf{\hat{x}},\mathbf{\hat{x}}^{T}]=\frac{i\Omega}{\kappa}.~
\label{commrel}%
\end{equation}

The coordinate transformations $\mathbf{\hat{x}}^{\prime}=$\textbf{$S$%
}$\mathbf{\hat{x}}$ that preserve the above commutation relations
form the
symplectic group, i.e. the group of real matrices such that \textbf{$S$%
}$\Omega$\textbf{$S$}$^{T}=\Omega$. There are essentially two
standard ways of grouping the quadrature operators, and the
definition of $\Omega$ changes accordingly. These are
\begin{equation}
\mathbf{\hat{x}}:=(\hat{q}_{1},\dots,\hat{q}_{n},\hat{p}_{1},\dots,\hat{p}%
_{n})^{T},\Omega:=%
\begin{pmatrix}
0 & 1\\
-1 & 0
\end{pmatrix}
\otimes\openone~, \label{group11}%
\end{equation}
where $\openone$ is the $n\times n$ identity matrix, or
\begin{equation}
\mathbf{\hat{x}}:=(\hat{q}_{1},\hat{p}_{1},\dots,\hat{q}_{n},\hat{p}_{n}%
)^{T},\Omega:=\bigoplus_{j=1}^{n}%
\begin{pmatrix}
0 & 1\\
-1 & 0
\end{pmatrix}
~. \label{group22}%
\end{equation}
All the formulae that we review here are independent of this
choice, provided that the grouping $\mathbf{\hat{x}}$ and matrix
$\Omega$ are chosen consistently.

Any multimode bosonic state $\rho$ can be described using
phase-space methods by means of the Wigner characteristic function
$\chi(\boldsymbol{\xi })=\mathrm{Tr}[\rho
e^{i\mathbf{\hat{x}}^{T}\Omega\boldsymbol{\xi}}]$. The state
$\rho$ is called Gaussian when $\chi(\boldsymbol{\xi})$ is
Gaussian~\cite{RMPwee}. For a Gaussian state, the density operator
${\rho}$ has a one-to-one correspondence with the first- and
second-order statistical moments of the state. These are the mean
value $\mathbf{\bar{x}}:=\langle
\mathbf{\hat{x}}\rangle_{{\rho}}=\mathrm{Tr}(\mathbf{\hat{x}}{\rho}%
)\in\mathbb{R}^{2n}$ and the covariance matrix (CM) \textbf{$V$},
with generic element
\begin{equation}
V_{kl}=\frac{1}{2}\langle\{\hat{x}_{k}-\bar{x}_{k},\hat{x}_{l}-\bar{x}%
_{l}\}\rangle_{{\rho}}~,
\end{equation}
where $\{,\}$ is the anticommutator.

According to Williamson's theorem, there exists a symplectic
matrix $\mathbf{S}$ such that~\cite{RMPwee}
\begin{equation}
\mathbf{V}=\mathbf{S}(\mathbf{D}\odot\mathbf{D})\mathbf{S}^{T},~~\mathbf{D}%
=\mathrm{diag}(v_{1},\dots,v_{n}), \label{WillDec}%
\end{equation}
where the $v_{j}$'s are called symplectic eigenvalues and satisfy $v_{j}%
\geq(2\kappa)^{-1}$. When $v_{j}=(2\kappa)^{-1}$ for all $j$ the
state is pure. With the canonical choice $\kappa=1$ this means
that a pure state is defined by $v_{j}=1/2$ for all $j$, while
with the choice $\kappa=1/2$ a pure state has $v_{j}=1$ for all
$j$. The dot operator $\odot$ in Eq.~(\ref{WillDec}) has been
introduced to make the notation uniform depending on the different
grouping rules of Eqs.~(\ref{group11}) and~(\ref{group22}). When
Eq.~(\ref{group11}) is employed the dot operator is defined as
$\mathbf{D}\odot\mathbf{D}:=\mathbf{D}\oplus\mathbf{D}=(v_{1},\dots
,v_{n},v_{1},\dots,v_{n})$, while when Eq.~(\ref{group22}) is
employed the dot operator is defined as
$\mathbf{D}\odot\mathbf{D}:=(v_{1},v_{1},\dots ,v_{n},v_{n})$.

Although the Wigner function formalism is a popular approach for
describing Gaussian quantum states~\cite{RMPwee}, quantities
normally appearing in quantum information theory can often be
computed more straightforwardly using an algebraic
approach~\cite{Banchi}. Any multi-mode Gaussian state $\rho
(\mathbf{V},\mathbf{\bar{x}})$ parameterized by the first- and
second-moments $\mathbf{\bar{x}}$ and $\mathbf{V}$ can be written
in the operator exponential form~\cite{Banchi} (see
also~\cite{Sorkin,Doklady})
\begin{equation}
\rho(\mathbf{V},\mathbf{\bar{x}})=\exp\left[
-\frac{\kappa}{2}(\mathbf{\hat
{x}}-\mathbf{\bar{x}})^{T}\mathbf{G}(\mathbf{\hat{x}}-\mathbf{\bar{x}%
})\right]  /Z_{\rho}~, \label{Gexpression}%
\end{equation}
where
\begin{equation}
Z_{\rho}=\det\left(  \kappa\mathbf{V}+\frac{i\Omega}{2}\right)
^{1/2},
\label{PartitionFunction}%
\end{equation}
and the \textit{Gibbs matrix} $\mathbf{G}$ is related to the CM
$\mathbf{V}$ by
\begin{equation}
\mathbf{G}=2i\Omega\,\coth^{-1}(2\kappa\mathbf{V}i\Omega),~~\mathbf{V}%
=\frac{1}{2\kappa}\coth\left(  \frac{i\Omega\mathbf{G}}{2}\right)
i\Omega.
\label{e.GtoV}%
\end{equation}
The above relations are basis independent and allow the direct
calculation of $\mathbf{G}$ from $\mathbf{V}$ without the need of
the symplectic diagonalization~\eqref{WillDec}. This is a
consequence of the \textquotedblleft symplectic
action\textquotedblright\ formalism that is discussed in the next
section. From the operator exponential form, we then show how to
compute quantities like the fidelity between Gaussian states, the
von Neumann entropy, the quantum relative entropy and its
variance.

\subsection{Symplectic action and its computation}

Given a function $f:\mathbb{R}\rightarrow\mathbb{R}$ we can extend
$f$ to map Hermitian operators to Hermitian operators in the
following way: let $M=UxU^{\dagger}$ be the spectral decomposition
of a Hermitian operator $M$, then $f(M):=Uf(x)U^{\dagger}$ where
$f(x)$ is a vector whose $j$-th element is $f(x_{j})$. For more
general operators $M$ that admit a decomposition $M=UxU^{-1}$,
with a possibly non-unitary $U$, we define an operator function as
$f(M) := Uf(x)U^{-1}$.

The symplectic action was introduced in
Ref.~\cite{spedalieri_limit_2013} to extend a function $f$ to any
operator with symplectic structure. More precisely, for a given
matrix $\mathbf{V}$ with symplectic diagonalization as in
Eq.~\eqref{WillDec}, the symplectic action $f_{\ast}$ on
$\mathbf{V}$ is defined by
\begin{equation}
f_{\ast}(\mathbf{V})=\mathbf{S}[f(\mathbf{D})\odot f(\mathbf{D})]\mathbf{S}%
^{T}, \label{SA}%
\end{equation}
where
$f(\mathbf{D})=\mathrm{diag}[f(v_{1}),f(v_{2}),\dots,f(v_{n})]$
acts as a standard matrix function. In Ref.~\cite{Banchi} it was
proven that, for any odd function $f(-x)=-f(x)$, the symplectic
action can be explicitly written as
\begin{equation}
f_{\ast}(\mathbf{V})=f(\mathbf{V}i\Omega)i\Omega. \label{simpleSA}%
\end{equation}
where $f(\mathbf{V}i\Omega)$ is a matrix function.

Matrix functions are part of most numerical libraries and symbolic
computer algebra systems, so their computation, either numerical
or analytical, can be easily done on a computer. This is an
advantage especially for symbolic calculations~\cite{Banchi,PLOB}.
On the other hand, for a full symplectic diagonalization, the
practical problem is not the computation of the
symplectic spectrum but the derivation of the symplectic matrix {$\mathbf{S}$%
}\ performing the diagonalization {$\mathbf{SVS}$}$^{T}$ into the
Williamson's form~\cite{RMPwee}. For this matrix {$\mathbf{S}$},
we know closed formulas only for specific types of two-mode
Gaussian states~\cite{QCB3} which appear in problems of quantum
sensing~\cite{SensingREV}, such as quantum
illumination~\cite{QIll2} and quantum reading~\cite{QR}.

The Gibbs exponential form \eqref{Gexpression} can be proven by
first noting that a single mode thermal state with diagonal CM
$\mathbf{V}=v\odot v$ can be written as an operator exponential
\begin{equation}
\rho=e^{-\frac{g}{2}\kappa(\hat{q}^{2}+\hat{p}^{2})}/Z_{\rho}~,
\end{equation}
with $g=2\coth^{-1}(2\kappa v)$ and extending the result to
multi-mode, possibly non-thermal states, via the symplectic
action.
Indeed, since $\rho\propto e^{-g\hat{a}^{\dagger}\hat{a}}$, we may
write
\begin{equation}
v:=\langle\hat{q}{^{2}}\rangle=\langle{\hat{p}^{2}}\rangle=\frac{\langle
{\hat{a}^{\dagger}\hat{a}}\rangle+1/2}{\kappa}=\frac{1}{2\kappa}\coth\frac
{g}{2}. \label{invFUN}%
\end{equation}
Following the same construction of Ref.~\cite{Banchi}, which was
done for $\kappa=1$, we get the final result of
Eq.~\eqref{Gexpression}.

\subsection{Fidelity between arbitrary Gaussian states}

The fidelity $F(\rho_{1},\rho_{2})$ quantifies the degree of
similarity between two quantum states $\rho_{1}$ and $\rho_{2}$.
It is a central tool is many areas of quantum information,
especially for quantum state discrimination which is a fundamental
process in any decoding process. For pure states, it is defined as
$F=|\langle{\psi_{1}}|\psi_{2}\rangle|^{2}$, while for mixed
states
it may be defined in terms of the trace norm $||O||:=\mathrm{Tr}%
|O|=\mathrm{Tr}\sqrt{O^{\dagger}O}$ as~\cite{Uhlmann}
\begin{equation}
F:=||\sqrt{\rho_{1}}\sqrt{\rho_{2}}||=\mathrm{Tr}\sqrt{\sqrt{\rho_{1}}\rho
_{2}\sqrt{\rho_{1}}}~. \label{e:fidelity}%
\end{equation}
A general closed-form for the fidelity between two arbitrary
multi-mode Gaussian states was derived in Ref.~\cite{Banchi}, thus
generalizing partial results known for single-mode
states~\cite{fidG2,fidG3,fidG4}, two-mode
states~\cite{marian_uhlmann_2012},
pure~\cite{spedalieri_limit_2013} or thermal
states~\cite{scutaru}.

Given two arbitrary multi-mode states with CMs $\mathbf{V}_{i}$
and first moments $\mathbf{\bar{x}}_{i}$, the fidelity $F$ can be
written as~\cite{Banchi}
\begin{equation}
F({\rho}_{1},{\rho}_{2})=\frac{F_{\mathrm{tot}}}{\sqrt[4]{\det\left[
\kappa(\mathbf{V}_{1}+\mathbf{V}_{2})\right]
}}e^{-\frac{1}{4}\delta
^{T}(\mathbf{V}_{1}+\mathbf{V}_{2})^{-1}\delta}, \label{genFIDmain}%
\end{equation}
where $\delta:=\mathbf{\bar{x}}_{2}-\mathbf{\bar{x}}_{1}$, while
the term $F_{\mathrm{tot}}$ only depends on $\mathbf{V}_{1}$ and
$\mathbf{V}_{2}$ and is easily computable from the auxiliary
matrix
\begin{equation}
\mathbf{V}_{\mathrm{aux}}=\Omega^{T}(\mathbf{V}_{1}+\mathbf{V}_{2}%
)^{-1}\left(  \frac{\Omega}{4\kappa^{2}}+\mathbf{V}_{2}\Omega\mathbf{V}%
_{1}\right)  ,
\end{equation}
as
\begin{equation}
F_{\mathrm{tot}}^{4}=\det\left[  2\kappa\left(
\sqrt{\openone+\frac
{(\mathbf{V}_{\mathrm{aux}}\Omega)^{-2}}{4\kappa^{2}}}+\openone\right)
\mathbf{V}_{\mathrm{aux}}\right]  ~. \label{Ftot0}%
\end{equation}
The general solution \eqref{genFIDmain} has been derived thanks to
the operator exponential form \eqref{Gexpression} that makes
straightforward the calculation of the operator square roots in
the fidelity \eqref{e:fidelity}. Indeed, using the Gibbs matrices
$\mathbf{G}_{i}$ of the two Gaussian states, it was found in
\cite{Banchi} that
\begin{equation}
F_{\mathrm{tot}}=\det\left(  \frac{e^{i\Omega\mathbf{G}_{\mathrm{tot}}%
/2}+\openone}{e^{i\Omega\mathbf{G}_{\mathrm{tot}}/2}-\openone}i\Omega\right)
^{\frac{1}{4}}~, \label{Ftot}%
\end{equation}
where
\begin{equation}
e^{i\Omega\mathbf{G}_{\mathrm{tot}}}=e^{i\Omega\mathbf{G}_{1}/2}%
e^{i\Omega\mathbf{G}_{2}}e^{i\Omega\mathbf{G}_{1}/2}~.
\end{equation}
The final form \eqref{genFIDmain} is then obtained by expressing
the above
matrix functions in terms of CMs. Note that the asymmetry of $\mathbf{V}%
_{\mathrm{aux}}$ upon exchanging the two states is only apparent
and comes from the apparent asymmetry in the definition of
Eq.~\eqref{e:fidelity}. One can check that the eigenvalues of
$\mathbf{V}_{\mathrm{{aux}}}\Omega$, and thus the determinant in
Eqs.~\eqref{Ftot0}, are invariant under such exchange.

As already mentioned, an efficient computation of the quantum
fidelity is crucial for solving problems of quantum state
discrimination~\cite{SensingREV,Chefles,QHT2}, where two
multi-mode Gaussian states must be optimally distinguished.
Consider $N$ copies of two multimode Gaussian states,
${\rho}_{1}^{\otimes N}$ and ${\rho}_{2}^{\otimes N}$, with
the same a priori probability. The minimum error probability $p_{\text{err}%
}(N)$ in their statistical discrimination is provided by the
Helstrom bound~\cite{Helstrom}, for which there is no closed form
for Gaussian states. Nonetheless, we may write a fidelity-based
bound~\cite{Fuchs,Banchi} as
\begin{equation}
\frac{1-\sqrt{1-\left[  F({\rho}_{1},{\rho}_{2})\right]
^{2N}}}{2}\leq p_{\text{err}}(N)\leq\frac{\left[
F({\rho}_{1},{\rho}_{2})\right]  ^{N}}{2}~.
\label{errPROB}%
\end{equation}

The fidelity can be expressed~\cite{FuchsCaves} as a minimization
over POVMs $E_{x}$ of the overlap between two classical
probability distributions $p_{i}=\mathrm{Tr}[\rho_{i}E_{x}]$
\begin{equation}
F(\rho_{1},\rho_{2})=\min_{\{E_{x}\}}\sqrt{\mathrm{Tr}[\rho_{1}E_{x}%
]\mathrm{Tr}[\rho_{2}E_{x}]}~.
\end{equation}
Calling $\tilde{E}_{x}$ the optimal POVM that achieves the minimum
of the above quantity, we see that the fidelity can be measured
with a single POVM without state tomography. As such we may write
\begin{equation}
p_{\text{err}}(N)\leq\frac{1}{2}\left(
\sqrt{\mathrm{Tr}[\rho_{1}\tilde
{E}_{x}]\mathrm{Tr}[\rho_{2}\tilde{E}_{x}]}\right)  ^{N}~.
\end{equation}
where $\tilde{E}_{x}$ is optimal for the bound, in the sense that
any other POVM provides a larger upper bound.
Recently~\cite{Changhun}, it has been shown that such optimal POVM
can be explicitly computed between any two multi-mode Gaussian
states. Indeed, it was found that the optimal POVM is formed by
the eigenbasis of the operator~\cite{Changhun}
\begin{equation}
\hat{M}\propto\hat{D}(\mathbf{\bar{x}}_{1})\exp\left[  -\frac{\kappa}%
{2}{\mathbf{\hat{x}}}^{\text{T}}\mathbf{G}_{M}{\mathbf{\hat{x}}}%
-v_{M}^{\text{T}}\mathbf{\hat{x}}\right]  \hat{D}^{\dagger}(\mathbf{\bar{x}%
}_{1}),
\end{equation}
where $\hat{D}$ is a displacement and $v_{M}=0$ when $\mathbf{\bar{x}}%
_{1}=\mathbf{\bar{x}}_{2}$, while the general case is provided in
\cite{Changhun}. On the other hand, the matrix $\mathbf{G}_{M}$ is
given by
\begin{equation}
e^{i\Omega\mathbf{G}_{M}}=e^{-i\Omega\mathbf{G}_{1}/2}\sqrt{e^{i\Omega
\mathbf{G}_{1}/2}e^{i\Omega\mathbf{G}_{2}}e^{i\Omega\mathbf{G}_{1}/2}%
}e^{-i\Omega\mathbf{G}_{1}/2}. \label{GMsol}%
\end{equation}
Based of this general multi-mode solution, it was found
in~\cite{Changhun} that, for single-mode Gaussian states, there
are only three possible kinds of optimal measurements, depending
on $\rho_{1}$ and $\rho_{2}$: number-resolving detection,
quadrature detection, or a projection onto the eigenbasis of
operator $\hat{q}\hat{p}+\hat{p}\hat{q}$.

\subsection{Entropic quantities}

Entropic quantities are widespread in quantum information theory,
and are employed to bound the performances of QKD protocols,
entanglement sharing and data compression, to name a few examples.
Here we provide some simple formula for the von Neumann entropy of
a Gaussian state and for the relative entropy
between two arbitrary Gaussian states $\rho_{1}(\mathbf{\bar{x}}%
_{1},\mathbf{V}_{1})$ and
$\rho_{2}(\mathbf{\bar{x}}_{2},\mathbf{V}_{2})$ directly in terms
of their first moments $\mathbf{\bar{x}}_{j}$ and covariance
matrices $\mathbf{V}_{j}$. The following results first appeared
in~\cite{PLOB}, where the operator exponential form
\eqref{Gexpression} was employed to explicitly evaluate operator
logarithms.

Consider two arbitrary multimode Gaussian states, $\rho_{1}(\mathbf{\bar{x}%
}_{1},\mathbf{V}_{1})$ and
$\rho_{2}(\mathbf{\bar{x}}_{2},\mathbf{V}_{2})$. Then, the
entropic functional
\begin{equation}
\Sigma:=-\mathrm{Tr}\left(  \rho_{1}\log_{2}\rho_{2}\right)
\label{e:Sigma_functional}%
\end{equation}
is given by~\cite[Theorem~7]{PLOB}%
\begin{align}
&  \Sigma(\mathbf{V}_{1},\mathbf{V}_{2},\delta) =\frac{1}{2\ln2}
\times\label{functional}\\
&  \left[  \ln\det\left(
\kappa\mathbf{V}_{2}+\frac{i\Omega}{2}\right)
+\kappa\mathrm{Tr}(\mathbf{V}_{1}\mathbf{G}_{2})+\kappa\delta^{T}%
\mathbf{G}_{2}\delta\right]  ,\nonumber
\end{align}
where $\delta:=\mathbf{\bar{x}}_{2}-\mathbf{\bar{x}}_{1}$ and
$\mathbf{G}_{j}$ are the Gibbs matrices, obtained from the
covariance matrices $\mathbf{V}_{j}$ from Eq.~(\ref{e.GtoV}).

From the above entropic functional, we may compute both the von
Neumann entropy and the quantum relative entropy. Indeed, from
~\eqref{functional},
the von Neumann entropy of a Gaussian state $\rho(\mathbf{\bar{x}}%
,\mathbf{V})$ is equal to%
\begin{equation}
S(\rho):=-\mathrm{Tr}\left(  \rho\log_{2}\rho\right)  =\Sigma(\mathbf{V}%
,\mathbf{V},0)~, \label{entroTH}%
\end{equation}
The functional $\Sigma(\mathbf{V},\mathbf{V},0)$ is a symplectic
invariant,
namely $\Sigma(\mathbf{SVS}^{T},\mathbf{SVS}^{T},0)=\Sigma(\mathbf{V}%
,\mathbf{V},0)$. As such, from the Williamson decomposition
\eqref{WillDec} we find that $S(\rho)$ only depends on the
symplectic eigenvalues $v_{j}$ of $\mathbf{V}$, that can be
written as
\begin{equation}
v_{j}=\frac{2\bar{n}_{j}+1}{2\kappa},
\end{equation}
where $\bar{n}_{j}$ are the mean number of photons in each mode.
The von Neumann entropy of an $m$-mode Gaussian state can then be
expressed as~\cite{Sorkin}
\begin{equation}
S(\rho)=\sum_{j=1}^{m}h(\bar{n}_{j}),
\end{equation}
where $h(x):=(x+1)\log_{2}(x+1)-x\log_{2}x$.

The entropic functional of Eq.~\eqref{e:Sigma_functional} also
provides a tool for writing the relative entropy between two
arbitrary Gaussian states
$\rho_{1}(\mathbf{\bar{x}}_{1},\mathbf{V}_{1})$ and $\rho_{2}(\mathbf{\bar{x}%
}_{2},\mathbf{V}_{2})$, in terms of their statistical moments.
Indeed, we may write~\cite{PLOB}
\begin{align}
S(\rho_{1}||\rho_{2})  &  :=\mathrm{Tr}\left[  \rho_{1}(\log_{2}\rho_{1}%
-\log_{2}\rho_{2})\right] \nonumber\\
&  =-S(\rho_{1})-\mathrm{Tr}\left(
\rho_{1}\log_{2}\rho_{2}\right)
\nonumber\\
&  =-\Sigma(\mathbf{V}_{1},\mathbf{V}_{1},0)+\Sigma(\mathbf{V}_{1}%
,\mathbf{V}_{2},\delta)~. \label{RelENTROPY}%
\end{align}
It is worth mentioning that the final result
Eq.~(\ref{RelENTROPY}) is expressed directly in terms of the
statistical moments of the two Gaussian states. There is no need
of resorting to full symplectic diagonalizations \eqref{WillDec}
as in previous formulations~\cite{Scheelm,Chenmain}. This is an
advantage because, while the computation of the symplectic
spectrum needed for the von Neumann entropy is easy to get, the
symplectic matrix $\mathbf{S}$ performing such a symplectic
diagonalization is known only in a few cases as for specific types
of two-mode Gaussian states~\cite{QCB3}. On the other hand, the
invariant matrix formulation shown Eq.~(\ref{RelENTROPY}) allows
one to bypass such complicate diagonalization and directly compute
the quantum relative entropy.

Finally, we consider the quantum relative entropy variance
\begin{equation}
V_{S}(\rho_{1}\Vert\rho_{2})=\mathrm{Tr}\left[  \rho_{1}\left(  \log_{2}%
\rho_{1}-\log_{2}\rho_{2}-S(\rho_{1}\Vert\rho_{2})\right)
^{2}\right]  ~.
\label{ReeVariance}%
\end{equation}
The relative entropy variance was introduced in
Ref.~\cite{ReeVar1,ReeVar2} to bound the capacity of quantum
channels and quantum hypothesis testing. Using the operator
exponential form \eqref{Gexpression} and the definitions
\eqref{e.GtoV}, one can show that, for any two Gaussian states
$\rho_{1}$ and $\rho_{2}$, the variance
$V_{S}(\rho_{1}\Vert\rho_{2})$ can be written in
terms of the states' first and second moments as%
\begin{equation}
V_{S}(\rho_{1}\Vert\rho_{2})=\frac{4\kappa^{2}\mathrm{Tr}[(\mathbf{V}%
_{1}\mathbf{\tilde{G}})^{2}]+\mathrm{Tr}[(\mathbf{\tilde{G}}\Omega
)^{2}]+\delta^{T}\mathbf{B}\delta}{2(2\ln2)^{2}}~, \label{ReeVarGaussian}%
\end{equation}
where $\mathbf{\tilde{G}}=\mathbf{G}_{1}-\mathbf{G}_{2}$, $\delta
=\mathbf{\bar{x}}_{1}-\mathbf{\bar{x}}_{2}$ and
$\mathbf{B}=8\kappa
^{2}\mathbf{G}_{2}\mathbf{V}_{1}\mathbf{G}_{2}$. The above formula
was first derived in Ref.~\cite{WildeREEVar} for $\kappa=1$. It
was then easily generalized to arbitrary $\kappa$\ in
Ref.~\cite{Spyros2}, where an alternative simplified proof was
presented by exploiting a trace formula from
Ref.~\cite{MonrasWick}.



\end{document}